\documentclass[twocolumn]{svjour3}
\pdfoutput=1 

\usepackage{preprintcover}  
\PreprintCoverPaperTitle{Electron and photon energy calibration \\ with the ATLAS detector using LHC Run 1 data}  
\PreprintIdNumber{CERN-PH-EP-2013-05}  
\PreprintCoverAbstract{
This paper presents the electron and photon energy calibration achieved with the ATLAS detector using about 25~fb$^{-1}$ of LHC
proton--proton collision data taken at centre-of-mass  energies of $\sqrt{s}=7$ and $8$~TeV. The reconstruction of electron and photon energies is optimised using multivariate algorithms. 
The response of the calorimeter layers is equalised in data and simulation, and the longitudinal
profile of the electromagnetic showers is exploited to estimate the
passive material in front of the calorimeter and reoptimise the
detector simulation. After all corrections, the $Z$ resonance is used
to set the absolute energy scale. For electrons from $Z$ decays, the achieved calibration is typically accurate to 0.05\% in most of the
detector acceptance, rising to 0.2\% in regions with large amounts of
passive material. The remaining inaccuracy is less than 0.2--1\% for electrons with a transverse energy
of 10~GeV, and is on average 0.3\% for photons. 
The detector resolution is determined with a relative inaccuracy of less than 10\% for electrons
and photons up to 60~GeV transverse energy, rising to 40\% for
transverse energies above 500 GeV.
}  
\PreprintJournalName{EPJC}  

\usepackage{float}
\usepackage{amsmath}
\usepackage{amssymb}
\usepackage{atlasphysics}
\usepackage{cite}
\usepackage{rotating}
\usepackage{subfigure}
\usepackage{url}
\usepackage{parskip}

\usepackage[hyperindex,breaklinks]{hyperref}

\newcommand{\wen}{\ensuremath{W \rightarrow e\nu}}

\newcommand{\zee}{\mbox{\ensuremath{Z \to ee }}}
\newcommand{\zmumu}{\mbox{\ensuremath{Z \to \mu\mu }}}
\newcommand{\jpsiee}{\mbox{\ensuremath{J/\psi \to ee }}}
\newcommand{\Zllgamma}{\mbox{\ensuremath{Z \to \ell\ell \gamma }}}
\newcommand{\Zmumugamma}{\mbox{\ensuremath{Z \to \mu\mu \gamma }}}

\newcommand{\mgg}{\ensuremath{m_{\gamma\gamma}}}

\def\Etrue{\ensuremath{E_{\mathrm{true}}}}
\def\Eacc{\ensuremath{E_{\mathrm{calo}}}}
\def\Emva{\ensuremath{E_{\mathrm{MVA}}}}
\def\Etacc{\ensuremath{\et^{\mathrm{calo}}}}

\def\Ettrue{\ensuremath{\et^{\mathrm{true}}}}
\def\etaCluster{\ensuremath{\eta_{\mathrm{cluster}}}}

\newcommand{\Eonetwo}{\mbox{\ensuremath{E_{1/2}}}}

\makeatletter
\def\blfootnote{\xdef\@thefnmark{}\@footnotetext}

\begin{document}

\title{Electron and photon energy calibration \\ with the ATLAS detector using LHC Run 1 data}

\author{The ATLAS Collaboration}
\institute{The complete list of authors is given at the end of the paper}

\maketitle

\begin{abstract}
  This paper presents the electron and photon energy calibration achieved with the ATLAS detector using about 25~fb$^{-1}$ of LHC
proton--proton collision data taken at centre-of-mass  energies of $\sqrt{s}=7$ and $8$~TeV. The reconstruction of electron and photon energies is optimised using multivariate algorithms. 
The response of the calorimeter layers is equalised in data and simulation, and the longitudinal
profile of the electromagnetic showers is exploited to estimate the
passive material in front of the calorimeter and reoptimise the
detector simulation. After all corrections, the $Z$ resonance is used
to set the absolute energy scale. For electrons from $Z$ decays, the achieved calibration is typically accurate to 0.05\% in most of the
detector acceptance, rising to 0.2\% in regions with large amounts of
passive material. The remaining inaccuracy is less than 0.2--1\% for electrons with a transverse energy
of 10~GeV, and is on average 0.3\% for photons. 
The detector resolution is determined with a relative inaccuracy of less than 10\% for electrons
and photons up to 60~GeV transverse energy, rising to 40\% for
transverse energies above 500 GeV.

\end{abstract}


\keywords{
  electron, 
  photon,
  calibration,
  electromagnetic shower,
  calorimeter response,
  energy scale,
  resolution  
} 



\section{Introduction} 
\label{sec:Introduction}

Precise calibration of the energy measurement of electrons and
photons is a fundamental input to many phy\-si\-cs measurements. In particular, after the discovery of the Higgs boson by the
ATLAS and CMS experiments~\cite{atlas_h_discovery,cms_h_discovery},
an accurate determination of its properties is of
primary importance. A precise measurement of the $W$ boson mass is
also a long-term goal of the LHC experiments, and requires an
excellent accuracy of the electron energy calibration.

A first electron and photon calibration analysis was performed using 40~pb$^{-1}$ of LHC
collision data taken in 2010 at a centre-of-mass energy $\sqrt{s}=7$~\TeV~\cite{perf2010}. The calibration of the
ATLAS liquid argon (LAr) calorimeter was primarily based on test-beam
measurements; only the absolute energy scale was set using the $Z$ boson
resonance. The uncertainty on the detector material upstream
of the LAr calorimeter, which is of primary importance in understanding
its response to incident electromagnetic particles, was estimated from
engineering drawings and a material survey during construction. The
achieved calibration was accurate to 0.5--1\% for electrons,
depending on pseudorapidity and energy.

This paper presents the calibration scheme develo\-ped for
precision measurements involving electrons and photons with
$|\eta|<2.47$\footnote{ATLAS uses a right-handed coordinate system with its origin at the 
nominal interaction point (IP) in the centre of the detector and the $z$-axis 
along the beam pipe. The $x$-axis points from the IP to the centre of the LHC 
ring, and the $y$-axis points upward. Cylindrical coordinates $(r,\phi)$ are 
used in the transverse plane, $\phi$ being the azimuthal angle around the 
beam pipe. The pseudorapidity is defined in terms of the polar angle 
$\theta$ as $\eta=-\ln\tan(\theta/2)$.} and mostly derived from collision data. It
comprises local corrections to the calorimeter energy measurement, and the
intercalibration of its longitudinal layers; a measurement of detector
material leading to an improved simulation; an
improved simulation-based calibration; and a measurement of the
absolute energy scale from $Z$ boson decays. The universality of the
energy scale is verified using $J/\psi\rightarrow ee$ and
$Z\rightarrow \ell\ell\gamma$ decays ($\ell=e,\mu$). The studies are primarily
based on 20.3 fb$^{-1}$ of proton--proton collision data collected in
2012 at $\sqrt{s}=8$~\TeV, and the algori\-thms are tested on 4.5
fb$^{-1}$ of data collected in 2011 at $\sqrt{s}=7$~\TeV. 

The paper is organised as follows. After an overview
of the energy reconstruction with the ATLAS LAr calorimeter in
Sect.~\ref{sec:egammareco}, the calibration procedure, the data
and simulated Monte Carlo (MC) samples used for this purpose are summarised in
Sects.~\ref{sec:Overview} and ~\ref{sec:datasim}. Section~\ref{sec:MCCalibration}
describes the simulation-based energy calibration
algorithm. Data-driven corrections to the energy measurement and to
the detector material budget are presented in 
Sects.~\ref{sec:uniformity} to \ref{sec:material}, and the absolute energy scale determination
from $Z$ boson decays is described in Sect.~\ref{sec:zeescales}. Systematic
uncertainties affecting the calibration and cross-checks of the
$Z$-based energy scale are given in
Sects.~\ref{sec:uncsum}\--\ref{sec:photoncheck}. The results of this
calibration procedure applied to the 2011 data sample are summarised
in Appendix~\ref{app:2011}. Uncertainties on the energy resolution are discussed in
Sect.~\ref{sec:resolution}, and the performance of an algorithm combining the
calorimeter energy measurement with the momentum measured in the
tracking detectors is presented in Sect.~\ref{sec:epcombination}. Section~\ref{sec:summary}
summarises the achieved results and concludes the paper.


\section{Electron and photon reconstruction and identification in ATLAS}
 \label{sec:egammareco}

\subsection{The ATLAS detector}

The ATLAS experiment \cite{ATLAS_detector} is a general-purpose particle
physics detector with a forward-backward symmetric cylindrical
geometry and near 4$\pi$ coverage in solid angle. The inner tracking
detector (ID) covers the pseudorapidity range $|\eta|<2.5$ and
consists of a silicon pixel detector, a silicon microstrip detector
(SCT), and a transition  radiation tracker (TRT) in the range
$|\eta|<2.0$. The ID is surrounded by a superconducting  solenoid
providing a 2~T magnetic field. The ID provides accurate reconstruction of
tracks from the primary proton--proton collision region and also
identifies  tracks from secondary vertices, permitting an efficient
reconstruction of photon conversions in the ID up to a radius of about
800~mm. 

The electromagnetic (EM) calorimeter is a LAr sampling calorimeter with an accordion geometry. It is divided into a
barrel section (EMB), covering the pseudorapidity region
$|\eta|<1.475$,\footnote{The EMB is split into two half-barrel modules
which cover the positive and negative $\eta$ regions.} and two endcap
sections (EMEC), covering $1.375<|\eta|<3.2$. The barrel and endcap sections are divided into 16 and 8
modules in $\phi$, respectively. The transition region between the EMB
and the EMEC, $1.37<|\eta|<1.52$, has a large amount of material in
front of the first active calorimeter layer ranging from 5 to almost 10 radiation lengths
($X_{0}$). A high voltage (HV) system generates
an electric field of about 1~kV/mm, which allows ionisation electrons to drift
in the LAr gap. In the EMB, 
the HV is constant along $\eta$, while in the EMEC, where the gap
varies continuously with radius, it is adjusted in ste\-ps along
$\eta$. The HV supply granularity is typically in sectors of $\Delta
\eta \times \Delta \phi = 0.2 \times 0.2$. Both the barrel and endcap
calorimeters are longitudinally segmented into three shower-depth layers for
$|\eta|<2.5$. The first one (L1), in the ranges $|\eta|<1.4$ and
$1.5<|\eta|<2.4$, has a thickness of about $4.4 X_{0}$ and is segmented into high-granularity strips in the $\eta$
direction, typically $0.003\times 0.1$ in $\Delta \eta \times \Delta \phi$ 
in EMB, sufficient to provide an event-by-event discrimination
between single photon showers and two overlapping showers coming from
the decay of neutral hadrons in jets \cite{PhotonReco}. The second layer (L2), which
collects most of the energy  deposited in the calorimeter by photon
and electron showers, has a thickness of about $17 X_{0}$  and a
granularity of $0.025\times 0.025$ in $\Delta \eta \times \Delta
\phi$. A third layer (L3), which has a granularity of $0.05\times
0.025$ in $\Delta \eta \times \Delta \phi$ and a depth of about
2$X_{0}$, is used to correct leakage beyond the EM calorimeter for 
high-energy showers. In front of the accordion calorimeter, a thin
presampler layer (PS), covering the pseudorapidity interval
$|\eta|<1.8$, is used to correct for energy loss upstream of the 
calorimeter. The PS consists of an active LAr layer with a thickness of
1.1 cm (0.5 cm) in the barrel (endcap) and has a granularity of $\Delta
\eta \times \Delta \phi = 0.025 \times 0.1$.

The hadronic calorimeter, surrounding the EM calorimeter,
consists of an iron/scintillator  tile calorimeter in the range
$|\eta|<1.7$ and two copper/LAr calorimeters  spanning $1.5<|\eta|<3.2$. The acceptance is extended by two copper/LAr and tungsten/LAr
forward calorimeters up to $|\eta| =$ 4.9. The forward calorimeters
also provide electron reconstruction capability, a feature that is not
discussed here.

The muon spectrometer, located beyond the calorimeters, consists of
three large air-core superconducting toroid systems with precision
tracking chambers providing accurate muon tracking for $|\eta|<2.7$
and fast detectors for triggering for $|\eta|<2.4$.

\subsection{Energy reconstruction in the electromagnetic calorimeter}
\label{subsec:lar_electronics}

Electrons and photons entering the LAr calorimeter develop EM showers
through their interaction with the lead absorbers. The EM showers ionise
the LAr in the gaps between the absorbers. The ionisation electrons
drift and induce an electrical signal on the electrodes which is
proportional to the energy deposited in the active 
volume of the calorimeter. The signal is brought via cables to the read-out Front
End Boards, where it is first amplified by a current-sensitive
pre-amplifier. In order to accommodate a large dynamic range, and to
optimise the total noise due to electronics and 
inelastic $pp$ collisions coming from previous bunch crossings (out-of-time pile-up),
the signal is shaped by a bipolar filter and
simultaneously amplified with three linear gains called low (LG),
medium (MG) and high (HG). For each channel, these three amplified
signals are sampled at a 40~MHz clock frequency and stored on a switched capacitor array, awaiting the level-1 trigger decision; upon receipt, the sample corresponding to the maximum amplitude of the physical pulse stored in MG is first digitised by a 12-bit analog-to-digital converter (ADC). Based on this sample, a hardware gain selection is used to choose the most suited gain. The samples of the chosen gain are digitised and routed via optical fibres to the read-out drivers. More
details on the ATLAS LAr calorimeter read-out and electronic
calibration are given in Refs.~\cite{LArElectronics} and
~\cite{LArReadiness}.

The total energy deposited in an EM calorimeter cell is reconstructed as
\begin{eqnarray}
  E_{\rm cell} = F_{\mathrm{\mu A\rightarrow MeV}} & \times & F_{\mathrm{DAC\rightarrow \mu A}} \\
  & \times & \frac{1}{\frac{M\mathrm {phys}}{M\mathrm {cali}}}\times G \times \sum_{j=1}^{\mathrm{N_{samples}}}{a_{j}(s_{j} - p)} \nonumber ,
\label{eq:ofreco}                                                                           
\end{eqnarray}  
where $s_{j}$ are the samples of the shaped ionisation signal
digitised in the selected electronic gain, measured in ADC counts in
$\mathrm{N_{samples}}$ time slices ($\mathrm{N_{samples}} = 5$) spaced
by 25 ns; \footnote{The delay between the event trigger and the time
  slices is optimised to ensure that the third sample is on average at
  the signal maximum in each read-out channel.} $p$ 
is the read-out electronic pedestal, measured for each gain in
dedicated calibration ru\-ns; the $a_{j}$ weights are the optimal
filtering coefficients (OFC) derived from the predicted shape of the
ionisation pulse and the noise autocorrelation, accounting for both
the electronic and the pile-up components \cite{Cleland94}. The cell gain 
$G$ is computed by injecting a known calibration signal and reconstructing the corresponding cell response. The factor
$\frac{M \mathrm {phys}}{M \mathrm {cali}}$, which quantifies the ratio of the maxima of the physical and calibration pulses corresponding to the same
input current, corrects the gain factor $G$ obtained with the calibration pulses to adapt it to physics-induced signals;
the factor $F_{\mathrm{DAC\rightarrow \mu A}}$ converts
digital-to-analog converter (DAC) counts set on the calibration board to a current in $\mu A$; the
factor $F_{\mathrm {\mu A \rightarrow MeV}}$ converts the ionisation current to the total deposited energy at the EM scale and is determined from test-beam
studies~\cite{Aharrouche:2010zz}. 

\subsection{Electron and photon reconstruction}
\label{sec:recoegamma}

The reconstruction of electrons and photons in the region
$|\eta|<2.47$ starts from energy deposits (clusters) in the EM
calorimeter. To reconstruct the EM clusters, the EM calorimeter is
divided into a grid of $N_{\eta} \times 
N_{\phi}$ towers of size $\Delta\eta \times \Delta\phi = 0.025\times 0.025$. Inside each of these elements, the energy of all cells
in all longitudinal layers is summed into the tower energy. These clusters are seeded by towers with total transverse energy above
2.5~GeV and searched for by a sliding-window algorithm~\cite{Clustering}, with a window size of 3 $\times$ 5 towers.

Clusters matched to a well-reconstructed ID track originating from a vertex found in the beam interaction region are classified as electrons. If the matched track is consistent with originating from a photon conversion and if in addition a conversion vertex is reconstructed, the corresponding candidates are considered as converted photons. They are classified as single-track or double-track conversions depending on the number of assigned electron-tracks. Clusters without matching
tracks are classified as unconverted photons \cite{PhotonReco}. 
The electron cluster is then rebuilt using an area of calorimeter
cells corresponding to 3$\times$7 and 5$\times$5 L2
cells \footnote{Only in L2 does the cell granularity correspond
  exactly to this tower size: the number of cells selected by the
  clustering algorithm in the other layers varies according to the
  position of the cluster barycentre in L2 \cite{Clustering}.} in the
EMB and EMEC respectively.  
For converted photons, the same 3$\times$7 cluster size is used in the barrel,
while a 3$\times$5 cluster is associated with unconverted photons due to their
smaller lateral size. A 5$\times$5 cluster size is used in the EMEC for converted and unconverted photons.
These lateral cluster sizes were optimised to take into account
the different overall energy distributions in the barrel and
endcap calorimeters while minimising the pile-up and noise contributions. 
The cluster energy is then determined by applying correction factors
computed by a calibration scheme based on the full detector simulation, which is described in Sect.~\ref{sec:MCCalibration}.

Photons and electrons reconstructed near regions of the
calorimeter affected by read-out or HV failures are 
rejected \cite{LArDQ}.

The relative energy resolution for these EM objects can be parameterised as follows:
\begin{equation}
\frac{\sigma}{E} = \frac{a}{\sqrt{E}} \oplus \frac{b}{E} \oplus c , 
\label{eq:resol}
\end{equation}
where $a$, $b$ and $c$ are $\eta$-dependent parameters; $a$ is the
sampling term, $b$ is the noise term, and $c$ is the constant term.
The sampling term contributes mostly at low energy; its design value is $(9\--10)\%/\sqrt{E [{\rm GeV}]}$ at low
$|\eta|$, and is expected to worsen as the amount of material in front of the
calorimeter increases at larger $|\eta|$. The noise term is about $350 \times \cosh\eta$ MeV for a $3\times 7$ cluster in $\eta \times \phi$ space in the barrel and for a mean number of interactions per bunch crossing $\mu =20$; it is dominated by the pile-up noise at high $\eta$. At higher energies the
relative energy resolution tends asymptotically to the constant term, $c$, which has a design value of 0.7\%. 

\subsection{Electron and photon identification}
\label{sec:IDegamma}

The clusters associated with electron and photon candidates must satisfy a set of 
identification criteria, requiring their longitudinal and transverse
profiles to be consistent with those expected for EM 
showers induced by such particles. 

Three reference sets of cut-based selections, labelled loose, medium and tight, 
have been defined for elect\-rons with increasing background rejection
power~\cite{Aad:2014fxa,ElectronID2012}. Shower shape variables in both the first and second layers of the
EM calorimeter and the fraction of energy deposited in the hadronic
calorimeter are used in the loose selection with additional
requirements on the associated track quality and track-cluster
matching. Tightened requirements on these discriminating variables are
added to the medium criteria together with a loose selection on the
transverse impact parameter and on the number of hits in the TRT associated with the track, and a measured hit in the
innermost layer of the pixel detector to discriminate against photon
conversions.\footnote{This cut is only applied when the traversed module is active} The tight selection adds a selection on the ratio of the candidate's reconstructed energy to its track momentum, $E/p$,
stricter requirements on the discriminating variables and TRT
information, and a veto on reconstructed photon conversion vertices
associated with the cluster.

The identification of photons is performed by applying cuts on shower shapes measured in the
first two longitudinal layers of the EM calorimeter 
and on the leakage into the hadronic calorimeter~\cite{PhotonID}. 

To further suppress background from hadronic decays, an isolation requirement is applied. 
The calorimeter isolation transverse energy
$\et^{\rm iso}$ is computed by summing the transverse energy of all
calorimeter cells in a cone of size $\Delta R = \sqrt{(\Delta\eta)^{2}+ (\Delta\phi)^{2}}$ around the candidate~\cite{Clustering}.
The isolation energy is corrected
by subtracting the estimated contributions from the photon or electron candidate itself and from the
underlying event and pile-up contributions using the technique proposed in Ref.~\cite{Salam} and implemented 
as described in Ref.~\cite{InclusivePhoton}.
A track isolation variable $\pt^{\rm iso}$ is also used for electrons and muons. It is built by
summing the transverse momenta of the tracks 
in a cone of size $\Delta R$ around the candidate, 
excluding the track associated with the candidate itself. The
tracks considered in the sum must come from the reconstructed vertex with the highest sum of all 
associated tracks and
must have at least four hits in either the pixel or SCT detector.


\section{Overview of the calibration procedure} 
\label{sec:Overview}

\begin{figure*}
  \centering
  \includegraphics[width=\textwidth]{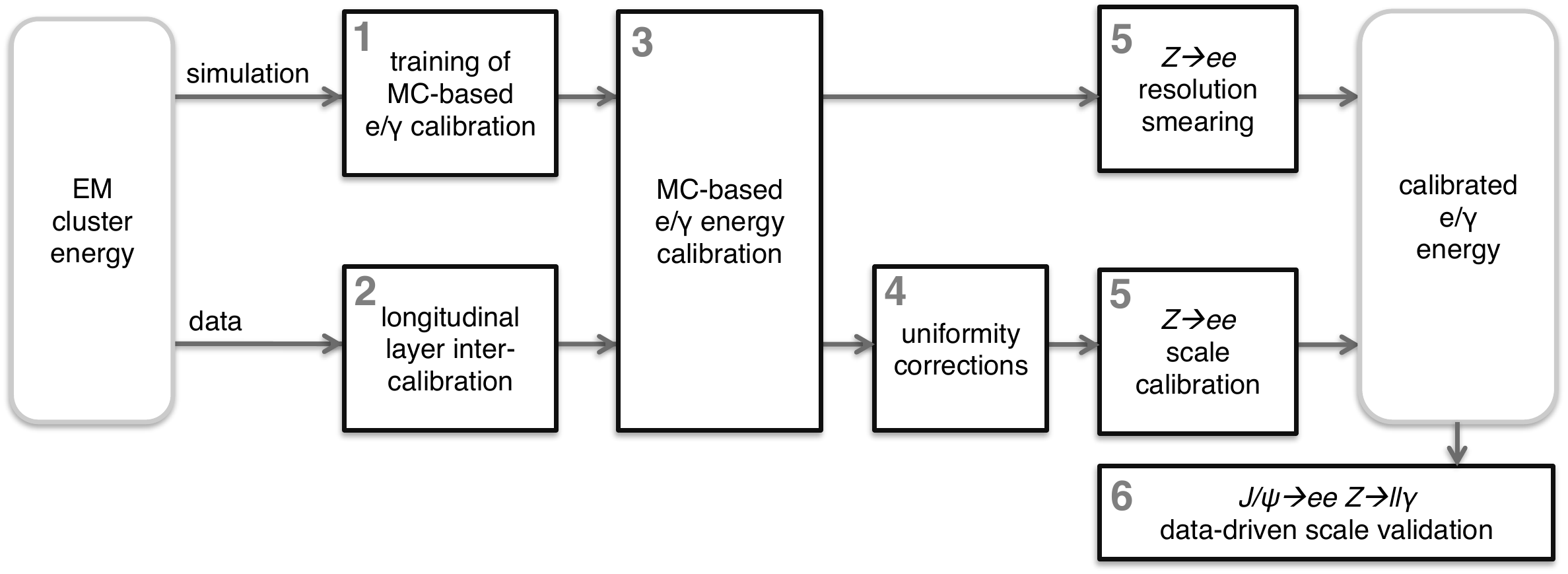}
  \caption{Schematic overview of the procedure used to calibrate the energy response of electrons and photons in ATLAS.}
  \label{fig:Flowchart}
\end{figure*}

The different steps in the procedure to calibrate the energy response
of electrons and photons described in this paper are summarised below,
with the item numbers referring to the calibration steps sketched
in Fig.~\ref{fig:Flowchart}. The references to their description in
the paper is also given.

The energy of an electron or photon candidate is built from
the energy of a cluster of cells in the EM calorimeter. 
The calibration proceeds as follows:

\begin{enumerate}

\item The EM cluster properties, including its longitudinal
  development, and additional information from the ATLAS inner
  tracking system, are calibrated to the original electron and photon
  energy in simulated MC samples using multivariate techniques (step
  1), which constitutes the core of the MC-based $e/\gamma$ response
  calibration (step 3). The calibration constants are determined using
  a multivariate algorithm (MVA)~\cite{TMVA}; its optimisation is
  performed separately for electrons, converted and unconverted
  photons. The MC samples used in the various analyses presented in
  this paper are detailed in Sect.~\ref{sec:datasim}, while the
  MC-based MVA calibration is described in
  Se\-ct.~\ref{sec:MCCalibration}. 

A prerequisite of this MC-based calibration is that the detector
geometry and the interactions of particles with matter are accurately
described in the simulation. The material distribution is
measured in data using the ratio of the first-layer energy to the second-layer energy in the longitudinally segmented EM calorimeter (\Eonetwo). Measuring
\Eonetwo\ in data with different samples (electrons and unconverted photons)
allows a precise determination of the amount of material in front of
the calorimeter and provides some sensitivity to its radial distribution as
described in Sect.~\ref{sec:material}. 

\item Since the EM calorimeter is longitudinally segmented, the scales
  of the different longitudinal layers have to be equalised in data
  with respect to simulation, prior to the determination of the
  overall energy scale, in order to ensure the correct
  extrapolation of the response in the full \pt\ range used in the
  various analyses (step 2). The procedure to measure the EM calorimeter layer
  scales is reviewed in Sect.~\ref{sec:samplingcalib}.  

\item The MC-based $e/\gamma$ response calibration is applied to the
  cluster energies reconstructed both from collision data and MC
  simulated samples (step 3). 

\item A set of corrections are implemented to account for response
  variations not included in the simulation in specific detector regions (step 4), e.g. non-optimal HV regions, geometric
  effects such as the inter-module widening (IMW) or biases associated with 
  the LAr calorimeter electronic calibration. These corrections
  are discussed in Sect.~\ref{sec:uniformity}, where the stability
  of the calorimeter response as a function of  $\phi$, time and
  pile-up is also presented. 

\item  The overall electron response in data is calibrated so that it agrees
  with the expectation from simulation, using a large 
  sample of $Z\rightarrow ee$ events as discussed in
  Sect.~\ref{sec:zeescales}. Per-electron scale factors are
  extracted and applied to electron and photon candidates in data
  (step 5). Using the same event sample it is found that the
  resolution in data is slightly worse than that in simulation, and
  appropriate corrections are derived and applied to simulation to
  match the data. The electron and photon calibration uncertainties are summarised in
  Sect.~\ref{sec:uncsum}. 

\item The calibrated electron energy scale is validated with electron
  candidates from \jpsiee\ events in data (step 6). 
  The scale
  dependence with $\eta$ and \pt, and its associated systematic
  uncertainties are summarised in Sect.~\ref{sec:electroncheck}. The
  scale factors extracted from $Z\rightarrow ee$ events are assumed to
  be valid also for photons, while photon-specific systematic
  uncertainties are applied, as discussed in
  Sect.~\ref{sec:photonsys}. This approach is validated with photon
  candidates from \Zllgamma\ events in data, and discussed
  in Sect.~\ref{sec:photoncheck}. 

\end{enumerate}

The determination of the electron and photon energy resolution, and
the associated uncertainties, are described in
Sect.~\ref{sec:resolution}. Finally, the potential for improving the electron
energy resolution, by combining the cluster energy with the momentum
measured by the ID, is described in
Sect.~\ref{sec:epcombination}.


\section{Collision data and simulated samples}
 \label{sec:datasim}

The results presented in this paper are primarily based on 20.3 fb$^{-1}$ of $pp$ collision data at
$\sqrt{s}$ = 8 \TeV, collected by ATLAS in 2012. The results of the
application of the same methods to 4.7 fb$^{-1}$ of $pp$ collision
data taken in 2011 at $\sqrt{s}$ = 7~\TeV\ are described in Appendix~\ref{app:2011}.

Table~\ref{tab:datamc} lists the kinematic selections applied to the
different calibration samples, the generators used and the
corresponding numbers of events in 2012 collision data. The average
electron transverse energy is around $\et^e\sim$ 40$\--$45~\GeV\ in the $W$ and $Z$
samples and $\et^e\sim$ 11~\GeV\ in the $J/\psi$ sample; for photons,
$\et^{\gamma}\sim$ 25, 100~\GeV\ in the \Zllgamma\ and $\gamma + X$
samples, respectively. The $W$ event 
selection relies on $\et^\mathrm{miss}$ and $\phi^\mathrm{miss}$,
respectively defined as the norm and azimuth of the total transverse
momentum imbalance of all reconstructed objects, and on the transverse
mass defined as $m_\mathrm{T} =
\sqrt{2\et^{e}\et^\mathrm{miss}(1-\cos\Delta\phi)}$ where $\Delta\phi = \phi^{e} - \phi^\mathrm{miss}$, $\phi^{e}$ being the azimuthal angle of the electron.  

\begin{table*}
  \begin{center}
    \begin{tabular}{llll}
      \hline\hline
      Process & Selections & $N_\mathrm{events}^\mathrm{data}$ & MC generator \\ 
      \hline
      
      $Z\rightarrow ee$               & $\et^e>27$~GeV, $|\eta^e|<2.47$ & 5.5 M & {\sc Powheg+Pythia} \\
      & $80<m_{ee}<100$~GeV & & \\[2mm]
      
      $W\rightarrow e\nu$ & $\et^e>30$~GeV, $|\eta^e|<2.47$ & 34 M & {\sc Powheg+Pythia} \\
      & $\et^\mathrm{miss} > 30$~GeV, $m_\mathrm{T} > 60 $~GeV & & \\[2mm]
      
      $J/\psi\rightarrow ee$          & $\et^e>5$~GeV, $|\eta^e|<2.47$      & 0.2 M & {\sc Pythia} \\
      & $2<m_{ee}<4$~GeV & & \\[2mm]
      
      $Z\rightarrow \mu\mu$           & $\pt^\mu>20$~GeV, $|\eta^\mu|<2.4$  & 7 M & {\sc Sherpa} \\
      & $60 < m_{\mu\mu} < 120$~GeV & & \\[2mm]
            
      $Z\rightarrow \ell\ell\gamma$,         & $\et^{\gamma}>15$~GeV, $|\eta^{\gamma}|<2.37$ & 20k (e) & {\sc Sherpa} \\
      large-angle & $\et^{e}>15$~GeV, $|\eta^{e}|<2.47$ & 40k ($\mu$)  & \\
      & $\pt^\mu>20$~GeV, $|\eta^\mu|<2.4$  & & \\
      & $45 < m_{\ell\ell} < 85$~GeV & & \\
      & $80 < m_{\ell\ell\gamma} < 120$~GeV & & \\
      & $\Delta R (\ell, \gamma) >0.4$  & & \\[2mm]

      $Z\rightarrow \mu\mu\gamma$,         & $\et^{\gamma}>7$~GeV,
      $|\eta^{\gamma}|<2.37$ & 120k & {\sc Sherpa} \\
      collinear & $\pt^\mu>20$~GeV, $|\eta^\mu|<2.4$  & & \\
      & $55 < m_{\mu\mu} < 89$~GeV & & \\
      & $66 < m_{\mu\mu\gamma} < 116$~GeV & & \\
      & $\Delta R (\mu, \gamma) <0.15$  & & \\[2mm]
      
      $\gamma + X$                    & $\et>120$~GeV, $|\eta^\gamma|<2.37$ & 3.1 M & {\sc Pythia} \\

      \hline\hline
    \end{tabular}
  \end{center}
  \caption{\label{tab:datamc}Summary of the processes used in the
    present calibration analysis, the kinematic selections, the numbers
    of events selected in data at $\sqrt{s}=8$~TeV (for 20.3 fb$^{-1}$) and the MC signal samples used.
    The region $1.37\leq |\eta|<1.52$ is excluded for photons.
  }
\end{table*}

The $J/\psi$ sample results from both direct production and $b\rightarrow J/\psi$ decays. Three different triggers are used for
this sample requiring a transverse energy of the leading lepton above
4, 9 and 14 GeV respectively. The trigger requirement significantly affects the
electron $\et$ distribution in this sample, which is not the case for
the other calibration samples.  

In the \Zllgamma\ sample, photons and electrons are required to have a
large-angle separation. A collinear sample in the
\Zmumugamma\ channel, where the photon is near one of the muons, is also selected. Isolation requirements are applied
to photons and leptons. In the large-angle sample, leptons are
required to have $p_{\rm T}^{\rm iso} (\Delta R =0.2) /p_{\rm T}^{\ell} <
0.1$; in addition electrons are required to satisfy $\et^{\rm iso} (\Delta R = 0.3)
/p_{\rm T}^{e} < 0.18$ while for photons $\et^{\rm iso} (\Delta R = 0.4) <
4$~\GeV. In the collinear sample, the same isolation cut is applied to
photons, but it is tightened for muons by applying $p_{\rm T}^{\rm iso} (\Delta R
=0.3) / p_{\rm T}^{\mu} < 0.15$.  

The measurements are compared to expectations from MC simulation. 
Comparisons between data and simulation are initially performed using the detector description originally used for most ATLAS analysis (for instance, in Ref.~\cite{atlas_h_discovery}), later refered to as the ``base'' simulation. The detector description resulting from the passive material determination described in Sec.~\ref{sec:material} is instead refered to as the ``improved'' simulation.
Large samples of \zee, \zmumu, $J/\psi \allowbreak \to ee$, \wen, \Zllgamma\ and $\gamma + X$  
events were generated with {\sc Sherpa}~\cite{sherpa} and {\sc
Powheg}~\cite{powheg1,powheg2,powheg3,powheg4} interfaced with {\sc
Pythia}~\cite{pythia8}. The generated events are processed through the
full ATLAS detector simulation~\cite{atlassimulation} based on {\sc
  Geant4}~\cite{geant4}. The size of the MC samples exceeds the
corresponding collision data samples by a factor of up to 10.  

For the optimisation of the MC-based $e/\gamma$ response calibration, a sample of 20 million single electrons,
and one of 40 million single photons are simulated. The \et\ distribution of such samples is tuned to cover
the range from 1 \GeV\ to 3 \TeV\, while maximising the statistics
between 7 and 100 \GeV.

For studies of systematic uncertainties related to the detector
description in simulation, samples with additional
passive material in front of the EM calorimeter are simulated, representing
different estimates of the possible amount of
material, based on studies using collision data
\cite{material1,material2,material3,material4,material5,material6}.  

Depending on the signal samples, backgrounds consist of $W\rightarrow \ell\nu$,
$Z\rightarrow \tau\tau$ and gauge boson pair production, simulated using
{\sc Powheg}; $b\bar b, c\bar c \rightarrow \mu +X$ simulated using
{\sc Pythia}; and $t\bar t$ production, simulated using {\sc Mc@nlo}
\cite{mcnlo}. For the \Zllgamma\ analysis, backgrounds from $Z$ production
in association with jets are si\-mu\-la\-ted using {\sc Sherpa}. 
Some background contributions are directly determined from data.

The MC events are simulated with additional interactions in the same or neighbouring bunch 
crossings to match the pile-up conditions during LHC operation, and are weighted to reproduce 
the distribution of the average number of interactions per bunch crossing in data.


\section{MC-based calibration for electrons and photons \label{sec:MCCalibration}}

Reconstructed electron and photon clusters are calibrated to correct for the energy lost in the material upstream of the calorimeter, the energy
deposited in the cells neighbouring the cluster in $\eta$ and $\phi$,
and the energy lost beyond the LAr calorimeter. Further corrections
are applied to correct for the response dependence as a function of the particle
impact point within the central cluster cell. The cluster-level
calibration constants are extracted from simulated electrons and
photons and strongly rely on the assumed amount of passive material in
front of the EM calorimeter. The simulation of the detector material uses the 
improvements described in Sect.~\ref{sec:material}.

The constants are determined using a multivariate algorithm, applied separately for
electrons, converted and unconverted photons in $\eta$ and \pt~bins.
The calibration method presented in this section supersedes
the procedure described in Refs.~\cite{Aad:2009wy} and~\cite{perf2010}, except for the transition
region $1.37\leq|\eta|<1.52$ where the initial calibration procedure is still used.

\subsection{Input variables}
\label{subsec:inputs}

The calibration procedure optimises the estimate of the true particle
energy at the interaction point ($E_\text{true}$) from the detector-level
observables. The algorithm uses cluster position measurements in the ATLAS
and EM calorimeter frames. The ATLAS coordinate system has its origin
at the nominal interaction point, with respect to which the
calorimeter is displaced by a few millimeters, while all calorimeter cells are in
their nominal position in the EM calorimeter frame. 

The quantities used for electrons and photons are the total energy
measured in the calorimeter, $\Eacc$; the ratio of the PS energy to the
calorimeter energy, $E_0/\Eacc$; the shower depth, defined as $X = \sum_i X_i
E_i/\sum_i E_i$, where $E_i$ and $X_i$ are the cluster energy and the
calorimeter thickness (in radiation lengths) in layer $i$; the cluster
barycentre pseudorapidity in the ATLAS coordinate system, $\etaCluster$; and the cluster barycentre in
$\eta$ and $\phi$ within the calorimeter frame. The variable $\etaCluster$ is included to account for the passive-material
variations in front of the calorimeter; the inclusion of the barycentre location in the calorimeter frame is important to accura\-tely correct for the increase of lateral energy leakage for particles that hit the
cell close to the edge, and for the sampling fraction variation as a
function of the particle impact point with respect to the calorimeter
absorbers. 

Photons are considered converted if the conversion radius $R_{\rm conv}$ is smaller than 800 mm. For these converted photons, $R_{\rm conv}$ is used as an additional input to the MVA only if the vectorial sum of the conversion track momenta, $\pt^{\mathrm{conv}}$, is above 3 GeV. In particular for conversions with both tracks containing at least one hit in either the pixel or SCT detector, further quantities are considered: the ratio $\pt^{\mathrm{conv}}/E_{\rm calo}$; and
the fraction of the conversion momentum carried by the
highest-\pt\ track, $\pt^{\mathrm{max}}/\pt^{\mathrm{conv}}$.

\subsection{Binning and linearity corrections}
\label{subsec:Binning}

To help the MVA optimise the energy response in different regions of
phase space, the sample is divided into bins of $|\etaCluster|$,
\Etacc, and according to the particle type (electron, unconverted
photon or converted photon). The binning is chosen to follow the known 
detector geometry variations and significant changes in the energy
response. A rectangular mesh of $10 \times 9$ bins in $|\etaCluster|
\times \Etacc$ is defined, and $2 \times 6$ bins are defined in
addition for the regions close to the edges of the two half-barrel 
modules: 
\begin{itemize}
\item[$\bullet$] $|\etaCluster|$: 0 - 0.05 - 0.65 - 0.8 - 1.0 - 1.2 - 1.37 ;
  1.52 - 1.55 - 1.74 - 1.82 - 2.0 - 2.2 - 2.47, where 1.37 - 1.52 is excluded and 0 - 0.05 and 1.52 - 1.55 are edge bins.
\item[$\bullet$] \Etacc\ (normal bins): 0 - 10 - 20 - 40 - 60 - 80 - 120 - 500 - 1000 and 5000 GeV.
\item[$\bullet$] \Etacc\ (edge bins): 0 - 25 - 50 - 100 - 500 - 1000 and 5000 GeV.
\end{itemize}
An independent optimisation is performed for each bin. 

Multivariate algorithms aim at optimising the energy response and
minimising the root mean square (RMS) resolution. The presence of
tails in the energy response results in remaining non-linearities
which are corrected by adjusting the peak position of the ratio of the
output energy \Emva\ to \Etrue\ to unity. These corrections range from
+2\% to +5\% depending on $\eta$ at $E_{\rm T}=$10 GeV, and rapidly
decrease to zero around 100 GeV.

\subsection{Performance}
\label{subsec:MVAPerformance}

\begin{figure*}
\centering
\includegraphics[width=0.49\textwidth]{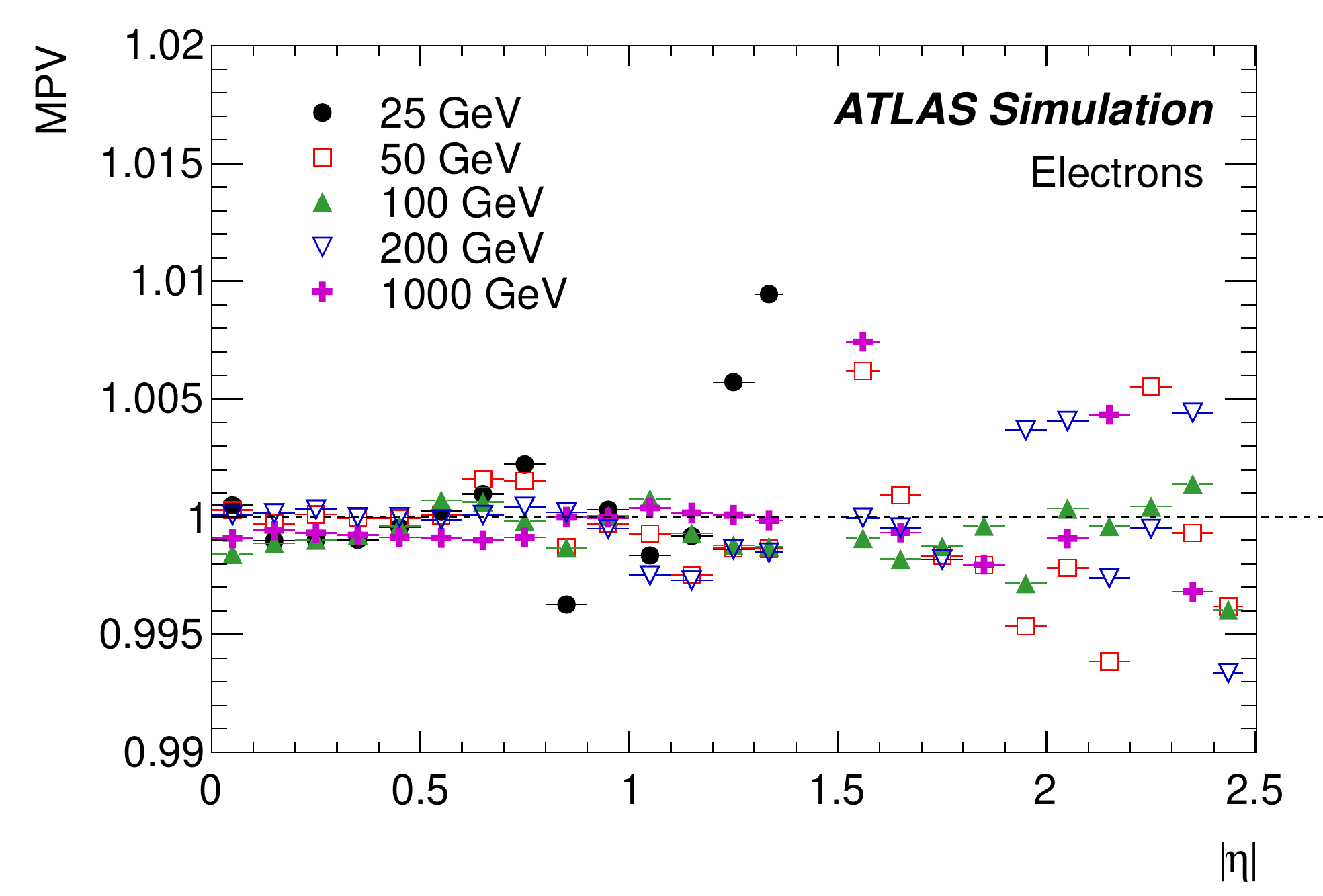}
\includegraphics[width=0.49\textwidth]{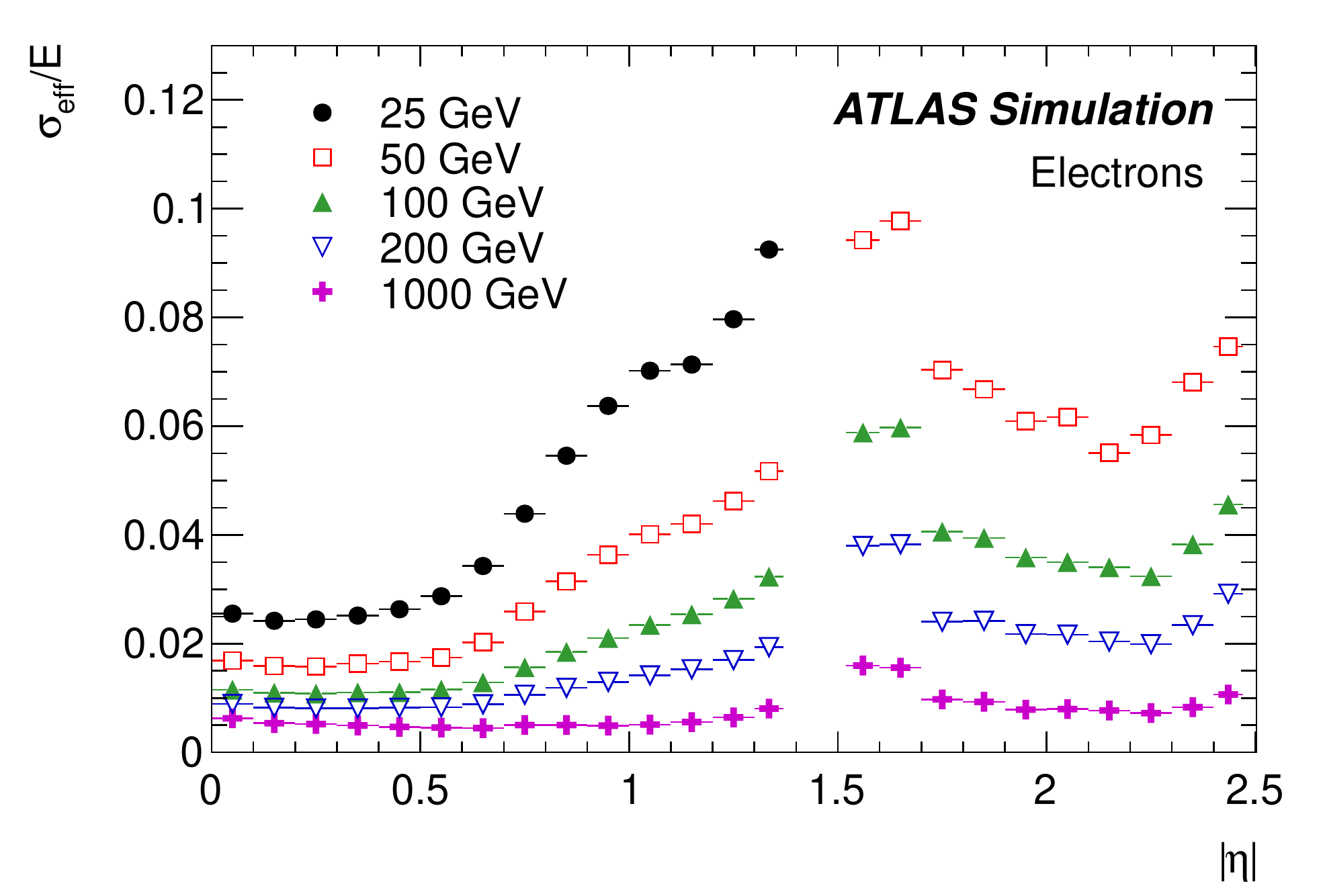}
\includegraphics[width=0.49\textwidth]{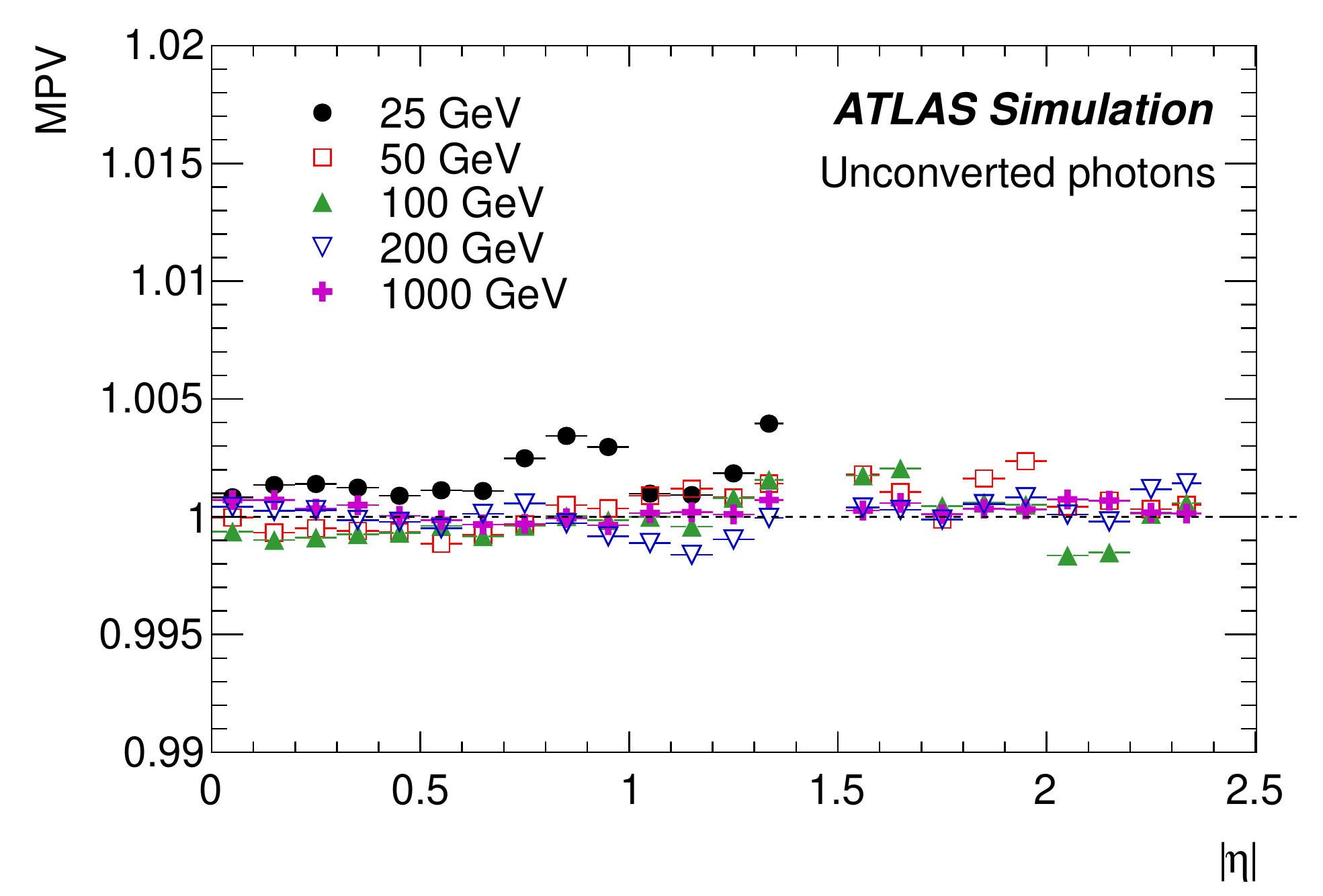}
\includegraphics[width=0.49\textwidth]{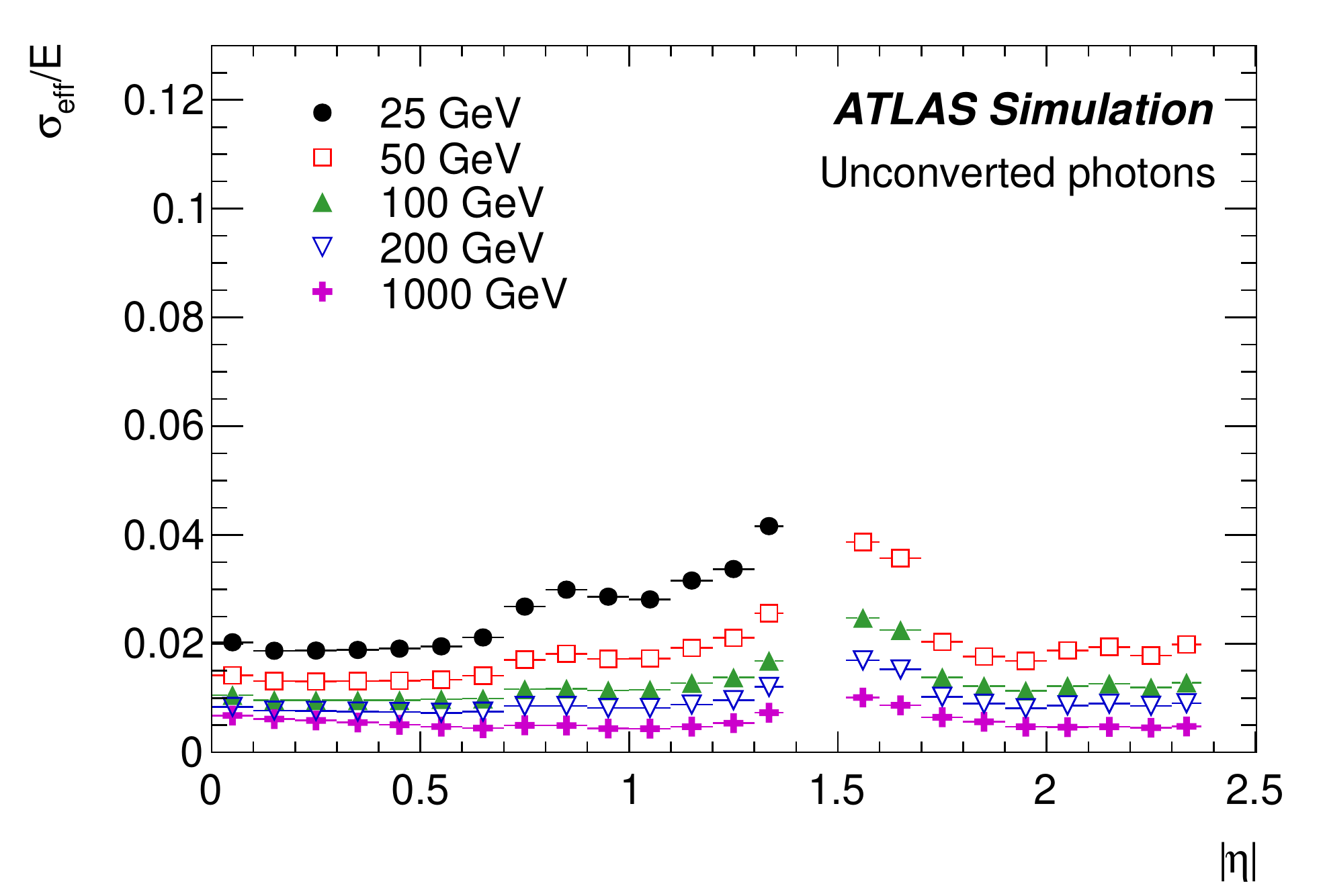}
\includegraphics[width=0.49\textwidth]{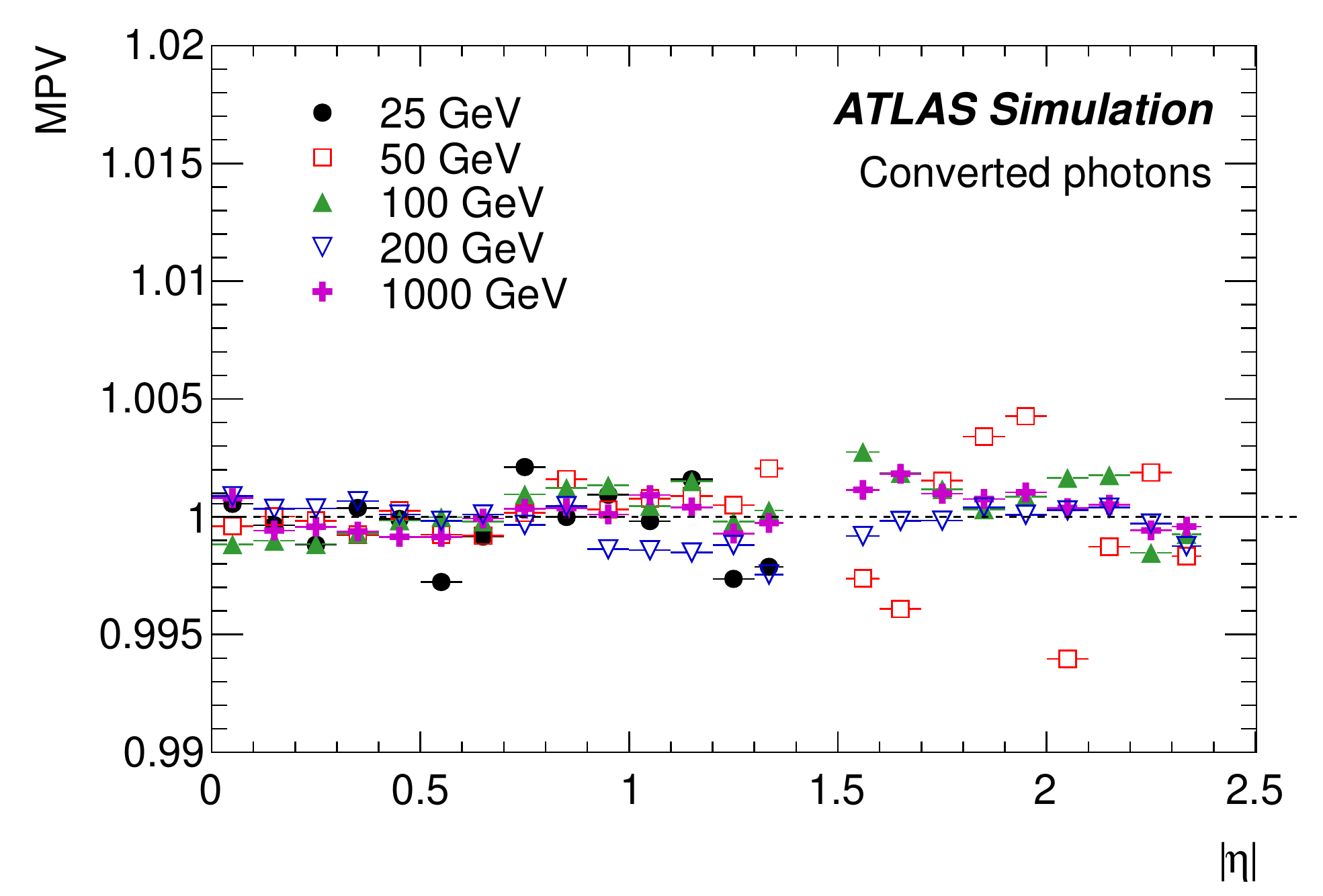}
\includegraphics[width=0.49\textwidth]{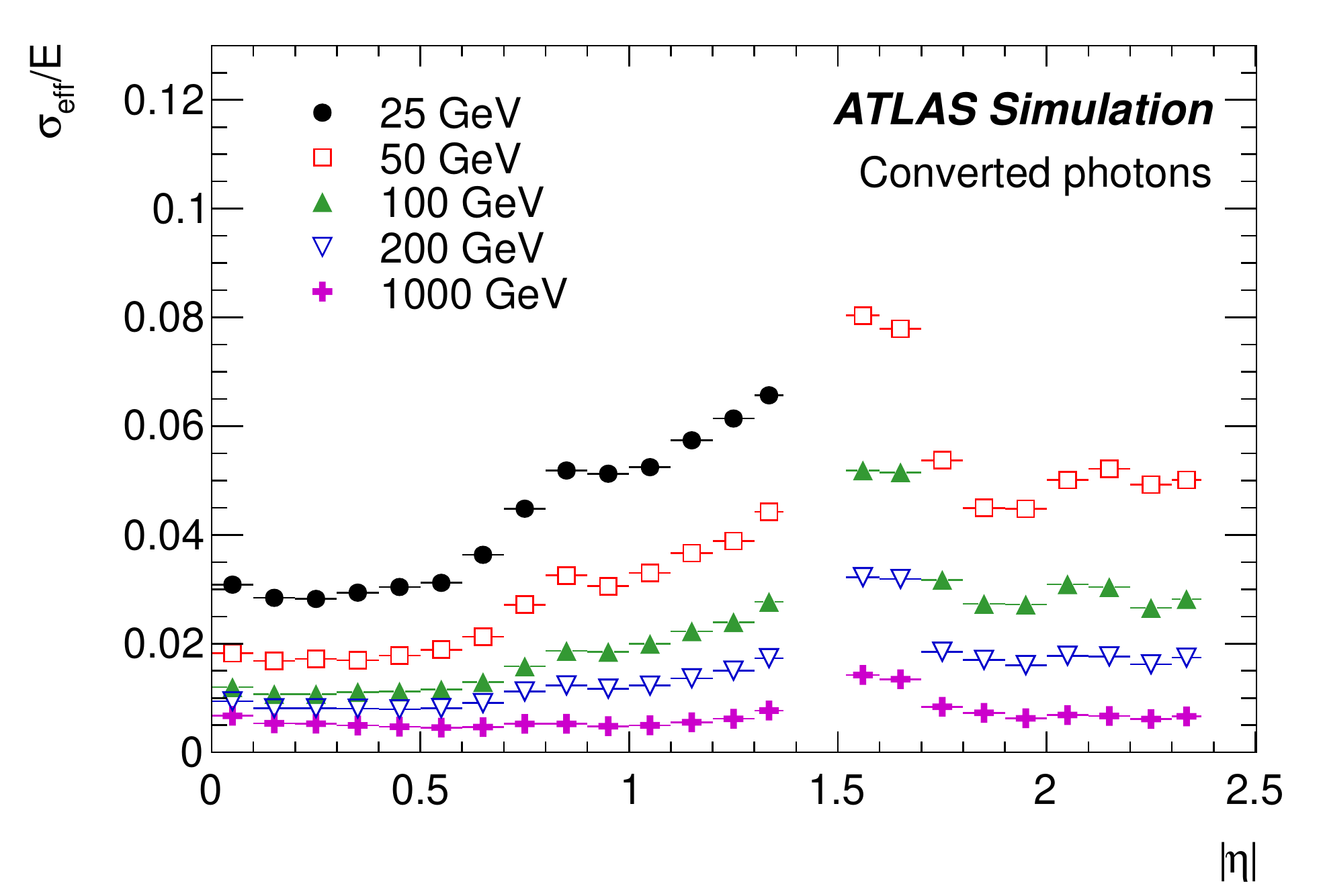}
\caption{Most probable value (MPV) of $E / \Etrue$ and relative effective resolution
  $\sigma_{\rm eff}/E$ as a function of $|\eta|$ for different energies,
  for electrons (top), unconverted photons (middle) and converted
  photons (bottom). The points at $E=$25~GeV are shown only for
  $|\eta|<1.37$, where they correspond to $E_{\rm T}>$10~GeV. \label{fig:MVA:single}}
\end{figure*}

The linearity and resolution of the MVA calibration are
illustrated in Fig.~\ref{fig:MVA:single}. The linearity is
defined as the deviation of the peak position of $E / \Etrue$ from unity as a function  
of \Ettrue, estimated by the most probable value (MPV) of a Gaussian function fitted
to the core of the distribution in each ($\Ettrue, |\eta|$) bin. The fits
are restricted to the range $[-1,+2]$ standard deviations. The
resolution $\sigma_\mathrm{eff}$ is defined as the interquartile range of $E/\Etrue$,
i.e. the interval excluding the first and the last quartiles of the
$E/\Etrue$ distribution in each bin, normalised to 1.35 standard
deviations, its equivalent definition for a normal
distribution. These estimators are chosen to reflect the influence of
energy tails.

The obtained MVA calibration non-linearity is everywhere below 0.3\%
for \Ettrue\ above 10~\gev, better than 1\% at lower transverse
energies, only reaching 2\% in localised regions for converted
photons. An improvement of more than a factor two compared to the
initial calibration is achieved, in particular in the high $|\eta|$
region. For the resolution, improvements of about 3\% to 10\% in the
barrel and 10\% to 15\% in the endcap are obtained for unconverted
photons. For converted photons in the same energy range, the resolution
is improved by typically 20\%. For electrons, improvements of a few
percent are obtained on average, except at
$1.52<|\eta|<1.8$ where they vary from 10\% to 30\% depending on
$\et$. While the resolution estimator used here reproduces the
expected sampling term resolution for unconverted photons
($\sigma/E\sim 0.1/\sqrt{E}$ on average), the worse resolution
obtained for electrons and converted photons reflects the
presence of significant energy tails induced by interactions with the material upstream of the calorimeter.

\begin{figure*}
\centering
\includegraphics[width=0.49\textwidth]{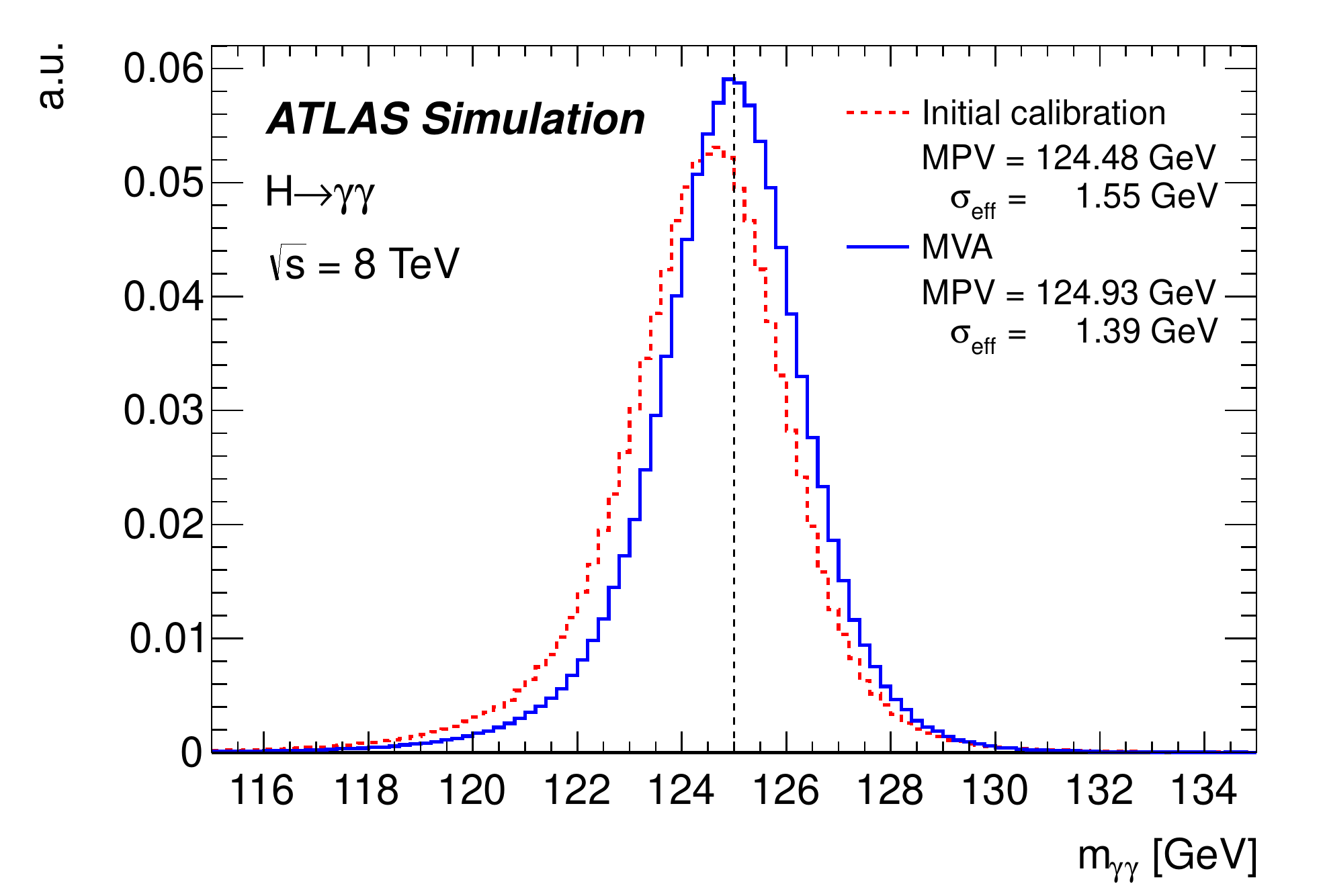}
\includegraphics[width=0.49\textwidth]{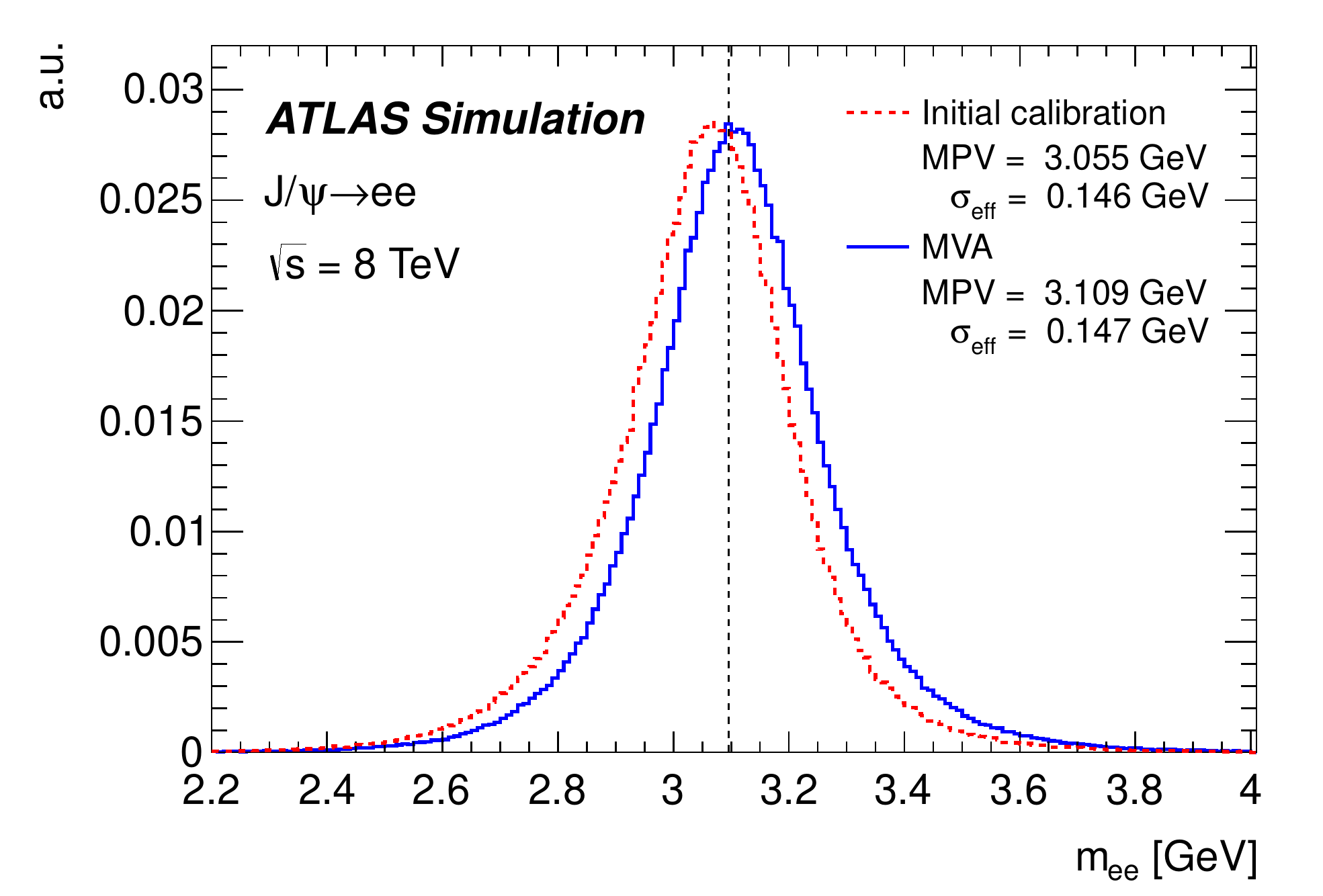}
\caption{Comparison of the diphoton invariant mass distributions, $m_{\gamma\gamma}$, for a simulated Standard Model Higgs boson with a
mass of 125 GeV, obtained with the initial calibration and with the
MVA calibration (left). The same comparison for the dielectron
invariant mass distributions, $m_{ee}$, for simulated \jpsiee~decays
(right). The vertical dashed lines indicate the simulated 
masses.\label{fig:MVA:performance}}
\end{figure*}

Fig.~\ref{fig:MVA:performance} compares the performance of the MVA
calibration with the initial calibration in simulated
$H\to\gamma\gamma$ ($m_{\mathrm{H}}=125$~GeV) and \jpsiee\ events. The invariant
mass resolution of the former improves by 10\% on average, with a maximum improvement of 15\% for converted
photons or in the barrel--endcap transition region. The latter reflects the expected linearity improvement; no
significant resolution improvement is obtained. 

\section{Uniformity and stability \label{sec:uniformity}}

Good uniformity of the EM calorimeter energy reconstruction across
pseudorapidity and azimuthal angle, and excellent stability of the response as a function
of time and pile-up conditions, are necessary to achieve optimal energy resolution
in data. They also constitute a prerequisite for the passive material
determination and energy scale measurement presented in 
Sects.~\ref{sec:samplingcalib}\--\ref{sec:zeescales}. The present
section describes a set of studies, based on the data collected at $\sqrt{s}=8\tev$, 
aiming to correct for local non-uniformities in the calorimeter response. 

The response uniformity is investigated using \textit{$E/p$} for electrons in
\wen\ events and the electron pair invariant mass in $Z$ boson
decays. Four classes of effects are discussed below. The stability
of the response as a function of $\phi$, time and pile-up is presented
after all corrections are applied.

\subsection{High-voltage inhomogeneities}

\begin{figure}
  \centering
  \includegraphics[width=\columnwidth]{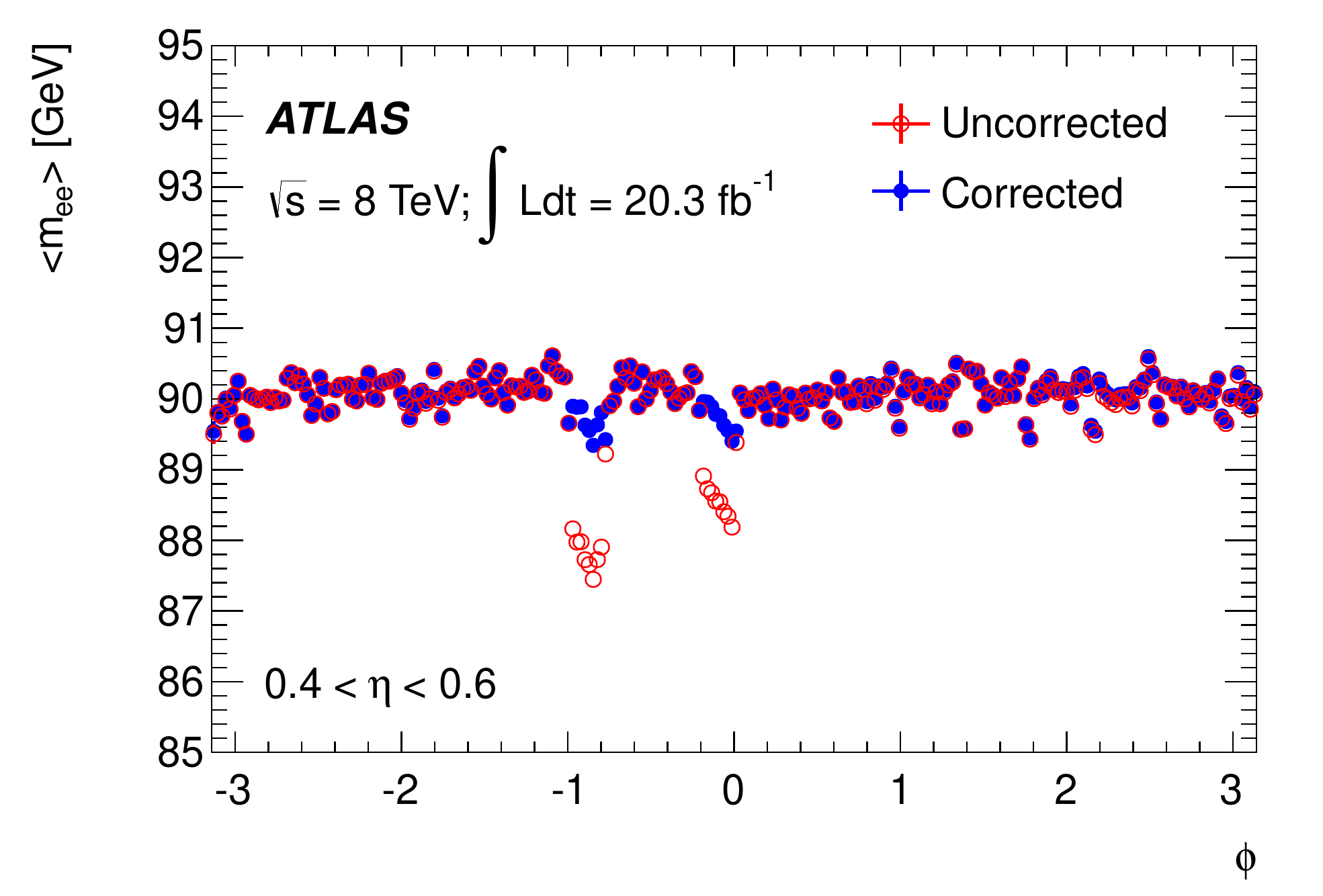}
  \caption{Average value of $m_{ee}$ as a function of the
    azimuthal position of the leading decay electron with $0.4<\eta<0.6$, before and 
    after the HV correction. The error bars include statistical uncertainties only.
    \label{fig:HVUniformityCorrections}
  }
\end{figure}

In a few sectors (of typical size $\Delta\eta\times\Delta\phi =
0.2\times 0.2$) of the EM calorimeter, the HV 
is set to a non-nominal value due to short circuits occurring in specific
LAr gaps \cite{LArDQ}. The effect of such modifications is first corrected at the reconstruction
level using the expected HV dependence of the response. 
The azimuthal profiles of the electron pair invariant mass in \zee~
events, however, show localised residual effects,
affecting less than 2\% of the total number of HV sectors in the EM calorimeter \cite{LArReadiness}.
An empirical correction is derived based on these profiles to restore
the azimuthal uniformity in the problematic sectors. The average
value of $m_{ee}$ as a function of the azimuthal position of its
leading decay electron, for $0.4<\eta<0.6$, is presented in
Fig.~\ref{fig:HVUniformityCorrections} before and after this
correction. In this example, two sectors are set to a non-nominal HV,
inducing a decrease of the response by about 2\% at $\phi \sim -1$ and
$\phi \sim 0$. After correction, the response is uniform. 

\subsection{Time dependence of the presampler response}

The nominal HV in the EM barrel PS is 2000 V. To limit the
increasing occurrence of sporadic electronics noise~\cite{perf2010} with increasing luminosity, the operating HV was
reduced to 1600 V during the 2011 run and until September
2012 (period P1). The HV was later further reduced to 1200 V, with some sectors brought down to 800 V (period P2).
As above, the non-nominal HV is at first compensated at the cell level
using a correction defined from the expected HV dependence of the PS
response. This correction is of the order of 8\% for P1 and 21\% for P2.  

\begin{figure}
  \centering
  \includegraphics[width=\columnwidth]{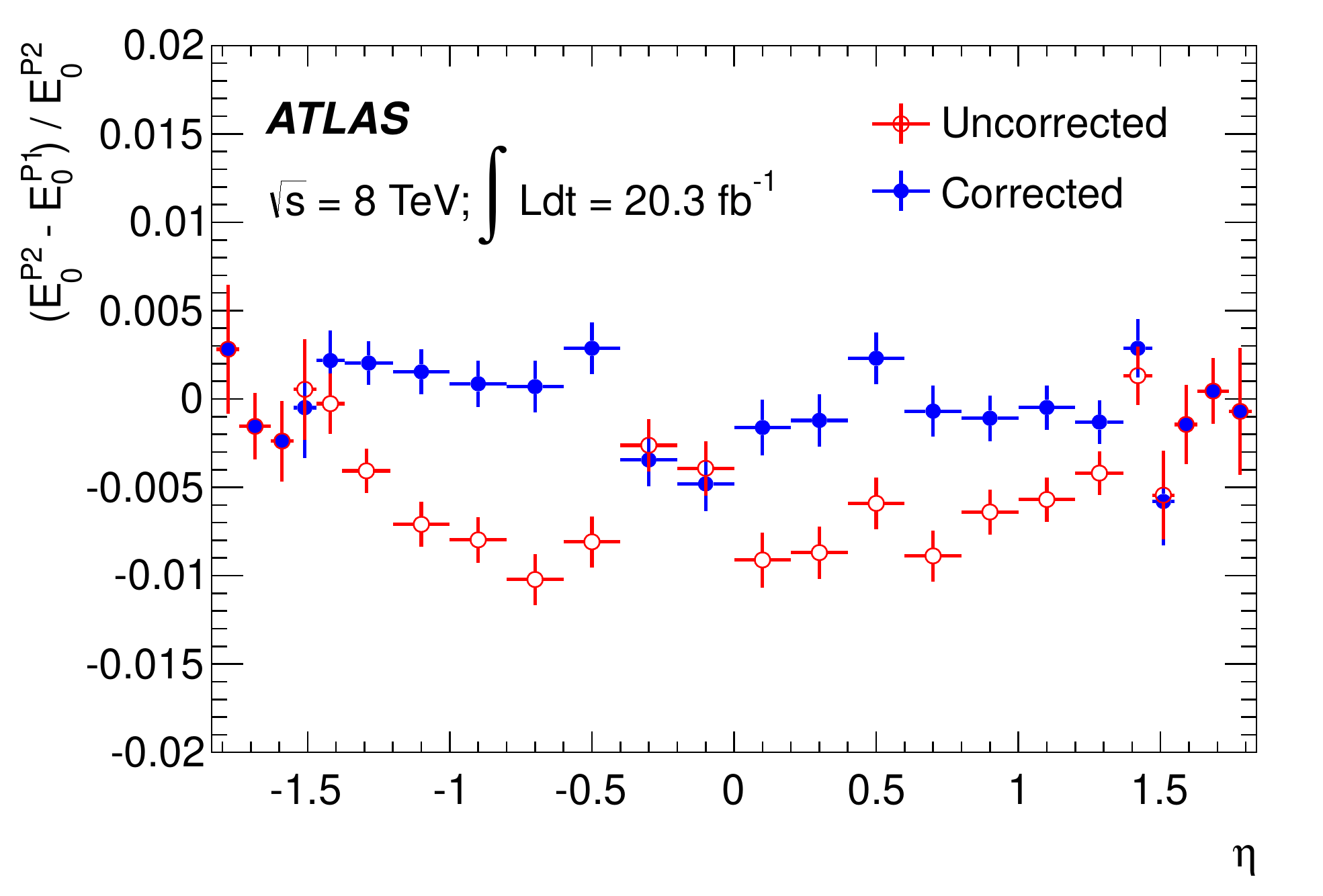}
  \caption{Relative difference in the raw PS energy response due to the change
    in HV settings, as a function of $\eta$, before and after
    correction of the residual HV dependence. The periods
    before and after the HV change are referred to as P1 and P2,
    respectively. The error bars include statistical uncertainties only.\label{fig:PSHVUniformityCorrections}
  }
\end{figure}

The accuracy of the correction is verified by comparing the
PS response for electrons from $Z\rightarrow ee$ data between P1 and P2; a residual
$\eta$-dependent variation of up to 1\% is observed. 
An additional empirical correction is applied to the PS energy at the cluster level,
reducing the bias to 0.4\% across $\eta$. The residual response bias
and its corrections are illustrated in Fig.~\ref{fig:PSHVUniformityCorrections}.

\subsection{Energy loss between the barrel calorimeter modules} 

\begin{figure}[!h]
\centering
\includegraphics[width=\columnwidth]{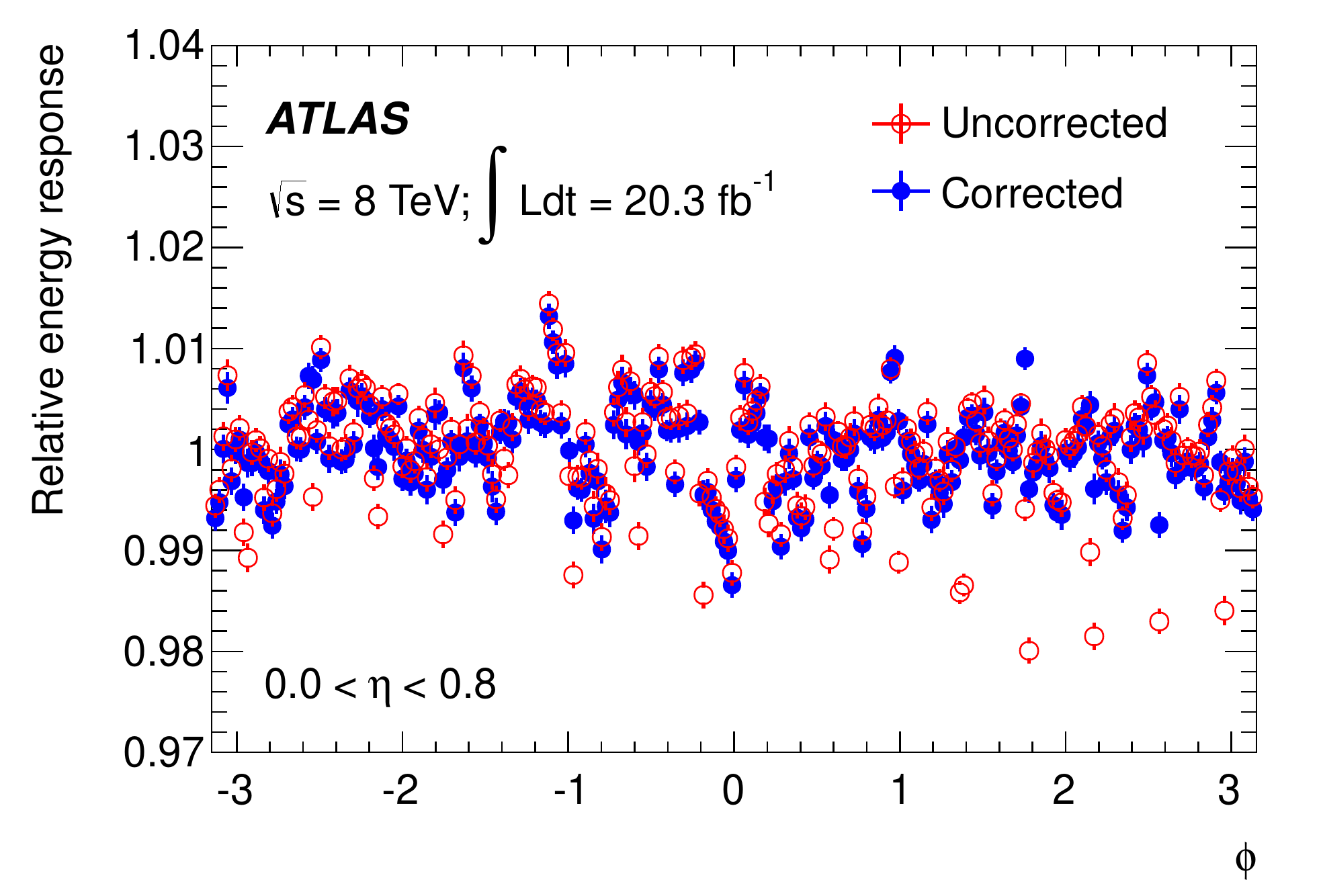}
\caption{Relative energy response of electrons as a function of $\phi$, before and after
  correction for the intermodule widening effect, for
  $0<\eta<0.8$. The relative energy scale is defined from the MPV of
  the \textit{$E/p$} distribution in $W$ events, normalised to its
  average over $\phi$. The error bars include statistical uncertainties only.}
\label{fig:IMWcorrections}
\end{figure}

When probing the energy response versus $\phi$ using the MPV of the \textit{$E/p$} distribution in \wen\ events in data, a $\pi/8$-periodical structure is
observed. The period and the location of the effect correspond to the
transitions between the barrel calorimeter modules. The size of the
modulation is $\sim$2\% in the $\phi>0$ region and $\sim$1\%
for $\phi<0$, and is interpreted as a gravity-induced widening 
of the inter-module gaps. The energy loss is adjusted with an
empirical function which is then used to correct the calorimeter
response. The effect of the inter-module widening and its correction
are shown in Fig.~\ref{fig:IMWcorrections}. This effect is not observed in the EMECs. 

\subsection{Energy response in high and medium gain\label{sec:unifGain}} 

\begin{figure}
\centering
\includegraphics[width=\columnwidth]{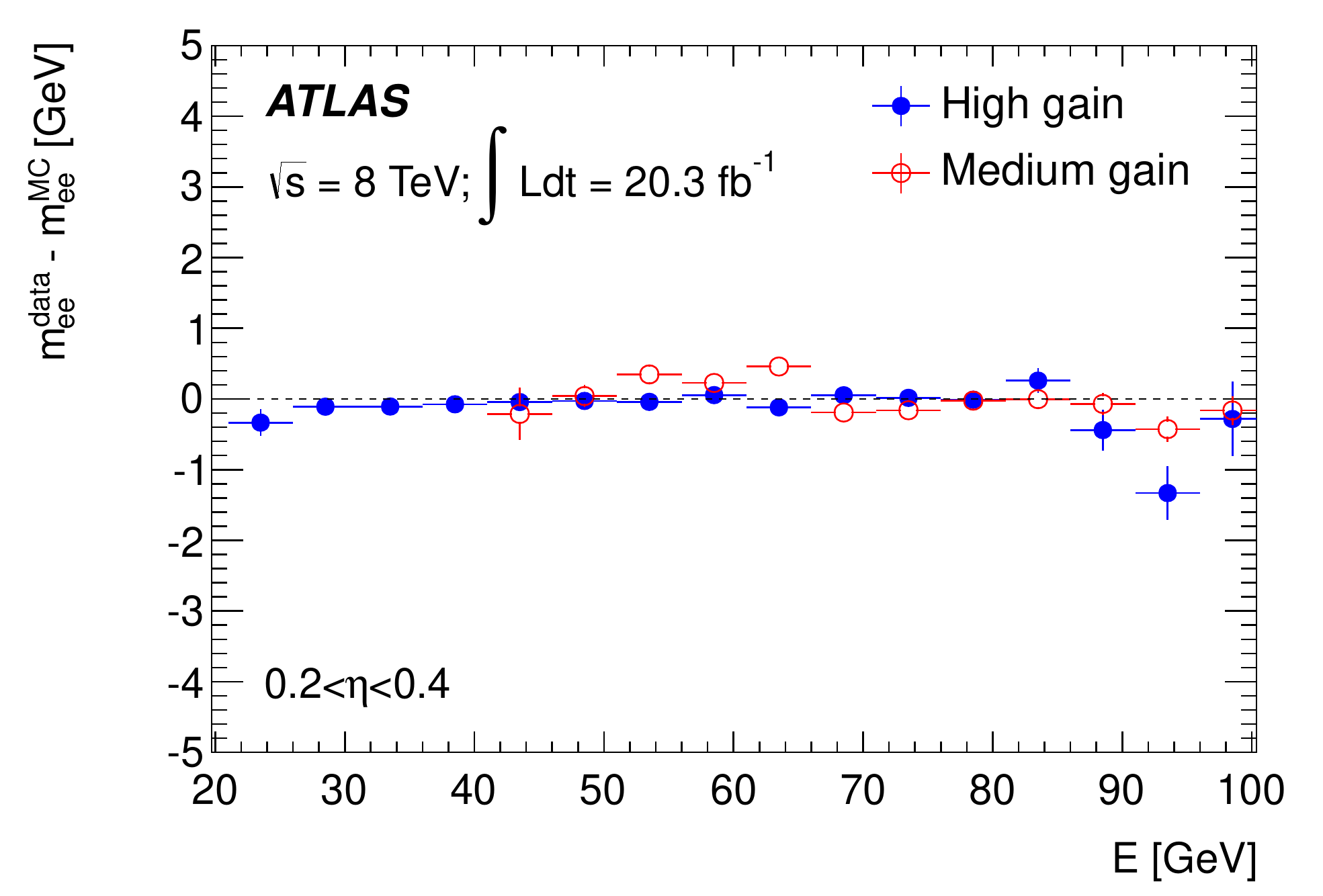} \\
\includegraphics[width=\columnwidth]{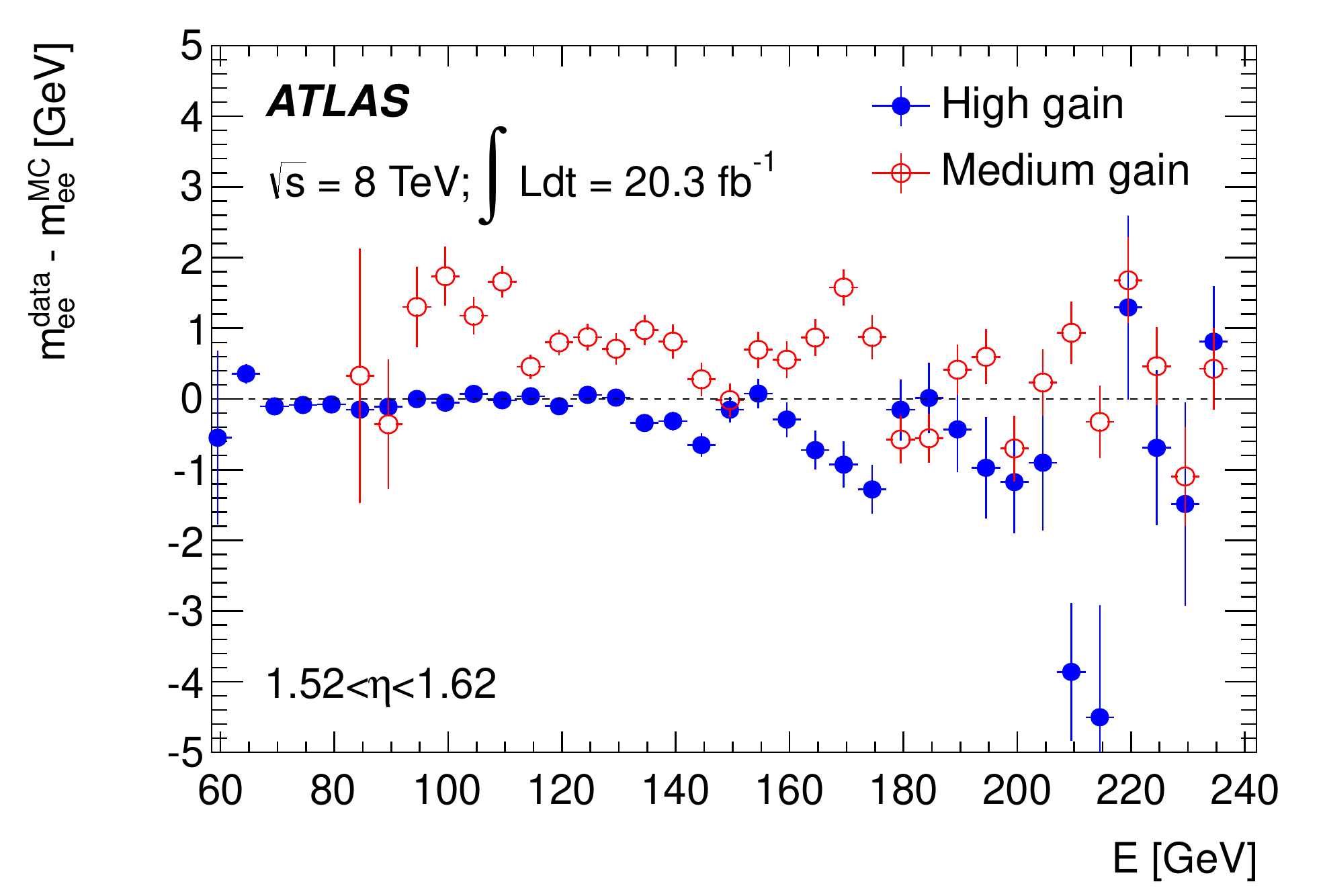}
\caption{Difference between data and simulation of the average reconstructed $Z$ boson
  mass as a function of the energy of one electron, for events
  where all cluster cells are in high gain from those where at least
  one cell is in medium gain, in a good region (top)
  and in a region with a significant bias (bottom). The error bars include statistical uncertainties only.} 
\label{fig:HGMG}
\end{figure}

To accommodate the wide range of expected energies in the calorimeter cells, the electronic signals are 
treated with three gains (see Sect.~\ref{subsec:lar_electronics}). In \zee~ events, used for
the absolute energy scale determination (see Sect.~\ref{sec:zeescales}),
most electron clusters have all their L2 cel\-ls recorded in HG. In
the case of $H\rightarrow \gamma \gamma$ ($m_{\mathrm{H}}=125$~GeV) for
example, roughly 1/3 of the events have a photon with at least one
cell in MG.

The reconstructed electron pair invariant mass is compared between
data and simulation as a function of the electron energy, for
events where all electron cluster cells in L2 are in HG and for those
where at least one cell is in MG. In most of the calorimeter, the
energy calibration is found to be gain independent within
uncertainties; however, a percent-level effect is seen in specific
$\eta$ regions (around $|\eta|\sim 0.6$ and $|\eta|\sim 1.6$). Two
example regions are illustrated in Fig.~\ref{fig:HGMG} for
$0.2<\eta<0.4$ and $1.52<\eta<1.62$. The observed effect is symmetric
in $\eta$. 

The observed gain dependence of the energy response is removed by applying
a correction defined from the data--MC difference of the energy response in HG
and MG, multiplied by the expected fraction of clusters with at least
one L2 cell in MG at given $\eta$ and $\et$. The LG case,
relevant only at very high energy, is assumed to have the same correction as the MG.

\subsection{Azimuthal non-uniformity and operational stability after corrections}

\begin{figure}
\centering
\includegraphics[width=\columnwidth]{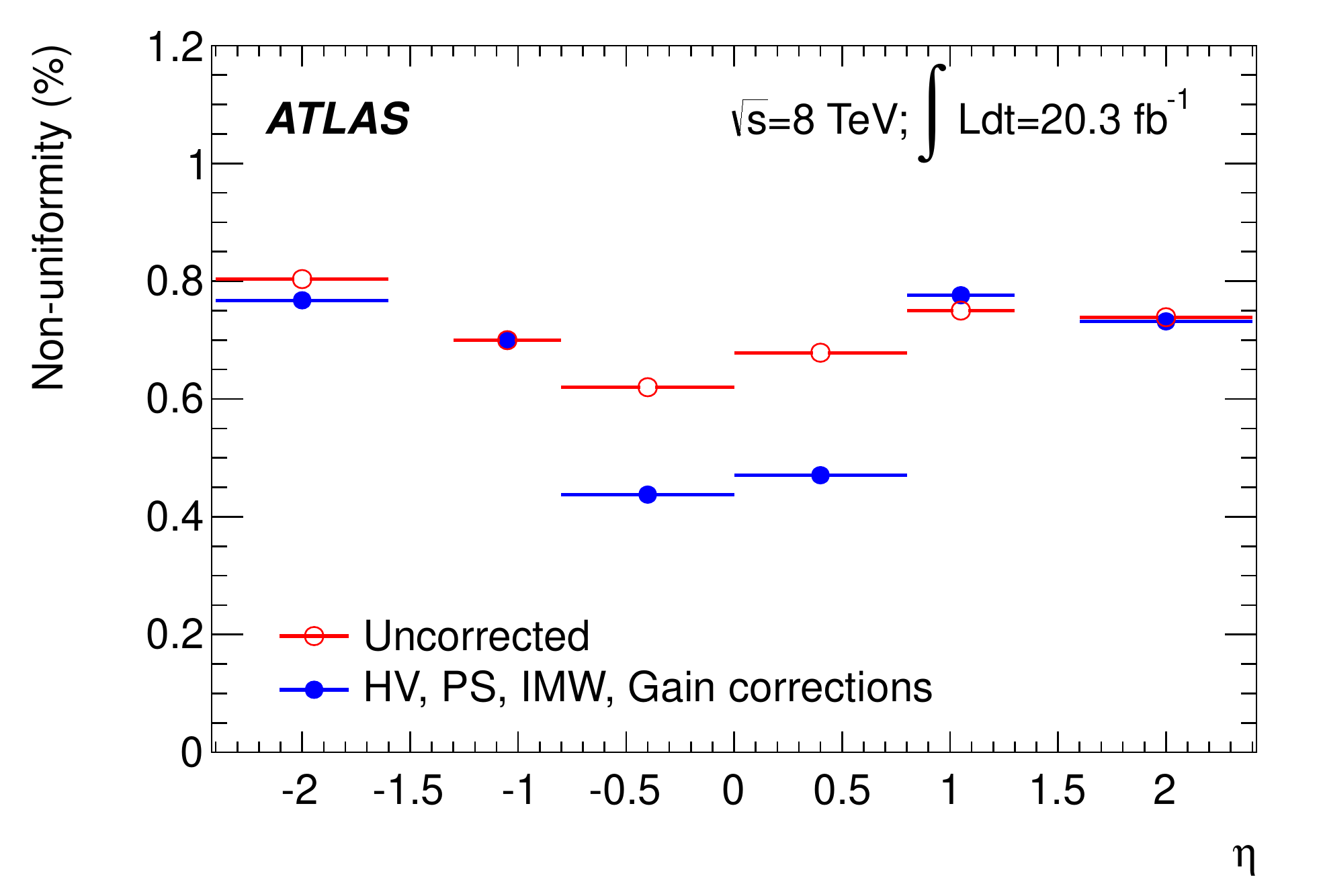}
\caption{Azimuthal non-uniformity of the energy response as a function of
  $\eta$, estimated from the electron pair invariant mass peak in
  \zee~ events.}  
\label{fig:phiuniformity}
\end{figure}

\begin{figure}
\centering
\includegraphics[width=\columnwidth]{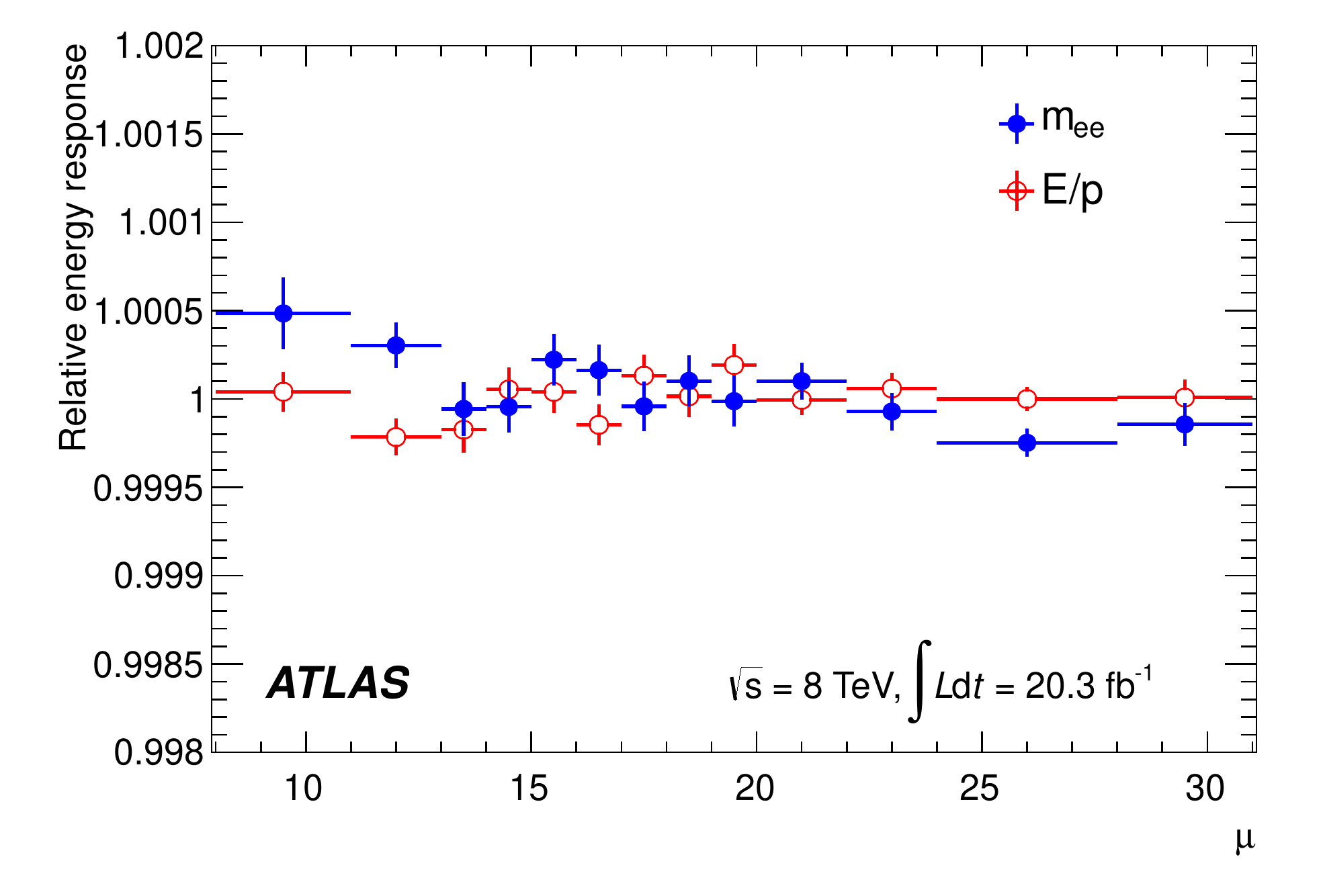}
\caption{Energy response as a function of $\mu$,
  normalised to its average. The energy response is probed using the
  peak position of the electron pair invariant mass peak in $Z$ events and the MPV of the 
  \textit{$E/p$} distribution in
  $W$ events, and $\mu$ is defined as the expected number of $pp$
  interactions per bunch crossing. The error bars include statistical uncertainties only.\label{fig:uni:vsPileUp}}
\end{figure}

\begin{figure}
  \centering
  \includegraphics[width=\columnwidth]{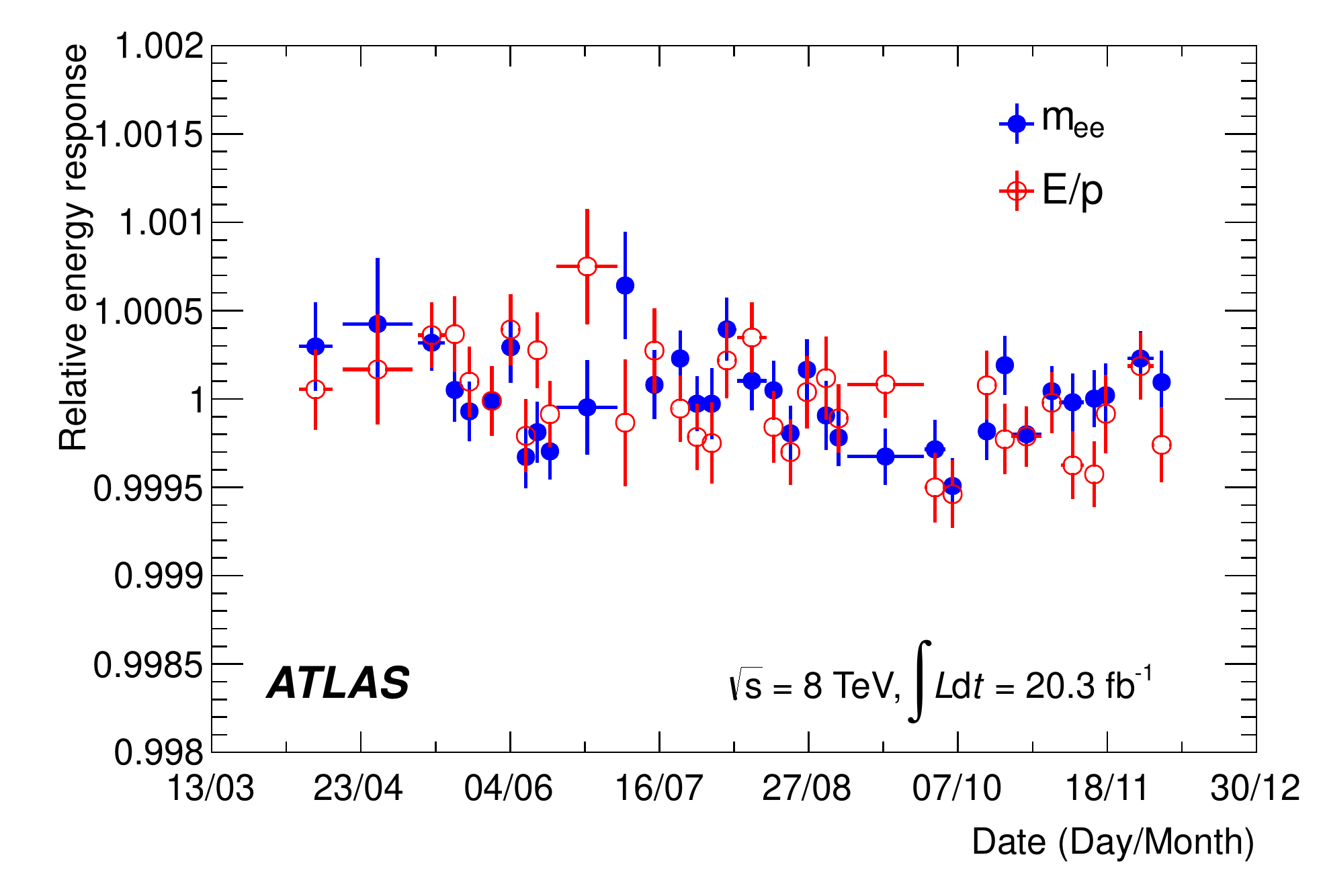}
  \caption{Energy response as a function of time, normalised to its average quantity. The energy response is probed using the
    peak position of the electron pair invariant mass peak in $Z$ events and the MPV of the 
    \textit{$E/p$} distribution in
    $W$ events; each point in time represents a recorded
    amount of data of around 100~pb$^{-1}$. The error bars include statistical uncertainties only.}
  \label{fig:uni:vsTime}
\end{figure}

\begin{figure}
\centering
\includegraphics[width=\columnwidth]{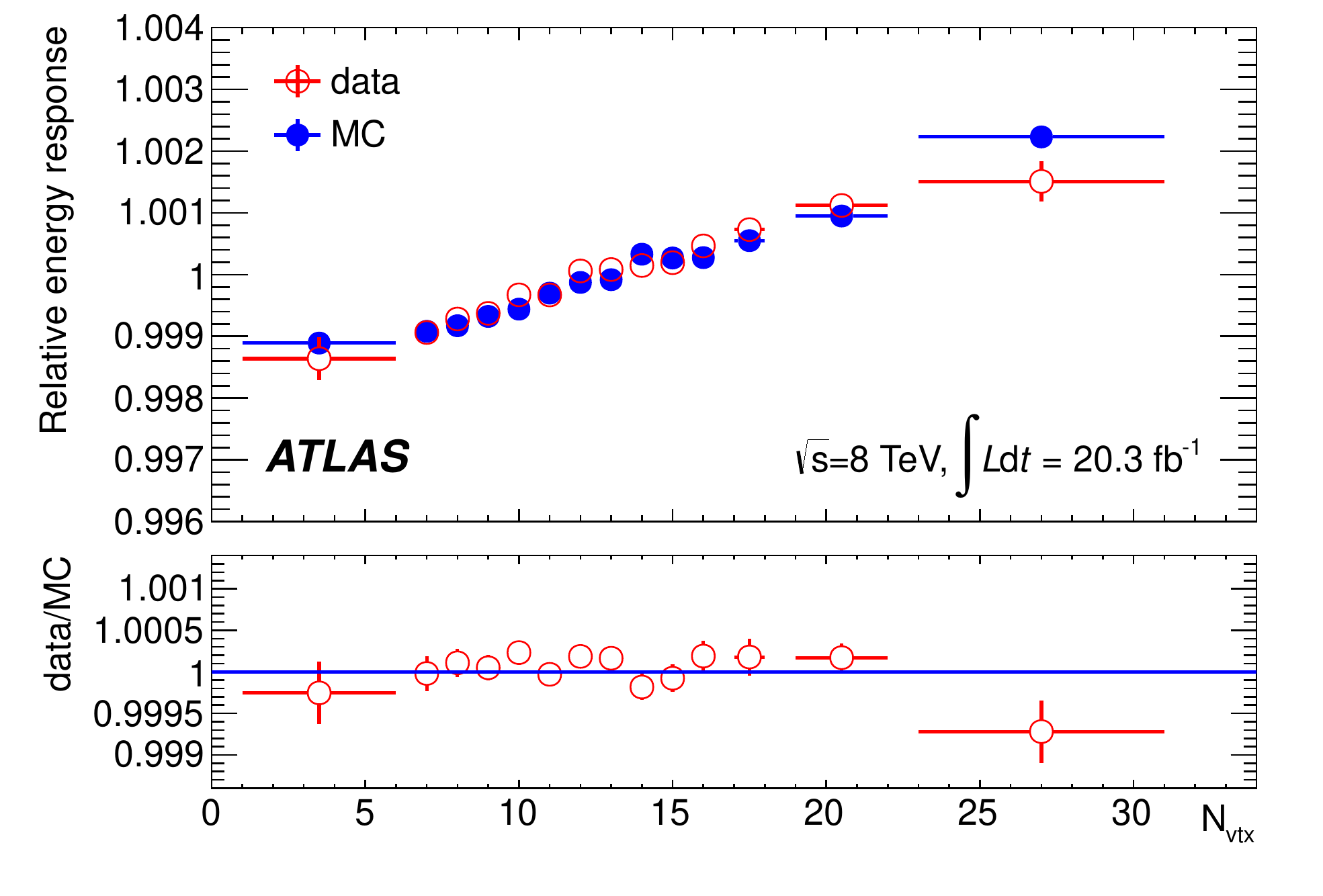}
\caption{Top: energy response as a function of $N_{\rm vtx}$,
  normalised to its average. The energy response is probed using the
  peak position of the electron pair invariant mass in $Z$
  events. Bottom: ratio of data to simulation. The error bars include statistical uncertainties only.\label{fig:uni:vsVertices}}
\end{figure}

The azimuthal non-uniformity before and after the corrections described above
is shown in Fig.~\ref{fig:phiuniformity}. This non-uniformity is defined as the RMS of the energy
 response versus $\phi$, probed with a granularity of $\Delta\phi=0.025$, after having subtracted the contribution from 
  the expected statistical fluctuations. The energy response is
probed using the electron pair invariant mass peak in \zee~
events, and the non-uniformity is defined from the RMS of the response
versus $\phi$, probed with a granularity of $\Delta\phi=0.025$,
corresponding to one cell in L2, and for coarse $\eta$ bins; the
contribution of the expected statistical fluctuations is subtracted
in quadrature. The result can be interpreted as the non-uniformity
contribution to the long-range resolution constant term.  A
non-uniformity of 0.45\% is achieved for $|\eta|<0.8$, and 0.75\% is
obtained in the rest of the calorimeter. 

The stability of the electron energy response as a function of
the mean number of interactions per bunch crossing ($\mu$), and as a function of time was measured using electrons from $Z$ boson decays. The
results presented in Figs.~\ref{fig:uni:vsPileUp} and~\ref{fig:uni:vsTime} show stability
at the level of $0.05\%$. The stability of the response as a function
of the  number of reconstructed collision vertices ($N_{\rm vtx}$) is
shown in Fig.~\ref{fig:uni:vsVertices}. Classifying events according
to $N_{\rm vtx}$, related to the actual number of interactions per bunch crossing, 
biases the pile-up activity of the colliding bunch
with respect to the average. In this case the compensation of the
pile-up contribution to the reconstructed energy by the bipolar
shaping becomes imperfect, giving rise to the observed slope. The description of this effect in the simulation is accurate to 0.05\%.


\section{Intercalibration of the LAr calorimeter layers \label{sec:samplingcalib}}

Corrections are needed in data to adjust residual effects not perfectly
accounted for by the cell electronic calibration discussed in
Sect.~\ref{subsec:lar_electronics}. 

The intercalibration of the first and second calorimeter layers uses
muons from $Z\rightarrow\mu\mu$ decays as probes, while the determination of the PS energy scale
exploits the PS energy distributions of electrons in data and
simulation, after effective corrections for possible mis-modelling of
the upstream passive material. The results are verified by a study of
the electron energy response as a function of shower depth. 

No dedicated intercalibration of the third EM longitudinal layer is carried out, as its contribution is
negligible in the energy range covered by the present studies.

\subsection{Intercalibration of the first and second calorimeter
  layers \label{sec:l1l2}}

Muon energy deposits in the calorimeter are insensitive to the 
amount of passive material upstream of the EM calorimeter and
constitute a direct probe of the energy response. The measured muon
energy is typically 60~MeV in L1 and about 210~MeV in L2, with a
signal-to-noise ratio of about three \cite{LArReadiness}. Muon energy deposits
are very localised, most of the energy being deposited in one or two
cells. Since the critical energy for muons interacting with the calorimeter is of the order of 100 GeV,
most muons from $Z\rightarrow\mu\mu$ decays are minimum ionising particles. 

The analysis uses muons from $Z\rightarrow\mu\mu$ decays, requiring
$\pt^\mu>25$~GeV. The calorimeter cells
crossed by the muon tracks are determined by extrapolating the muon
tracks to each layer of the calorimeter, taking into account the
geometry of the calorimeter and the residual magnetic field seen by 
the muon along its path in the calorimeter. In L1, the
muon signal is estimated by summing the energies measured in three
adjacent cells along $\eta$, centred around the cell of highest
energy among the few cells around the extrapolated track. In L2, due to the accordion geometry, the
energy is most often shared between two adjacent cells along $\phi$;
hence the signal is estimated from the sum of the highest energy cell
and its most energetic neighbour in $\phi$.

\begin{figure}
  \centering
  \includegraphics[width=\columnwidth]{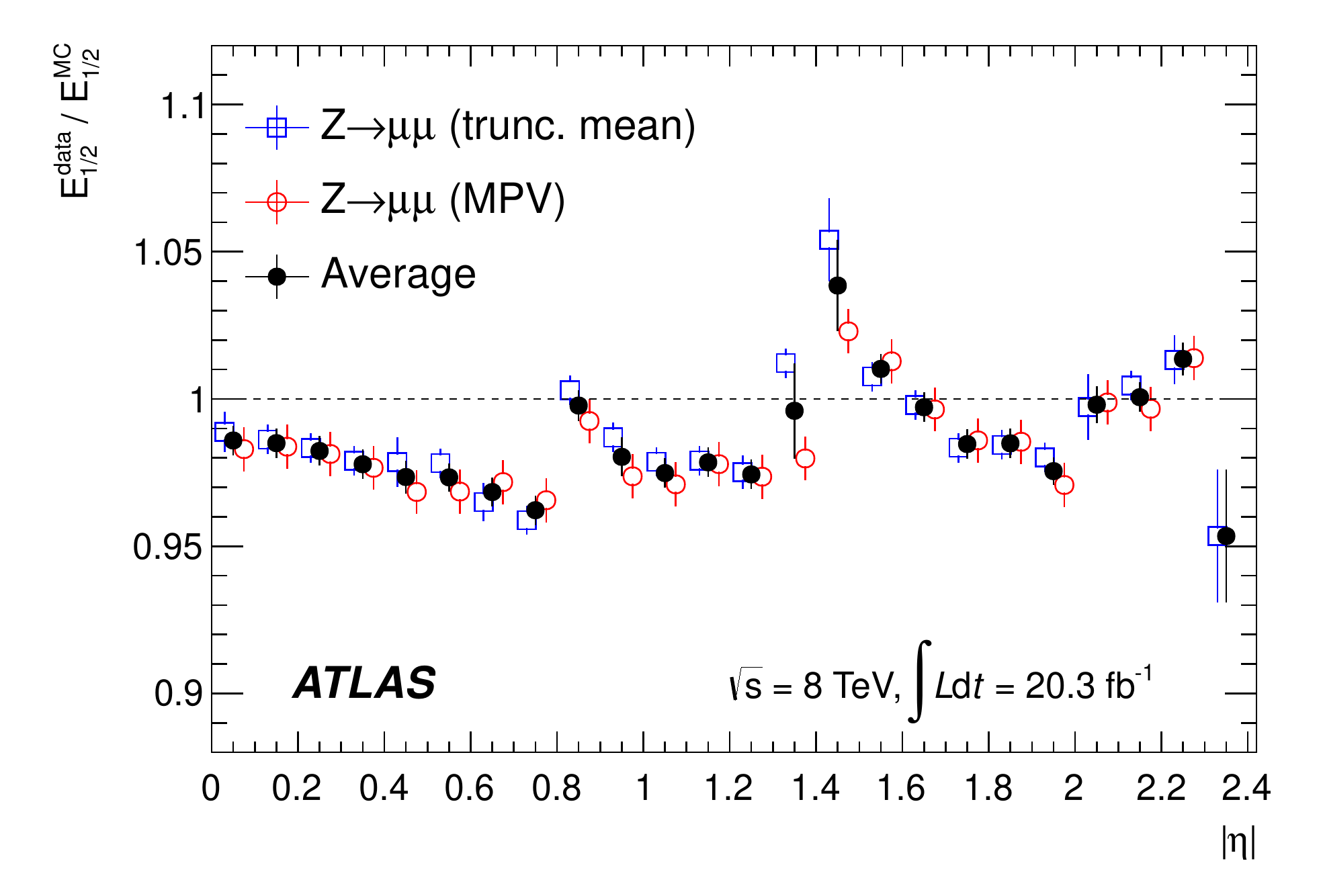}
  \caption{Ratio $\langle E_{1/2} \rangle^{\mathrm{data}}/ \langle E_{1/2} \rangle^{\mathrm{MC}}$ as a function of $|\eta|$, as obtained from the peak
    position of muon energy deposits in L1 and L2, and from the mean of
    these energy deposits computed in an interval containing 90\% of
    the distribution. The error bars represent the total uncertainty specific to the
    $Z\rightarrow\mu\mu$ analysis.\label{fig:E12muons}} 
\end{figure}

The observed muon energy distribution in each layer is given by the
convolution of a Landau distribution describing the energy deposit,
and a Gaussian distribution corresponding to the electronic noise. The
MPV of the deposited energy is extracted using an analytical fit
with the convolution model, or is alternatively estimated using a truncated mean, by defining the
interval as the smallest one containing 90\% of the energy
distribution. Denoting $\langle E_{1/2} \rangle$ the ratio of the MPVs
in L1 and L2, the intercalibration result is defined as $\alpha_{1/2}
=\langle E_{1/2} \rangle^{\mathrm{data}}/ \langle E_{1/2} \rangle^{\mathrm{MC}}$. The
central value of $\alpha_{1/2}$ is given by the average of the two
methods; the difference is used to define its systematic 
uncertainty. The statistical uncertainty is negligible. The result is
illustrated in Fig.~\ref{fig:E12muons}. All features are observed to
be symmetric within uncertainties with respect to $\eta=0$, and are therefore shown as
a function of $|\eta|$. In the barrel, a negative bias of about 3\%  is observed; it shows a falling structure from
$|\eta|=0$ to 0.8 and from $|\eta|=0.8$ to 1.4, with a positive step
at the boundary between these regions. In the endcap,
$\alpha_{1/2}\sim 1$ on average, but its behaviour across
pseudorapidity is not uniform. 

The intercalibration of the calorimeter layers with muons relies on
the proper modelling in the simulation of the induced ionisation
current by muons in each calorimeter layer. The following sources of
uncertainty are considered: 

\begin{itemize} 

\item[$\bullet$] uncertainty in the exact path length traversed by muons, related
  to uncertainty in the geometry of the read-out cells; 

\item[$\bullet$] uncertainty in the effect of the reduced electric field at the
  transition between the different calorimeter layers; 

\item[$\bullet$] uncertainty in the modelling of the conversion of deposited
  energy to ionisation current due to variations in the electric field
  following the accordion structure of the absorbers and electrodes; 
 
\item[$\bullet$] uncertainty in the cross-talk between different calorimeter
  cells (between L1 cells, between L1 and L2 cells and between L2
  cells)~\cite{Aharrouche:2007nk} which affects the measured energy
  for muons (using three cells in L1 and two cells in L2). 

\end{itemize}

These uncertainties are evaluated by implementing the
corresponding changes in the simulation. The resulting uncertainty on
the relative calibration of L1 and L2 rises from 1\% to 1.5\% with the
pseudorapidity in the barrel and is 1.5\% in the endcap. 

These uncertainties are also propagated to uncertainties on the
modelling of $E_{1/2}$ for electrons and phot\-ons, as this variable is
used in Sect.~\ref{sec:material} for the passive-material
determination. For this modelling, the difference between data and
simulation in the description of lateral EM shower shape is also
taken into account, as it affects L2 more than L1. 

In addition, the HG response in L1 is found to be sensitive to
the pile-up-dependent optimisation of the OFC, for $1.8<|\eta|<2.3$,
with an uncertainty rising from 1\% to 5\% in this region. Since in this
region most high-\et\ EM showers have their highest energy cell in L1 recorded in MG,
this additional uncertainty is accounted for when applying the
muon-based calibration to electrons or photons.

The L1/L2 calibration bias $\alpha_{1/2}$ discussed in this section is
removed by applying an $|\eta|$-dependent correction to the layer
intercalibration in data. The correction can be applied to the energy measured either in
L1 (by defining $E_1^{\rm corr} = E_1 \,/\, \alpha_{1/2}$) or in L2
($E_2^{\rm corr} = E_2 \times \alpha_{1/2}$). The latter option is
chosen, as a direct comparison of $E_2$ in data and simulation shows
that the pattern vs $|\eta|$ observed in Fig.~\ref{fig:E12muons} is localised in
L2. After all other corrections discussed in the rest of the paper are applied, and in particular the overall energy scale
correction discussed in Sect.~\ref{sec:zeescales}, the calibrated
particle energy is unaffected by this choice.

\subsection{Presampler energy scale \label{sec:psscale}}

The presampler energy scale $\alpha_\mathrm{PS}$ is determined from the ratio of PS energies
in data and MC simulation and estimated using electrons from $W$ and $Z$
decays. Before this ratio can be interpreted in terms of an energy
scale, the effects of passive-material mis-modelling must be taken
into account, as an inaccurate passive-material description in the detector affects the electron shower development and hence the 
predicted PS energy distributions with respect to the data,
resulting in an apparent energy scale bias. This is
addressed by exploiting the expected correlation between $E_{1/2}$ and
$E_0$ for electrons, at a given $\eta$ value, under variations of the passive
material upstream of the PS. 

\begin{figure*}
  \centering
  \includegraphics[width=\columnwidth]{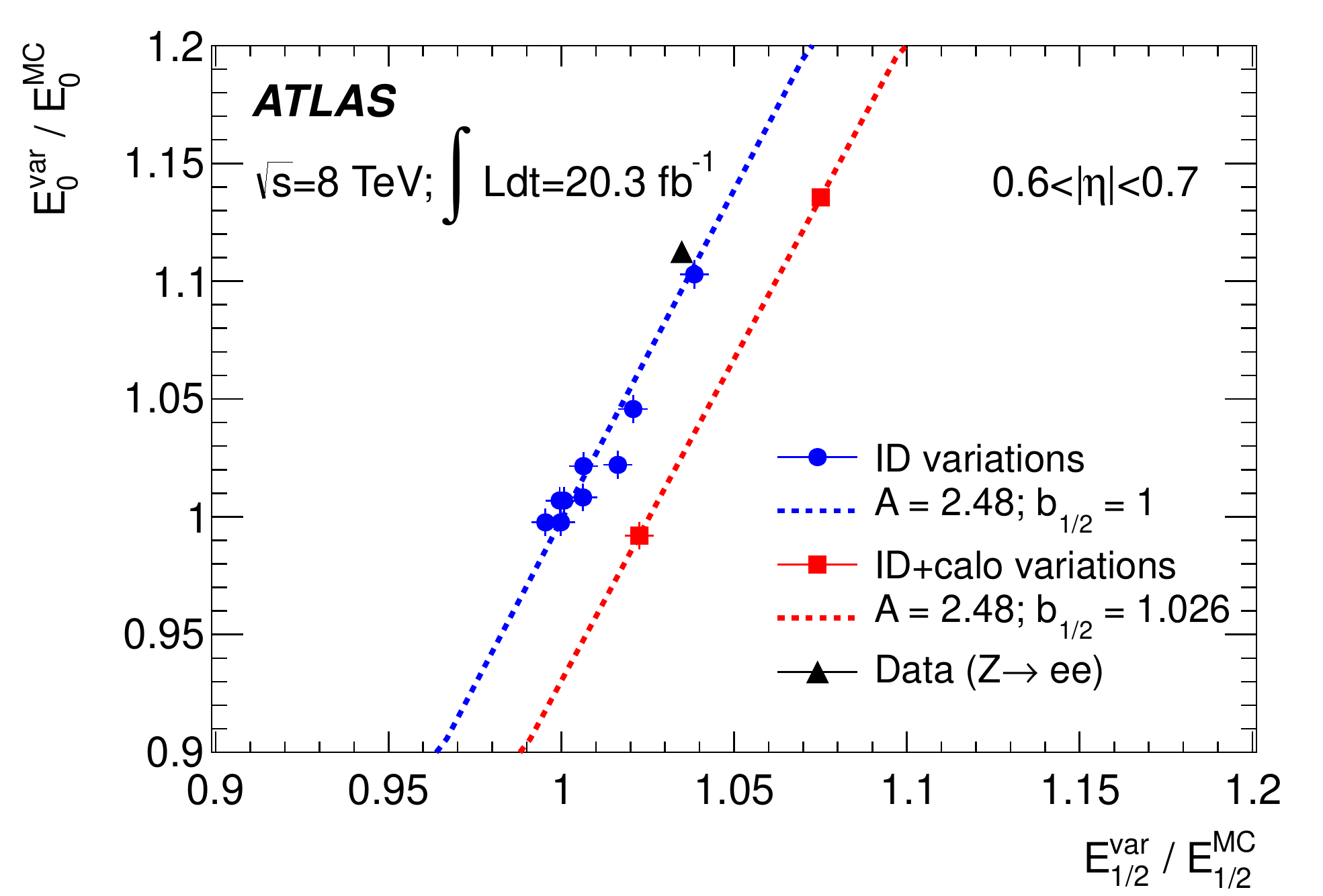}
  \includegraphics[width=\columnwidth]{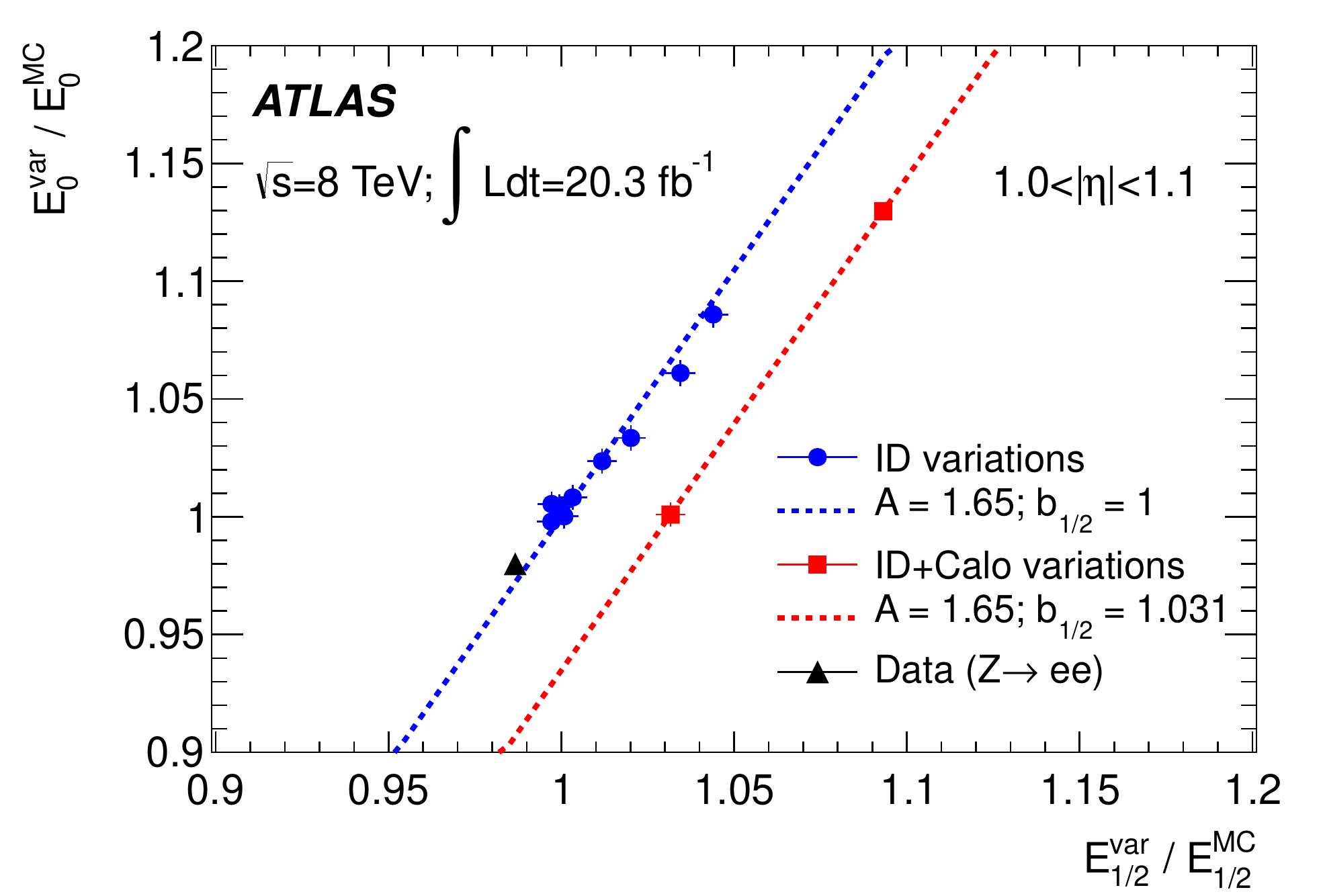}       
  \caption{Examples of correlation between $E_0$ and $E_{1/2}$ ratios under material variations 
    upstream of the calorimeter in the simulated sample, and their corresponding linear parameterisations,
    for $0.6<|\eta|<0.7$ (left) and
    $1.0<|\eta|<1.1$ (right).
    ID material variations refer to
    additions of up to 15\%$X_0$ inside the ID volume
    (circles). Calorimeter variations correspond to 
    5\%$X_0$ added between the PS and L1, separately or in addition
    to ID material variations (squares). The predictions of $E_0$ and $E_{1/2}$
    in the simulation variations, $E_0^{\mathrm{var}}$ and $E_{1/2}^{\mathrm{var}}$,  are normalised to their values
    predicted by the nominal simulation,  $E_0^{\mathrm{MC}}$ and $E_{1/2}^{\mathrm{MC}}$. The triangle shows the
    values obtained from $Z\rightarrow ee$ data, after L1/L2
    calibration correction. The errors bars are statistical only.\label{fig:E0E12Correlation}} 
\end{figure*}

To study this correlation, a set of detector
material variations is implemented in simulation, increasing the passive material in
the various detector sub-systems upstream of the PS (ID, services,
cryostat) and within the calorimeter volume between the PS and L1.  
The results are illustrated in
Fig.~\ref{fig:E0E12Correlation}. Simulations with additional passive
material upstream of the PS result in an earlier shower and
simultaneously increase the PS activity and $E_{1/2}$; a linear
correlation between these observables is observed. Simulations also
including passive material between the PS and L1 exhibit the same
slope of $E_0$ versus $E_{1/2}$, but with an offset along $E_{1/2}$ as material additions after
the PS can not affect the PS activity, but generate earlier showers in
the calorimeter. The following linear parameterisation describes the
impact of upstream passive-material variations on $E_0$ and $E_{1/2}$: 

\begin{equation}
\frac{E^{\rm var}_0(\eta)}{E^{\rm MC}_0(\eta)} =
1 + A(\eta) \,\,\, \left( \frac{E^{\rm var}_{1/2}(\eta)}{E^{\rm MC}_{1/2}(\eta) \,\,
  b_{1/2}(\eta)} - 1 \right),
\label{eq:e0matcorr}
\end{equation} 

where $E^{\rm MC}_0$ and $E^{\rm MC}_{1/2}$ are the predicted values
of $E_0$ and $E_{1/2}$ in the nominal simulation, and $E^{\rm var}_0$
and $E^{\rm var}_{1/2}$ their values in the varied simulations. The simulation samples described above predict
$A=2.48\pm 0.09$ for $|\eta|<0.8$, and $A=1.65\pm 0.05$ for
$|\eta|>0.8$. Assuming correct L1/L2 calibration, $b_{1/2}$
parameterises the remaining potential mis-modelling of
$E_{1/2}$ for effects unrelated to the material upstream of the PS (such as an imperfect
description of the passive material between the PS and L1 and a
possible mis-modelling of the cross-talk between L1 and L2); by definition 
$b_{1/2}\equiv 1$ in the absence of bias.

Correlating the data/MC ratios of $E_0$ and $E_{1/2}$ thus
approximately removes the impact of local material variations on the
former, and provides a corrected prediction for this quantity,

\begin{equation}
\frac{E^\mathrm{corr}_0(\eta)}{E^\mathrm{MC}_0(\eta)} =
1 + A(\eta) \,\,\, \left( \frac{E^\mathrm{data}_{1/2}(\eta)}{E^\mathrm{MC}_{1/2}(\eta)
  \,\, b_{1/2}(\eta)} - 1 \right),
\label{eq:PScorr}
\end{equation}

where $E^\mathrm{corr}_0(\eta)$ corresponds to the amount of expected
PS energy in the simulation, corrected for local material bias
via $E^{\mathrm{data}}_{1/2}(\eta)$ and $b_{1/2}(\eta)$. Finally, the PS energy scale is defined by 

\begin{equation}
\alpha_\mathrm{PS}(\eta) = \frac{E^\mathrm{data}_0(\eta)}{E^\mathrm{corr}_0(\eta)}.
\label{eq:PSscaleEq}
\end{equation}

The offset $b_{1/2}$ is probed using a sample of unconverted
photons selected from radiative $Z$ decays and inclusive photon
production, and defined as $b_{1/2}\equiv
E^{\mathrm{data}}_{1/2}/E^{\rm MC}_{1/2}$ for this sample. In addition to the identification criteria summarised in
Sect.~\ref{sec:egammareco}, the unconverted photon sample should
satisfy $E_0<500$~MeV to limit the probability that a conversion
occurred between the end of the ID and the PS. It is verified using
simulation that this cut indeed minimises the sensitivity of this
sample to material variations upstream of the PS, and that $E_{1/2}$
modelling uncertainties from material after the PS or cross-talk
between L1 and L2 affect electrons and photons in a similar way, so
that this photon sample probes $b_{1/2}$ for electrons with an
inaccuracy of less than 1--2\% depending on pseudorapidity. 

\begin{figure}
  \centering
  \includegraphics[width=\columnwidth]{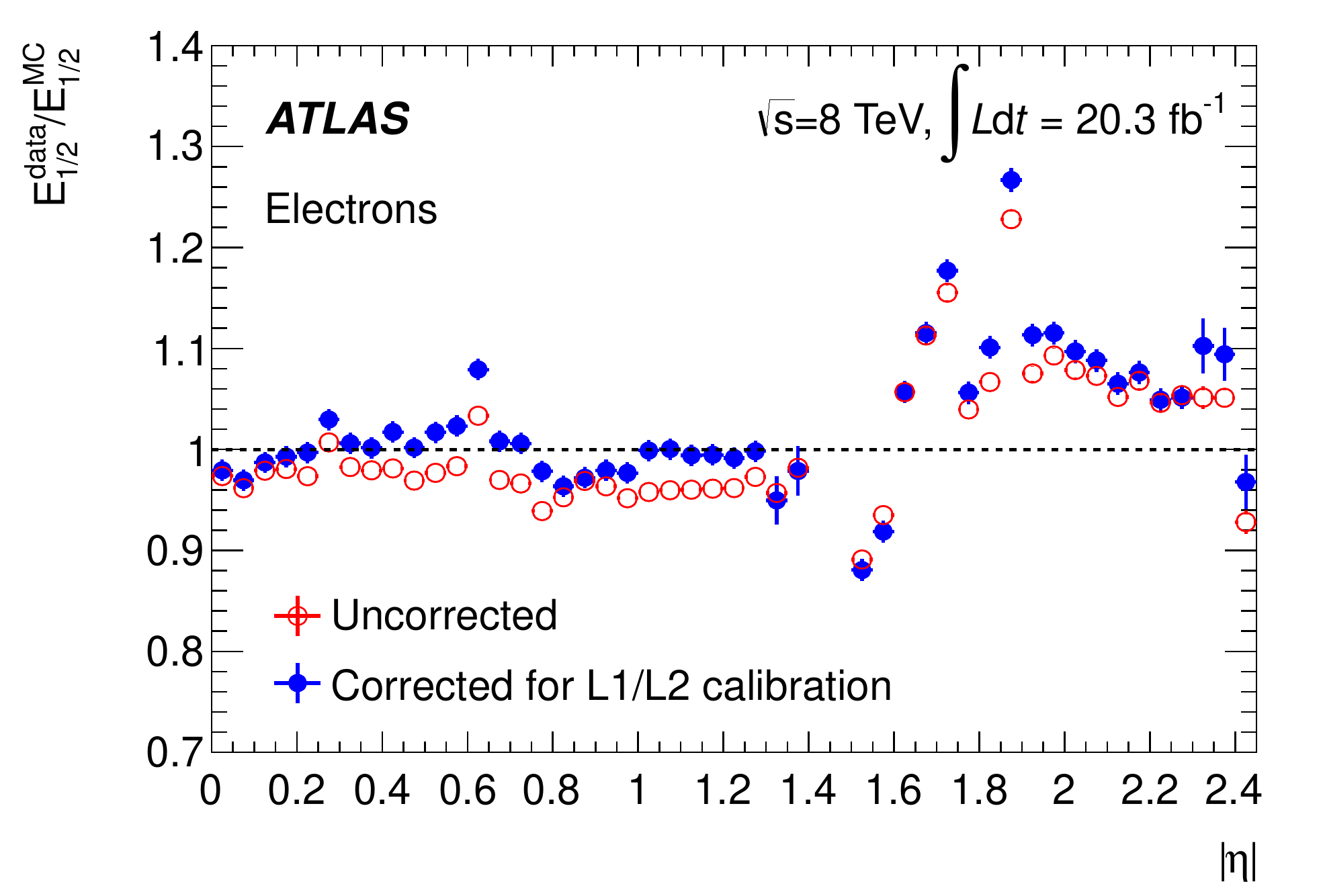} \\
  \includegraphics[width=\columnwidth]{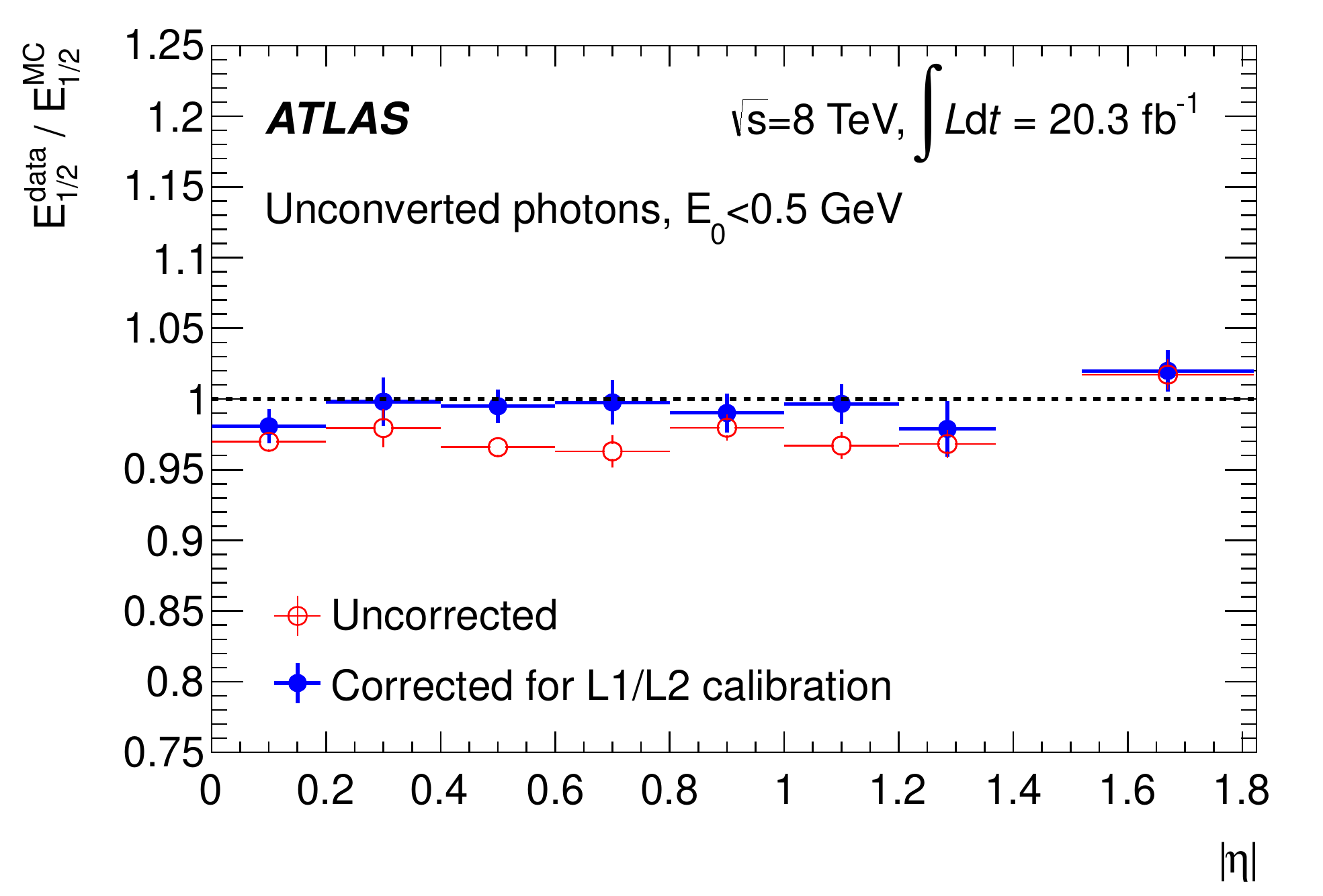}
  \caption{Top: ratio $E_{1/2}^{\mathrm{data}}/E_{1/2}^{\mathrm{MC}}$, for
    electrons from $W$ and $Z$ decays. Bottom: $b_{1/2}$, defined as
    $E^{\mathrm{data}}_{1/2}/E^{\rm MC}_{1/2}$ for unconverted photons
    with $E_0<500$~MeV. Both observables are shown as a function 
    of $|\eta|$, before and after the L1/L2 calibration
    corrections. The errors bars on the uncorrected points are
    statistical only; after corrections, the error bars also include
    systematic uncertainties related to the L1/L2 calibration.} 
  \label{fig:b12summary}
\end{figure}

Fig.~\ref{fig:b12summary} shows the comparison of $E_{1/2}$ between data
and simulation for electrons and for the unconverted photon sample, before and after the L1/L2
calibration correction described in Sect.~\ref{sec:l1l2}. Before this calibration correction, the ratio of data to MC simulation for electrons
and photons is on average below one by 3\% in the barrel. After calibration
corrections, $b_{1/2}$ is everywhere close to one, which suggests that
there is no significant material mis-modelling downstream of the PS. In
contrast, the electron data in the endcap show a residual positive
bias of about 7\% on average, indicating a discrepancy in the description of the material. An explicit passive-material measurement using these data
is performed in Sect.~\ref{sec:material}.  

\begin{figure}
  \centering
  \includegraphics[width=\columnwidth]{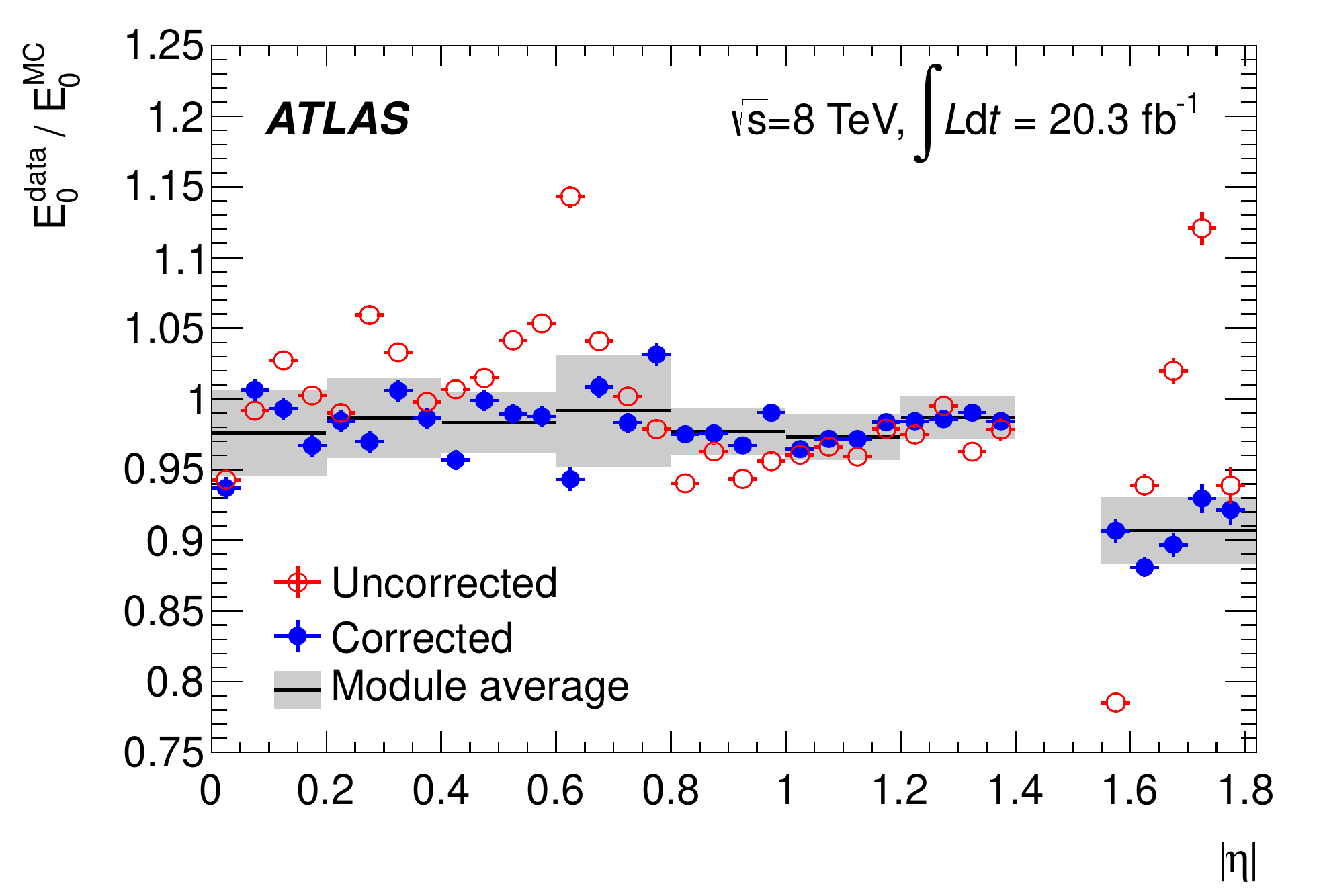}
  \caption{Ratio of the average PS energies, $E_0^{\mathrm{data}}/E_0^{\mathrm{MC}}$, for electrons in data and
    simulation as a function of $|\eta|$, before and after corrections
    for $b_{1/2}$ and material upstream of the PS. The full lines with
    shaded bands represent the PS energy scale as a function of $|\eta|$, $\alpha_{\mathrm{PS}}(\eta)$, and its uncertainty.\label{fig:alphaps2012}} 
\end{figure}

Figure~\ref{fig:alphaps2012} summarises the PS scale calculated according to
Eqs.~\eqref{eq:PScorr} and \eqref{eq:PSscaleEq} and
Fig.~\ref{fig:b12summary}, from which the corrected values are used
as input to the calculation. The material 
corrections based on Eq.~\eqref{eq:PScorr} visibly reduce the
variations of $E_0^{\rm data}/E_0^{\rm corr}$ versus $\eta$
compared to $E_0^{\mathrm{data}}/E_0^{\rm MC}$, 
especially in the regions $0.6<|\eta|<0.8$ and
$1.52<|\eta|<1.82$. After this correction, the PS energy scale $\alpha_\mathrm{PS}$ is defined by
averaging $E_0^{\mathrm{data}}/E_0^{\rm corr}$ over intervals corresponding to the PS module
size ($\Delta\eta=0.2$ in the barrel, $\Delta\eta=0.3$ in the
endcap). 
As it is located in the transition region, the correction to the PS energy scale for the module covering $1.4<|\eta|<1.55$ is not addressed by this analysis. For particles entering this region,
$\alpha_\mathrm{PS}$ and its uncertainty are taken from the closest
range among $1.2<|\eta|<1.4$ and $1.52<|\eta|<1.82$.

The measured PS energy scale $\alpha_\mathrm{PS}$ defines a correction factor
that is applied to the data. Uncertainties affecting its determination arise from the
statistical and systematic uncertainties affecting $b_{1/2}$ and $A$,
and from the residual variations of $E_0^{\mathrm{data}}/E_0^{\rm corr}$ within a PS
module, which indicates that the material correction via
Eq.~\eqref{eq:PScorr} is only approximate. The statistical uncertainty on
$E_0^{\mathrm{data}}/E_0^{\mathrm{MC}}$ and
$E_{1/2}^{\mathrm{data}}/E_{1/2}^{\mathrm{MC}}$ from the electron
samples is negligible. The PS scale measurement is accurate to 2--3\%, depending on pseudorapidity. 

\subsection{Layer intercalibration cross-check}

\begin{figure}
  \centering
  \includegraphics[width=\columnwidth]{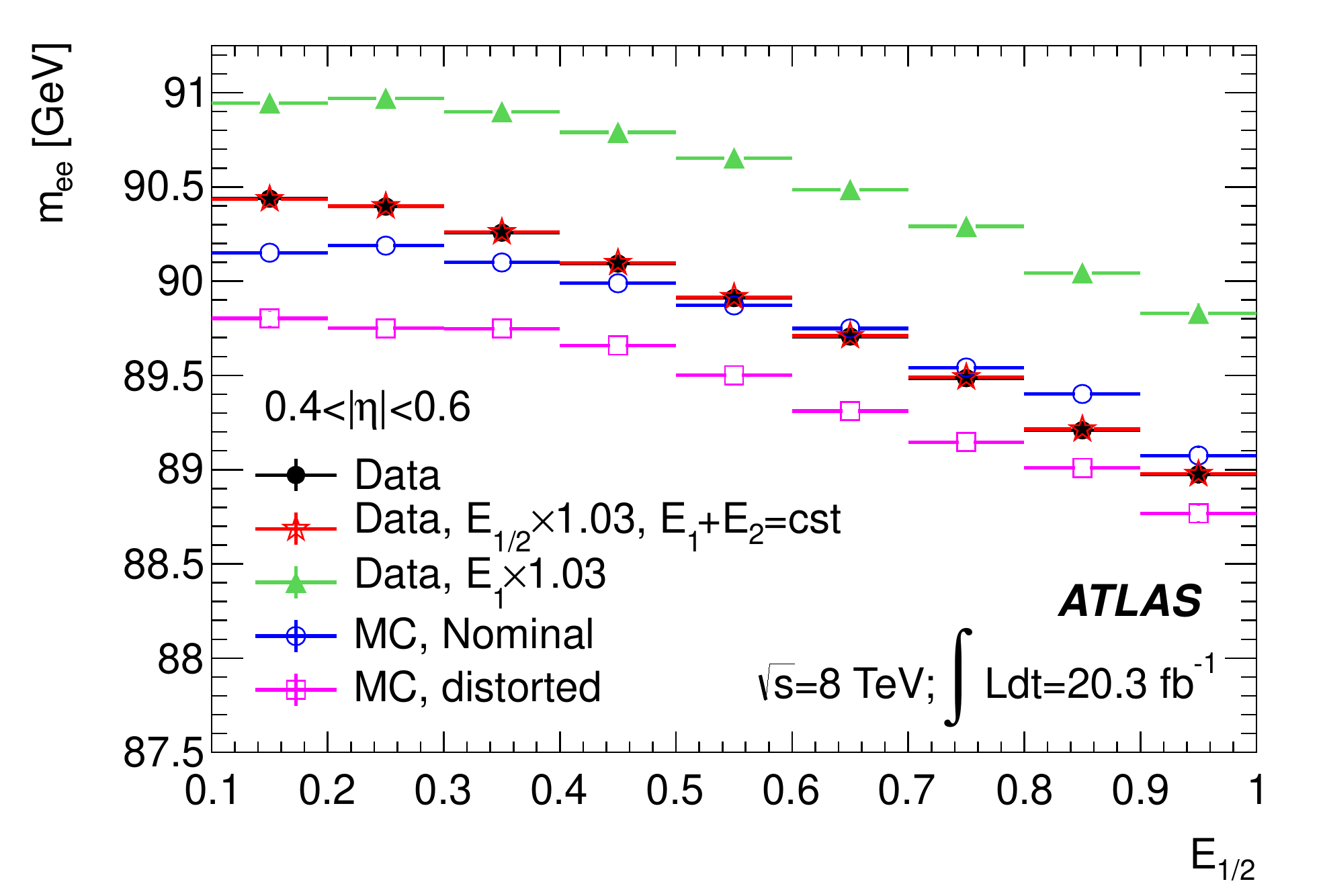} \\
  \includegraphics[width=\columnwidth]{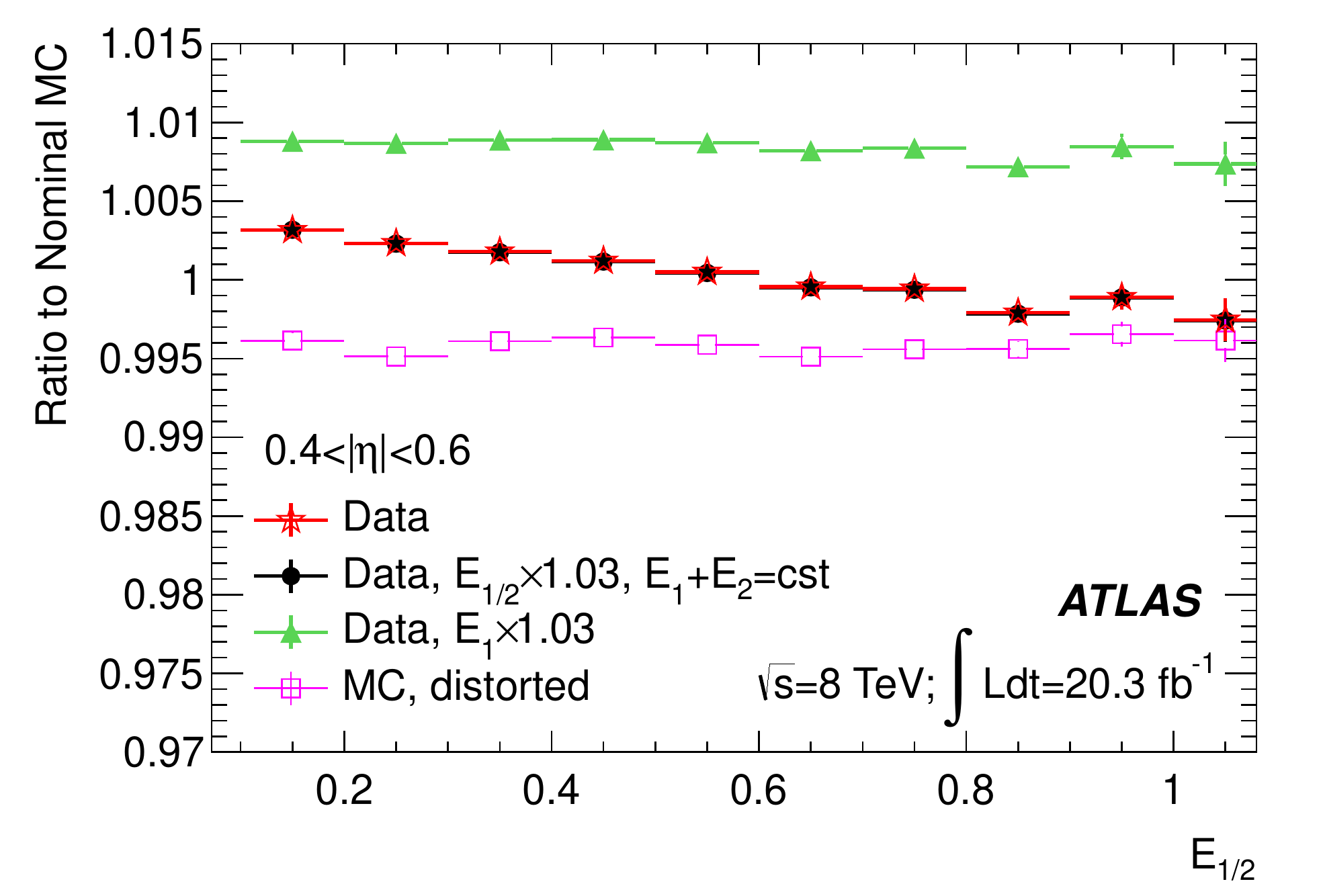}
    \caption{Top: $m_{ee}$ as a function of
      $E_{1/2}$ for the data (points), the nominal simulation (open
      circles), the simulation with additional material (open squares), the data with a 3\% scaling 
      of $E_1$ (triangles), and the data with a 3\% scaling of $E_1$ with
      $E_1+E_2$ kept constant (open stars). Bottom: ratios of the
      curves shown in the top plot to the nominal simulation. 
      The plots show values for electrons in the pseudorapidity bin $0.4<|\eta|<0.6$. The error
      bars include statistical uncertainties only.}  
    \label{fig:Zpeak_vs_E12}
\end{figure}

The dependence of the electron energy response on shower depth allows
a direct extraction of $\alpha_{1/2}$ for EM showers, providing a test
of the baseline approach described in
Sect.~\ref{sec:l1l2}. Figure~\ref{fig:Zpeak_vs_E12} shows the
correlation between the invariant mass of electron pairs from $Z\rightarrow
ee$ decays and $E_{1/2}$ for data and simulation, in the
representative bin $0.4<|\eta|<0.6$. The PS scale
corrections determined in Sect.~\ref{sec:psscale} are applied. 

The ratio between data and the nominal simulation is not constant versus $E_{1/2}$. A constant 
data-to-simulation ratio is recovered by rescaling the L1 response in 
data and recomputing the invariant mass accordingly, adjusting
$\alpha_{1/2}$ to maximise the compatibility of the ratio with a
horizontal line. This procedure is applied to derive $\alpha_{1/2}$ as a
function of $|\eta|$, and the optimum is determined by $\chi^2$
minimisation. 

\begin{figure}
  \centering
  \includegraphics[width=\columnwidth]{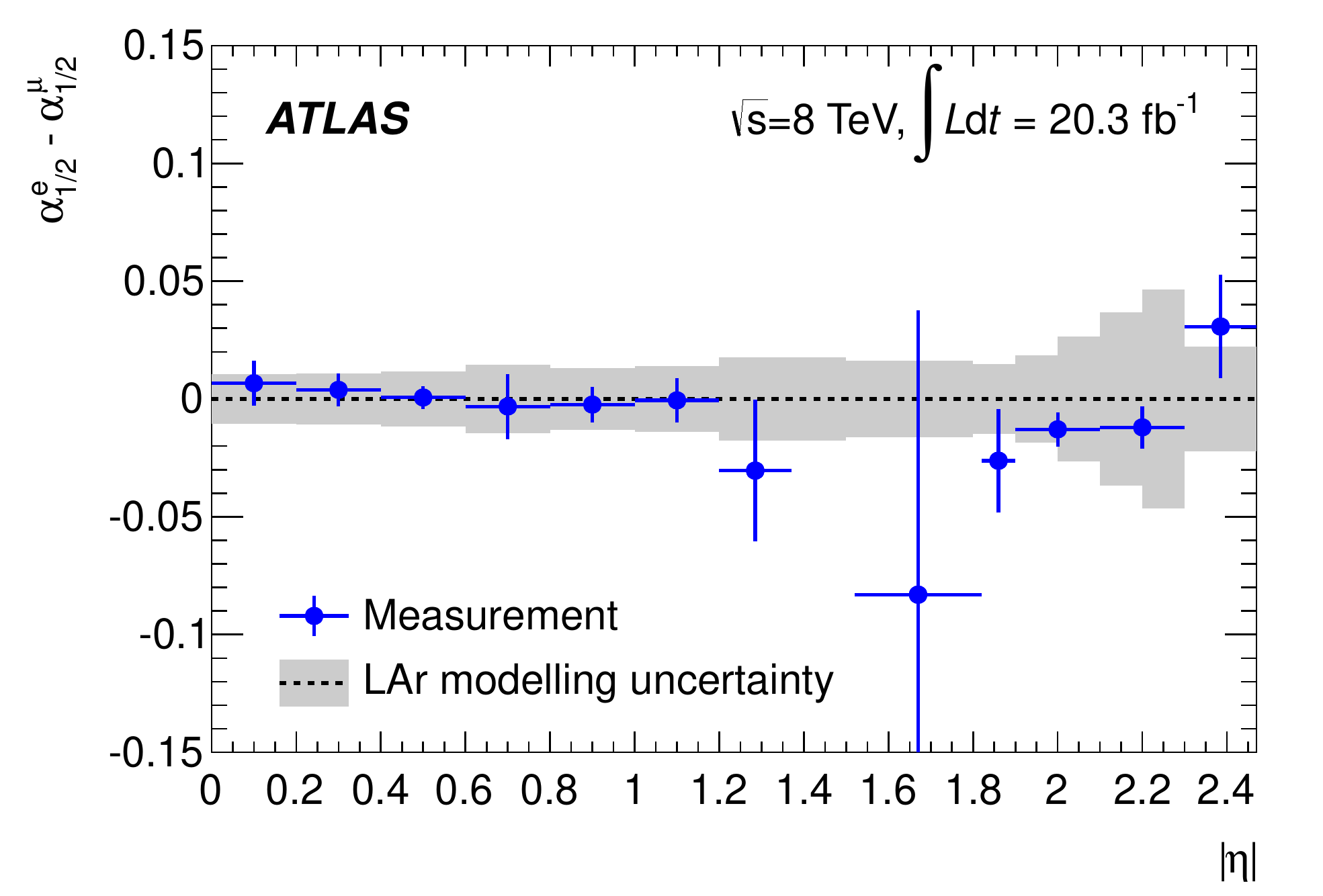}
  \caption{Difference between the electron- and muon-based L1/L2 calibration
    results, denoted $\alpha_{1/2}^e$ and $\alpha_{1/2}^\mu$ respectively, as a function of $|\eta|$. The uncertainty band reflects the systematic
    uncertainties affecting the muon result; the error bars represent
    the uncertainty on the electron result. \label{fig:E12difference}} 
\end{figure}

The difference between the values of  $\alpha_{1/2}$ obtained with this procedure and with the muon-based L1/L2
calibration are shown in Fig.~\ref{fig:E12difference} as a function of $|\eta|$. Good
compatibility in the full pseudorapidity range is observed, confirming
the validity of the muon-based calibration. For $1.2<|\eta|<1.37$ and 
$1.52<|\eta|<1.82$, the $E_{1/2}$ distributions for electrons in data and simulation differ significantly regardless of $\alpha_{1/2}$, leading to poor
convergence of the minimisation procedure and enhanced uncertainties
in these bins. 

The uncertainties on the electron measurement include systematic
contributions from detector material mis-modelling and from
uncertainties on the cross-talk between L1 and L2. To test the
influence of passive material,  a $Z\rightarrow ee$ sample with
20$\--$35\%$X_0$ additional material, depending on $|\eta|$, is simulated
and treated as the data. The $\alpha_{1/2}$ values extracted from this
sample represent a conservative passive-material contribution to the
uncertainty on $\alpha_{1/2}$, and contribute about 0.5\% on average,
except for $1.37<|\eta|<1.82$ where the uncertainty is 1$\--$2\%. The
influence of cross-talk is probed by rescaling the L1 response in data,
requiring in addition that the sum of the L1 and L2 energies be
constant. Such variations have no impact on the data/MC ratio and the
contribution of this effect is negligible. These systematic variations
are also illustrated in Fig.~\ref{fig:Zpeak_vs_E12}, for
$0.4<|\eta|<0.6$.

\section{Passive-material determination \label{sec:material}}

After L1/L2 calibration corrections, the $E_{1/2}$
distribution observed for EM showers in the data can be used to quantify the
amount of detector material upstream of the active calorimeter. Higher
values of $E_{1/2}$ in data would indicate earlier shower development,
and hence a local excess of material in comparison with the simulation. Although
$E_{1/2}(\eta)$ is intrinsically a measure of the material integral in
front of the calorimeter at a given pseudorapidity, the study is performed for
different categories of EM showers (electrons, and unconverted photons
without PS activity), providing partial information on the distance
from the beam axis at which the material is probed. 

The detector material categories can be grouped under ID material; cryostat material
(``Cryo''), designating material located  between the maximal
conversion radius and the PS; and calorimeter material (``Calo''), for
passive material located between the PS and L1.

\subsection{Methodology~\label{sec:mat:method}}

\begin{figure*}
	\centering
        \includegraphics[width=0.7\textwidth]{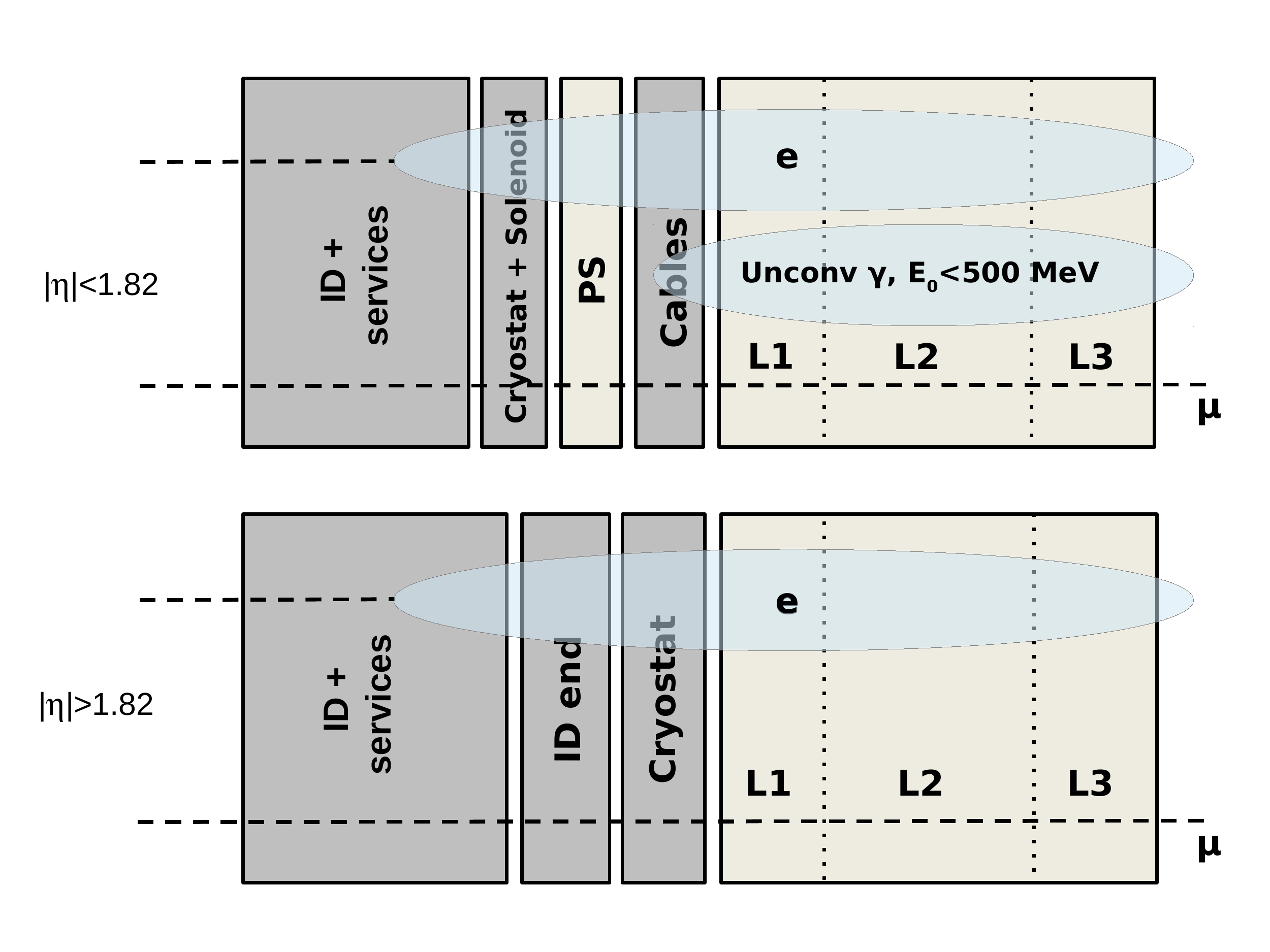}
	\caption{Sketch of EM shower development for the different particle categories described in the text, for $|\eta|<1.82$ (top) 
          and $|\eta|>1.82$ (bottom).
          The interaction point is located to the left of the figure.
          \label{fig:E12drawing}}
\end{figure*}

Electrons are sensitive to all detector material crossed along their
trajectory, from the interaction point up to L1; unconverted
photons are insensitive to the ID material upstream of the conversion
radius. Within the PS acceptance ($|\eta|<1.82$), a veto on the PS
activity can be required to minimise the probability that a conversion
happened in front of the PS, making such phot\-ons specifically
sensitive to passive material between the PS and L1. The shower
development for these different types of particles 
is sketched in Fig.~\ref{fig:E12drawing}. 

The sensitivity of
$E_{1/2}$ for these probes of detector material is evaluated using simulated
samples including the following variations:
\begin{itemize}
\item[$\bullet$] +5\% relative scaling of the ID inactive material;
\item[$\bullet$] +10\%$X_0 \times \cosh\eta$ in front of the barrel calorimeter;
\item[$\bullet$] +5\%$X_0 \times \cosh\eta$ between the barrel PS and L1;
\item[$\bullet$] +15\%$X_0 \times \tanh\eta$ at the end of  the active SCT and TRT endcap detectors;
\item[$\bullet$] +15\%$X_0 \times \tanh\eta$ at the end of the ID volume, in  front of the EMEC cryostat;
\item[$\bullet$] +30\%$X_0 \times \tanh\eta$  in front of the endcap calorimeter, for $1.5<|\eta|<1.82$;
\item[$\bullet$] +5\%$X_0 \times \tanh\eta$ between the endcap PS and L1.
\end{itemize}
The material additions in the barrel are placed at constant radius,
and their thickness is constant as a function of $z$, hence the material
seen by particles coming from the interaction point increases with
pseudorapidity as indicated above. Similarly, the material additions in the endcap are
placed at constant $z$ and have constant thickness as a function of radius.

\begin{figure}
  \centering
  \includegraphics[width=\columnwidth]{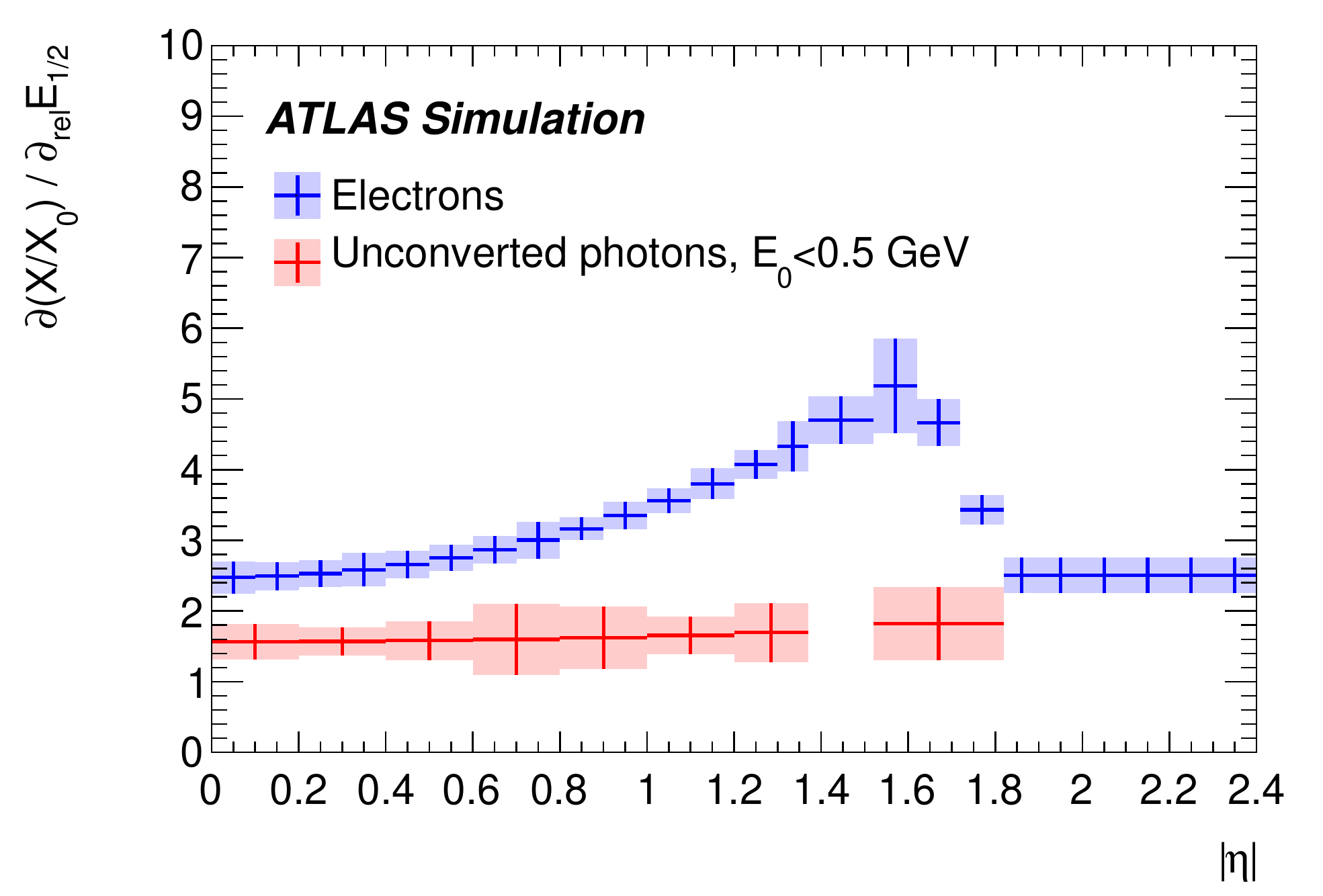}
  \caption{Sensitivity factor $\frac{\partial X/X_0}{\partial_{\rm rel} E_{1/2}}$ 
    as a function of $|\eta|$, for material variations upstream of
    the PS for electrons, and for variations between the PS and L1
    for unconverted photons with $E_0<500$~MeV. The shaded bands
    represent the systematic uncertainty due to the dependence of
    this quantity on the location of the material additions. \label{fig:x0sensElectrons}}  
\end{figure}

For each category and in a given $|\eta|$ region, the amo\-unt of
additional material $X$, expressed in terms of $X_0$, is normalised to the relative shift induced in
$E_{1/2}$ for electrons or photons respectively, obtaining a
sensitivity factor $\frac{\partial   X/X_0}{\partial_{\rm rel}
  E_{1/2}}$. Figure~\ref{fig:x0sensElectrons} shows the sensitivity
curve obtained from the various material distortions upstream of the
PS, for electrons. The behaviour is approxima\-tely universal, and
parameterised as a single curve. At small $\eta$, a 1\%
relative change in $E_{1/2}$ corresponds to about 2.5\%$X_0$. The
sensitivity of unconverted photons with $E_0<500$~MeV to material
between the PS and L1 is also shown; a 1\% relative change in
$E_{1/2}$ corresponds to about 1.5\%$X_0$, independently of $\eta$.

This factor is scaled by the observed relative difference
$\Delta E_{1/2}^\mathrm{data}$ of $E_{1/2}$ between data and
simulation after calibration corrections (see
Fig.~\ref{fig:b12summary}), yielding an estimate of the passive-material offset with  respect to the nominal simulation: 
\begin{equation}
\Delta X / X_0 = \Delta E_{1/2}^\mathrm{data} \left(\frac{\partial
  X/X_0}{\partial_{\rm rel} E_{1/2}}\right).
\end{equation}
The uncertainty on the material measurement receives contributions
from $\Delta E_{1/2}^\mathrm{data}$, reflecting the residual L1/\allowbreak L2 calibration
uncertainty discussed in the previous section, and from
$\frac{\partial X/X_0}{\partial_{\rm rel} E_{1/2}}$. The intrinsic EM
shower development modelling accuracy contributes to the latter; this
item is evaluated by simulating high-$\et$ electron samples and varying
the associated {\sc Geant4} options to test refinements in the
theoretical description of bremsstrahlung and photon conversion cross
sections, as well as alternative electron multiple scattering models,
and found to be $\sim$1\%. The residual sensitivity differences between
the various material configurations contributes a systematic
uncertainty of $\sim$10\% to the parameterisation.

Two categories of detector material are probed for $|\eta|<1.82$: the
integral between the interaction point and the PS, i.e. the sum of
ID and cryostat material; and calorimeter material between the PS and
L1. The former is obtained by comparing $E_{1/2}$ in the electron and
unconverted photon data samples in order to subtract, from the
electron probe, the influence of material after the PS. The latter is
obtained by comparing $E_{1/2}$ for unconverted photons between data and
simulation. For $|\eta|>1.82$, only the total amount of material up to L1 is measured, by comparing $E_{1/2}$ for electrons in data and simulation. 

The ID material is considered known $a$~$priori$, with an accuracy of 5\% from detailed monitoring and weighing during the construction
and installation ~\cite{ATLAS_detector}. Studies using $K_S^0$ decays, secondary hadronic
interactions and photon conversions were also
performed~\cite{material3,material2}, with no indication of ID
material mis-modelling larger than 5\%. The ID material accuracy is
combined with the measured material integral to derive an estimate for
the cryostat material. The calorimeter material is measured without
external inputs.

\begin{figure*}
  \centering
  \includegraphics[width=0.49\textwidth]{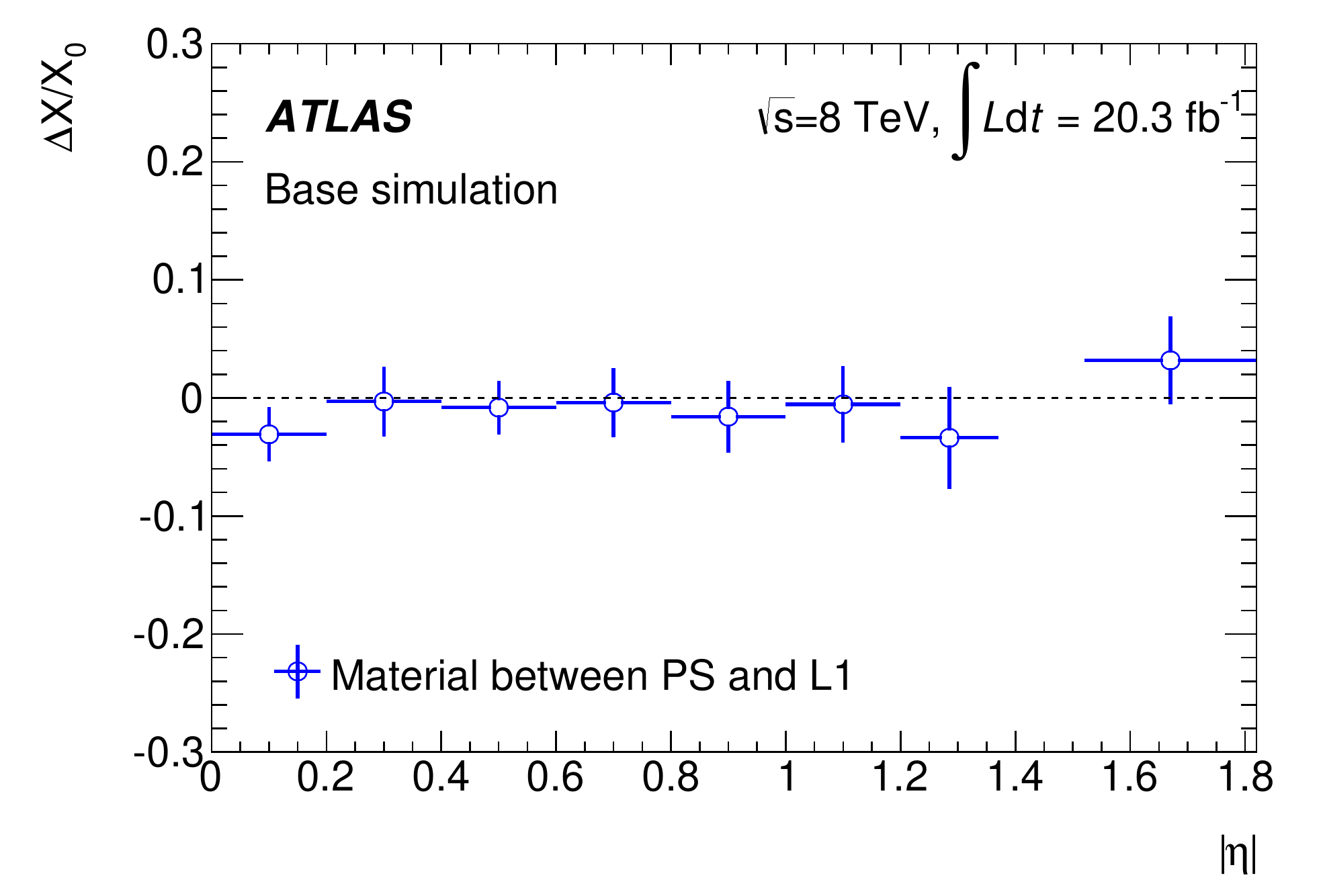}
  \includegraphics[width=0.49\textwidth]{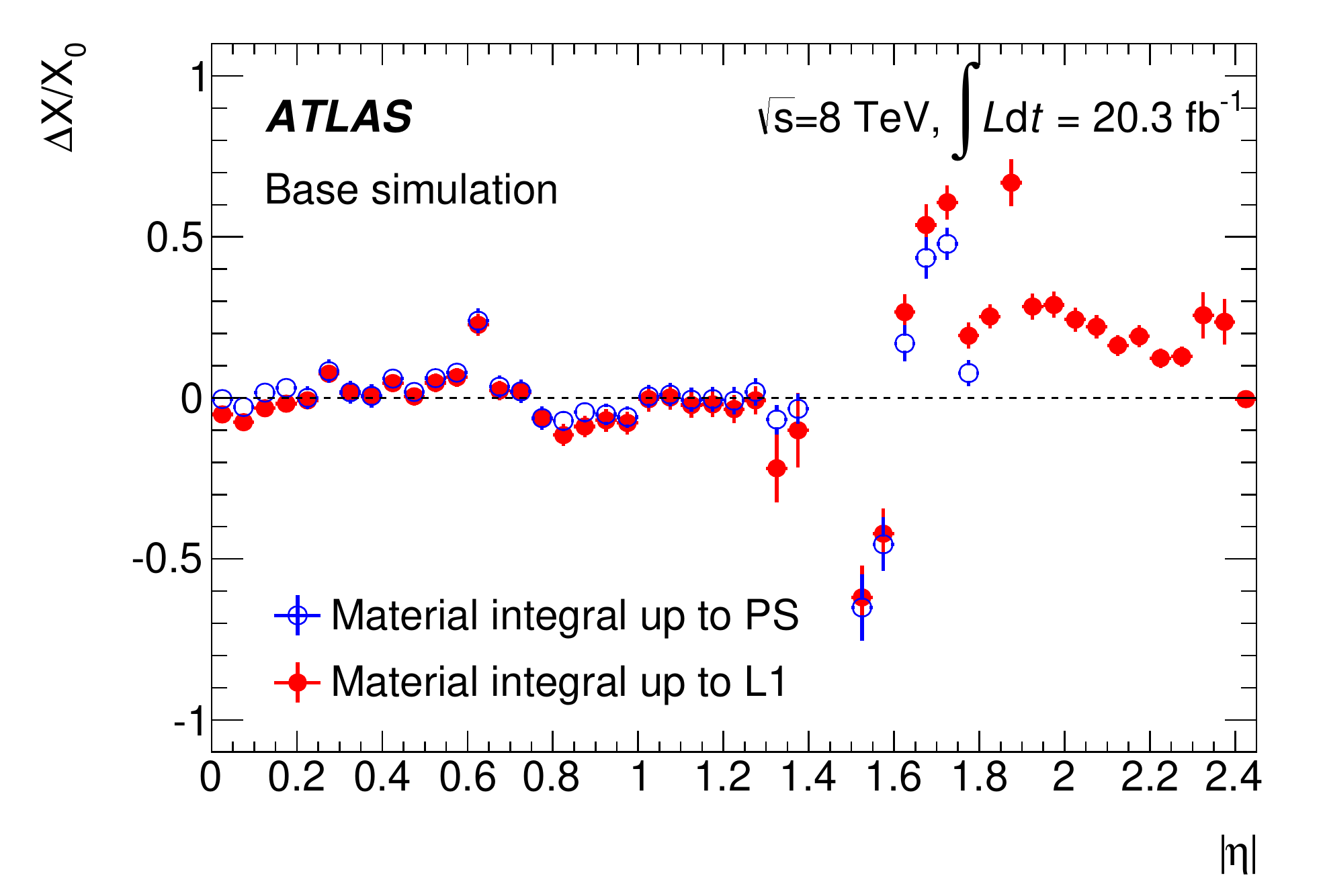}
  \caption{Difference between the material estimate, $\Delta X/X_0$, from data and the
    nominal base simulation as a function of $|\eta|$. Left: calorimeter material estimate
    obtained from data/MC comparisons of $E_{1/2}$ for unconverted
    photons, after calibration corrections. Right:
    integrated estimate up to L1, obtained from data/MC
    comparisons for electrons, after calibration corrections;
    integrated estimate up to the PS, obtained by comparing electron
    and unconverted photon data. The error bars include statistical
    and systematic uncertainties. \label{fig:matoldgeo}}
\end{figure*}

\begin{figure*}
  \centering
  \includegraphics[width=0.49\textwidth]{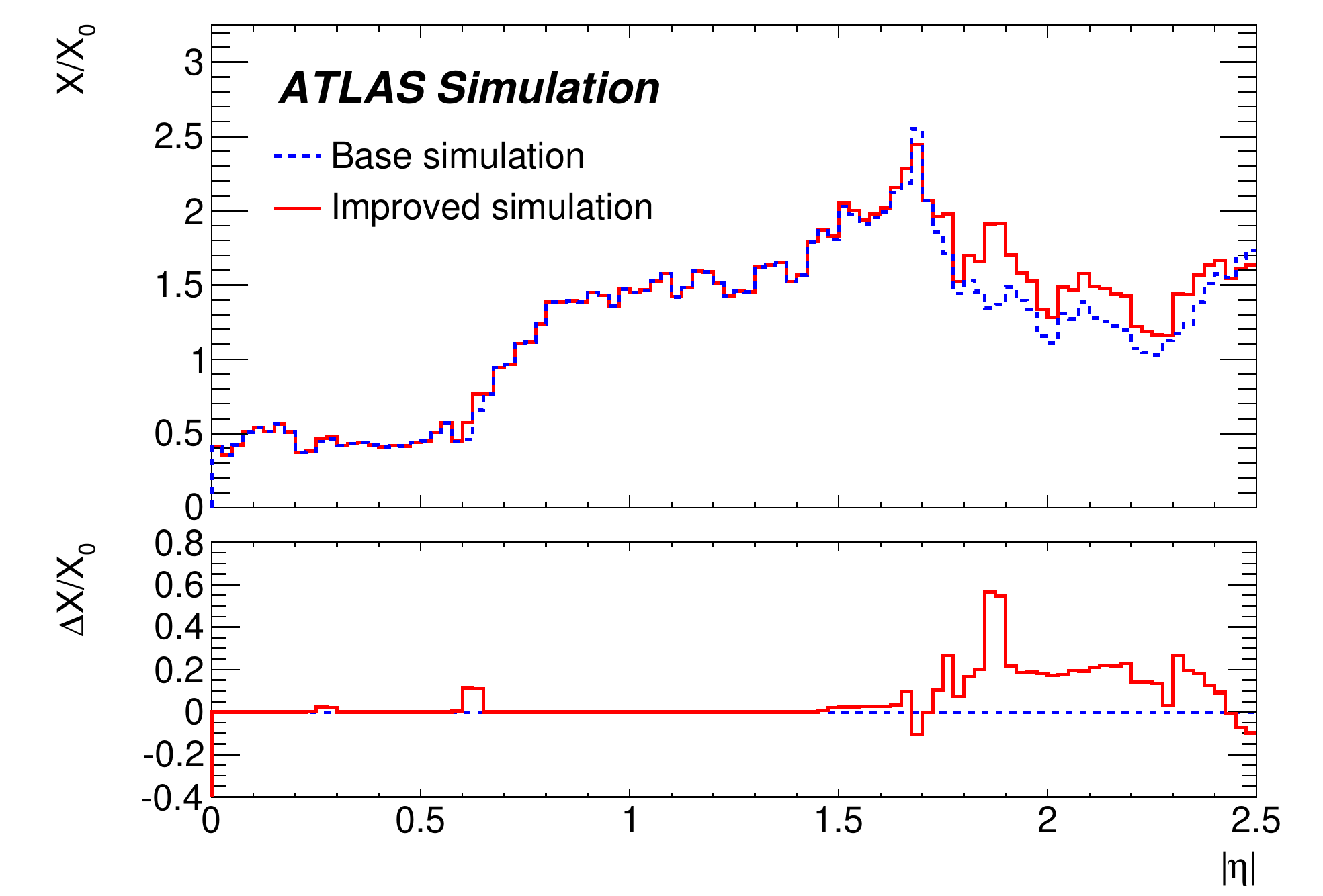}
  \includegraphics[width=0.49\textwidth]{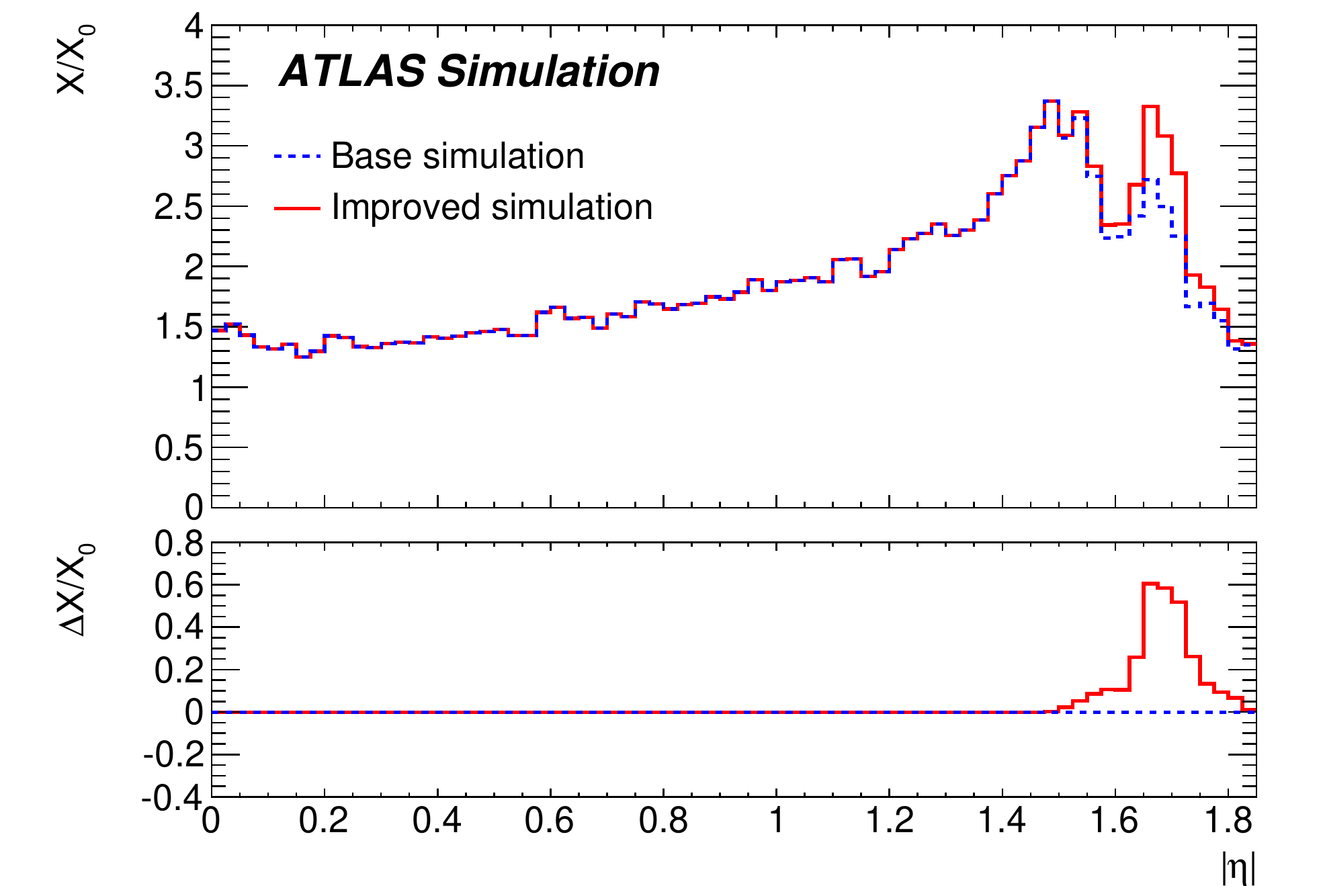}
  \caption{Amount of material traversed by a particle, $X/X_0$, as a function of $|\eta|$, for the base
    simulation and including the corrections based on calorimeter
    measurements, up to the ID boundaries (left), and between
    the ID boundaries and the PS (right). The lower panels indicate the
    difference between the improved and the base
    simulations.\label{fig:newgeoDiff}}   
\end{figure*}

\begin{figure*}
  \centering
  \includegraphics[width=0.49\textwidth]{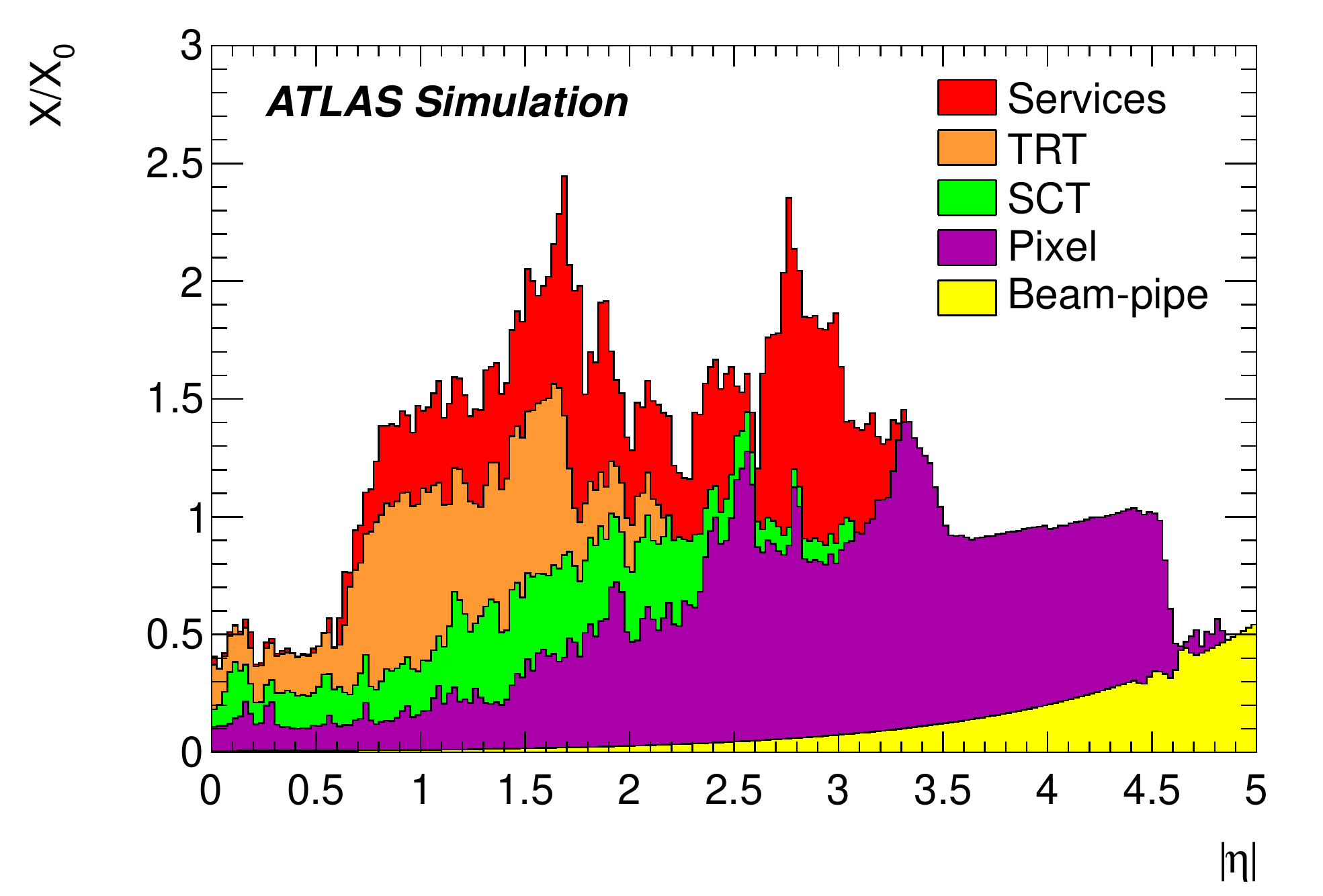}
  \includegraphics[width=0.49\textwidth]{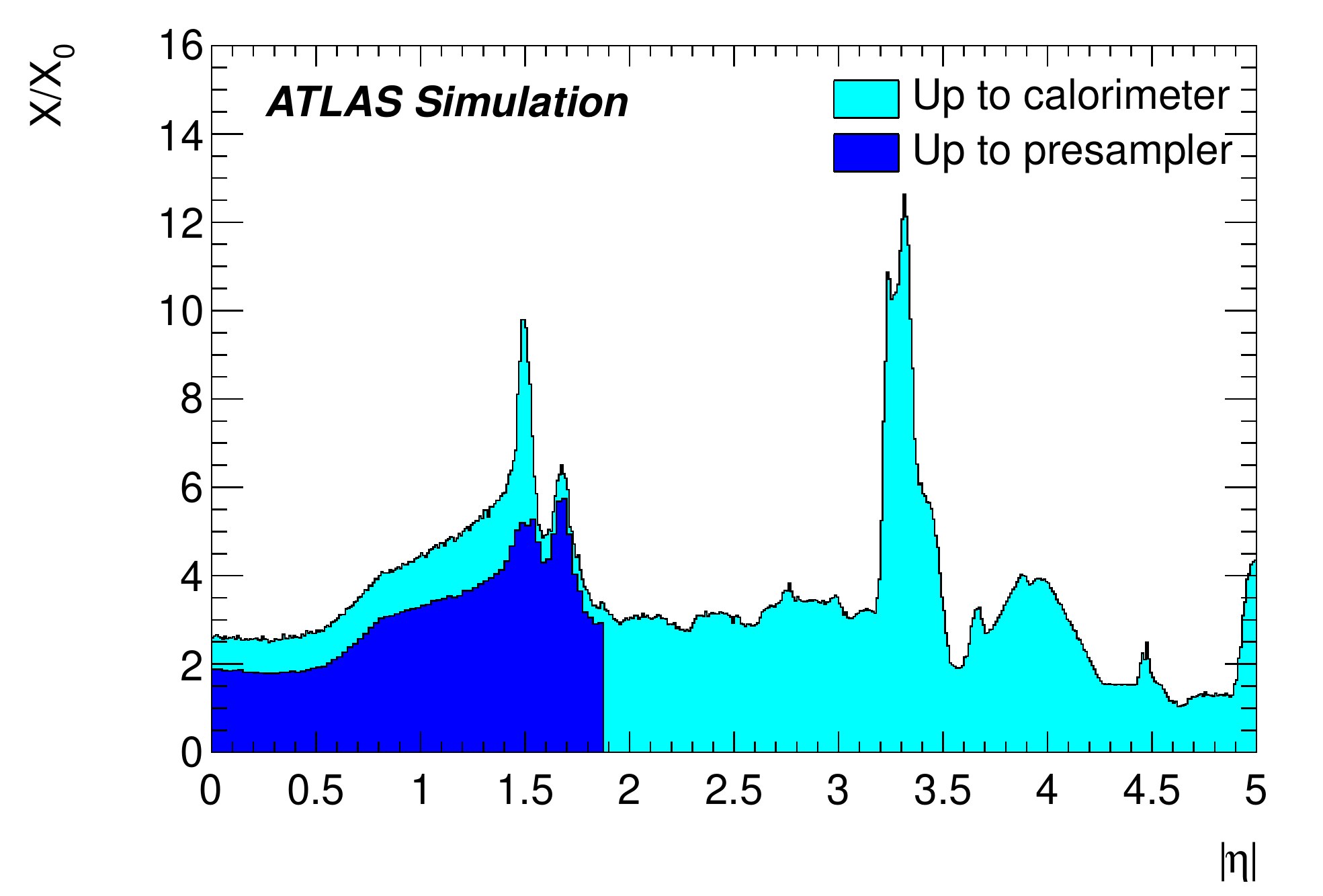}
  \caption{Amount of material traversed by a particle, $X/X_0$, as a function of
    $|\eta|$, in the improved simulation, up to the ID boundaries
    (left), and up to the PS and the EM calorimeter (right). The
    contributions of the different detector elements, including the
    services and thermal enclosures are shown separately by filled
    colour areas. \label{fig:newgeoTotal}} 
\end{figure*}

\subsection{Material determination}

The difference between the calorimeter material estimate from data and simulation, as obtained from the comparison of $E_{1/2}$ for unconverted photons after calibration corrections, is summarised in
Fig.~\ref{fig:matoldgeo} (left). The statistical uncertainty in the measurement is dominated by the
size of the unconverted photon control samples. Systematic
uncertainties come from the accuracy of the material sensitivity
calibration, the finite size of the MC sample, and from the residual
sensitivity of unconverted photons to passive-material variations
upstream of the PS. After calibration corrections, no
significant bias remains in $E_{1/2}$ (see
Fig.~\ref{fig:b12summary}, bottom), translating into material discrepancies of at
most 0.03$X_0$. The measurement accuracy is about
0.03$X_0$. 

In Fig.~\ref{fig:matoldgeo} (right), the data--MC material difference integrated up to L1, denoted $\Delta X_{\rm L1}$ and
expressed in units of $X_0$, results from the combination of the observed
$E_{1/2}$ profile for electrons after calibration correct\-ions (see
Fig.~\ref{fig:b12summary}, top) and the
corresponding sensitivity curve (see Fig.~\ref{fig:x0sensElectrons}). In the barrel, moderate features are
observed, primarily a 0.2$X_0$ excess at $|\eta|=0.6$, and a slight 
--0.1$X_0$ deficit between $0.8<|\eta|<1$. In the endcap, the measurement
is characterised by very strong excesses, up to 0.6$\--$0.7$X_0$, in the
region $1.65<|\eta|<1.75$, and around $|\eta|=1.9$ because of an incomplete description of SCT cooling pipes. In the remaining
part of the endcap, an overall bias of about 0.2$X_0$ is observed. In
contrast, a deficit of about --0.5$X_0$ is observed within
$1.55<|\eta|<1.6$. The material bias integrated up to the PS, $\Delta X_{\rm PS}$, is obtained
after subtracting, from the above, material contributions located
after the PS, i.e. $\Delta X_{\rm Calo}$. This is derived by
comparing the electron and unconverted photon data. The features
observed within the PS acceptance are very similar to $\Delta X_{\rm L1}$,
which indicates that the material biases are located upstream of the
PS. For both integrated estimates, the measurement accuracy ranges
from about 0.04$X_0$ to 0.06$X_0$.

\subsection{Improvements to the ATLAS material simulation}

This section presents the detector simulation
improvements implemented following the results obtained in the
previous section. Given the absence of significant biases 
in $\Delta X_{\rm Calo}$, the data suggest the need to implement in the simulation material modifications upstream of the PS. Most of the discrepancies correspond
to areas with a large amount of material from services between the ID active area and the calorimeter cryostat. The corrections were implemented in an effective way,
adding material in the most discrepant areas and in amounts corresponding to the measurement. The
modifications to the detector material description are illustrated in
Fig.~\ref{fig:newgeoDiff}. The total amount of detector material
within the ID boundaries, and up to the active calorimeter are
illustrated in Fig.~\ref{fig:newgeoTotal} for the improved
simulation. 

\begin{figure}
  \centering
  \includegraphics[width=\columnwidth]{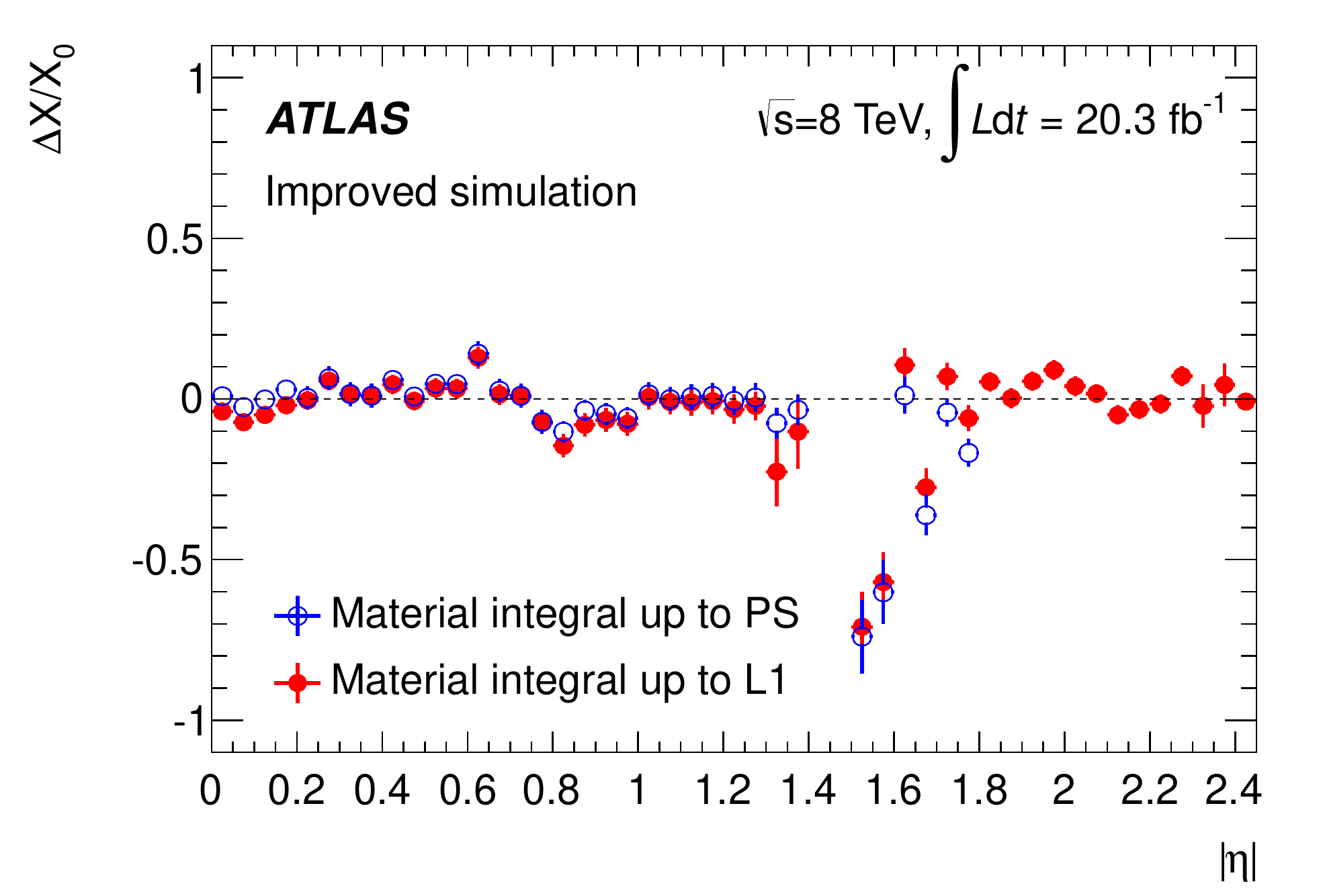}
  \caption{Difference between the material estimate, $\Delta X/X_0$, from data and the
    improved simulation as a function of $|\eta|$. The integrated material estimate up to L1 is obtained from data/MC comparisons for
    electrons, after L1/L2 calibration corrections; the
    integrated estimate up to the PS is obtained by comparing electron
    and unconverted photon data. The error bars include statistical
    and systematic uncertainties.\label{fig:e12newgeo}}
\end{figure}

\begin{figure}
  \centering
  \includegraphics[width=\columnwidth]{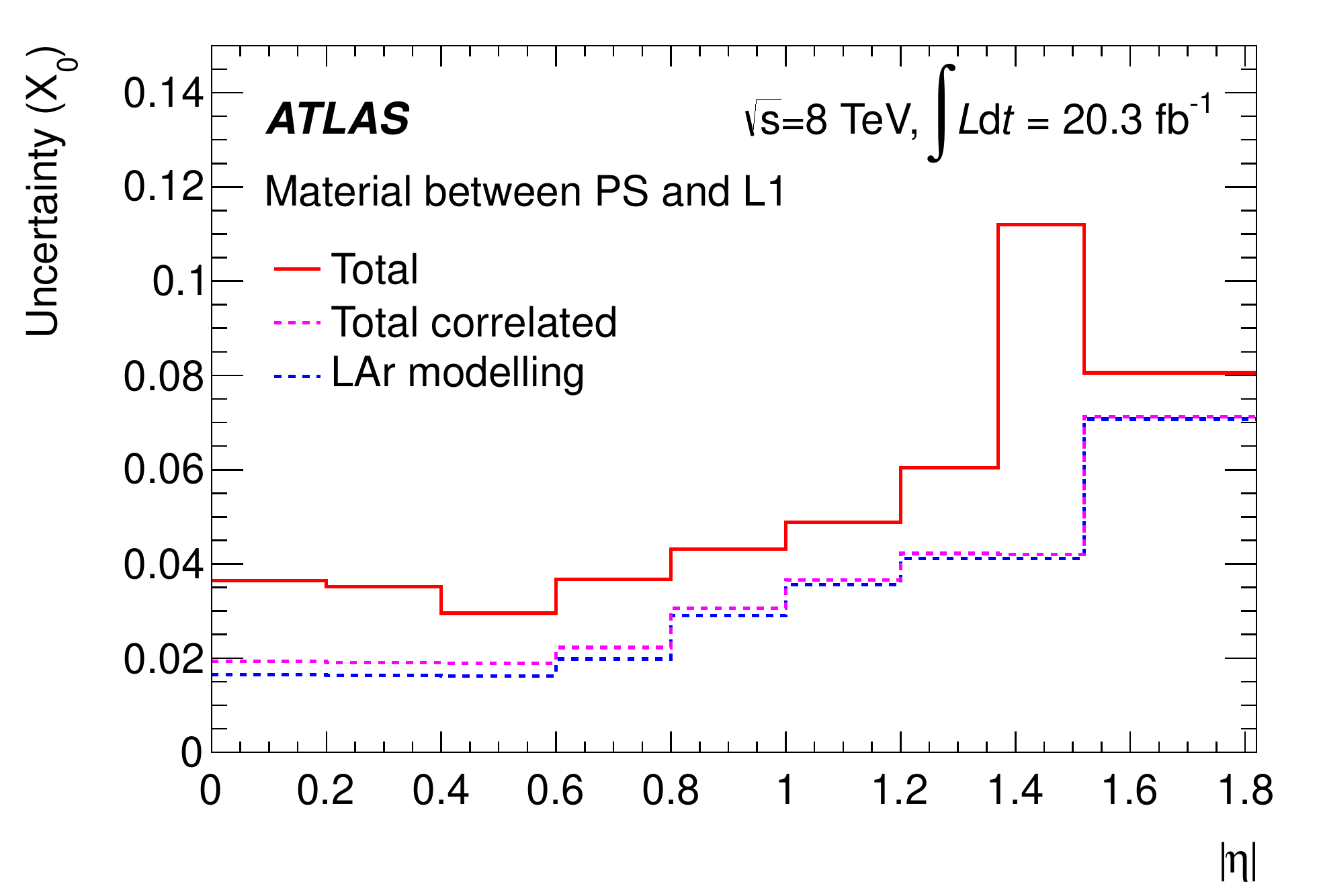} \\
  \includegraphics[width=\columnwidth]{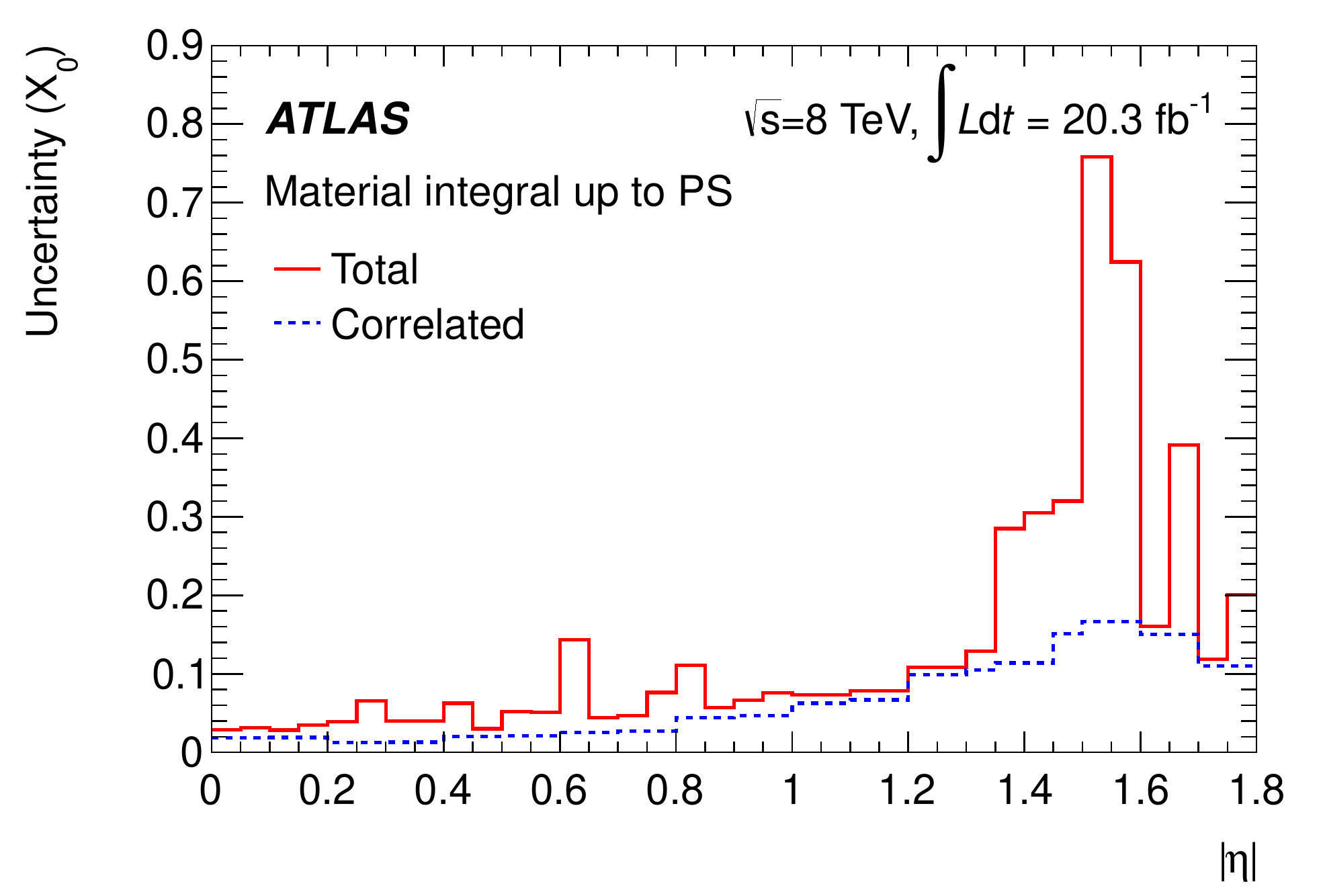} \\
  \includegraphics[width=\columnwidth]{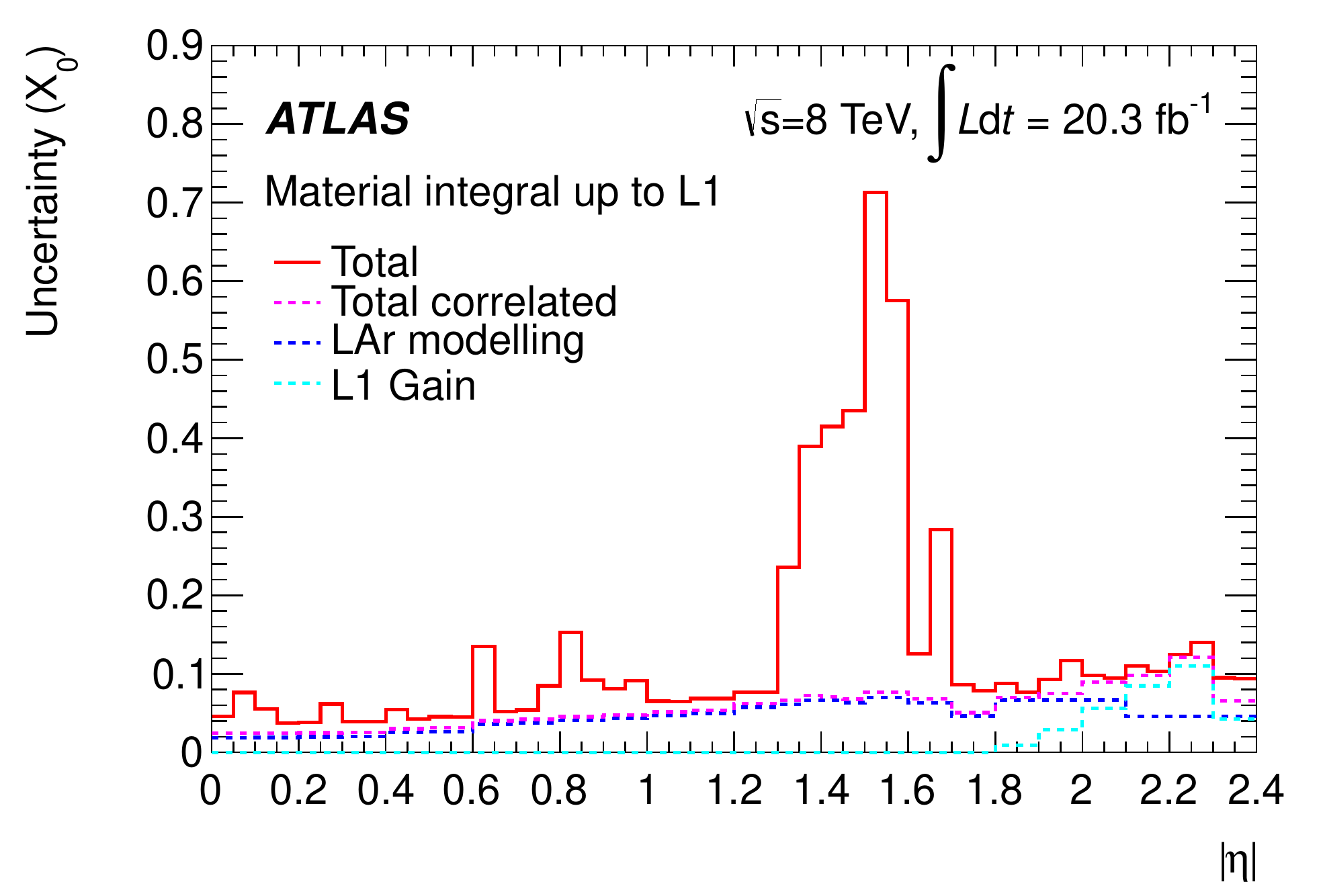}
  \caption{From top to bottom : uncertainty on the calorimeter material estimate, $\Delta
    X_{\rm Calo}$; on the material integral up to the PS, $\Delta
    X_{\rm PS}$; and on the material integral up to L1, $\Delta
    X_{\rm L1}$. The LAr $E_{1/2}$ modelling, {\sc Geant4} and L1 gain systematic
    uncertainties are assumed correlated across $\eta$. The remaining
    part of the uncertainty is data driven and considered
    uncorrelated.\label{fig:materrnewgeo1}} 
\end{figure}

After implementation and validation of the improved simulation, the
$Z\rightarrow ee$ samples were resimulated, and the $E_{1/2}$ data/MC
comparisons repeated. The difference between the material estimate from data and the improved simulation is summarised in Fig.~\ref{fig:e12newgeo}. As can be seen, the improved simulation behaves as
expected in most of the acceptance: the overall discrepancy in the
endcap has disappeared, as well as the strong peak around
$|\eta|=1.9$. The deficit within $1.5<|\eta|<1.6$ remains, as it has
not been addressed. In the barrel, the excess at $|\eta|=0.6$ has been
halved. The residual passive-material uncertainties in this improved 
simulation are presented in Fig.~\ref{fig:materrnewgeo1}. Where no
significant excess or deficit remains, the measurement uncertainty is
given by the L1/L2 calibration uncertainty, and the sensitivity curves' systematic uncertainties. When the residual discrepancy is
larger than the measurement uncertainty, the size of the discrepancy
is taken as the final uncertainty. No measurement was performed for $1.37<|\eta|<1.52$;
in this region the uncertainty on the material upstream of L1 is
estimated to be $\sim 0.4X_0$, following
Ref.~\cite{perf2010}. The $E_{1/2}$ modelling systematic uncertainties   
summarised in Sect.~\ref{sec:samplingcalib} are considered
correlated across $\eta$, separately in the barrel and endcap
calorimeters. Among these, the L1 gain systematic uncertainty only affects the
measurement of the material integral up to L1, for $|\eta|>1.8$. The
{\sc Geant4} systematic uncertainties are fully correlated across
$\eta$. The data-driven components of the material determination are
estimated in bins of size $\Delta\eta=0.2$, and the corresponding
uncertainties are uncorrelated beyond this range.

The MC-based energy calibration described in
Sect.~\ref{sec:MCCalibration} is applied using the new
detector description, to reoptimise the energy response in the endcap,
where significant amounts of passive material were added in the simulation. The
resulting MC calibration constitutes, together with the energy corrections
described in Sec\-ts.~\ref{sec:uniformity} and~\ref{sec:samplingcalib},
the basis of the absolute scale determination presented in the next section.


\section{Energy scale and resolution determination with electrons from $Z\rightarrow ee$ decays \label{sec:zeescales}}

As shown in Sect.~\ref{sec:uniformity}, no significant
mis-calibration is observed as a function of $\phi$, in a given 
$\eta$ region. The residual non-uniformity is at the level of 0.75\% or better
as illustrated in Fig.~\ref{fig:phiuniformity}, matching the design
constant term of 0.7\%. Furthermore, the energy response is shown in
Figs.~\ref{fig:uni:vsPileUp} and \ref{fig:uni:vsTime} to be stable with time at the level
of 0.05\%. Consequently, the absolute scale determination is
carried out as a function of $\eta$ only, and independently of time and
azimuth.

\subsection{Methodology\label{sec:zmeth}}

\begin{figure*}
  \centering
  \includegraphics[width=0.49\textwidth]{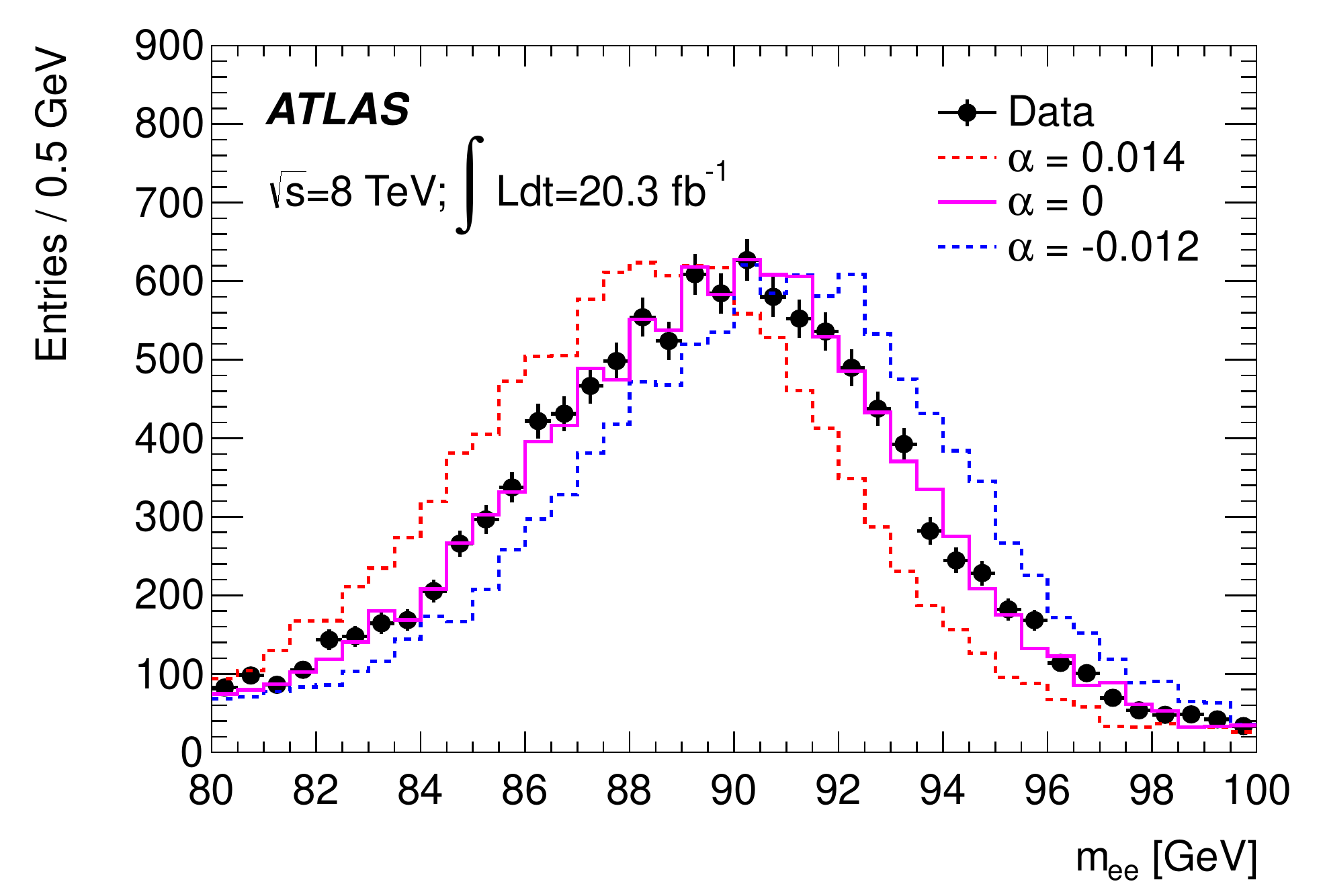}
  \includegraphics[width=0.49\textwidth]{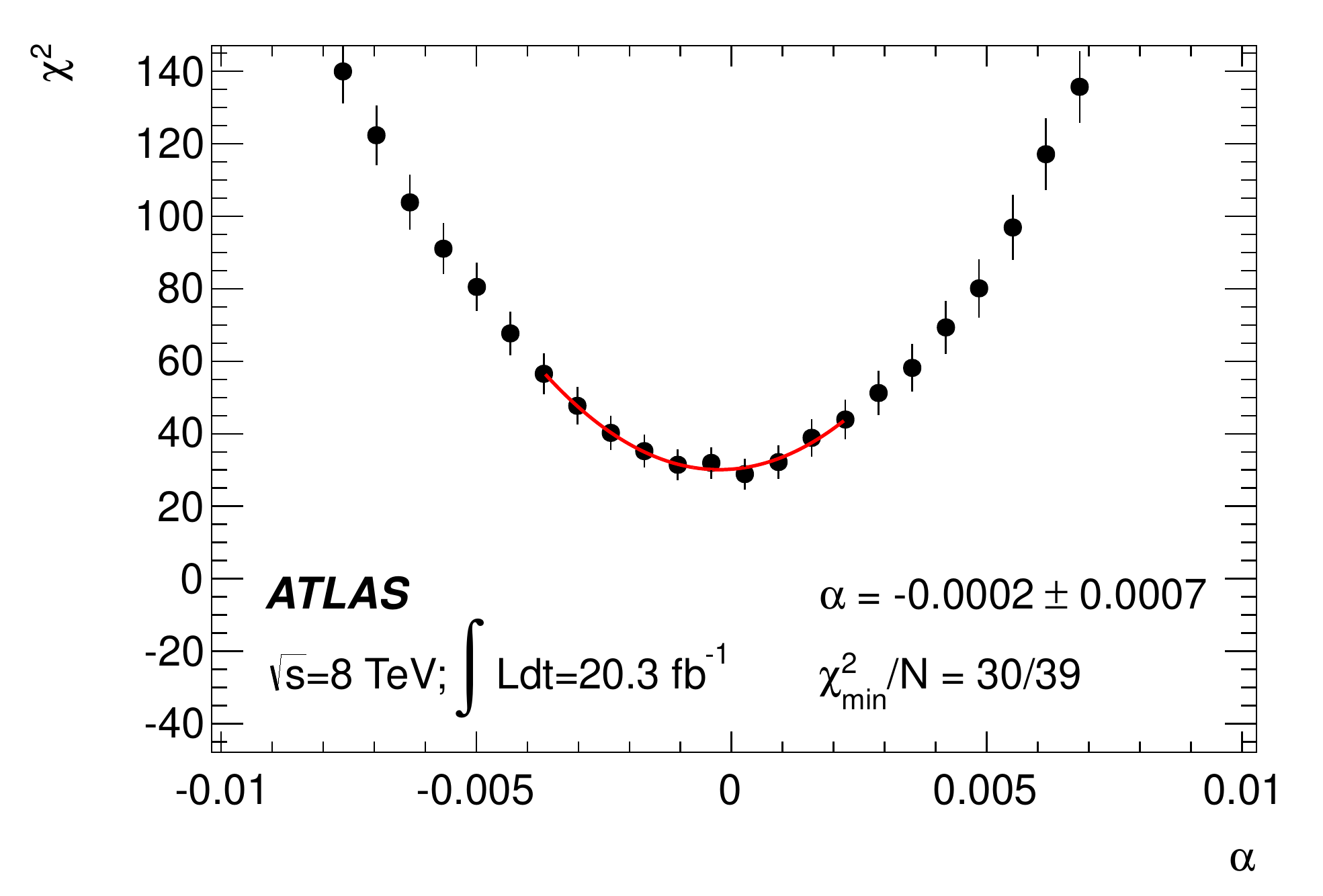} \\
  \includegraphics[width=0.49\textwidth]{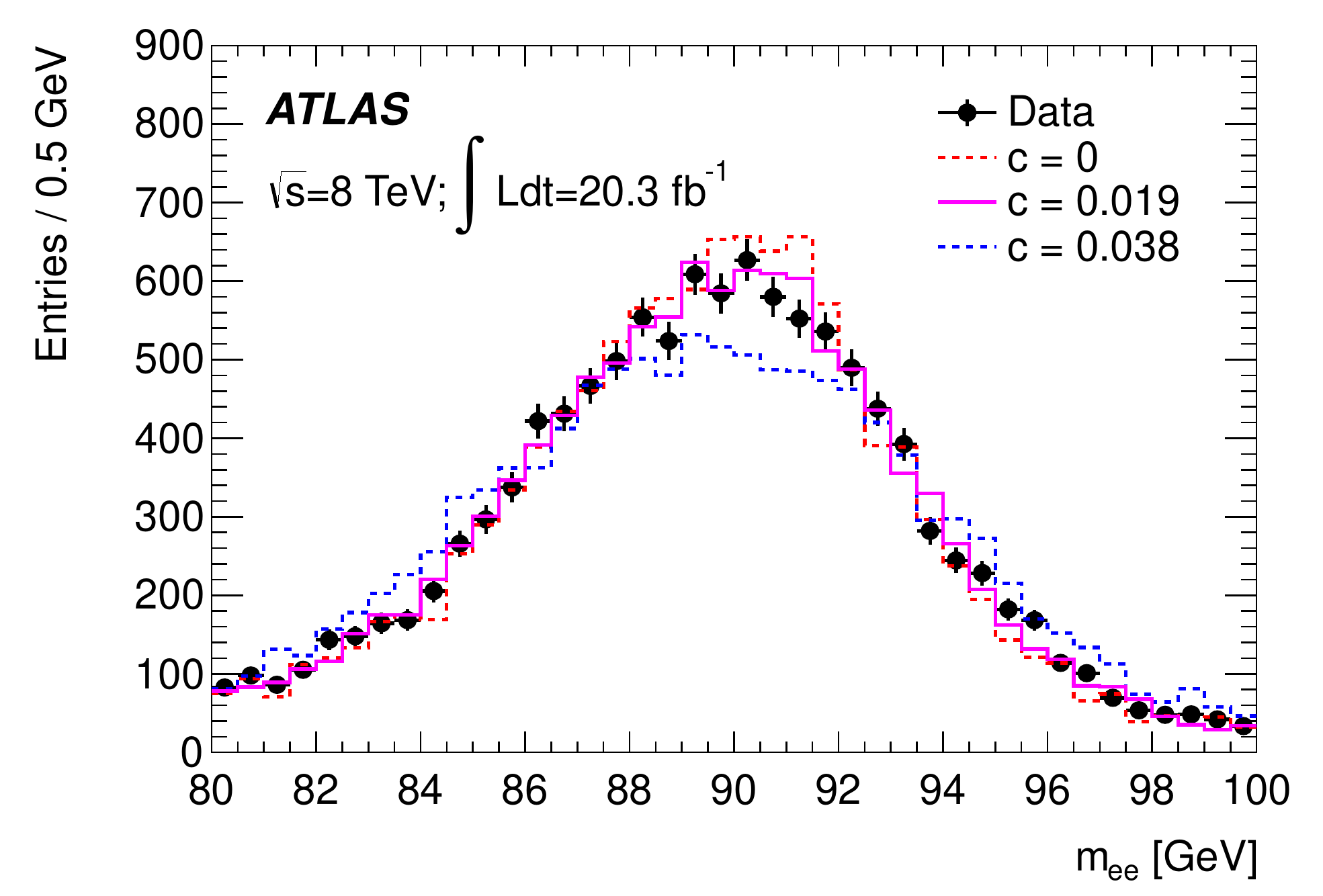}
  \includegraphics[width=0.49\textwidth]{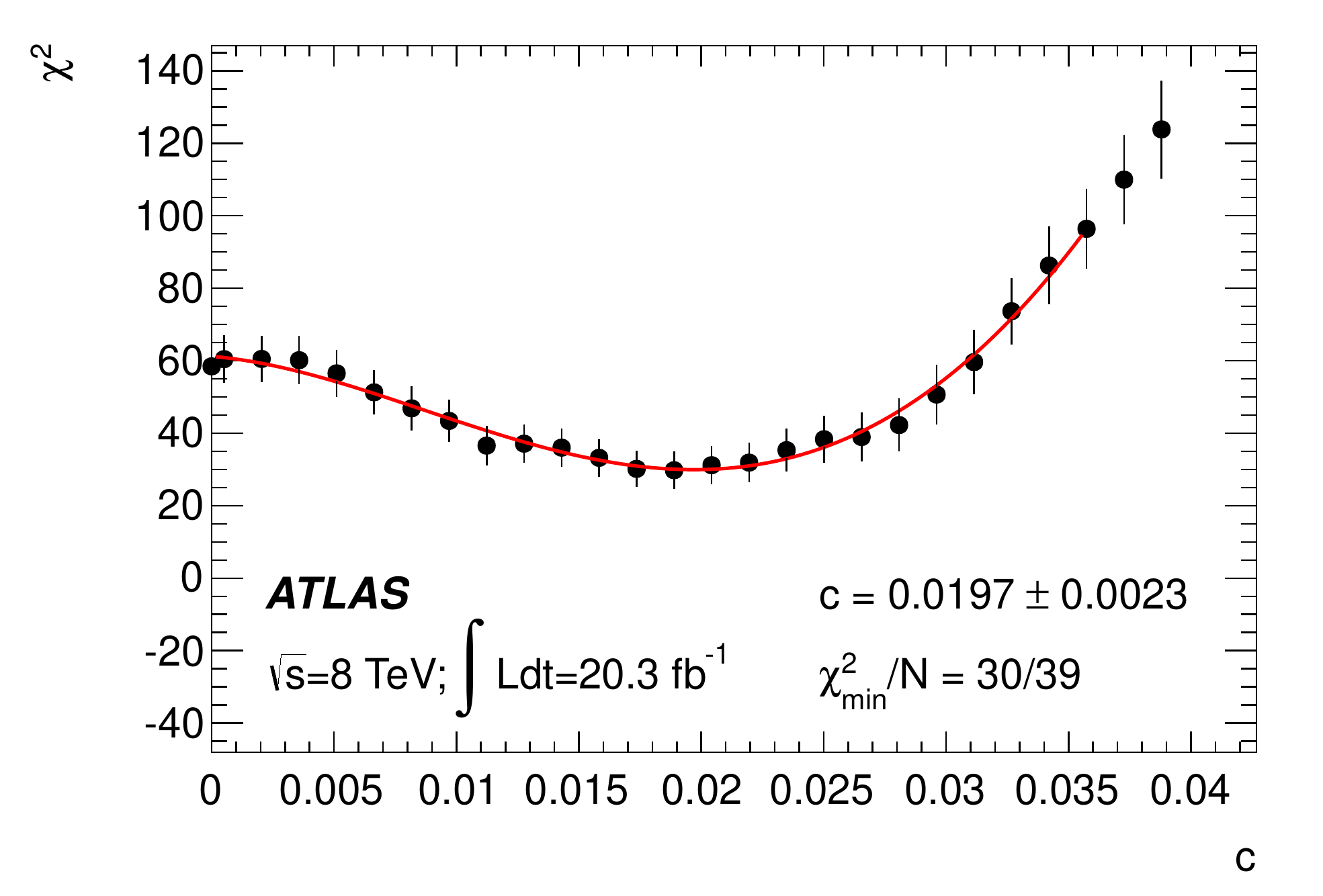}
  \caption{The electron pair invariant mass distribution, $m_{ee}$, for $Z\rightarrow ee$ candidates in data compared to MC templates for events with one electron within $1.63<\eta<1.74$, and the other within $2.3<\eta<2.4$. Top left: templates of $m_{ee}$ for different values of
    $\alpha$, for a fixed value of
    $c$. Bottom left: templates of $m_{ee}$ for different values of
    $c$, for a fixed value of
    $\alpha$. Top right: $\chi ^{2}$ as a function of
    $\alpha$, for the energy scale fit. Bottom right: $\chi ^{2}$ as a function of
    $c$, for the resolution fit. \label{fig:Tmeth}}
\end{figure*}

The energy mis-calibration is defined as the difference in response
between data and simulation, and is parameterised as follows:
\begin{equation}
  E^{\rm data}=E^{\rm MC}(1+\alpha_i)
  \label{eq:bias}
\end{equation}
where $E^{\rm data}$ and $E^{\rm MC}$ are the electron energy in data and simulation, and $\alpha_i$ represents the 
departure from optimal calibration, in a given pseudorapidity bin
labelled $i$. For $Z\rightarrow ee$ decays, the effect of electron
mis-calibration on the electron
pair invariant mass is  
\begin{eqnarray}
  \nonumber m_{ij}^{\rm data} &=&  m_{ij}^{\rm MC} (1+\alpha_{ij}), \\
  \alpha_{ij} &\sim& \frac{(\alpha_{i} + \alpha_{j})}{2},
  \label{eq:massequality}
\end{eqnarray}
neglecting second-order terms and assuming that the angle between the two
electrons is perfectly known; $m_{ij}^{\rm data}$ and $m_{ij}^{\rm MC}$ are the
invariant mass in data and simulation for an electron pair
reconstructed in pseudorapidity bins $i$ and $j$, and 
$\alpha_{ij}$ the induced shift of the mass peak. Electron resolution
corrections are derived under the assumption that the resolution curve 
is well modelled by the simulation up to a Gaussian constant term    
\begin{equation}
  \left(\frac{\sigma_E}{E}\right)^{\rm data} = \left(\frac{\sigma_E}{E}\right)^{\rm MC} \,\,\, \oplus \,\,\, c.
  \label{eq:smearingcorr}
\end{equation}
The sampling term is assumed to be known to $10\%$ from test-beam studies~\cite{Aharrouche:2006nf}. 
For each $(\eta_i,\eta_j)$ category, the relative electron and
invariant mass resolutions satisfy 
\begin{eqnarray}
  \nonumber \left(\frac{\sigma_m}{m}\right)_{ij}^{\rm data} & = &
  \left(\frac{\sigma_m}{m}\right)_{ij}^{\rm MC} \oplus c_{ij} \\
  & = & \frac{1}{2} \left[ \left(\frac{\sigma_E}{E}\right)_{i}^{\rm MC} \oplus
  c_{i} \oplus \left(\frac{\sigma_E}{E}\right)_{j}^{\rm MC} \oplus
  c_{j} \right], \nonumber \\
  c_{ij} &=& \frac{(c_{i} \oplus c_{j})}{2},
  \label{eq:cij}
\end{eqnarray}
where $c_{ij}$ is the relative invariant mass resolution correction for
$(\eta_i,\eta_j)$. To determine the $\alpha$ and $c$ parameters, histograms of the invariant mass are created from the simulation, including energy scale and 
resolution perturbations to the reconstruction-level quantities, in a range
covering the expected uncertainty in narrow steps. The templates are
built separately for the electron pseudorapidity
configurations $(\eta_i,\eta_j)$ and constitute a two-dimensional grid
along ($\alpha_{ij}$, $c_{ij}$). The data are categorised 
accordingly. The optimal values, uncertainties and correlations of
$\alpha_{ij}$ and $c_{ij}$ are obtained by $\chi^2$ minimisation, as
illustrated in Fig.~\ref{fig:Tmeth}. The individual
electron energy scales and resolution corrections are obtained by
solving the system given by Eqs.~\eqref{eq:massequality} and~\eqref{eq:cij}.

An alternative method relies on the same kinematical relations, but
replaces the templates by a parameterisation of the Monte Carlo
distributions, and performs a likelihood fit to the energy
scales. The parameterisation is based on the convolution of a
Breit--Wigner function and a Gaussian distribution; the energy
scales are determined by minimising with respect to $\alpha_{ij}$ the following likelihood:
\begin{equation}
  - \ln L_{\mathrm{tot}} = \sum^{N_{\mathrm{events}}}_{k=1} -\ln L_{ij}\left(\frac{m_{k}}{1+\frac{\alpha_{i}+\alpha_{j}}{2}}\right) , 
  \label{eq:likelihood}
\end{equation}
where 0$<i,j<N_{\mathrm{regions}}$, $N_{\mathrm{regions}}$ is the number of regions
considered for the calibration, $N_{\mathrm{events}}$ is the total number of
selected events and $L_{ij}(M|\alpha_i,\alpha_j)$ is the probability density function
(PDF) quantifying the compatibility of an event with the expected
$Z$~boson line shape at the reconstruction level, when the two
electrons fall in regions $i$ and $j$.

\subsection{Results and uncertainties}

The methods described above are applied to a sample of $Z\rightarrow
ee$ decays, selecting two electrons in the final state,
satisfying $\et>27$~GeV, $|\eta|<2.47$, $80<m_{ee}<100$~GeV and medium ID selection criteria. The
$\eta$ categories are defined as bins of size $\Delta\eta=0.2$ in the
barrel; a more complicated structure is defined in the endcap,
according to the PS acceptance boundaries and the $\eta$-dependent
HV settings. 

\begin{figure}
  \centering
  \includegraphics[width=\columnwidth]{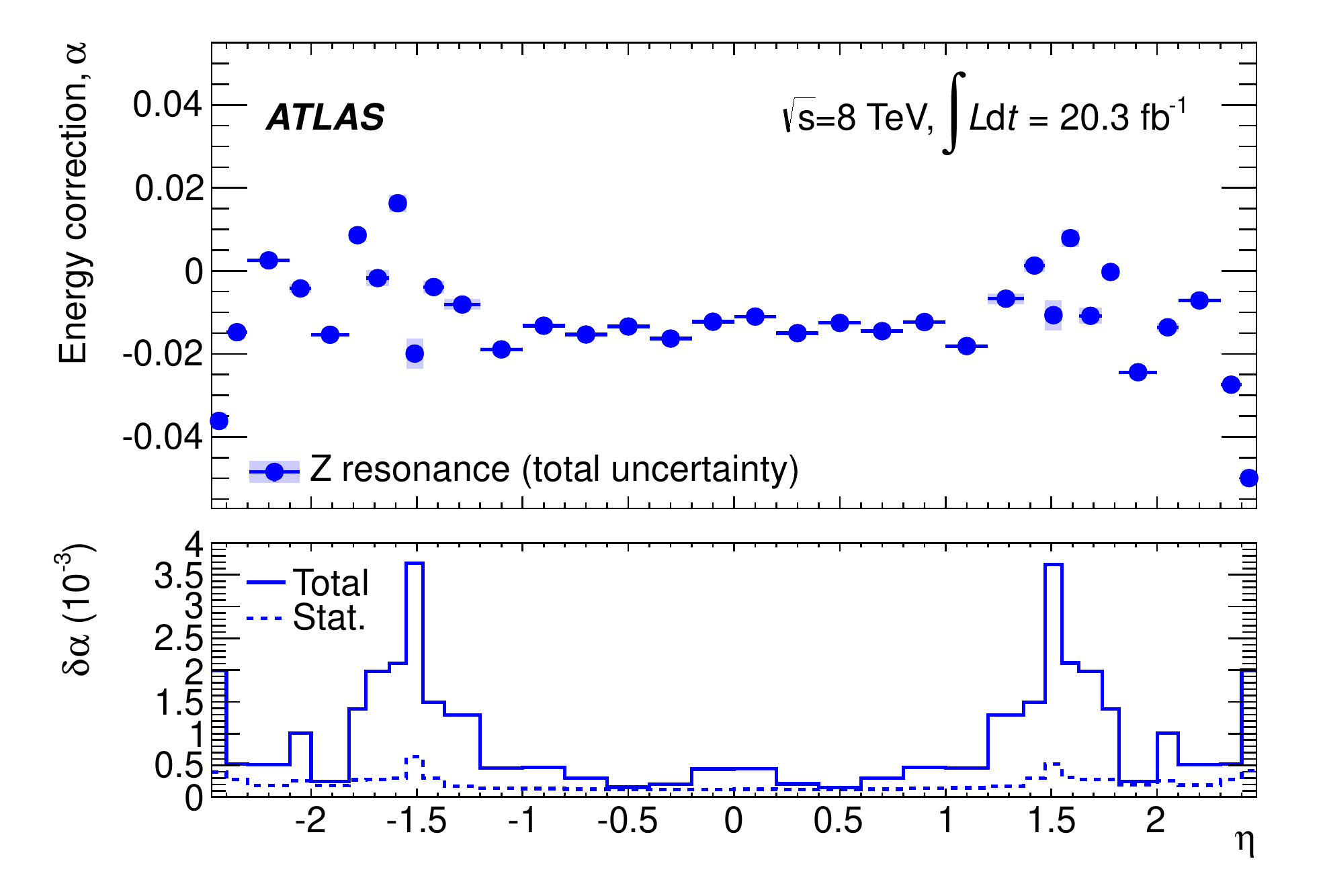}
  \caption{Top: energy scale corrections $\alpha$, defined as
    $E^{\rm data} = (1+\alpha)E^{\rm MC}$ and derived from $Z\rightarrow
    ee$ events using the template method, as a function of $\eta$. The corrections are defined
    with respect to the 2010 calibration scheme, and after uniformity
    and layer calibration corrections. The error bands include
    statistical and systematic uncertainties. Bottom: statistical and 
    total energy scale uncertainties, $\delta \alpha$ as a function of $\eta$.\label{fig:finalplota}} 
\end{figure}    

\begin{figure}
  \centering
  \includegraphics[width=\columnwidth]{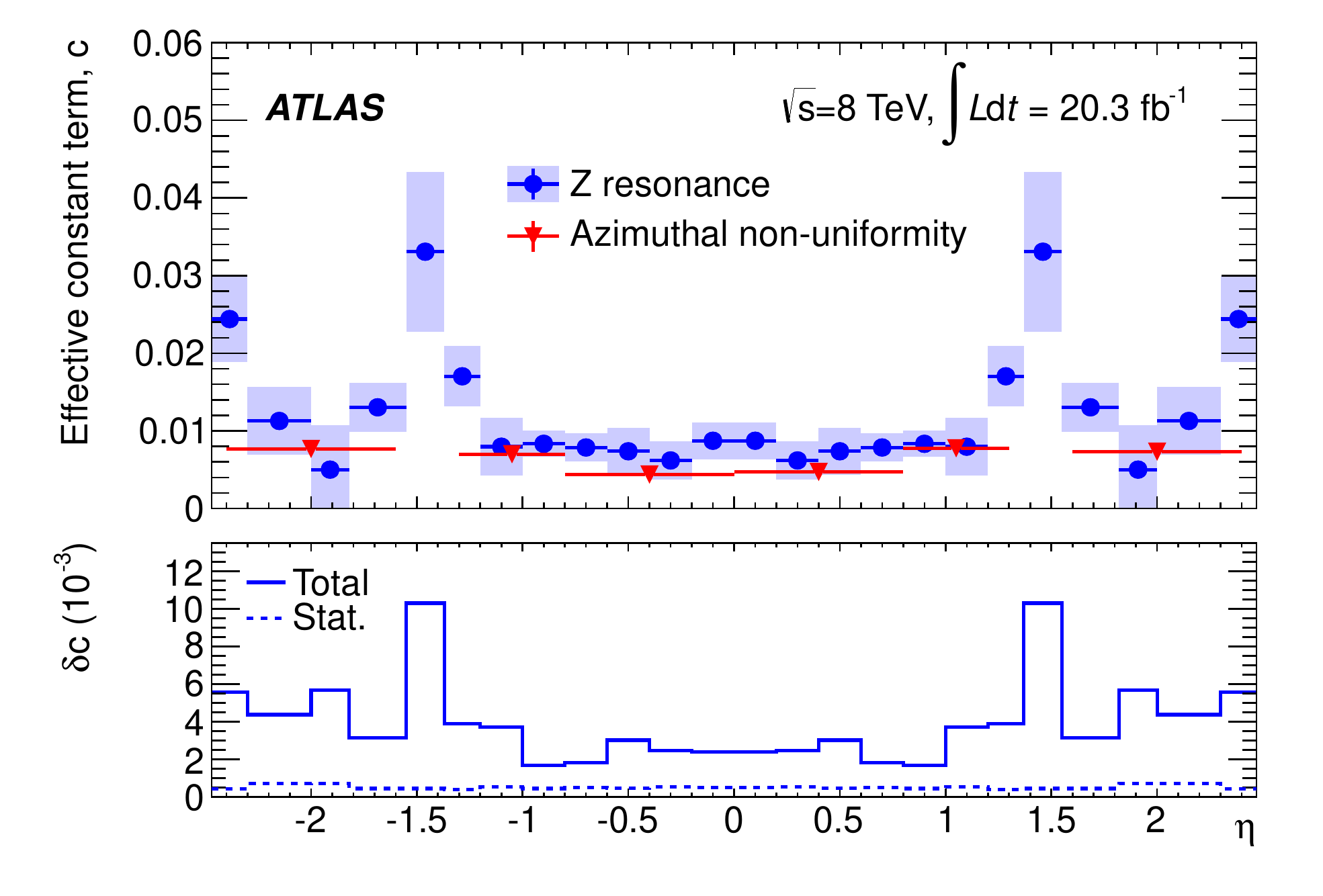}
  \caption{Top: effective constant term corrections $c$ derived from $Z\rightarrow
    ee$ events using the template method, as a function of $\eta$. The values of $c$ are
    symmetrised with respect to $\eta=0$. The error bands include
    statistical and systematic uncertainties. The contribution from the
    response uniformity derived in Sect.~\ref{sec:uniformity} of the paper is included for
    comparison and is not symmetrised. Bottom: statistical and total uncertainties on
    $c$, $\delta c$ as a function of $\eta$.\label{fig:finalplotc}}
\end{figure}

The results are summarised in Figs.~\ref{fig:finalplota}
and~\ref{fig:finalplotc}, where the energy scale and the effective constant term corrections are illustrated as a function of $\eta$. The energy scale factors are defined with
respect to the 2010 calibration results~\cite{perf2010}, which were
implemented as cell-level energy corrections in coarse $\eta$ bins;
the values obtained here reflect the reoptimisation of the OFC
coefficients for the 2012 pile-up conditions, and the uniformity and
layer calibration corrections performed ahead of the absolute scale
determination and discussed in the previous sections. They are found to be symmetric with respect to $\eta=0$
in the barrel; in the endcaps, similar patterns are observed for $\eta>0$
and $\eta<0$, up to an overall shift of about 1\% resulting from the
slightly different LAr temperature in the two cryostats. The energy scale determination is 
accurate to $0.3\times 10^{-3}$ for $|\eta|<1.37$, $2\times 10^{-3}$ for
$1.37<|\eta|<1.82$ and $0.5\times 10^{-3}$ for $|\eta|>1.82$.  The
resolution corrections are about 0.8\% on average in the barrel, and
about 1\% in the endcap, and are determined to be accurate on average to 0.3\% and 0.5\%, respectively. At given $|\eta|$, the values of $c$
are found to be statistically compatible for $\eta>0$ and $\eta<0$
and are symmetrised in Fig.~\ref{fig:finalplotc}. The uncertainty contributions are
detailed below.

The agreement of the energy scale between the template and likelihood
methods is good, with an average difference of $\delta\alpha=3.8 \times
10^{-4}$ when the full $\eta$ range is considered, and of $1.3 \times
10^{-4}$ when the transition region ($1.37<|\eta|<1.55$) is
discarded. The fitted constant terms agree within $\delta c=1.2 \times 10^{-3}$ between the two methods.

The intrinsic accuracy of the template method is tested by injecting
known energy scale and resolution distortions into a simulation
sample that is treated as the data. The corresponding $\alpha$ and $c$
corrections are then derived by comparing the modified simulation to the
templates, and compared to the injected values. No bias is found
beyond the statistical accuracy of the test; systematic uncertainties
of $\delta\alpha=0.5 \times 10^{-4}$ for the energy scales, and  $\delta c=1.1 \times 10^{-3}$ for
the resolution corrections are assigned.

The sensitivity of the result to the event select\-ions is studied by
varying the electron identification criteria and the mass window used for
the fit. Repeating the energy correction
determination using electrons with tight ID selection criteria gives
an average difference $\delta\alpha=1.2 \times 10^{-4}$ for the energy scales, and  $\delta c=1.1 \times 10^{-3}$ for
the resolution corrections. Uncertainties on the efficiency
corrections for trigger, identification and reconstruction can distort
the invariant mass distribution and lead to a total uncertainty of
about $\delta\alpha=0.4 \times 10^{-4}$ on the energy scales and $\delta c=0.3 \times
10^{-3}$ on the resolution corrections. The impact of the choice of
mass window is on average 
$\delta\alpha=0.9 \times 10^{-4}$ and $\delta c=0.9 \times
10^{-3}$. 

The dedicated tracking algorithm used for elect\-rons provides
momentum measurements at the interaction point and at the outer
radius of the ID, denoted by ($q/p)^{\rm IP}$ and ($q/p)^{\rm
  out}$ respectively. The momentum lost by bremsstrahlung is quantified by defining $f_{\rm brem} =
1-(q/p)^{\rm IP}/(q/p)^{\rm out}$, where values close to 0 select
electrons which have lost a small fraction of their
momentum. Repeating the analysis requiring $f_{\rm brem}<0.3$ selects an electron sample with less bremsstrahlung than the inclusive
sample, with an efficiency of about 50\%. The impact of this selection is
$\delta\alpha=6 \times 10^{-4}$ and $\delta c=1.5 \times 10^{-3}$.

Uncertainties induced by the general modelling of the signal process
(pile-up, interaction point distribution, theoretical description of the
$Z$ lineshape) contribute $\delta\alpha=0.4 \times 10^{-4}$ and
$\delta c=0.5 \times 10^{-3}$.

Electroweak, top and multijet backgrounds constitute about 0.13\% of the
selected $Z$ boson sample. To propagate the corresponding uncertainty,
the normalisation of the electroweak and top backgrounds is varied
within the theoretical cross-section uncertainties, which are as large as 10\% depending on the channel, with an impact of $\delta\alpha=0.3
\times 10^{-4}$ and $\delta c=0.4 \times 10^{-3}$. The multijet background
fraction is estimated by comparing the electron isolation distribution
observed after\- all selections with the expected distributions for
signal and multijet production~\cite{Aad:2014fxa}. The signal distribution is determined
from the simulation, while the multijet distribution is determined
from a jet-enriched sample obtained by selecting electron pairs passing
only the loose identification criterion. The relative uncertainty of
this determination is 50\% and contributes $\delta\alpha=0.2
\times 10^{-4}$ and $\delta c=0.1 \times 10^{-3}$.

The uncertainties quoted above are averages; the values depend on
pseudorapidity, with larger values in regions with a large amount of
material upstream of the calorimeter and in the transition region between the
barrel and endcap. The stability of the corrections with the energy is
discussed in Sect.~\ref{sec:electroncheck}.

\subsection{Data/MC comparison after corrections}

\begin{figure}
  \centering
  \includegraphics[width=\columnwidth]{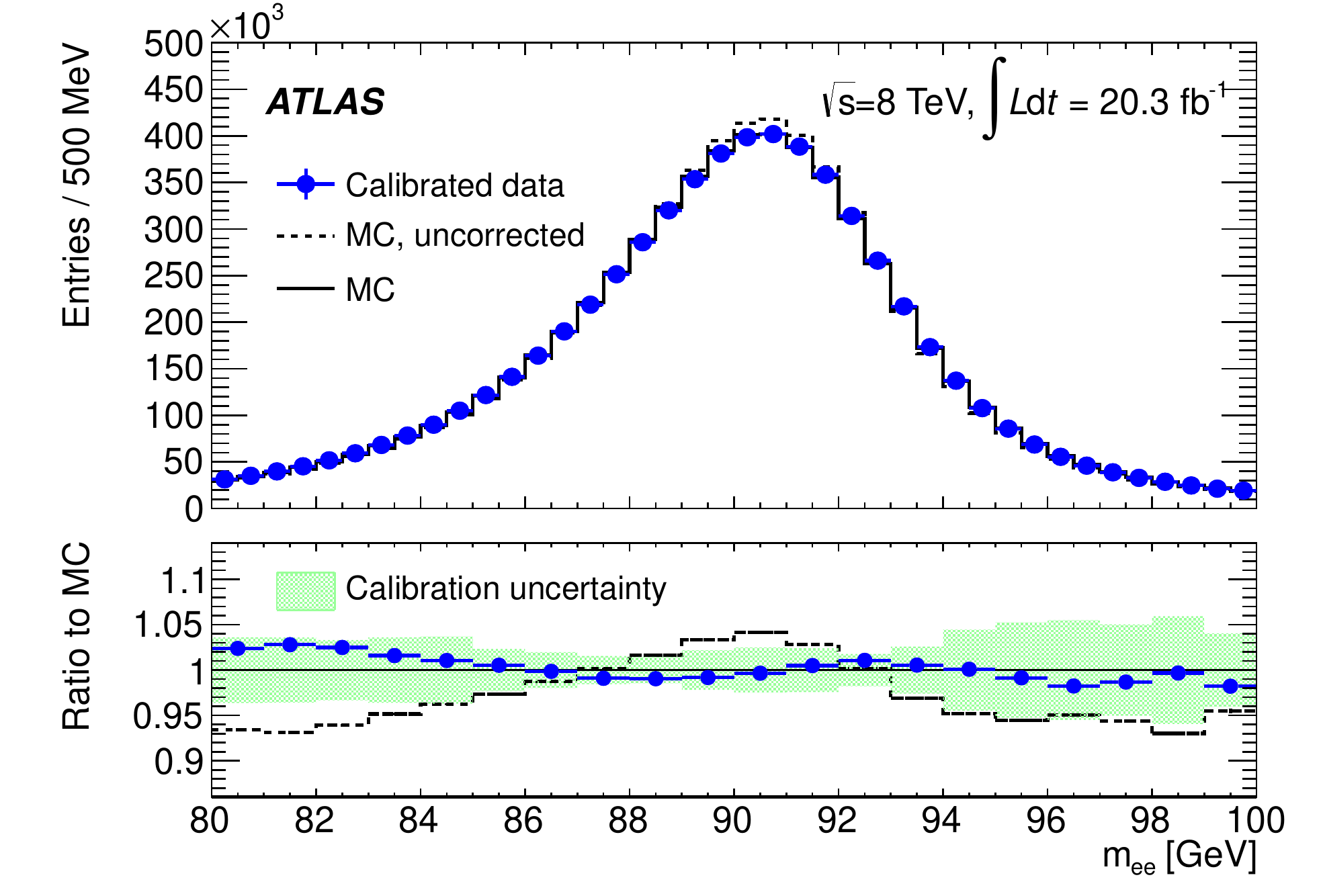}
  \caption{Top: electron pair invariant mass distribution for
    $Z\rightarrow ee$ decays in data and improved simulation. Energy scale
    corrections are applied to the data. The improved simulation is shown
    before and after energy resolution corrections, and is normalised to the number of
    events in data. Bottom: ratio of the data and uncorrected MC
    distributions to the corrected MC distribution with the calibration uncertainty band. 
    \label{fig:zeefinalresults}}
\end{figure}

After all corrections, the dielectron mass distribution in data and simulation agree at the level of 1\%
in the mass range $80<m_{ee}<100$~GeV, rising to 2\% towards the low
end of the interval. The energy scale adjustement, followed by the resolution correction, are the main causes of the improved agreement with respect to that obtained with the previous calibration and simulation \cite{perf2010}. 
The jet, electroweak and top backgrounds contribute about 1.5\% near
$m_{ee}=80$~GeV and $m_{ee}=100$~GeV. Figure~\ref{fig:zeefinalresults}
shows the dielectron mass distribution for the data corrected with the
energy scale factors and 
for the MC simulation with and without the resolution corrections. In
addition the ratios of the corrected data and uncorrected MC
distributions to the corrected MC distribution are illustrated
together with the final calibration uncertainty. 

A slight excess persists at low mass, indicating that the energy tails in
the data are not entirely modelled by the simulation, even after the
calibration and detector geometry improvements described above. However, as shown in Fig.~\ref{fig:zeefinalresults}, this discrepancy lies within the
quoted passive-material uncertainty. Its impact on the energy scale
and resolution corrections is covered by the systematic variations
described in the previous section.


\section{Summary of uncertainties common to electrons and photons\label{sec:uncsum}}

\newcommand{\etZ}{\et^{e (Z\to ee)}}

The calibration uncertainty for electrons from $Z$ boson decays is determined, at
given pseudorapidity and for $\langle\etZ\rangle\sim 40$~GeV, by the accuracy of the $Z$-based calibration described in the previous
section. Other effects are generally energy and particle-type
dependent and can be written as follows:
\begin{equation}
\delta E_{i}^{e,\gamma}(\et,\eta) = \Delta E_{i}^{e,\gamma}(\et,\eta) - \Delta E_{i}^{e}(\langle\etZ\rangle,\eta)
\end{equation}
A given source of uncertainty $i$ changes the energy scale by $\Delta
E_{i}^{e,\gamma}(\et,\eta)$, which is a function of $\et$ and $\eta$
and depends on particle type. The $Z$-based effective calibration absorbs the effect
for electrons with $\et=\langle\etZ\rangle$ and leaves the residual uncertainty $\delta
E_{i}^{e,\gamma}(\et,\eta)$. Because of this subtraction, $\delta
E_{i}^{e,\gamma}(\et,\eta)$ can change sign as a function of \et.

\begin{figure*}
  \centering
  \includegraphics[width=0.49\textwidth]{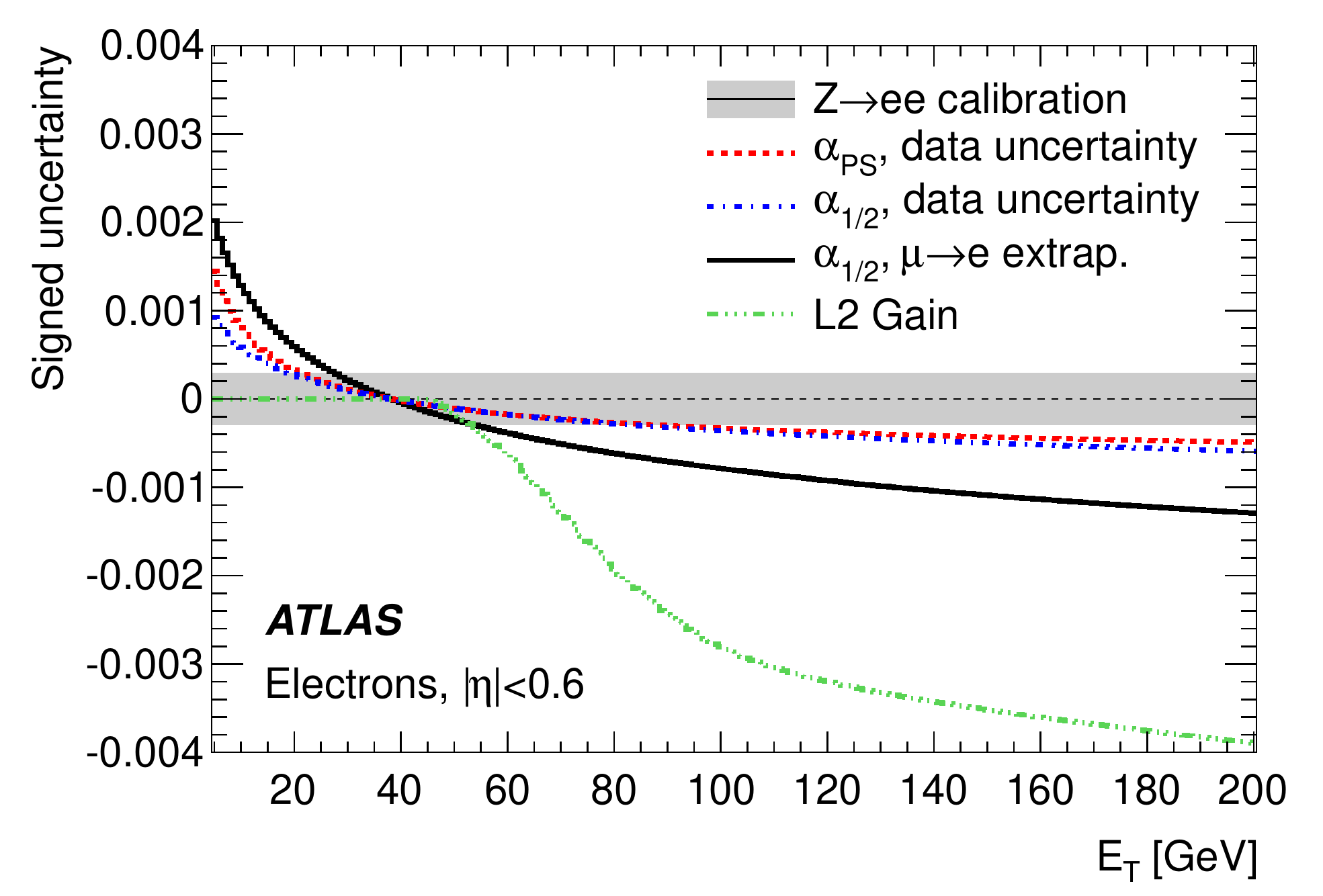}
  \includegraphics[width=0.49\textwidth]{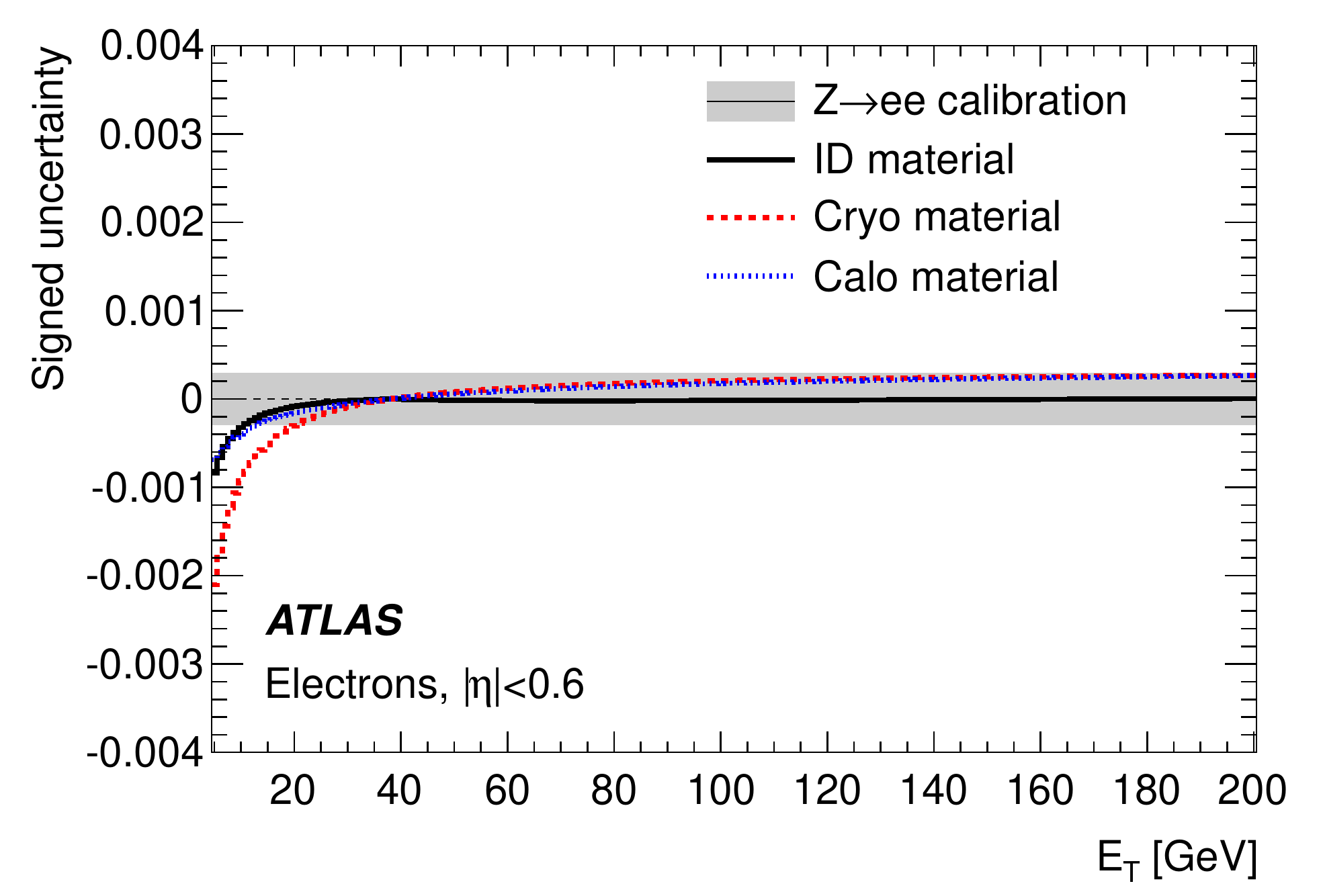} \\
  \includegraphics[width=0.49\textwidth]{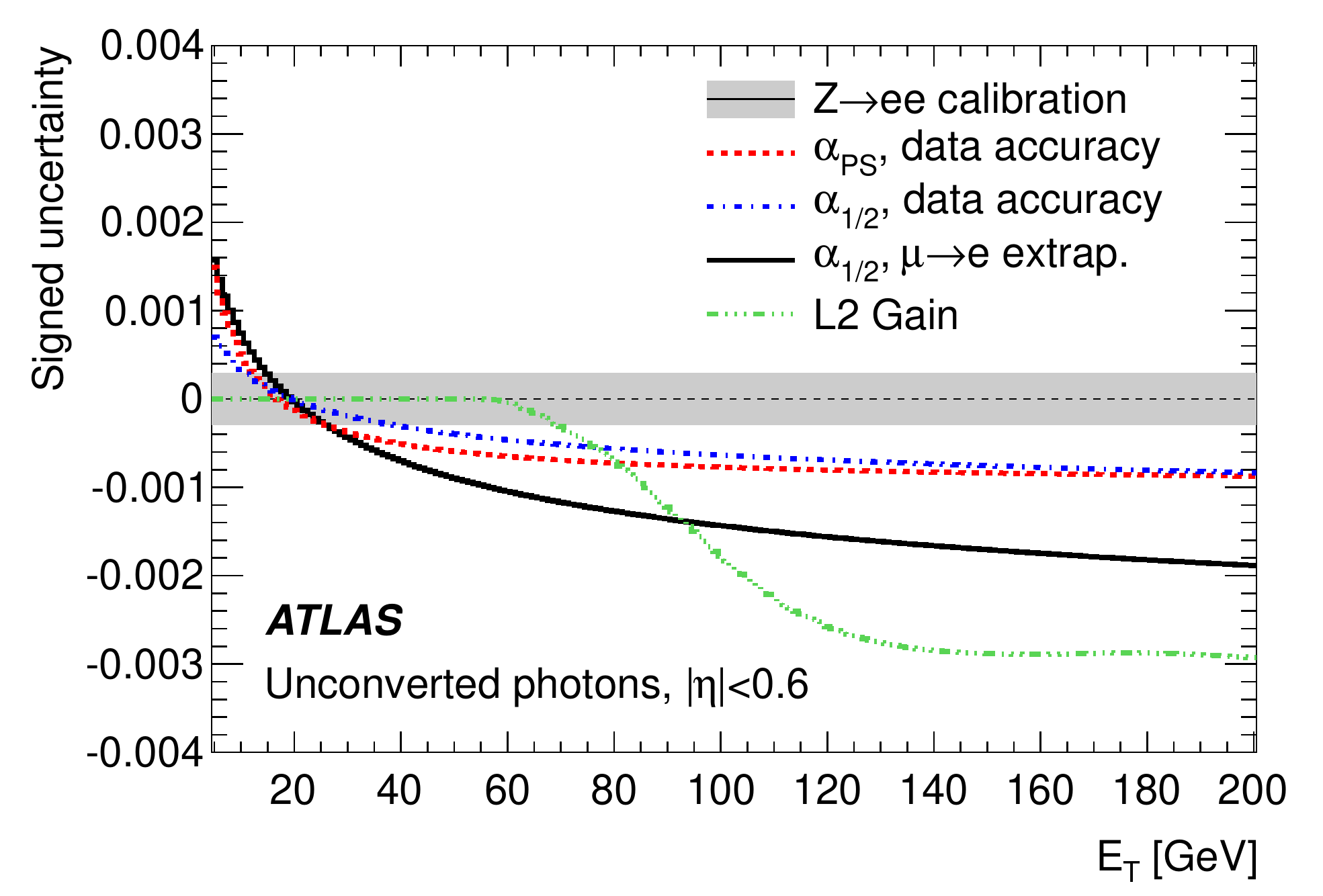}
  \includegraphics[width=0.49\textwidth]{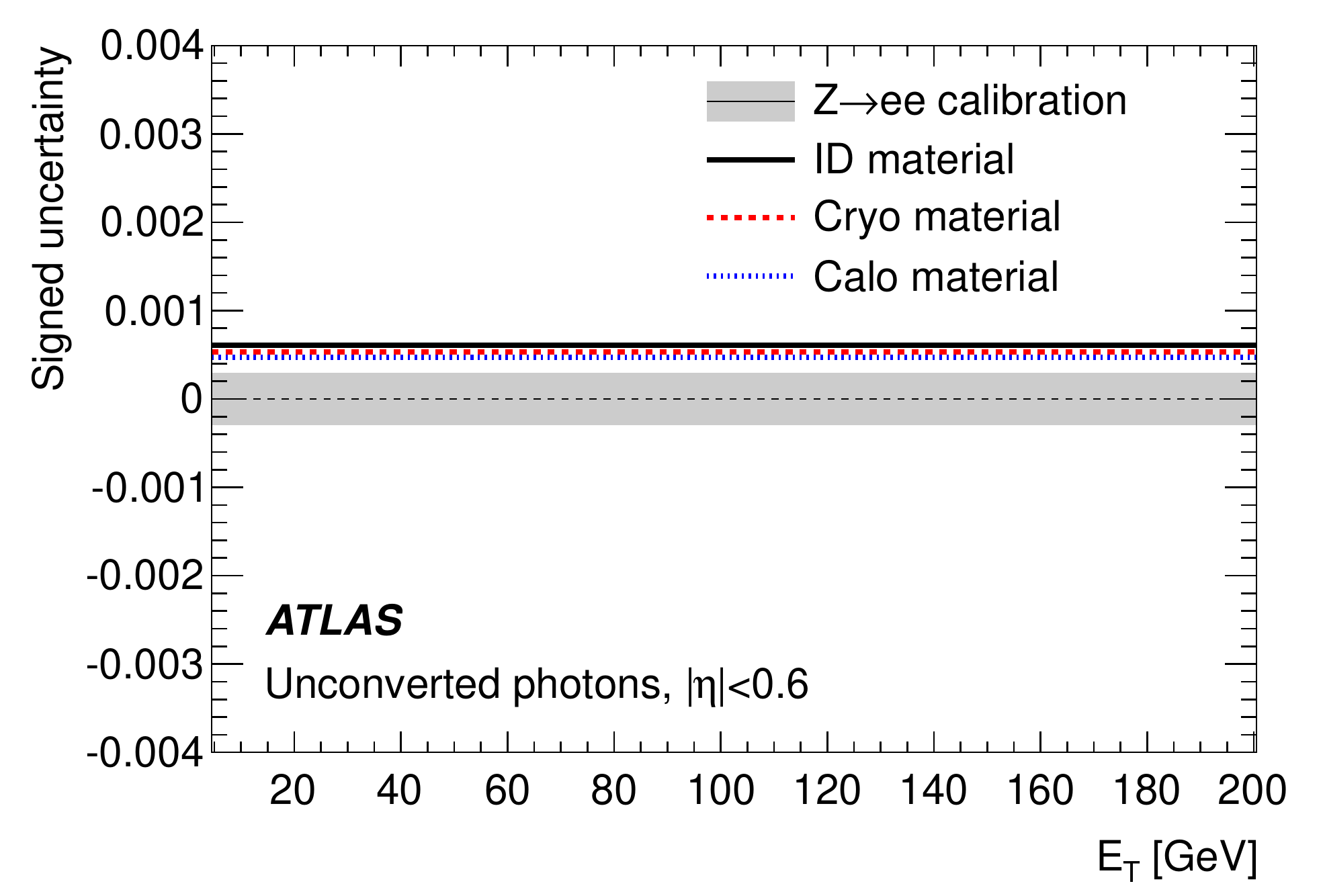}
  \caption{The $E_{\rm T}$ dependence of the dominant calibration systematic
    uncertainties for $|\eta|<0.6$, for electrons and unconverted
    photons. The curves show the effect on the energy scale of upward variations of the
    listed sources of uncertainty; depending on source and particle
    type, the effect can change sign as a function of \et. Left:
    uncertainties from the relative response of L1 and L2 (data-driven uncertainties of the PS and L1/L2
    response measurements; uncertainty in the application of the
    muon-based measurement to electrons; L2 HG/MG uncertainty). Right: passive-material
    uncertainties (ID, cryostat and calorimeter material).\label{fig:uncsumElectrons}}
\end{figure*}

The gain dependence of the energy response is measured in data by
comparing the $Z$ peak position for electron clusters with all L2 cells
recorded in HG to electrons with at least one cell recorded in
MG. The probability of an electron or photon cluster to
contain at least one MG cell increases smoothly with energy in a given
$\eta$ range: it is   
parameterised using simulated $Z\rightarrow ee$ and
$H\rightarrow\gamma\gamma$ samples, and is then validated using
$Z\rightarrow ee$ events from data. The parameterisation is used to correct for the effect in data, and 
the full size of the correction is taken as uncertainty. 
As a separate effect, the energy pedestal of electron and photon
clusters is verified by comparing pile-up only events in data and
simulation. An average offset of about 15~MeV is found, depending on
$\eta$. The induced energy non-linearity is negligible at high energy
but can reach 0.1\% for $\et\sim 10$~GeV and is counted as a separate
systematic uncertainty.

The impact of the PS and L1/L2 calibration uncertainties on the
reconstructed particle energy depends on the fractional energies of the cluster carried in those layers, $f_{\rm PS}$ and $f_{\rm L2}$. The energy and
pseudorapidity dependence of $f_{\rm PS}$ and $f_{\rm L2}$ is
parameterised for electrons and photons using the simulation, and the
uncertainty is given by the product of these energy fractions and the
corresponding layer calibration uncertainties, at given $\et$ and
$\eta$. An additional uncertainty is assigned to the intrinsic
accuracy of EM shower development simulation, by varying physics
modelling options in {\sc Geant4}. 

The ID, Cryo and Calo material uncertainties are propagated after comparing
the energy response in samples simulated with modified and nominal
detector mate\-rial. The simulation modifications follow the description in
Sect.~\ref{sec:mat:method}; the corresponding response
differences are scaled to the actual material measurement
uncertainties, yielding the energy response uncertainties.

The dominant sources of uncertainty are illustrated in Fig.~\ref{fig:uncsumElectrons} for
electrons and unconverted photons, for $|\eta|<0.6$ and $5<E_{\rm T}<200$~GeV. All curves
correspond to an upward variation of the considered source of
uncertainty by one standard deviation; the effect of this variation
can change sign as a function of \et, as discussed above. For electrons, the layer
calibration uncertainties reach about 0.15\% at low \et, are minimal
for $\et\sim 40$~GeV, and rise to about 0.05\% towards high
energy. For unconverted photons, the layer uncertainties are largest
at high energy, reaching about 0.1\%. The uncertainty related to the
HG/MG transition contributes above $\et\sim 50$~GeV, and reaches 0.3\%
at high transverse energy. A deficit of passive material in 
the simulation induces a drop in the energy scale towards low \et; in
this pseudorapidity region, the drop is largest for cryostat material
and reaches --0.3\% at $\et=5$~GeV. After the $Z$-based calibration, the same
material deficit induces an overestimate of the energy response for
unconverted photons by about 0.05\%, regardless of the passive
material type. The energy dependence of this effect is negligible for
unconverted photons and the corresponding uncertainties are considered independent of \et. 

Other sources of uncertainty are subleading and not discussed here
explicitly. All sources are considered as independent and their
sum in quadrature defines the total uncertainty at given $\et$ and 
$\eta$. Tables~\ref{tab:systelec1}$\--$\ref{tab:systconv} show the
uncertainty contributions discussed above as a function of
pseudorapidity, and fixed values of \et. For electrons, $\et=$ 11~GeV and
40~GeV are typical of $J/\psi\rightarrow ee$ and $Z\rightarrow ee$
decays, respectively; $\et=$ 200~GeV illustrates the asymptotical behaviour of the
uncertainty. For photons, $\et=$ 60~GeV is typical for $H\rightarrow\gamma\gamma$ decays with
$m_{\mathrm{H}}\sim 125$~GeV.

\begin{table*}
  \centering
  \begin{tabular}{lcccccc}
    \hline
    \hline
    &  \multicolumn{6}{c}{Electrons, $E_{\rm T}=11$~GeV} \\
 $|\eta|$ range                       & 0--0.6 & 0.6--1 & 1--1.37 & 1.37--1.55 & 1.55--1.82 & 1.82--2.47 \\
 \hline
 $Z\rightarrow ee$ calibration      & 0.03 & 0.04 & 0.08 & 0.22 & 0.22 & 0.05 \\ 
 Gain, pedestal                     & 0.09 & 0.09 & 0.07 & 0.00 & 0.09 & 0.24 \\ 
 Layer calibration                  & 0.15 & 0.19 & 0.14 & 0.16 & 0.13 & 0.19 \\ 
 ID material                        & 0.03 & 0.03 & 0.31 & 0.88 & 0.33 & 0.10 \\ 
 Other material                     & 0.12 & 0.38 & 0.58 & 0.20 & 1.00 & 0.15 \\ 
 \hline                      
 Total                              & 0.22 & 0.44 & 0.69 & 0.94 & 1.09 & 0.36 \\
 \hline
 \hline
  \end{tabular}
  \caption{Summary of energy scale systematic uncertainty contributions from
    sources common to electrons and photons, estimated for
    electrons with $E_{\rm T}=11$~GeV, in \%.\label{tab:systelec1}}
\end{table*}

\begin{table*}
  \centering
  \begin{tabular}{lcccccc}
    \hline
    \hline
    &  \multicolumn{6}{c}{Electrons, $E_{\rm T}=40$~GeV} \\
 $|\eta|$ range                       & 0--0.6 & 0.6--1 & 1--1.37 & 1.37--1.55 & 1.55--1.82 & 1.82--2.47 \\
 \hline
 $Z\rightarrow ee$ calibration     & 0.03 & 0.04 & 0.08 & 0.22 & 0.22 & 0.05 \\ 
 Gain, pedestal                    & 0.00 & 0.00 & 0.00 & 0.00 & 0.02 & 0.01 \\ 
 Layer calibration                 & 0.01 & 0.01 & 0.01 & 0.00 & 0.00 & 0.01 \\ 
 ID material                       & 0.00 & 0.00 & 0.01 & 0.00 & 0.00 & 0.00 \\ 
 Other material                    & 0.00 & 0.01 & 0.02 & 0.00 & 0.02 & 0.00 \\ 
 \hline                      
 Total                             & 0.03 & 0.04 & 0.08 & 0.22 & 0.22 & 0.05 \\ 
 \hline
 \hline
  \end{tabular}
  \caption{Summary of energy scale systematic uncertainty contributions from
    sources common to electrons and photons, estimated for
    electrons with $E_{\rm T}=40$~GeV, in \%.\label{tab:systelec2}}
\end{table*}

\begin{table*}
  \centering
  \begin{tabular}{lcccccc}
    \hline
    \hline
    &  \multicolumn{6}{c}{Electrons, $E_{\rm T}=200$~GeV} \\
 $|\eta|$ range                       & 0--0.6 & 0.6--1 & 1--1.37 & 1.37--1.55 & 1.55--1.82 & 1.82--2.47 \\
 \hline
 $Z\rightarrow ee$ calibration      & 0.03 & 0.04 & 0.08 & 0.22 & 0.22 & 0.05 \\ 
 Gain, pedestal                     & 0.21 & 0.36 & 0.40 & 0.00 & 2.14 & 0.61 \\ 
 Layer calibration                  & 0.15 & 0.18 & 0.16 & 0.18 & 0.18 & 0.21 \\ 
 ID material                        & 0.00 & 0.06 & 0.07 & 0.78 & 0.17 & 0.06 \\ 
 Other material                     & 0.06 & 0.17 & 0.35 & 0.20 & 0.63 & 0.07 \\ 
 \hline                      
 Total                              & 0.27 & 0.45 & 0.57 & 0.85 & 2.25 & 0.65 \\ 
 \hline
 \hline
  \end{tabular}
  \caption{Summary of energy scale systematic uncertainty contributions from
    sources common to electrons and photons, estimated for
    electrons with $E_{\rm T}=200$~GeV, in \%.\label{tab:systelec3}}
\end{table*}

\begin{table*}
  \centering
  \begin{tabular}{lccccc}
    \hline
    \hline
    &  \multicolumn{5}{c}{Unconverted photons, $E_{\rm T}=60$~GeV} \\
 $|\eta|$ range                       & 0--0.6 & 0.6--1 & 1--1.37 & 1.55--1.82 & 1.82--2.47 \\
 \hline
 $Z\rightarrow ee$ calibration      & 0.03 & 0.04 & 0.08 & 0.16 & 0.05 \\ 
 Gain, pedestal                     & 0.03 & 0.02 & 0.01 & 0.89 & 0.55 \\ 
 Layer calibration                  & 0.15 & 0.20 & 0.20 & 0.25 & 0.26 \\ 
 ID material                        & 0.06 & 0.12 & 0.19 & 0.07 & 0.12 \\ 
 Other material                     & 0.09 & 0.17 & 0.40 & 0.96 & 0.09 \\ 
 \hline                      
 Total                              & 0.19 & 0.31 & 0.50 & 1.35 & 0.63 \\
 \hline
 \hline
  \end{tabular}
  \caption{Summary of energy scale systematic uncertainty contributions from
    sources common to electrons and photons, estimated for
    unconverted photons with $E_{\rm T}=60$~GeV, in \%.\label{tab:systunc}}
\end{table*}

\begin{table*}
  \centering
  \begin{tabular}{lccccc}
    \hline
    \hline
    &  \multicolumn{5}{c}{Converted photons, $E_{\rm T}=60$~GeV} \\
 $|\eta|$ range                       & 0--0.6 & 0.6--1 & 1--1.37 & 1.55--1.82 & 1.82--2.47 \\
 \hline
 $Z\rightarrow ee$ calibration     & 0.03 & 0.04 & 0.08 & 0.22 & 0.05 \\ 
 Gain, pedestal                    & 0.03 & 0.02 & 0.01 & 0.86 & 0.06 \\ 
 Layer calibration                 & 0.03 & 0.05 & 0.06 & 0.11 & 0.05 \\ 
 ID material                       & 0.03 & 0.10 & 0.09 & 0.15 & 0.05 \\ 
 Other material                    & 0.03 & 0.04 & 0.13 & 0.37 & 0.05 \\ 
 \hline                      
 Total                             & 0.18 & 0.34 & 0.50 & 1.00 & 0.23 \\
 \hline
 \hline
  \end{tabular}
  \caption{Summary of energy scale systematic uncertainty contributions from
    sources common to electrons and photons, estimated for
    converted photons with $E_{\rm T}=60$~GeV, in \%.\label{tab:systconv}}
\end{table*}

The uncertainty model developed and summarised here is expected to be
valid up to $\et\sim 500$~GeV. At these energies, the contribution of the
third calorimeter layer to the energy measurement is enhanced, and a
significant fraction of electrons and photons are recorded in low
gain. These aspects are not addressed in this paper.


\section{Electron calibration cross-checks \label{sec:electroncheck}}

At this point, the electron energy scale is fully specified; after all
corrections are applied, the energy response is expected to be linear
and uniform. The objective of the present section is to test the
extrapolation to different $\et$ regimes using additional probes.

\subsection{Energy scale from $J/\psi\rightarrow ee$\label{sec:jpsi}}

The results presented here are based on the \jpsiee\ sample described
together with the selection criteria in Table~\ref{tab:datamc}, and consisting of about 185K
events. After selections \cite{Aad:2014fxa}, the average electron transverse energy is
$\et=11$~GeV, making this sample a useful test of the energy response
in the low energy range where electrons from $Z$ decays are not available.

The invariant mass distribution of the events selected in the data sample
shows a sizeable background contribution and a hint of the $\psi(2S)$
resonance. To disentangle these contributions, the Monte
Carlo \Jpsi~signal is parameterised by an empirical function taking
into account the Gaussian core of the peak and the non-Gaussian
tails. The parameterised signal is used as input for a fit to the data
to extract the combinatorial background, parameterised as a second-order
polynomial, and the $\psi(2S)$ contribution. The \Jpsi~peak position
is unconstrained, and the $\psi(2S)$ resonance is assumed to be identical
to that of the \Jpsi~signal, up to a scaling of the mass and the
corresponding expected change in resolution.

The events are categorised as a function of electron pseudorapidity as in
Sect.~\ref{sec:zeescales}. For each ($\eta_i , \eta_j$) category,
the electron pair invariant mass PDF, $L_{ij}(m_{ee})$, is built from
the fitted signal and background components. The electron energy scale
factors $\alpha_i$ are extracted using a simultaneous fit using the
likelihood function given in Eq.~\eqref{eq:likelihood}. 

The statistical uncertainty on the electron energy scale extraction
amounts to 0.1\% to 0.2\%, depending on pseudorapidity. The main source
of systematic uncertainty is induced by the imperfect modelling of the
electron isolation. The $J/\psi$ sample results from direct production and from 
$b\rightarrow J/\psi$ decays; in the latter process, the electrons are 
produced in the vicinity of jets, which contribute to the measured electron cluster
energy. The uncertainty on the relative fractions of the two
processes and on the modelling of the jet contribution to the
electron energy contributes an uncertainty of~0.2\%.

\begin{figure}
  \centering
  \includegraphics[width=\columnwidth]{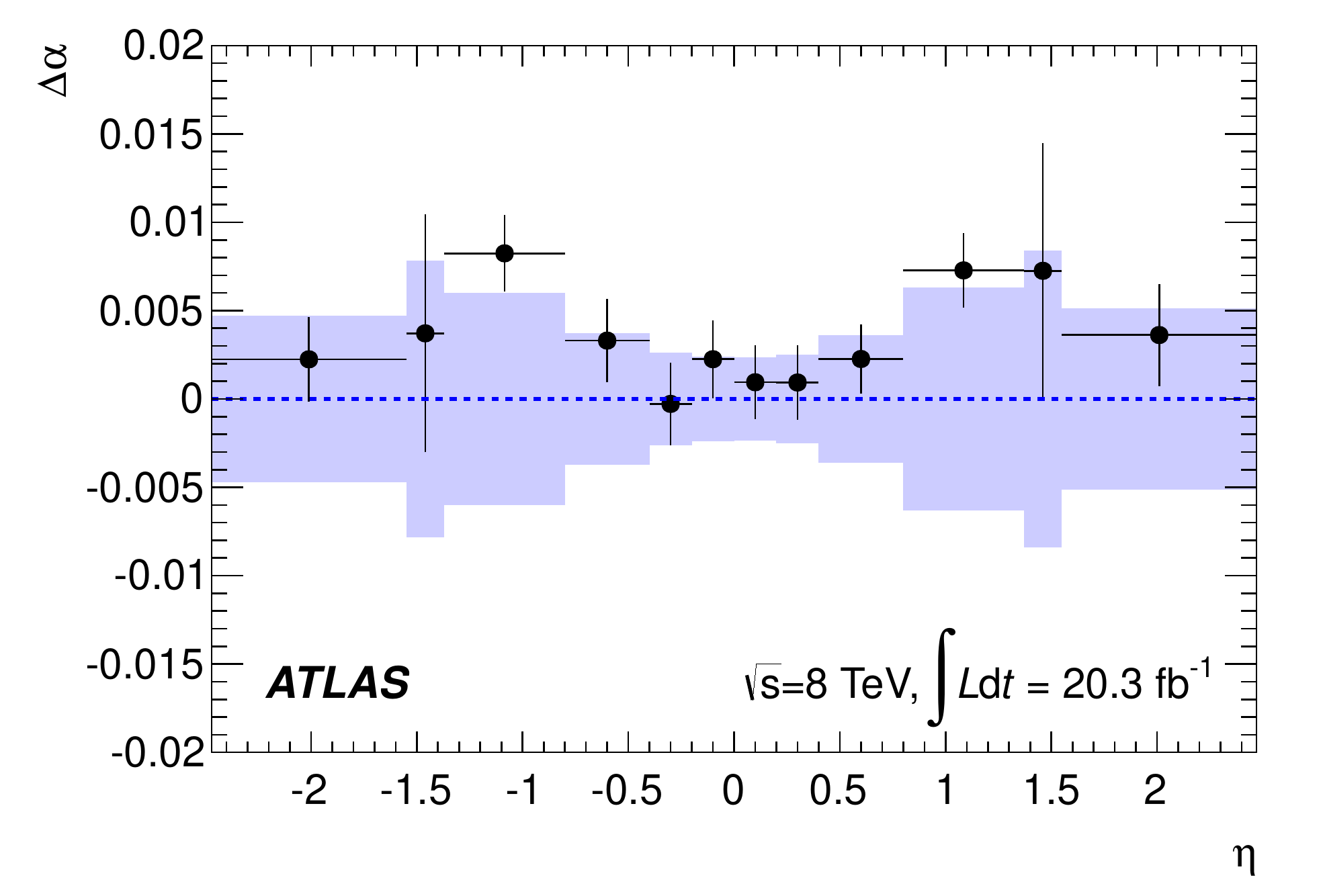}
  \caption{Energy scale factors $\Delta\alpha$ obtained after $Z$-
  based calibration from the \Jpsi~sample, as a function
    of the electron pseudorapidity. The
    band represents the calibration systematic uncertainty. The error bars on the data points
    represent the total uncertainty specific to the
    $J/\psi\rightarrow ee$ analysis.
    \label{fig:jpsi_z_uniformity}}
\end{figure}

\begin{figure}
  \centering
  \includegraphics[width=\columnwidth]{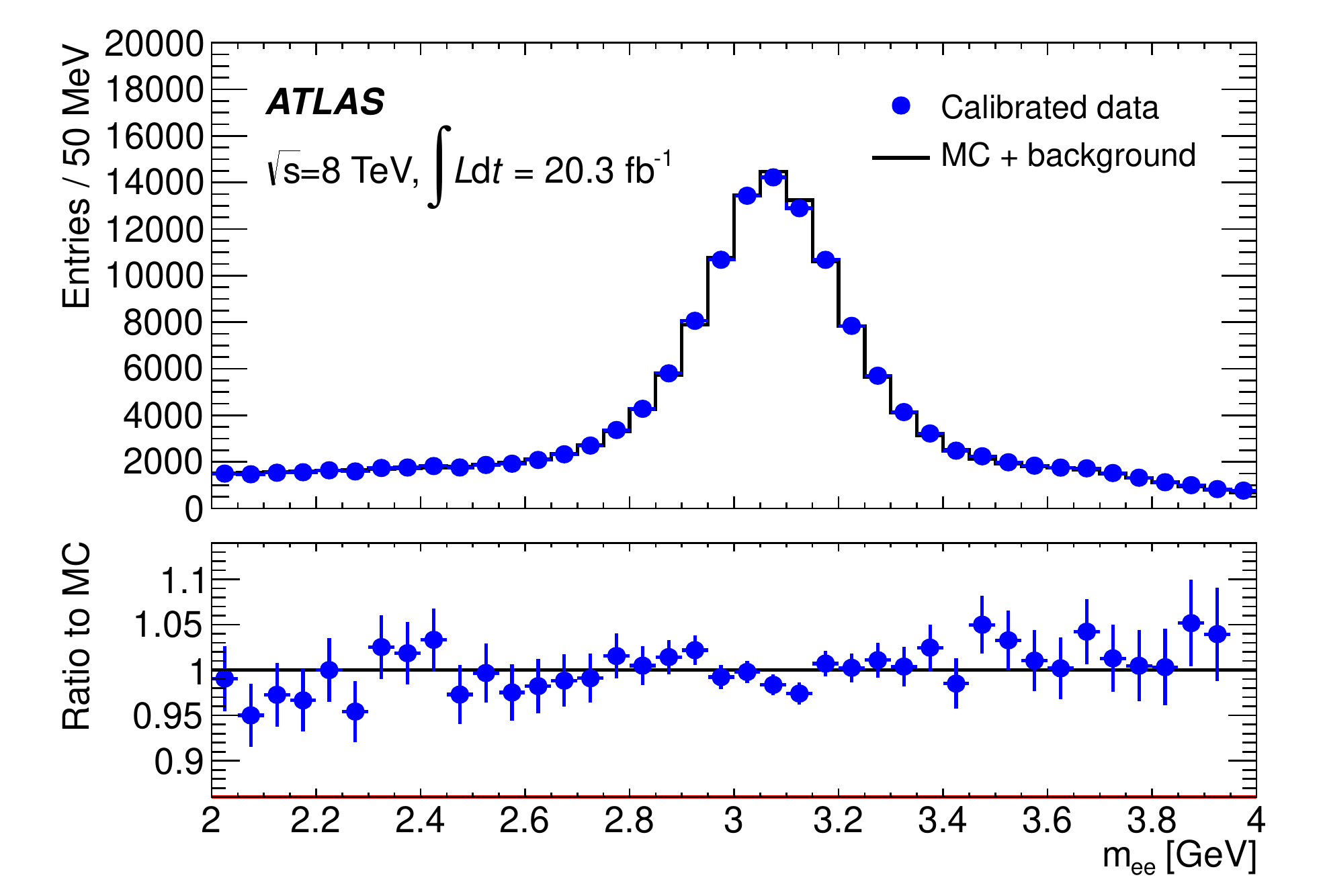}
  \caption{Invariant mass distribution of the electron pair corresponding to the $J/\psi$ selection in data and simulation, after the energy
  corrections described in the text are applied to the electron candidates in data. 
  A background component is determined from data and added
  to the simulated $J/\psi$ resonance. \label{fig:jpsi_datamc}.}
\end{figure}

The fit results are shown in Fig.~\ref{fig:jpsi_z_uniformity}, and are
compared to the expected uncertainties, composed of $Z$ scale
uncertainties, PS and L1/L2 intercalibration, and passive-material
uncertainties extrapolated to $\et=11$~GeV. The uncertainties on the
energy scales determined from the $J/\psi$ sample include the
contributions discussed above. Satisfactory agreement is obtained,
although the $J/\psi$ results tend to be higher than the $Z$ ones by
about one standard deviation. The residual differences in the central
region can be explained by an imperfect calibration of the cell
response, associated to a difference of the read-out pedestals in
physics and  calibration runs. This small bias is understood as
resulting from a different setting of the electronics
configuration used in both cases. In the region $1<|\eta|<1.82$, the differences are most
probably related to residual uncertainties in the detector material
description. Figure~\ref{fig:jpsi_datamc} shows the electron pair
invariant mass distribution in data and MC simulation, and the data/MC
ratio as a function of $m_{ee}$ after energy corrections. The
corrected data and the simulation agree within uncertainties across
the mass window used in the analysis.

\subsection{Energy linearity in $Z$ events and overall electron calibration accuracy \label{sec:electronscale}}

\begin{figure*}
  \centering
  \includegraphics[width=\columnwidth]{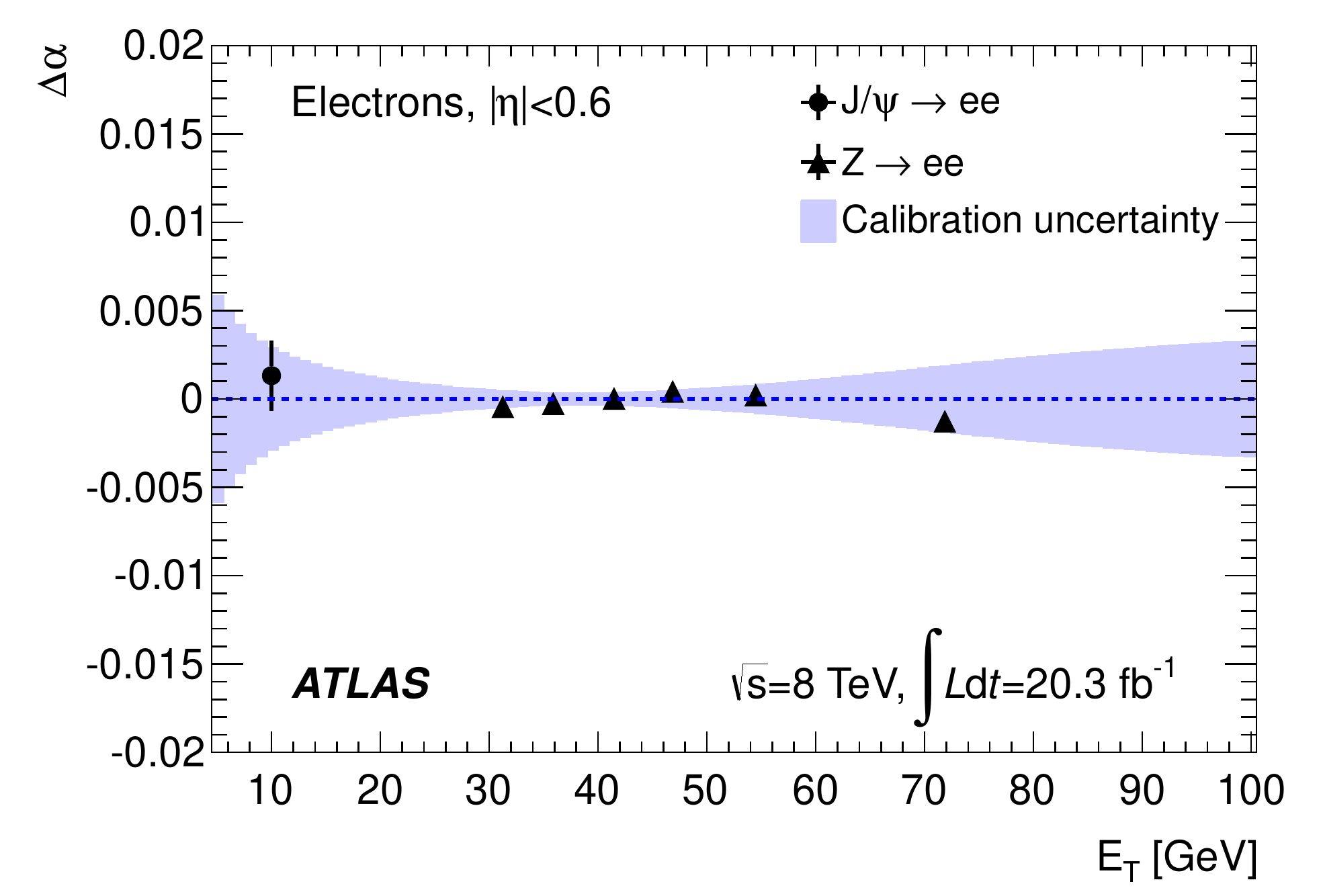}
  \includegraphics[width=\columnwidth]{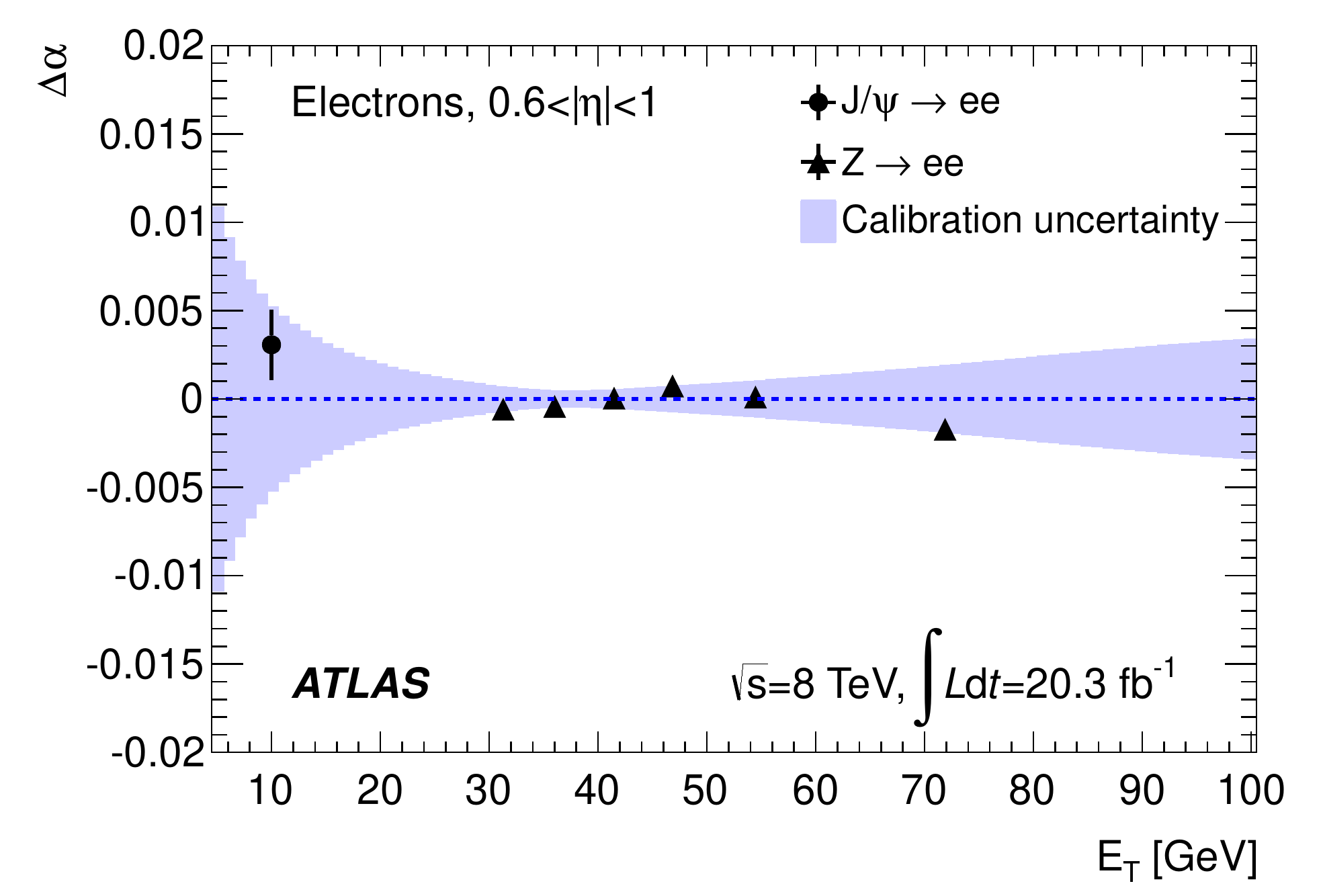} \\
  \includegraphics[width=\columnwidth]{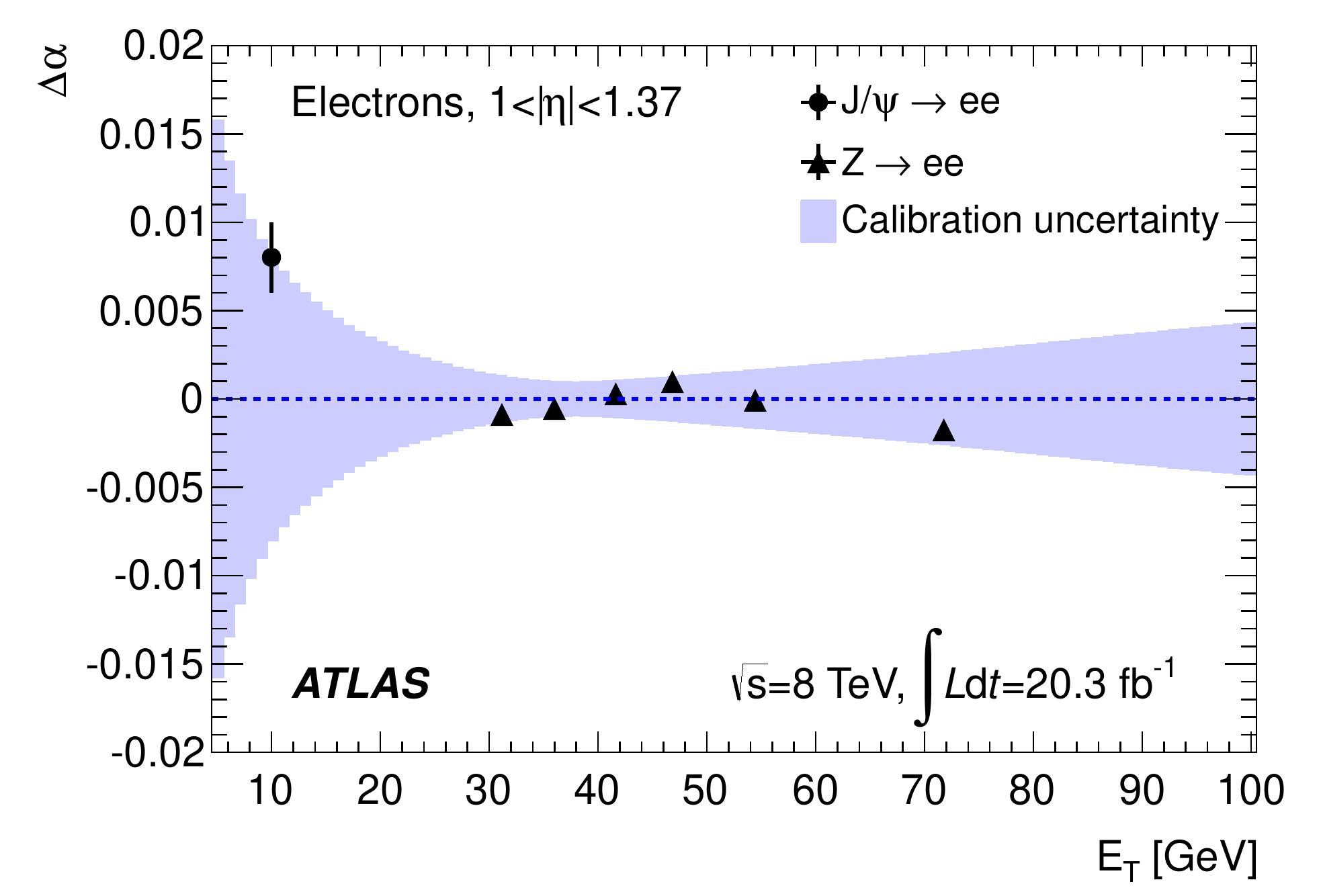}
  \includegraphics[width=\columnwidth]{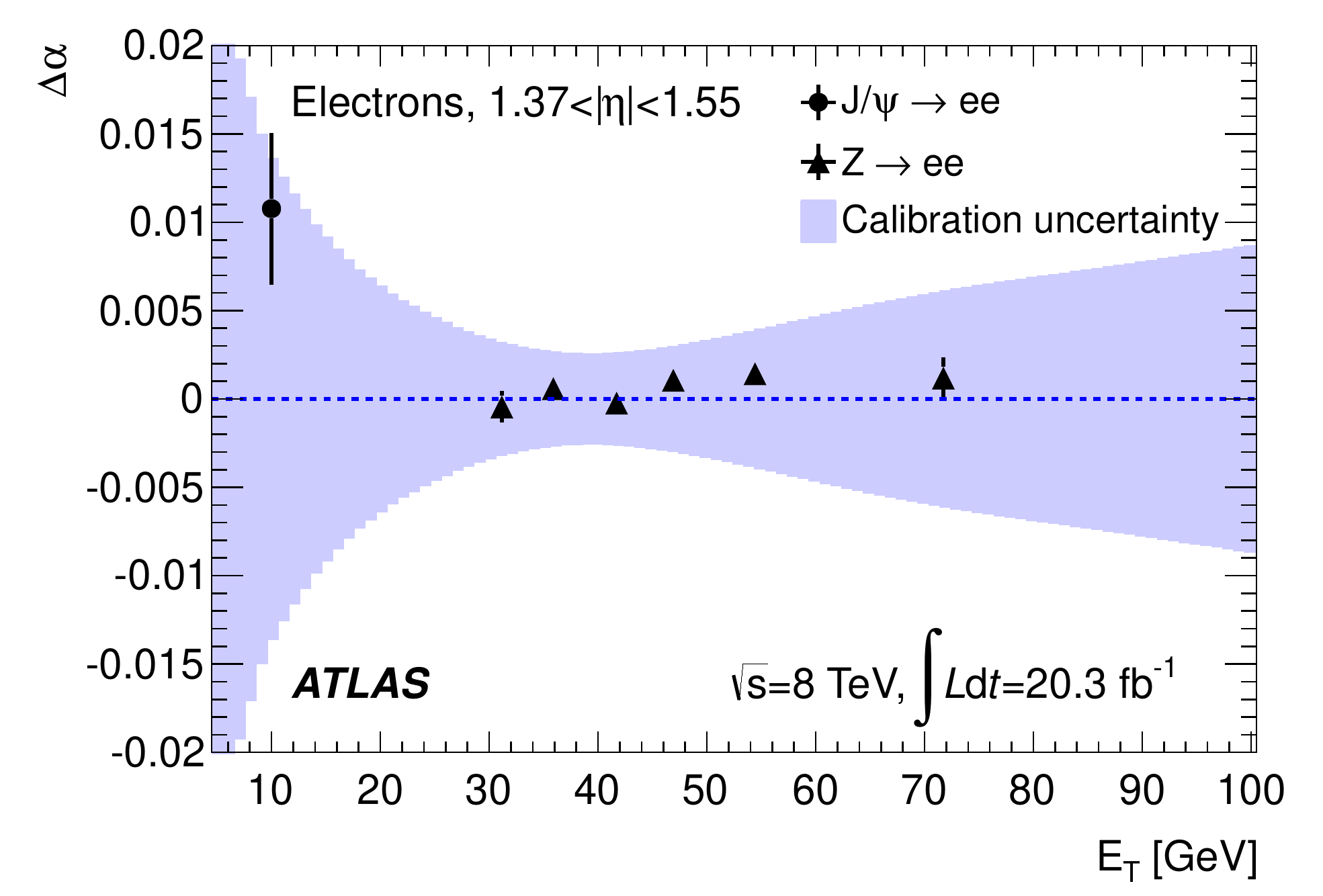} \\
  \includegraphics[width=\columnwidth]{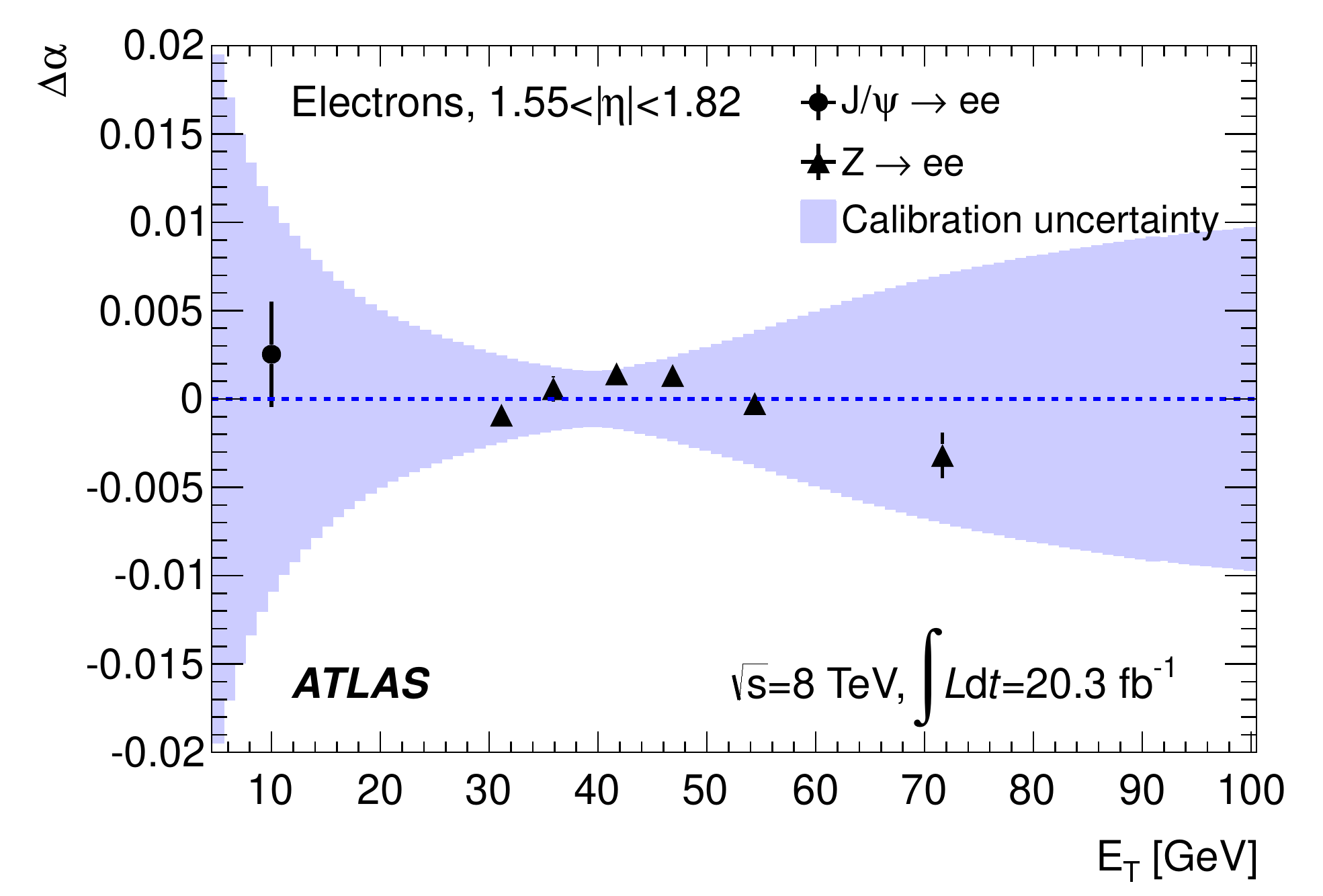} 
  \includegraphics[width=\columnwidth]{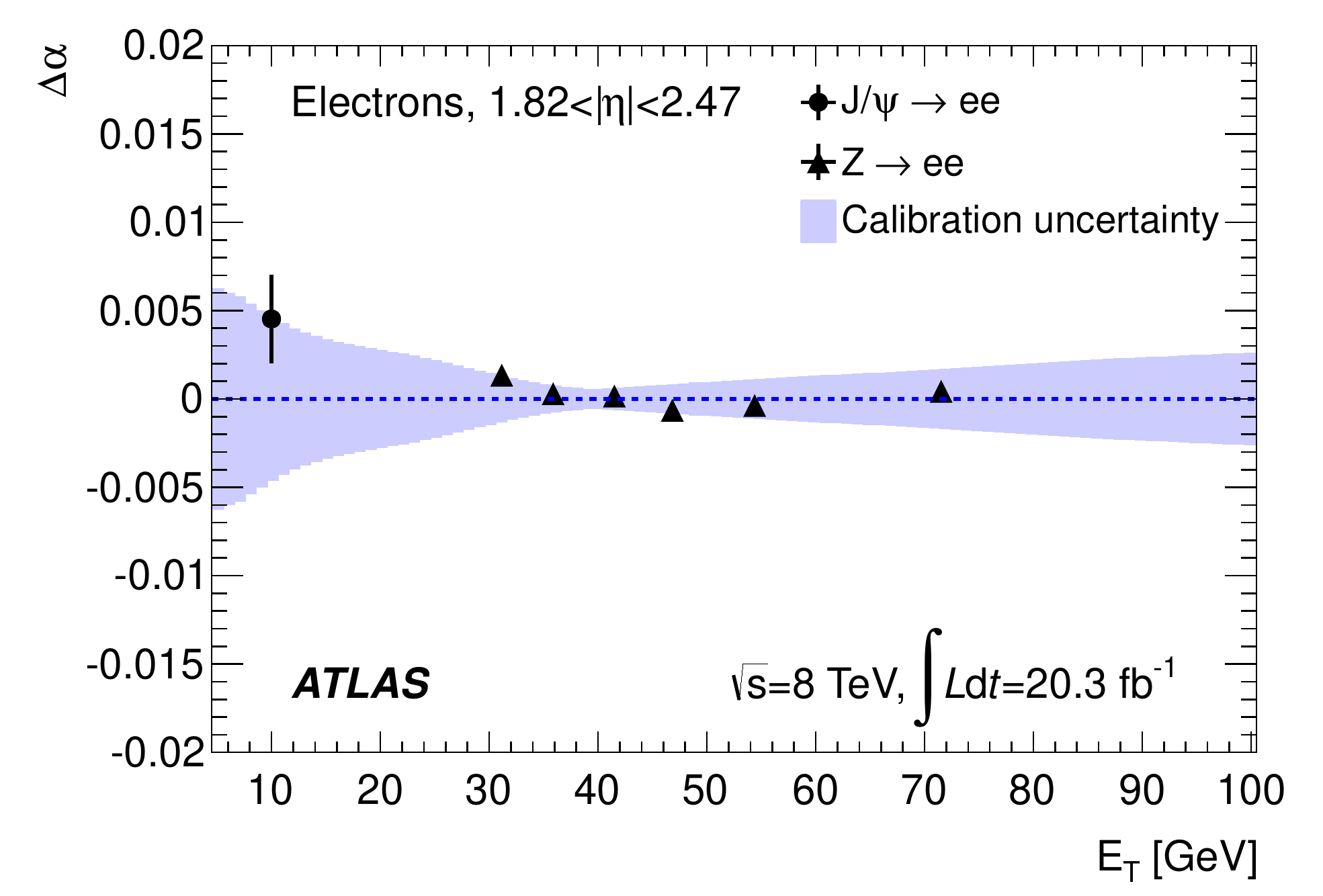}
  \caption{Energy scale factors $\Delta\alpha$ obtained after $Z$-based calibration from the \Jpsi\ and the
  \et-dependent $Z\rightarrow ee$ analyses, as function of \et\ in different pseudorapidity bins. The band represents the calibration
  systematic uncertainty. The error bars on the data points represent the total
  uncertainty specific to the cross-checking analyses.\label{fig:jpsi_lin}}
\end{figure*}

Finally, a study of the energy dependence of the calibration is
performed. The $Z\rightarrow ee$ and $J/\psi\rightarrow ee$ analyses of
Sects.~\ref{sec:zeescales},~\ref{sec:jpsi} are repeated, now 
categorising the electrons in broad intervals of $|\eta|$, with the following boundaries:
\begin{itemize}
\item $|\eta|$ : 0 - 0.6 - 1 - 1.37 - 1.55 - 1.82 - 2.47.
\end{itemize}
In addition, the $Z\rightarrow ee$ sample is subdivided in electron \et\ intervals:
\begin{itemize}
\item $\et$ : 27 - 35 - 42 - 50 - 100~GeV.
\end{itemize}
The analysis is performed after applying all corrections derived
above, so that the energy scale corrections derived here are expected to be close
to zero and constant. The results are shown in
Fig.~\ref{fig:jpsi_lin}. In all cases, the resulting energy scales
lie within the calibration systematic uncertainty envelopes. 

\section{Photon-specific uncertainties \label{sec:photonsys}}

\subsection{Conversion reconstruction inefficiency and fake conversions}
\label{sec:convFR}

  \begin{table*}[t]
    \centering
    \begin{tabular}{ccccc}
      \hline\hline
      Uncertainty & $|\eta| < 0.6$ & $0.6 \leq |\eta| < 1.37$ & $1.52 \leq |\eta| <1.81$ & $1.81 \leq |\eta| <2.37$ \\
      \hline
     Inefficiency & $0.02$ & $0.03$ & $0.10$ & $0.02$ \\
     Fake Rate & $0.01$ & $0.06$ & $0.06$ & $0.03$ \\ 
     \hline\hline
    \end{tabular}
    \caption{Impact on the energy scale of unconverted (converted) photons from the additional inefficiency (fake rate) in four pseudorapidity bins, in \%.}
    \label{tab:convFR}
  \end{table*}

  \begin{table*}
    \centering	
    \begin{tabular}{cccc}
      \hline\hline
      Particle type & $|\eta| < 0.8$ & $0.8 \leq |\eta| < 1.37$ & $1.52 \leq |\eta| < 2.37$ \\
      \hline
      $\Delta(\gamma-e)$, converted    & $0.16 \pm 0.11$ & $0.46 \pm 0.10$ & $0.19 \pm 0.10$ \\
      $\Delta(\gamma-e)$, unconverted  & $0.03 \pm 0.04$ & $0.10 \pm 0.06$ & $0.05 \pm 0.04$  \\
      \hline\hline
    \end{tabular}
    \caption{Difference between out-of-cluster energy loss for electrons and photons, $\Delta(\gamma-e)$, in \%.}  
    \label{tab:LeakagePhoton:Results}
  \end{table*}

The fraction of photons that convert to electron--positron pairs before
reaching the calorimeter is directly connected to the amount of
material upstream. The efficiency to reconstruct the corresponding
tracks and match them to clusters is close to unity for convers\-ions in
the innermost layers of the detector and drops at larger radii,
in the region instrumented by the TRT. The energy of true converted
photons reconstructed as unconverted is typically underestimated by
about 2\%, depending on \pt, $\eta$ and the radius of the
conversion. On the other hand, wrong associations between tracks
induced by pile-up interactions or fake tracks and clusters lead to
``fake conversions'' that induce around 2\% overestimation of the
energy. This effect is also more frequent for tracks reconstructed in
the TRT due to the imprecise measurement of $\eta$. 

Both effects impact the absolute photon energy scale if the
efficiency and fake rates are imperfectly described by the simulation. Systematic uncertainties associated with these quantities were estimated by
comparing the conversion rates in data and MC simulation, and using a template
method based on the ratio \Eonetwo. The latter exhibits a different
behaviour for photons that do or do not convert before the calorimeter
and therefore is sensitive to the ``true'' conversion status of the
photon. By combining the true and reconstructed conversion status, one
can determine the reconstruction efficiency and fake rate in data. The
method can also provide an estimate of the material upstream and could
be used in the future to constrain this quantity. 

The study was performed in four $|\eta|$ bins (0--0.6, 0.6--1.37,
1.52--1.81, 1.81--2.37) using the same event selection as typically adopted by the
$H \rightarrow \gamma\gamma$ analysis \cite{mHpaper}. Events with two photon
candidates satisfying tight identification criteria based on
calorimeter shower shapes and both track- and calorimeter-based
isolation were considered. The ratio of the transverse momenta of the
photons to the diphoton invariant mass, $\pt/\mgg$, was required to
be above 0.35 and 0.25 for the leading and subleading photons,
respectively, with $105$~GeV~$< \mgg <$~160~GeV.  
The study was limited to the leading photon of each event in order to
limit contamination by jets misidentified as photons, estimated to
be $\sim 10$\%. This jet background is subtracted in each \Eonetwo\ bin using a sideband method based on the identification and isolation
criteria \cite{InclusivePhoton}. The contribution of Drell--Yan events was estimated to be
$\sim 0.3$\% using MC simulation. Systematic uncertainties associated with imperfect knowledge of \Eonetwo\ and the material upstream were propagated to
the templates and the expected conversion rate in each bin. 

The results point to inefficiencies and fake rates that exceed by up
to a few percent the  predictions from MC simulation. The impact on the energy
scale of unconverted (converted) photons from the additional
inefficiency (fake rate) is shown in Table~\ref{tab:convFR}. It is typically around few $10^{-4}$ depending on $\pt$ and $\eta$, and reaches $10^{-3}$ in the bin $1.52 < |\eta| < 1.81$ around $\et = 60$~GeV. 

\subsection{Lateral leakage mis-modelling}

Electrons and photons deposit about 6\% of their energy outside of the
cluster used in the reconstruction, depending on pseudorapidity and
particle type. Although this effect is to first order taken into
account by the MC calibration, a calibration bias could appear in the case of
imperfect modelling of lateral shower development, as the energy loss
would be different in data and simulation. 

The energy scale factors obtained from the $Z$-based effective
calibration described in Sect.~\ref{sec:zeescales} absorb such
discrepancies. The energy dependence of the difference between data and simulation for electron lateral leakage was investigated and found to be negligible. Therefore the only component yet to be determined is the
mis-modelling difference between electrons and photons.

The $Z\rightarrow \ell\ell\gamma$ and $Z \rightarrow ee$ samples are used to estimate this difference. The energy of the decay electrons and radiative photons in the nominal cluster size is compared to the
energy found in a larger window of size $\Delta\eta\times\Delta\phi = 7 \times 11$ cells defined around it, using only cells in L2. The normalised difference between electrons and photons, 
\begin{eqnarray}
\Delta(\gamma-e) & = & \left(\frac{E_{7\times 11} - E_\mathrm{nom}}{E_\mathrm{nom}}\right)^\mathrm{data} \nonumber \\
                 & & \qquad - \left(\frac{E_{7\times 11} - E_\mathrm{nom}}{E_\mathrm{nom}}\right)^\mathrm{MC} \quad,
\end{eqnarray}
is estimated for three pseudorapidity intervals, separating unconverted and converted
photons.

The effect of the mis-modelling of photon conversion reconstruction is
tested by correcting the fraction of converted photons in simulation
according to the results of Sect.~\ref{sec:convFR}, and
$\Delta(\gamma-e)$ is evaluated with and without this
correction. Table \ref{tab:LeakagePhoton:Results}  shows the most
conservative values obtained for each bin and photon type according to
this procedure. 

The most significant effects found are $(0.46\pm 0.10)$\% for converted photons within $0.8<|\eta|<1.37$, and $(0.10 \pm 0.06)$\% for unconverted photons in
the same pseudorapidity bin. The systematic uncertainty on the photon
energy scale related to this effect is defined in each bin as the
larger of $\Delta(\gamma-e)$ and its uncertainty. This uncertainty is
assumed to not depend on the photon transverse energy.


\section{Photon calibration cross-checks \label{sec:photoncheck}}

\begin{figure*}
  \centering
  \includegraphics[width=0.49\textwidth]{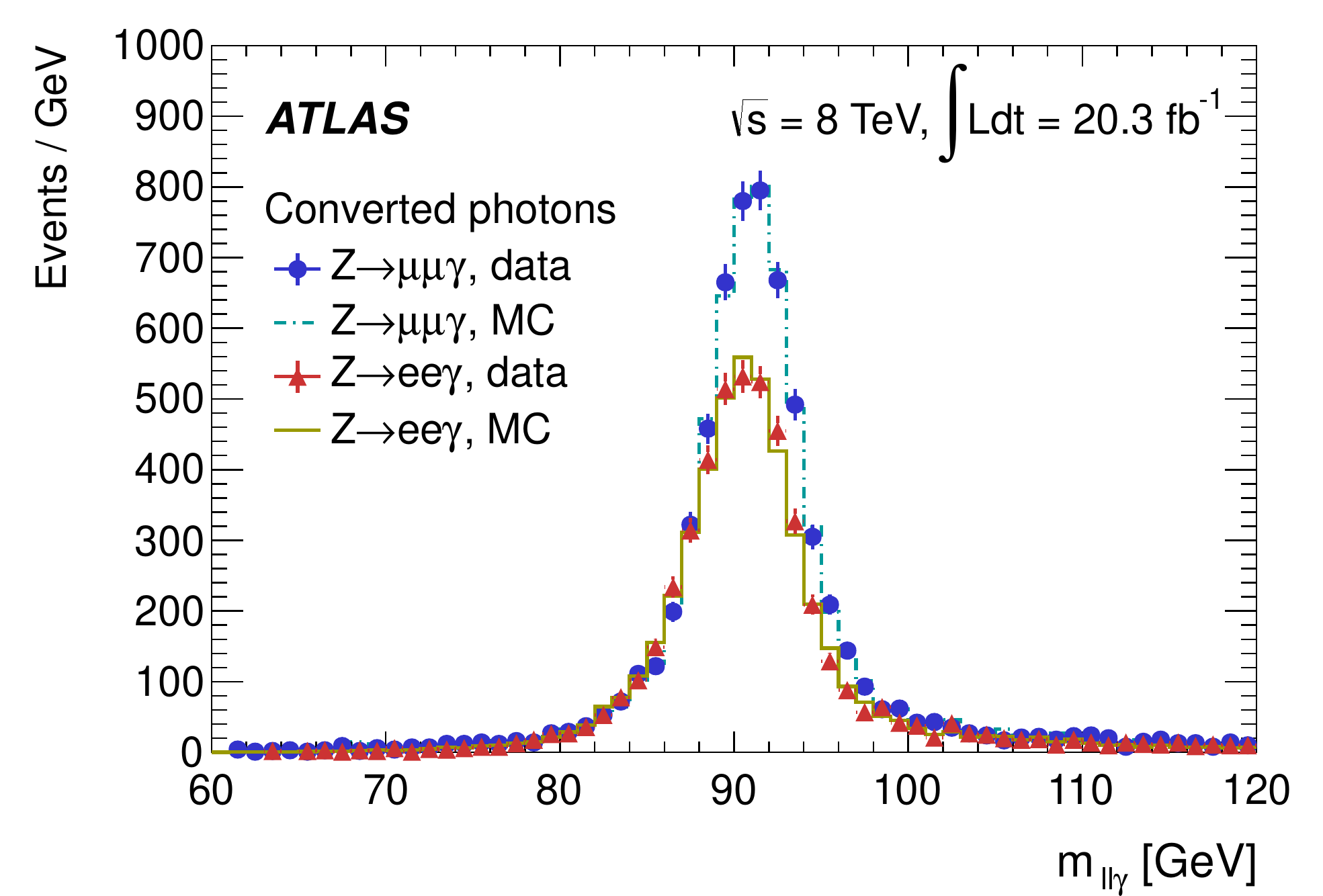}
  \includegraphics[width=0.49\textwidth]{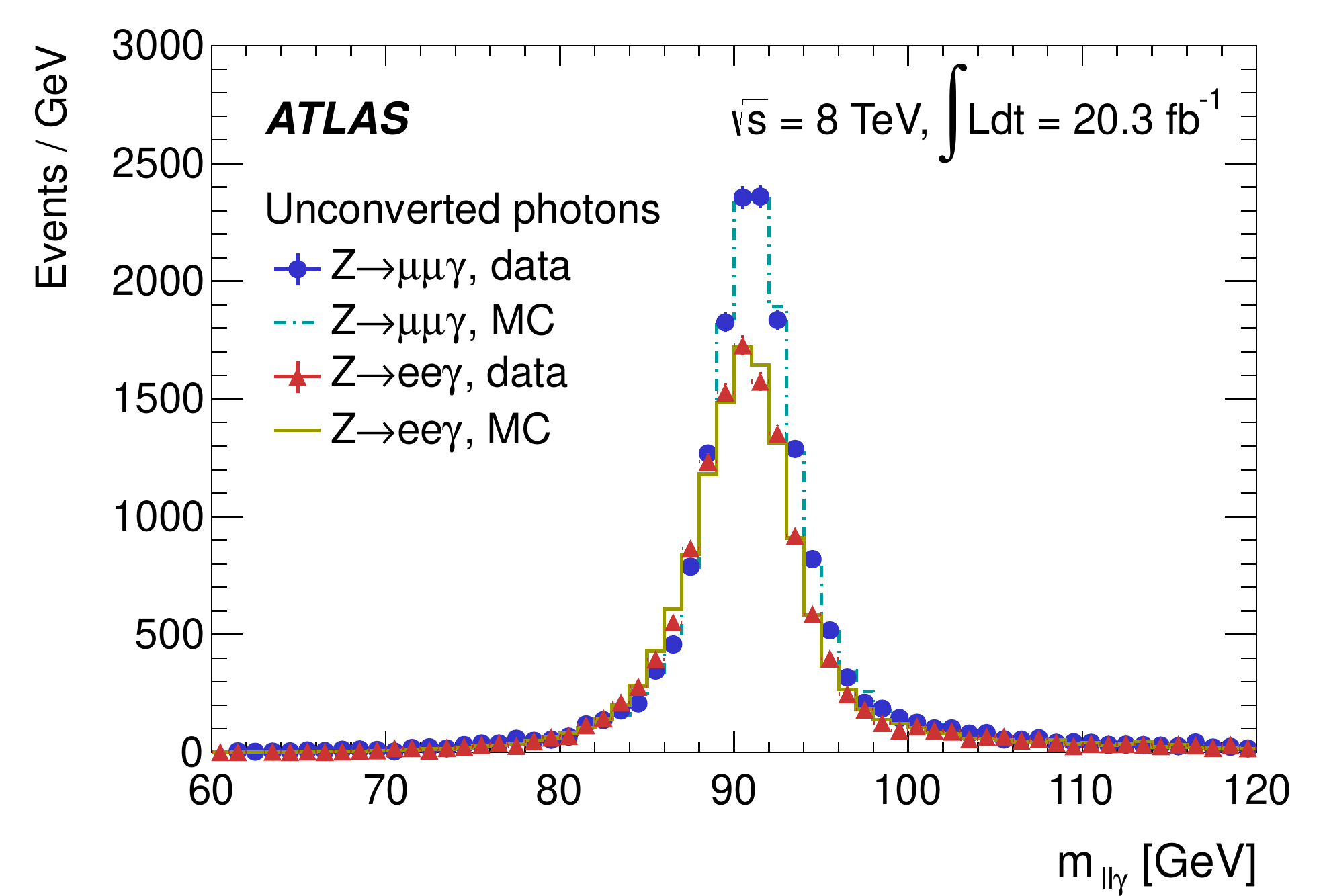}
  \caption{Invariant mass distributions in data and simulation, for large-angle
    $Z\rightarrow\ell\ell\gamma$ events with converted (left) and
    unconverted photons (right) in the electron and muon
    channels, as described in the legend, for $\Delta R(\ell,\gamma)>0.4$ and $E_{\rm T}^\gamma>15$~GeV. Energy corrections are applied. The MC simulation is normalised to
    the number of events in data. \label{fig:RadiativeMassDist}} 
\end{figure*}

The energy scale factors extracted in Sect.~\ref{sec:zeescales}
are expected to be valid for electrons and photons. Their universality
is tested using photons from radiative $Z$ decays in the electron and
muon channels, separately for unconverted, one-track and two-track converted photons. 
A first selection requires a large-angle separation between the radiative
photon and the leptons and is applied in the electron and muon
channels, and for converted and unconverted photons; a sample of
unconverted collinear photons is also selected in the muon
channel \cite{ATLAS:2012vua}. The event selection is detailed in Table~\ref{tab:datamc}; the photon purity in the various samples is estimated to range
between 97\% and 99\%. 
Figure~\ref{fig:RadiativeMassDist} shows the three-body invariant mass distributions for converted and unconverted
photons in data and MC simulation for large-angle $Z\rightarrow\ell\ell\gamma$ events in the electron and muon channels after all energy corrections are applied. 

Residual mis-calibrations between data and MC simulation are parameterised
following $E_i^{\rm data}=(1+\alpha_i) E_i^{\rm MC}$, similar to the procedure applied to electrons from $Z$ decays in Sect.~\ref{sec:zeescales}. Here, $E_i^{\rm MC}$ and $E_i^{\rm data}$ are
photon energies in region $i$ for MC simulation and data respectively, and
$\alpha_{i}$ measures the residual photon energy mis-calibration. 
For each $\alpha_{i}$ applied to data, the three-body invariant mass $m(\ell\ell\gamma(\alpha_i))_{\rm data}$ is recomputed; its agreement with MC simulation is quantified using a double ratio method,

\begin{equation}
R(\alpha_i)= \frac{\left<m(\ell\ell\gamma(\alpha_i))_{\rm data}\right>/\left<m(\ell\ell)_{\rm data}\right>}{\left<m(\ell\ell\gamma)_{\rm MC}\right>/\left<m(\ell\ell)_{\rm MC}\right>},
\end{equation}

 where $\left<m(\ell\ell\gamma)\right>$ and $\left<m(\ell\ell)\right>$
 are the mean values of the three-body and two-body invariant masses
 in the radiative and non-radiative samples, respectively. Taking the ratio
 of $\left<m(\ell\ell\gamma(\alpha_i))\right>$ to $\left<m(\ell\ell)\right>$ in the $R(\alpha_i)$ numerator suppresses the lepton
 energy scale uncertainties; normalising this ratio to the MC 
 expectation removes possible biases due to the different lepton
 kinematics in \Zll~and \Zllgamma~events. The value of $\alpha_i$ that
 provides the best agreement in the distributions with $R(\alpha_i)=1$
 defines the photon energy scale. The photon energy scales are
 separately derived for non-collinear $Z\rightarrow ee\gamma$ and
 \Zmumugamma~events for both unconverted and converted photons, while
 collinear \Zmumugamma~events are only used for unconverted
 photons. The energy scales from the different event topologies are
 then combined.

\begin{figure*}
  \centering
  \includegraphics[width=0.49\textwidth]{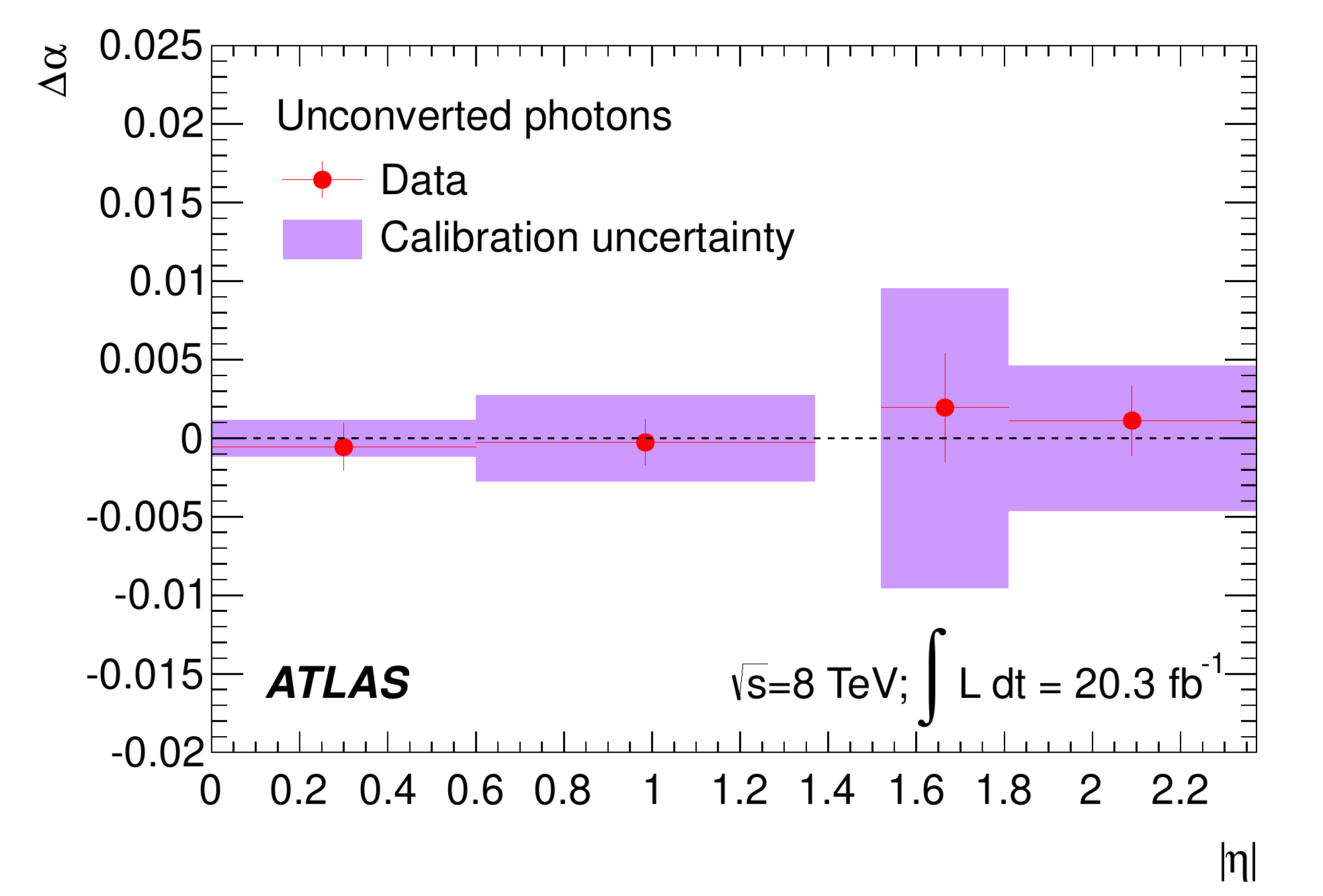}
  \includegraphics[width=0.49\textwidth]{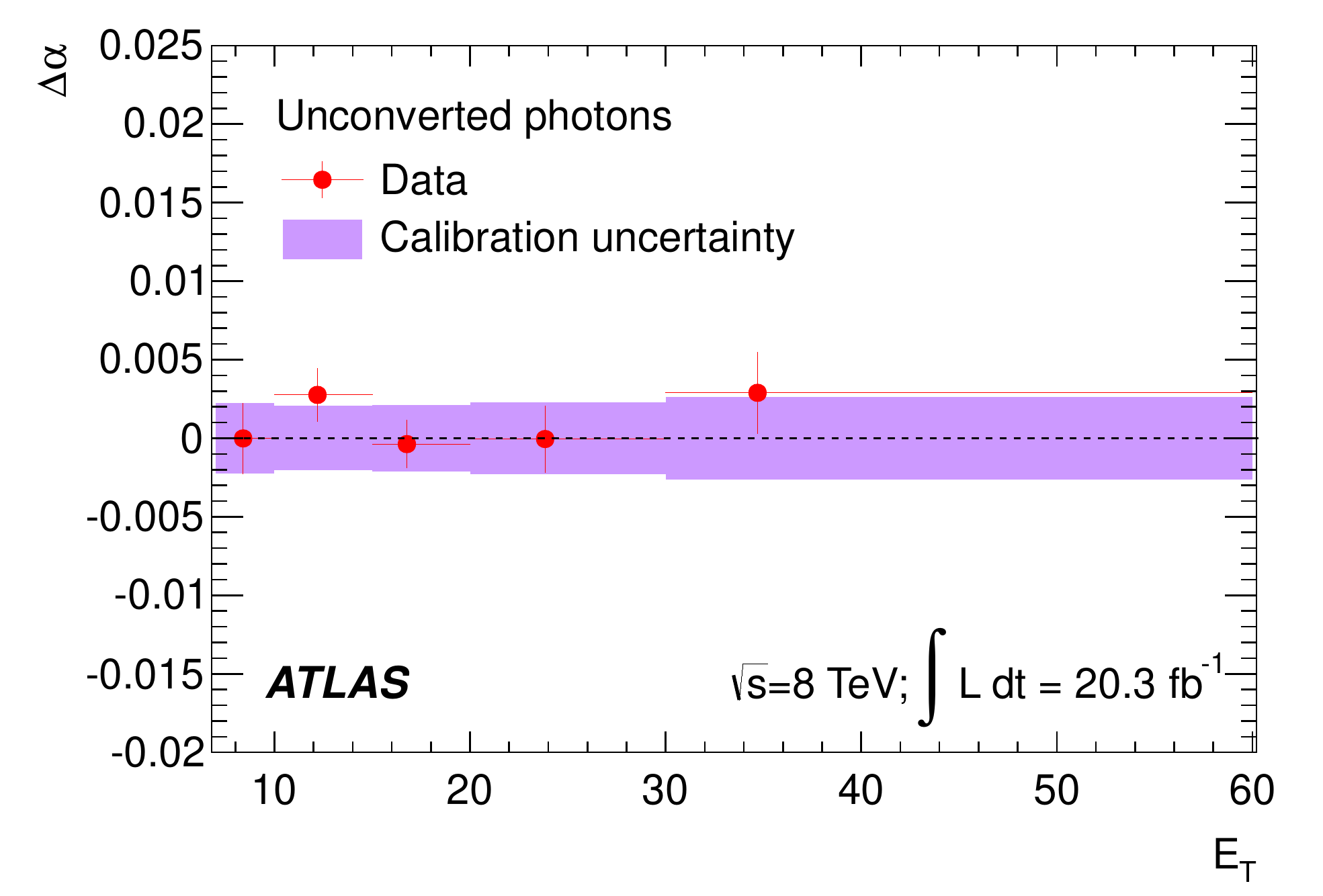} \\
  \includegraphics[width=0.49\textwidth]{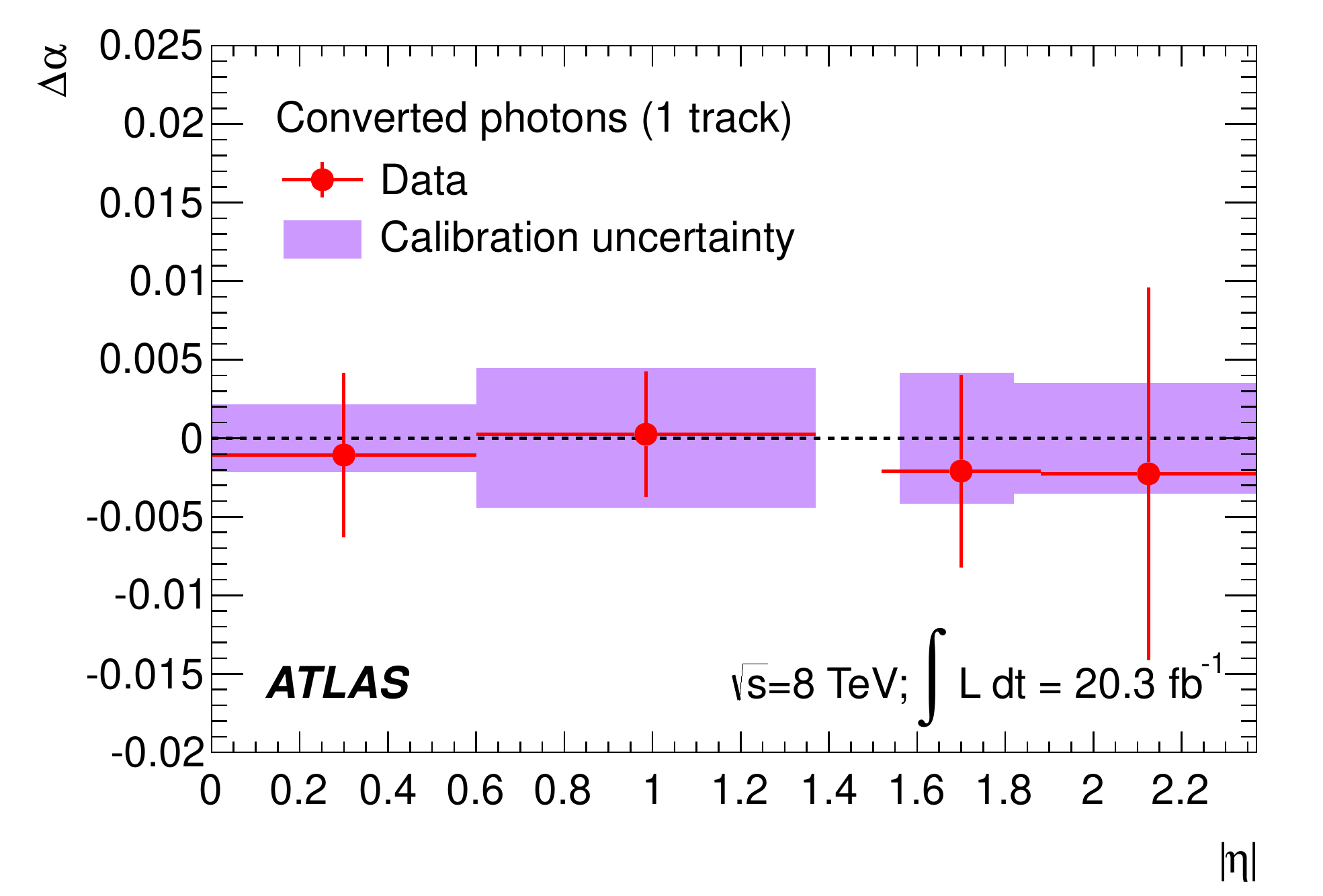}
  \includegraphics[width=0.49\textwidth]{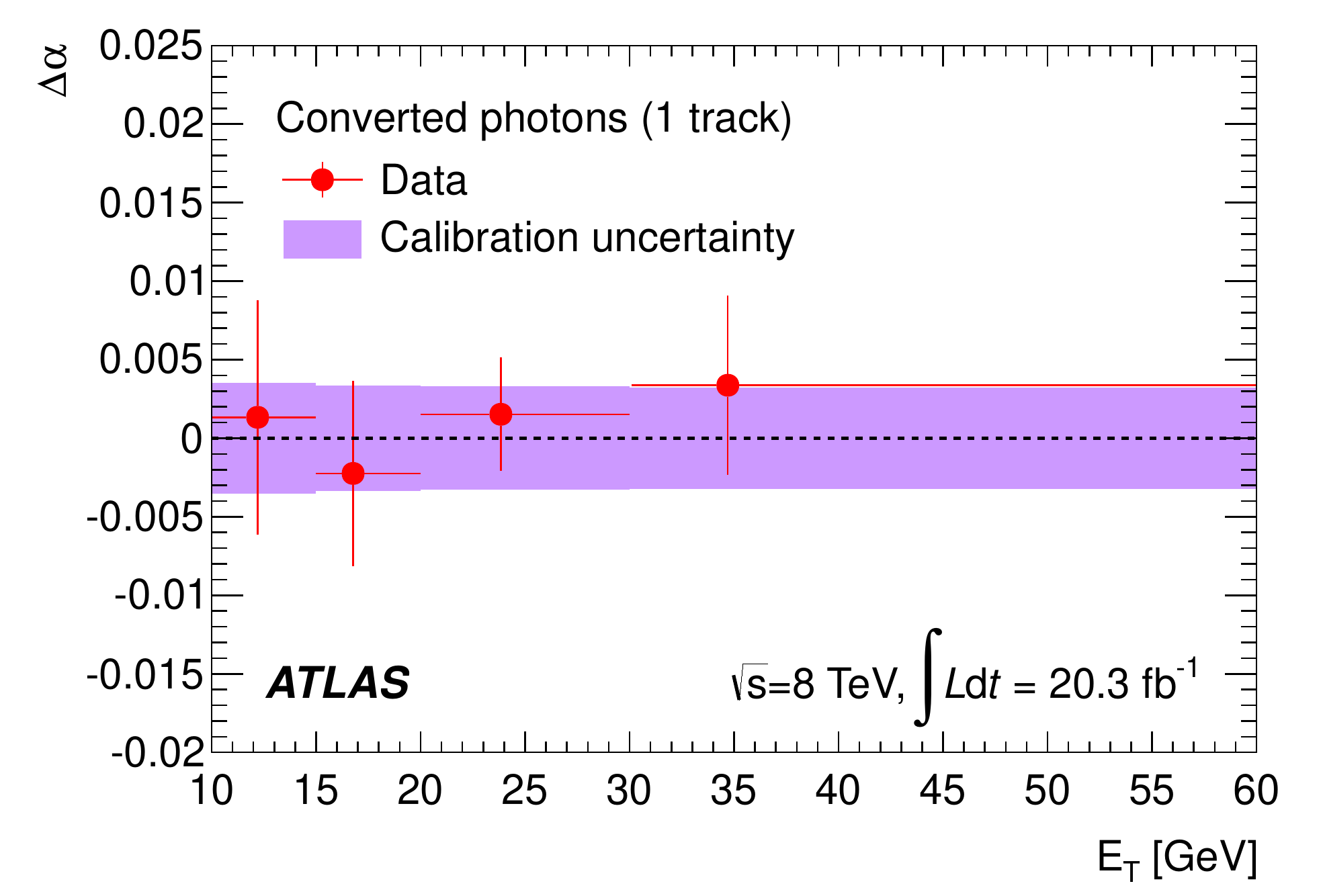} \\
  \includegraphics[width=0.49\textwidth]{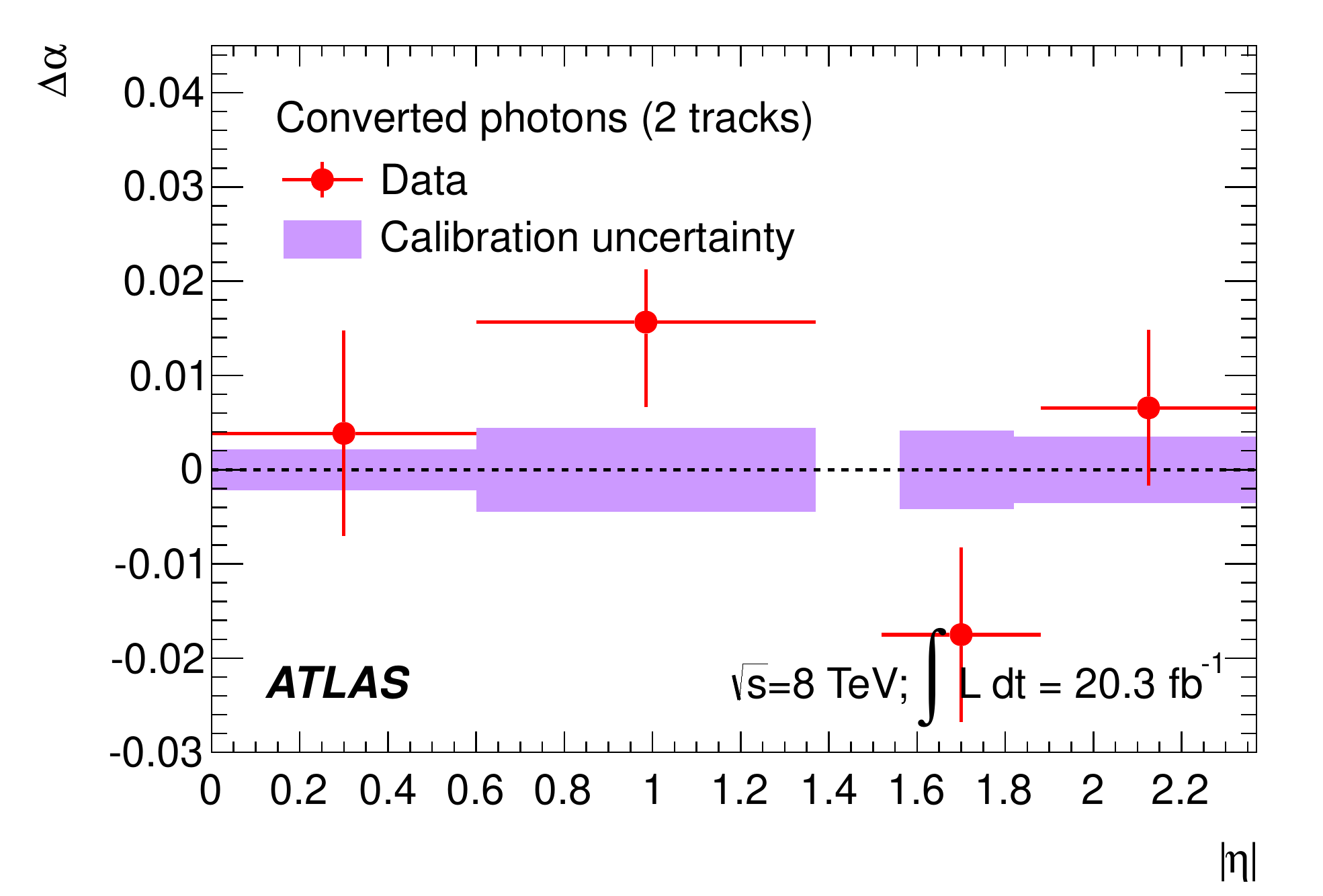} 
  \includegraphics[width=0.49\textwidth]{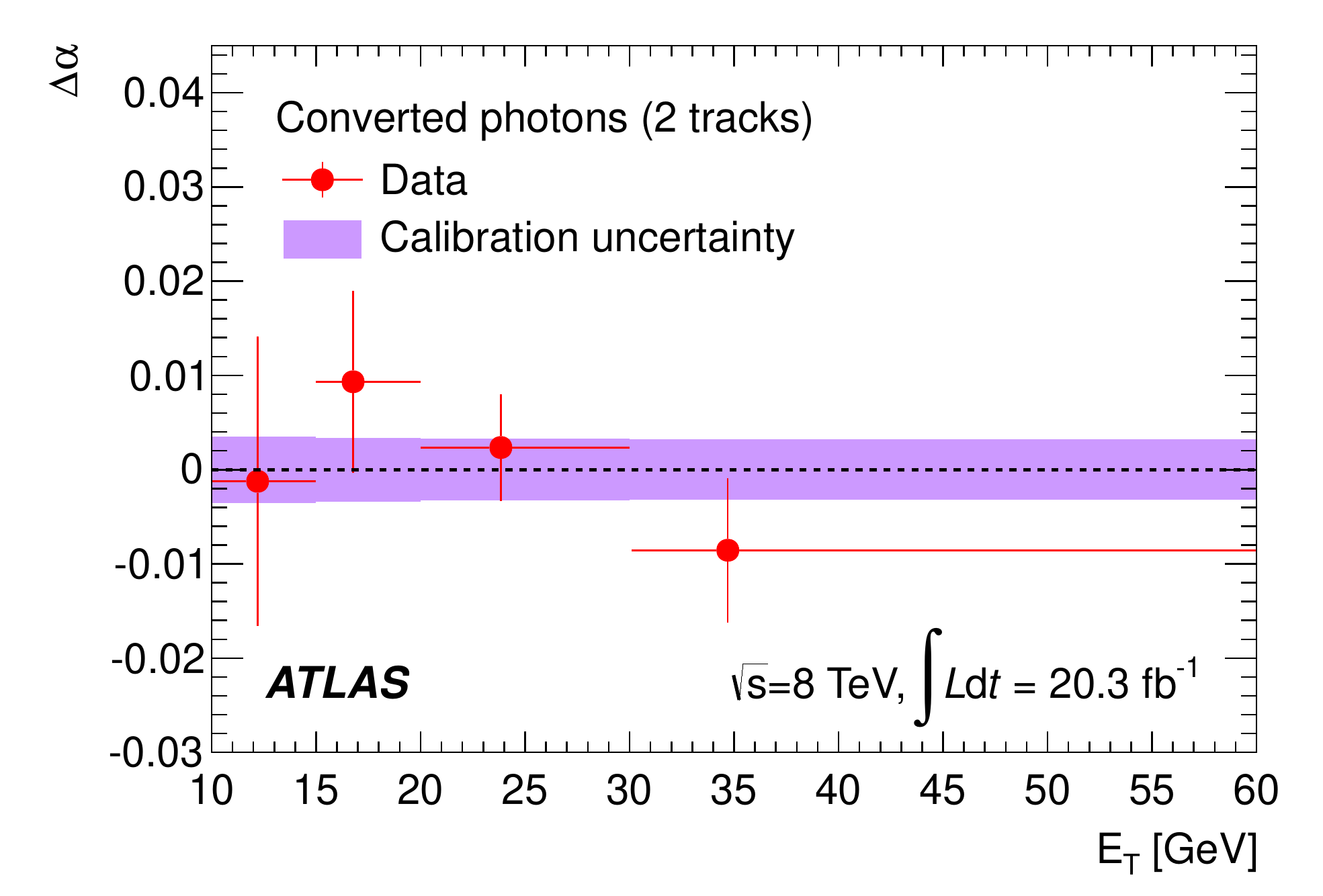}
  \caption{Combined photon energy scale factors $\Delta\alpha$ obtained
    after $Z$-based calibration as a function of $|\eta|$ (left) and
    $E_{\mathrm{T}}$ (right), for unconverted, one-track converted and two-track converted
    photons. The band represents the calibration systematic 
    uncertainty. The error bars on the data points
    represent the total uncertainty specific to the
    $Z\rightarrow\ell\ell\gamma$ analyses.\label{fig:CombFSRetaUnconv}}
\end{figure*}

Several sources of systematic uncertainty are considered in this study: background contamination, fit range, muon momentum scale and resolution in the \zmumu~ channel and electron energy scale and resolution in the \zee~ channel. 
The total systematic uncertainty is of the order of $0.1$\% while the statistical
uncertainty ranges between 0.2\% and 1.5\% depending on the pseudorapidity and on the photon conversion type.
%
%
Figure~\ref{fig:CombFSRetaUnconv} shows the combined photon energy scales as a
function of both $\eta$ and \et, separately for unconverted, single- and double-track converted photons. The bands around zero represent
the calibration systematic uncertainty, including contributions
discussed in Sects.~\ref{sec:uncsum} and~\ref{sec:photonsys}. The measured photon energy
scales agree with the expectation within uncertainties.

\section{Resolution accuracy\label{sec:resolution}}

The main way to probe the resolution in data is provided
by the study of the $Z$ resonance width, which provides a constraint on the total
resolution at given $\eta$ and for $\langle\etZ\rangle\sim 40$~GeV, the average
transverse energy of electrons from $Z$ decays. The resolution
corrections $c$ are derived in Sect.~\ref{sec:zeescales} as an
effective constant term to be added in quadrature to the expected
resolution. However, as is the case for the energy scales, $c$ absorbs the
potential mis-modelling of the resolution sampling term, the electronics noise 
term, the asymptotic resolution at high energy, and the effect of passive material upstream of the
calorimeter. Uncertainties related to these sources thus reappear when
considering different energies or particle types. 

\begin{figure*}
       \centering
       \includegraphics[width=\columnwidth]{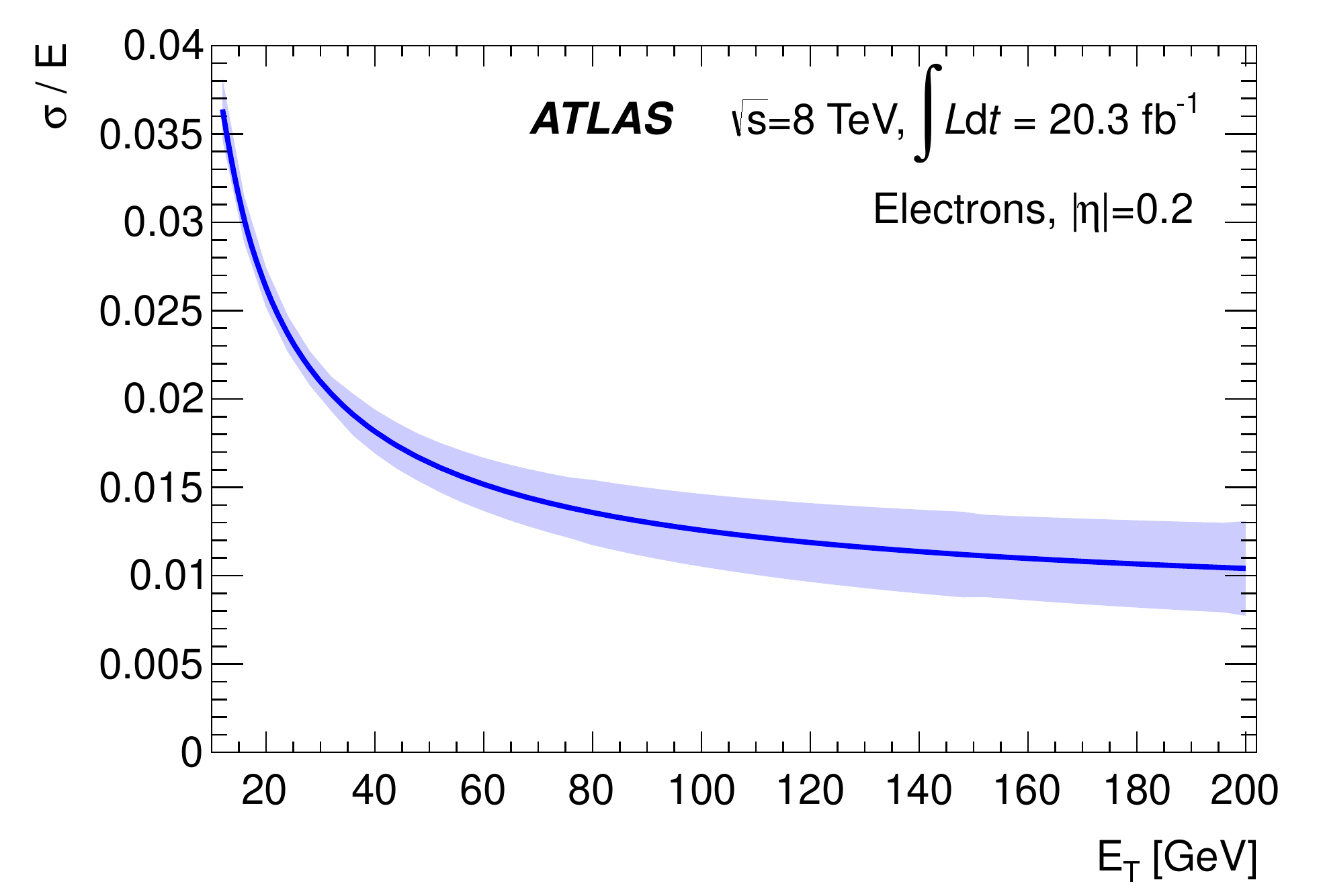}
       \includegraphics[width=\columnwidth]{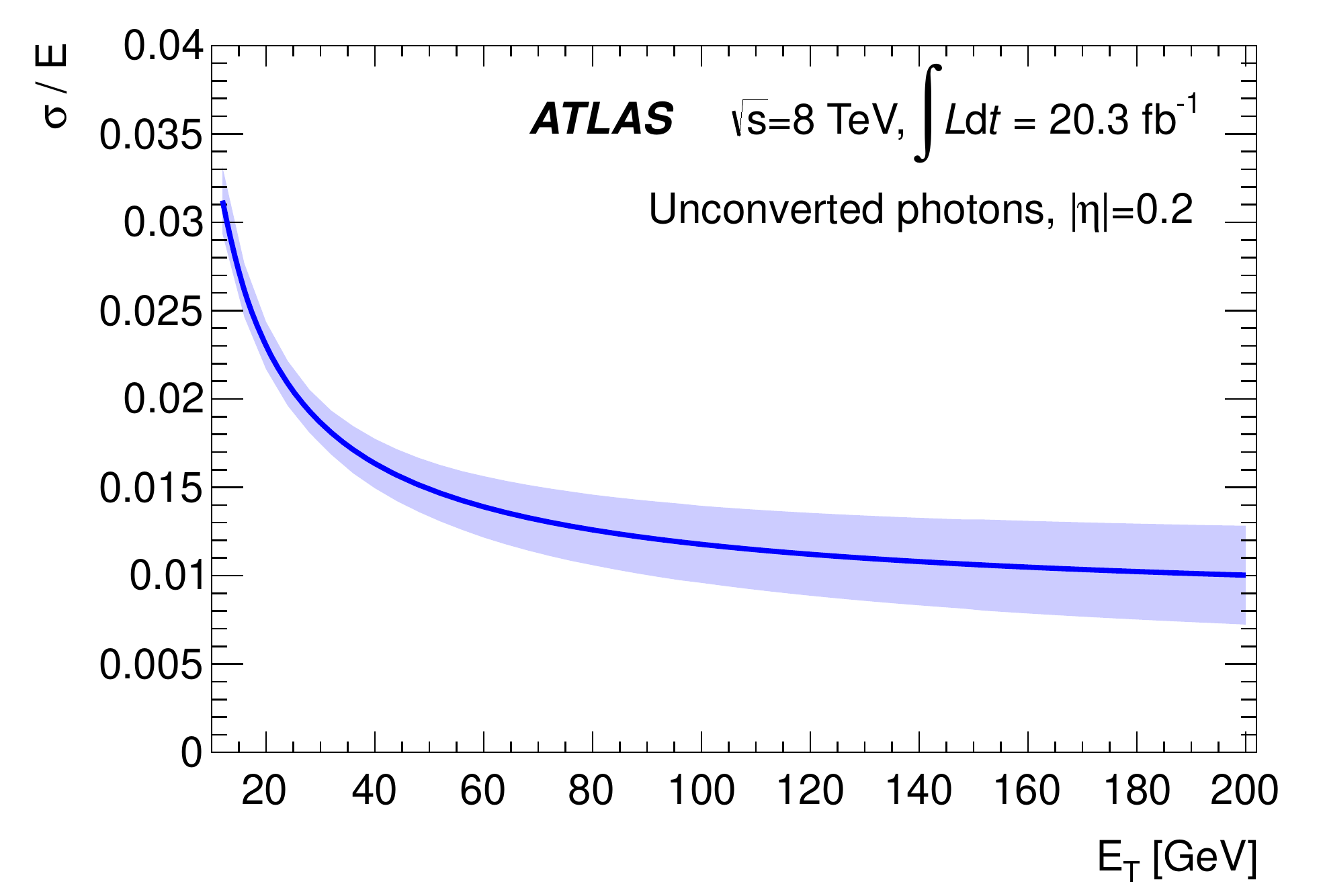}
       \caption{Resolution curve and its uncertainty as a function of $\et$ for
         electrons (left) and unconverted photons (right) with $|\eta|=0.2$.}
       \label{fig:resolutioncurve_eta0-0.4}
\end{figure*}

\begin{figure*}
       \centering
       \includegraphics[width=\columnwidth]{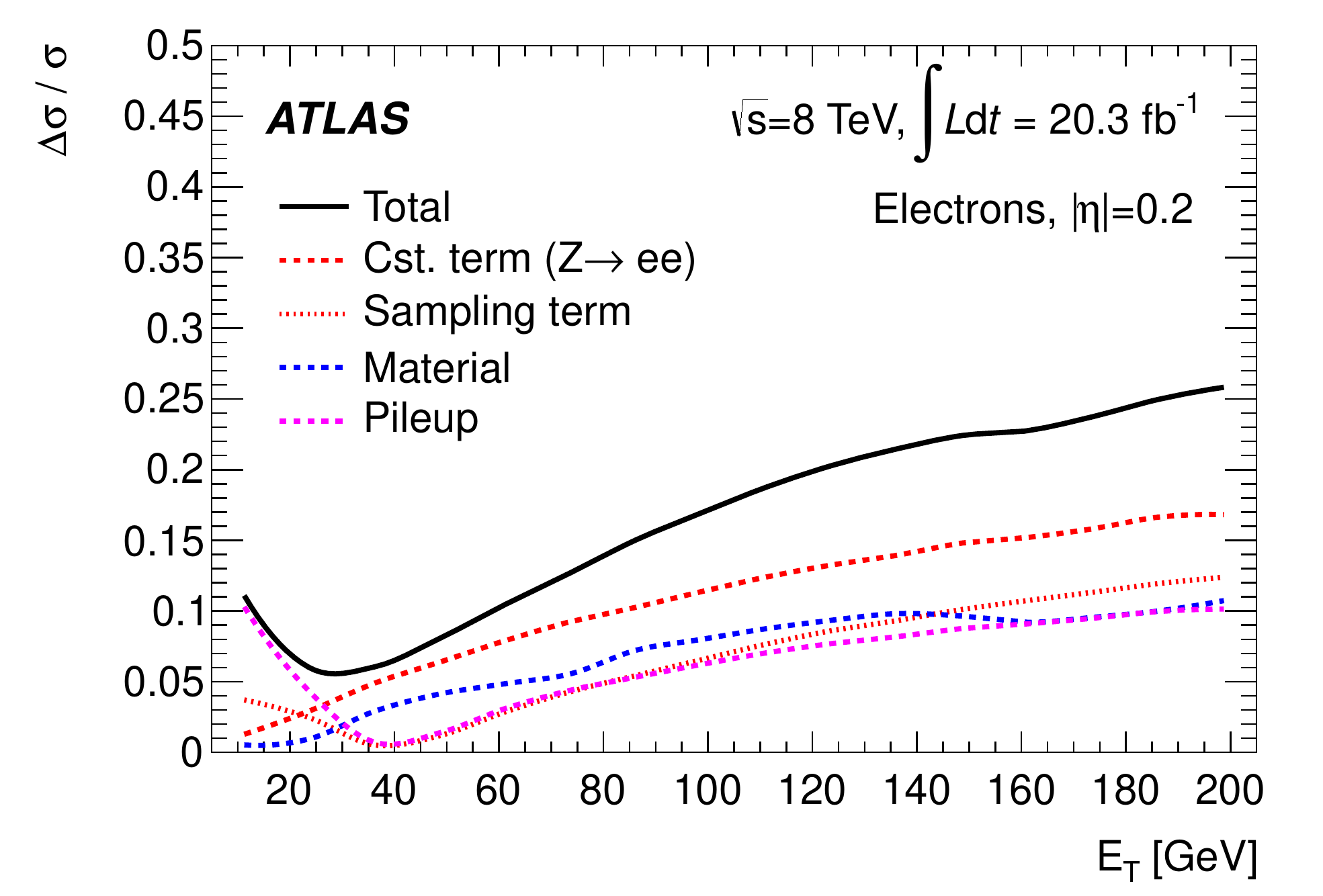}
       \includegraphics[width=\columnwidth]{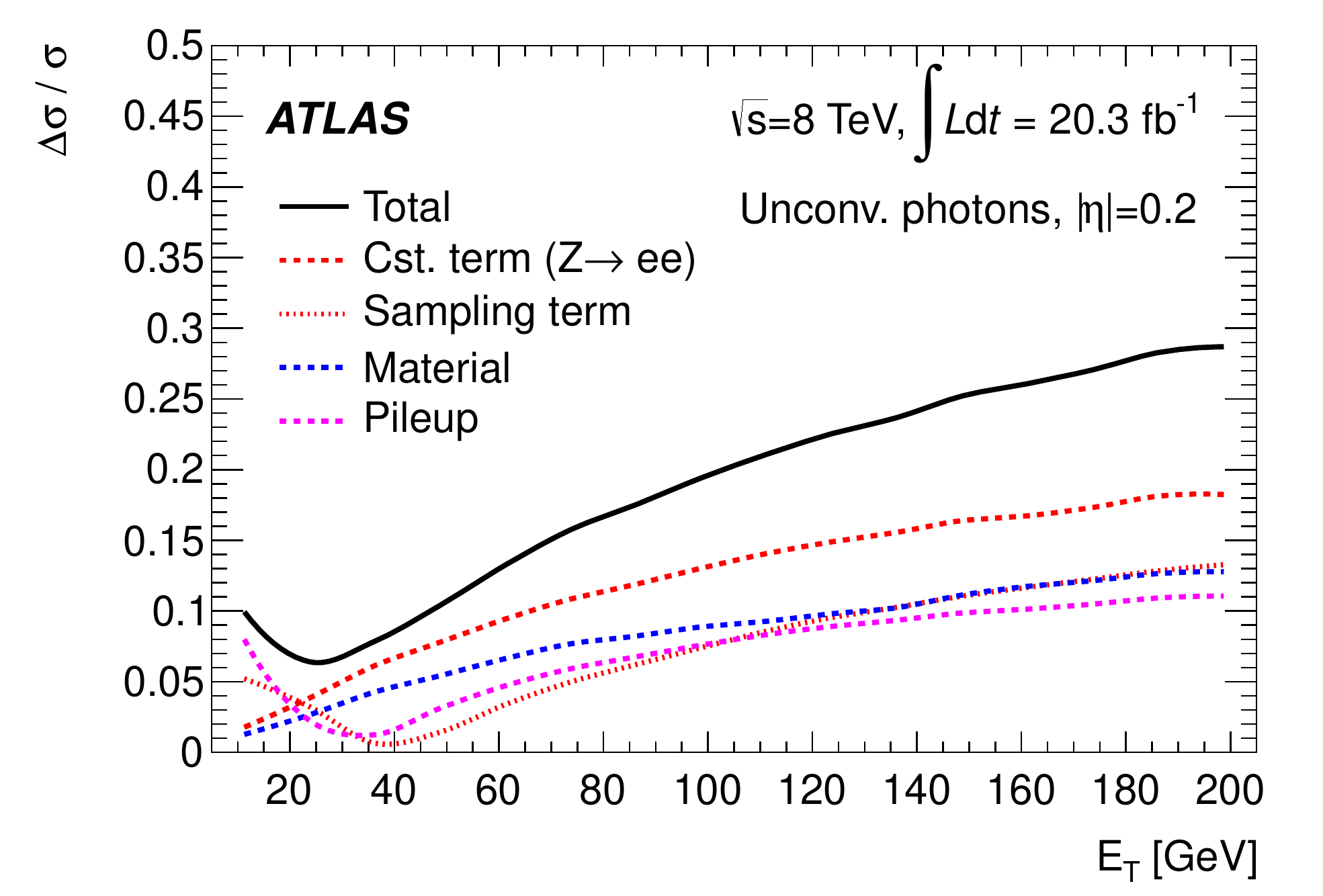}
       \caption{Contributions of the different uncertainties to the
         relative resolution uncertainty as a function of $\et$ for
         electrons (left) and unconverted photons (right) with $|\eta|=0.2$.}
       \label{fig:resolutionuncertainty_eta0-0.4}
\end{figure*}

The calorimeter intrinsic resolution,
defined as the expected resolution in the absence of upstream material
and with perfect response uniformity, is a function of $\eta$ and
scales as 1/$\sqrt{E}$. A 10\% uncertainty is assumed from test-beam
studies~\cite{Aharrouche:2006nf} and from simulation predictions obtained by varying the
physics modelling options in {\sc Geant4}. Taking into account that the
resolution is constrained at $\langle\etZ\rangle\sim 40$~GeV, the
uncertainty $\Delta_\mathrm{int}$ on the squared resolution 
induced by a 10\% relative increase in the intrinsic resolution is:
\begin{eqnarray}
  \Delta_\mathrm{int}(E,\eta) & = &  (1.1^2 -1) \times   \left[\sigma^2_\mathrm{int}(E,\eta) - \right. \nonumber  \\
    & & \sigma^2_\mathrm{int}(\langle\etZ\rangle   \times  \left. \cosh\eta, \eta) \right]
\label{eq:resolsyst1}
\end{eqnarray} 
where $\sigma_\mathrm{int}(E,\eta)$ is the intrinsic
resolution (in un\-its of GeV), and $\Delta_\mathrm{int}$ can be positive or negative depending on
particle type and energy. Equation~\eqref{eq:resolsyst1} is obtained by varying the
sampling term by 10\% in Eq.~\eqref{eq:resol} and requiring that the total
resolution is unchanged at $\et\sim 40$~GeV.

The resolution noise term scales as $1/\et$ for pile-up noise and $1/E$ for electronic noise, and mostly matters at low
energy. It receives contributions from the read-out electronics and
pile-up. A measurement of the total noise affecting electron and photon
clusters is performed by comparing pile-up-only events in data and simulation. Random clusters of the size used for
electrons and photons are drawn over $\eta,\phi$, and the
noise $\sigma_\mathrm{noise}$ is defined as the spread of the cluster
transverse energy distribution. The
noise systematic uncertainty $\delta_\mathrm{noise}$ is defined as the
difference in quadrature between $\sigma_\mathrm{noise}$ in data and simulation. An
uncertainty of $\delta_\mathrm{noise}=100$~MeV is 
found across $\eta$, apart for $1.52<|\eta|<1.82$ where
$\pm200$~MeV is appropriate. Its impact is

\begin{equation}
\frac{\Delta_\mathrm{noise}(E,\eta)}{{\et}^2} \, = \left(\frac{\delta_\mathrm{noise}(\eta)}{\et}\right)^2 \, - \, 
\left(\frac{\delta_\mathrm{noise}(\eta)}{\langle\etZ\rangle}\right)^2 .
\end{equation} 
 
The impact of detector material uncertainty on the resolution is
treated as follows. Assuming the ID, cryostat and calorimeter material
uncertainties discussed in Se\-ct.~\ref{sec:material}, $Z\rightarrow
ee$ samples are simulated with corresponding geometry distortions. The
distorted geometry samples are used as pseudo-data, and the impact
of the additional passive material on the reconstructed $Z\rightarrow
ee$ invariant mass distribution is calculated as in
Sect.~\ref{sec:zmeth}, yielding a material-induced constant 
term correction. This correction is applied to simulated
particles in the nominal geometry; subtracting the result in
quadrature from the actual particle resolution obtained in the
distorted simulation yields the material contribution to the
resolution uncertainty. The impact is
\begin{equation}
\Delta_\mathrm{mat}(E,\eta)= \sigma^2_\mathrm{dist}(E,\eta) - \sigma^2_\mathrm{nom}(E,\eta) - c_\mathrm{dist}^2 (\eta) \, ,
\end{equation}
where $c_\mathrm{dist}$ is the material-induced resolution correction
obtained as described above. The resolutions
$\sigma_\mathrm{dist}(E,\eta)$ and $\sigma_\mathrm{nom}(E,\eta)$
are parameterised for electrons and photons separately; hence $\Delta_\mathrm{mat}(E,\eta)$
depends on particle type, energy and $\eta$.

A given uncertainty source $i$ ($i$ = $c$, int, mat, noise)
contributes to a change $\Delta_i$ in the
squared resolution. Its contribution to the total resolution uncertainty is
\begin{equation}
\delta\sigma_i = \sqrt{\sigma_0^2 + \Delta_i} - \sigma_0 \, ,
\end{equation}
where $\sigma_0(E,\eta)$ is the nominal energy resolution. The
$\delta\sigma_i$ summed in quadrature give the total resolution uncertainty.

\begin{figure*}
  \centering
  \includegraphics[width=\columnwidth]{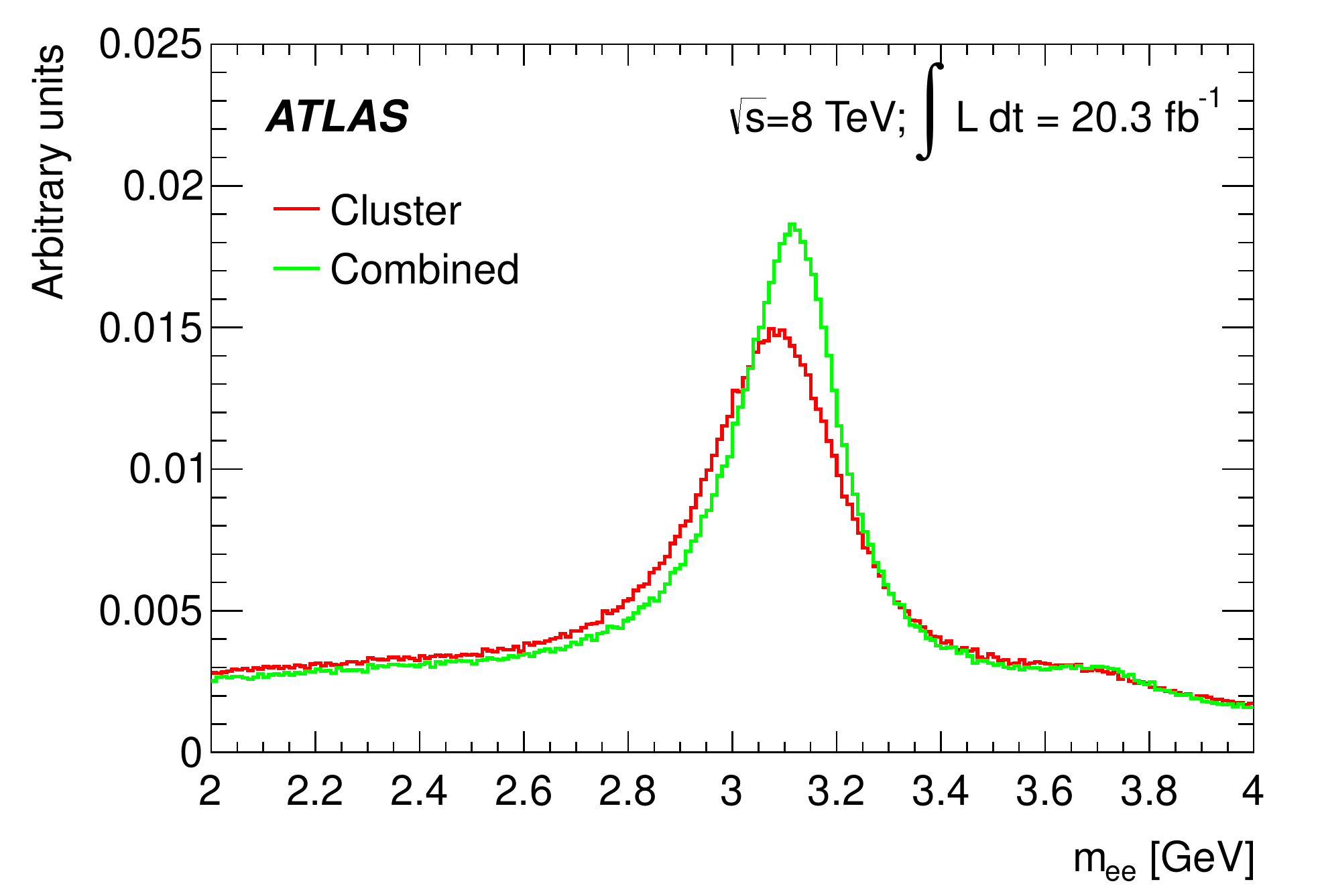}
  \includegraphics[width=\columnwidth]{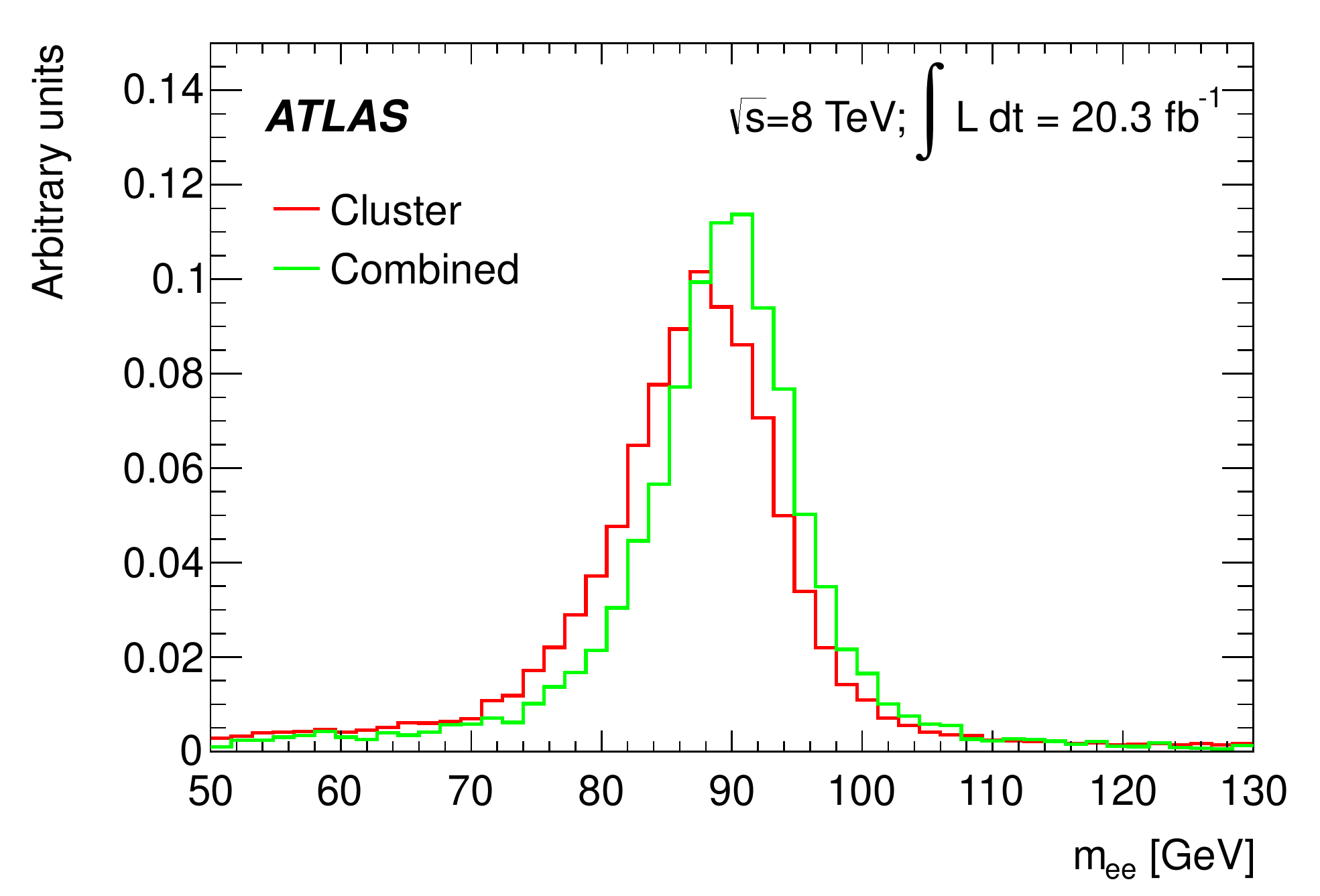}
  \caption{The electron pair invariant mass distribution, as reconstructed from data using either the calibrated cluster energies, or the combination of the cluster energy and the track momentum. 
    Left: $J/\psi\rightarrow ee$ selection. 
    Right: $Z\rightarrow ee$ selection, with one electron candidate in $1.37<|\eta|<1.52$.}
  \label{fig:epperfdata}
\end{figure*}

The resolution curve is shown for electrons and unconverted photons
in Fig.~\ref{fig:resolutioncurve_eta0-0.4}, as a function
of energy for $|\eta|=0.2$. The different contributions to
the resolution uncertainty are shown in
Fig.~\ref{fig:resolutionuncertainty_eta0-0.4}. The relative
uncertainty is 
minimal for electrons at $\et=40$~GeV, where the measurement of $c$
translates into an uncertainty of 5\%. At higher
transverse energy, the sampling term and detector material
contributions are significant; at low energy, the pile-up
contribution dominates. For unconverted photons, the uncertainty is about 10\% for $\et=40$~GeV. 

The asymptotic resolution uncertainty is given by the accuracy of the
$c$ constants, which are determined at $\et\sim 40$~GeV. At high
energy, the increased contribution of the third sampling and the
turn-on of low-gain amplification (mentioned in
Section~\ref{sec:uncsum}) are not expected to generate energy 
fluctuations that invalidate this model. 


\section{Energy--momentum combination \label{sec:epcombination}}

Combining the track momentum and cluster energy measurements improves
the electron energy resolution, in particular for electrons of 
low energy. For electrons, the momentum provided by the ID
provides the best measurement for low-\pt~particles, whereas the
calorimeter energy measurement is superior at high \pt. The
combination also improves the electron resolution in the transition
region of the electromagnetic calorimeter, $1.37 < |\eta| <
1.52$. This method is not applied for $|\eta|>1.52$.

The method relies on the ID momentum and calorimeter energy response
functions determined from simulation. Electrons are categorised
according to transverse momentum ($7 < \pt < 15 \GeV$, $15 < \pt < 30
\GeV$, $\pt > 30 \GeV$) and pseudorapidity ($|\eta| < 0.8$, $0.8 <
|\eta| < 1.37$, $1.37 < |\eta| < 1.52$, $|\eta|>1.52$). For each
category, the distributions of $\pt^\mathrm{reco}/\pt^\mathrm{true}$ and
$\et^\mathrm{reco}/\pt^\mathrm{true}$ are parameterised using Crystal
Ball functions. The following likelihood is then maximised for a given
electron candidate:  
\begin{equation}
\mathcal{L}(x) =  f_\mathrm{ID}\left(\frac{\pt}{x}\right) \cdot f_\mathrm{Calo} \left( \frac{\et}{x} \right),
\end{equation} 
where $f_\mathrm{ID}$ and $f_\mathrm{Calo}$ represent the ID and EM calorimeter response
functions, and $x$ represents the combined transverse momentum. The
combined transverse momentum is given by the value of $x$ for which $\mathcal{L}(x)$ is maximal.

The results are illustrated in Fig.~\ref{fig:epperfdata}. In the low
$\et$ range, the performance of the combination algorithm is
assessed using $J/\psi$ events. The invariant mass resolution 
is improved by about 20\% and the low-mass tails are significantly
reduced. A significant improvement is also obtained for $Z$ decay
electrons in the calorimeter barrel--endcap transition region where the
tracking information compensates for the locally poor energy
measurement. 

Energy--momentum combination is applied in measurements involving 
final states with low-\et\ electrons, such as Higgs boson decays to
four leptons \cite{mHpaper}. The systematic uncertainty on the combined momentum is
given by the cluster energy scale uncertainty summarised
in Sect.~\ref{sec:uncsum}, and by the momentum scale
uncertainty. The latter is assessed using $J/\psi\rightarrow ee$ events
reconstructed using ID information only. Comparing the position of the
electron pair invariant mass peak in data and simulation yields a
systematic uncertainty ranging from about 0.1\% near $\eta=0$ to about
1\% in the barrel--endcap transition region. The cluster energy and
track momentum systematic uncertainties are combined in quadrature to obtain the total uncertainty on the combined energy--momentum scale.


\section{Summary\label{sec:summary}}

The calibration procedure for electron and photon energy measurement
with the ATLAS detector is presented using LHC Run 1 proton--proton
collision data corresponding to a total integrated luminosity of
about 25 fb$^{-1}$ taken at centre-of-mass energies of $\sqrt{s}=7$~TeV and $\sqrt{s}=8$~TeV.
The calorimeter energy measurement is optimised on  
simulation using MVA techniques, improving the energy resolution with respect to the previous calibration approach \cite{perf2010} by about 10\% (20\%) for unconverted (converted) photons; for electrons,
the improvement ranges from a few percent in most of the acceptance
up to 30\% in the region with the largest amount of material upstream
of the active calorimeter, $1.52<|\eta|<1.82$.

The calorimeter energy response in data is stable at
the level of 0.05\% as a function of time and pile-up. After
corrections for local mechanical and high-voltage defects, the
azimuthal non-uniformity is less than 0.5\% in the barrel, and less
than 0.75\% in the endcap, meeting the original design goals. The
relative response of the calorimeter layers is analysed, notably
correcting a bias of about 3\% in the barrel. After calibration
corrections, the shower depth is used to probe the amount of material upstream of
the calorimeter with a typical accuracy of 3$\--$10\%$X_0$, depending on
pseudorapidity; the detector material description is adjusted
accordingly. The MVA calibration is optimised with the improved simulation,
and the calorimeter absolute energy scale is determined using electron pairs from $Z$ boson decays.

This procedure yields definite predictions for the energy dependence
of the electron and photon calibration and its uncertainty. The
uncertainty for electrons at $\et\sim 40$~GeV is on average 0.04\% for
$|\eta|<1.37$, 0.2\% for $1.37<|\eta|<1.82$ and 0.05\% for
$|\eta|>1.82$. At $\et\sim 
11$~GeV, the electron response uncertainty ranges from 0.4\% to 1\%
for $|\eta|<1.37$, is about 1.1\% for $1.37<|\eta|<1.82$, and again 0.4\%
for $|\eta|>1.82$. The photon energy scale uncertainty is typically 0.2\% to 0.3\%
for $|\eta|<1.37$ and $|\eta|>1.82$; for $1.52<|\eta|<1.82$, the
uncertainty is 0.9\% and 0.4\% for unconverted and converted photons,
respectively. Outside of this range, similar accuracy is achieved for converted and  
unconverted photons, and the energy dependence is weak. The electron
and photon calibration is confirmed using independent resonances
provided by $J/\psi$ events and $Z$ boson radiative decays. The
present energy scale uncertainty model is expected to be valid up to
$\et\sim 500$~GeV.

Finally, the relative uncertainty on the energy resolution is better than 10\% for
$\et<50$~GeV, and asymptotically rises to about 40\% at high
energy. For analyses involving low-$E_{\mathrm {T}}$ electrons, an energy--momentum
combination algorithm is defined, which improves the electron energy resolution obtained from the calorimeter cluster by about 20\% for $E_{\mathrm {T}}<30$~GeV and $|\eta|<1.52$.

The present results form the basis of ATLAS precision measurements
using electrons and photons in LHC Run-1 data.


\section*{Acknowledgements}

We thank CERN for the very successful operation of the LHC, as well as the
support staff from our institutions without whom ATLAS could not be
operated efficiently.

We acknowledge the support of ANPCyT, Argentina; YerPhI, Armenia; ARC,
Australia; BMWFW and FWF, Austria; ANAS, Azerbaijan; SSTC, Belarus;
CNPq and FAPESP, Brazil; NSERC, NRC and CFI, Canada; CERN; CONICYT,
Chile; CAS, MOST and NSFC, China; COLCIENCIAS, Colombia; MSMT CR, MPO
CR and VSC CR, Czech Republic; DNRF, DNSRC and Lundbeck Foundation,
Denmark; EPLANET, ERC and NSRF, European Union; IN2P3-CNRS,
CEA-DSM/IRFU, France; GNSF, Georgia; BMBF, DFG, HGF, MPG and AvH
Foundation, Germany; GSRT and NSRF, Greece; ISF, MINERVA, GIF, I-CORE
and Benoziyo Center, Israel; INFN, Italy; MEXT and JSPS, Japan; CNRST,
Morocco; FOM and NWO, Netherlands; BRF and RCN, Norway; MNiSW and NCN,
Poland; GRICES and FCT, Portugal; MNE/IFA, Romania; MES of Russia and
ROSATOM, Russian Federation; JINR; MSTD, Serbia; MSSR, Slovakia; ARRS
and MIZ\v{S}, Slovenia; DST/NRF, South Africa; MINECO, Spain; SRC and
Wallenberg Foundation, Sweden; SER, SNSF and Cantons of Bern and
Geneva, Switzerland; NSC, Taiwan; TAEK, Turkey; STFC, the Royal
Society and Leverhulme Trust, United Kingdom; DOE and NSF, United
States of America.

The crucial computing support from all WLCG partners is acknowledged
gratefully, in particular from CERN and the ATLAS Tier-1 facilities at
TRIUMF (Canada), NDGF (Denmark, Norway, Sweden), CC-IN2P3 (France),
KIT/GridKA (Germany), INFN-CNAF (Italy), NL-T1 (Netherlands), PIC
(Spain), ASGC (Taiwan), RAL (UK) and BNL (USA) and in the Tier-2
facilities worldwide.

\clearpage
\appendix
\section{Calibration results using 2011 data\label{app:2011}}

The calibration procedure described in this paper was applied to
4.7~fb$^{-1}$ of data collected at 7~TeV, and the same cross-check
measurements are carried out. The event selections are
modified to account for the different trigger conditions, yielding a
sample of about 1.1M $Z\rightarrow ee$ events, 100K $J/\psi\rightarrow
ee$ events and 90K $Z\rightarrow \ell\ell\gamma$ events.

\begin{figure}
  \centering
  \includegraphics[width=\columnwidth]{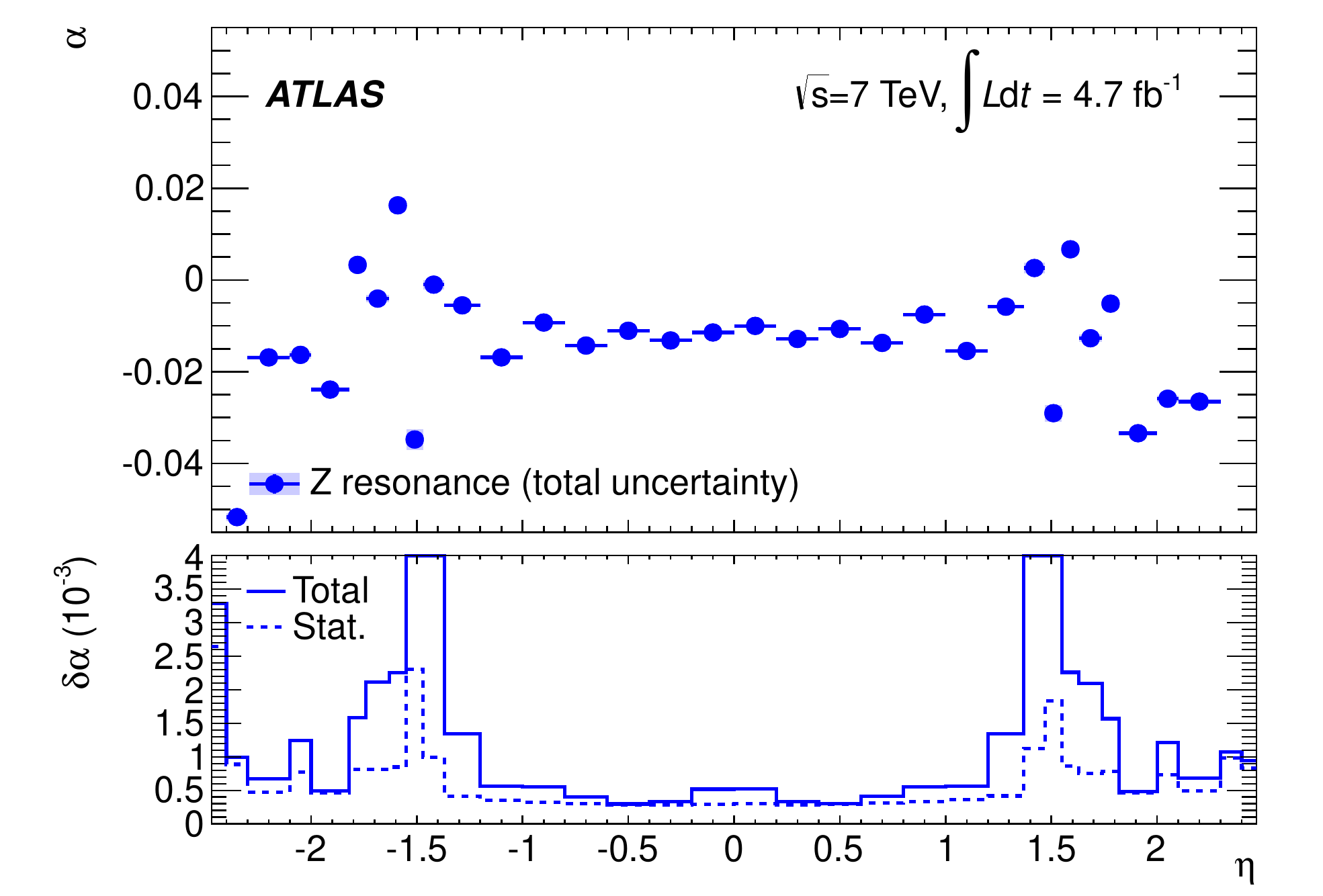}
  \caption{Top: 2011 energy scale corrections $\alpha$ derived from $Z\rightarrow
    ee$ events with respect to the 2010 calibration scheme~\cite{perf2010}, as a function of $\eta$. Bottom: statistical and
    total energy scale uncertainties.\label{fig:finalplota2011}} 
\end{figure}

\begin{figure}%
  \centering
  \includegraphics[width=\columnwidth]{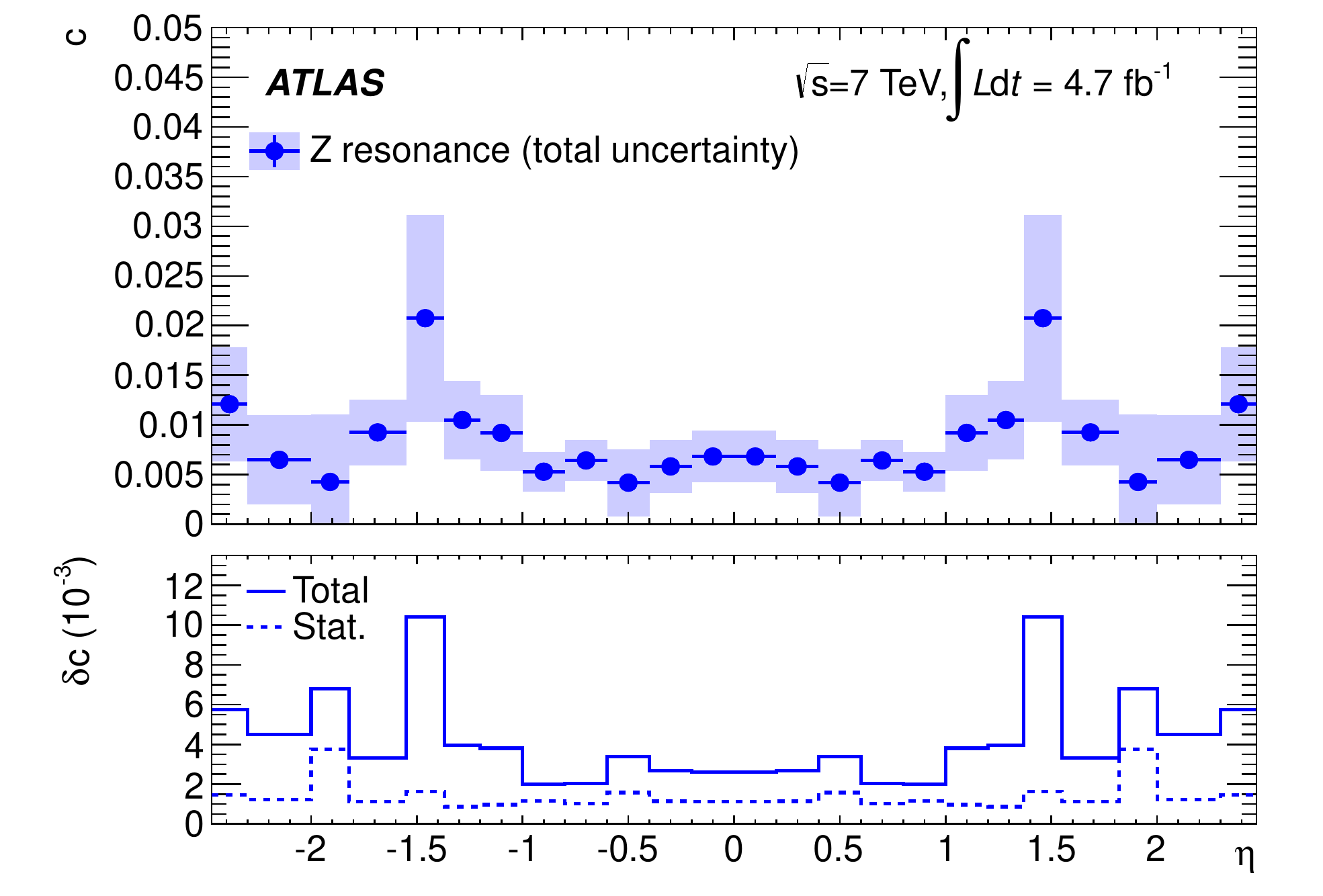}
  \caption{Top: 2011 constant term resolution corrections $c$ derived from $Z\rightarrow
    ee$ events, as a function of $\eta$. Bottom: statistical and total uncertainties on
    $c$.\label{fig:finalplotc2011}}
\end{figure}

The intercalibration constants and effective smearing corrections are
shown in Figs.~\ref{fig:finalplota2011} and~\ref{fig:finalplotc2011}. The
binning in pseudorapidity is identical to that used in the 8~TeV data
analysis. The features observed with the 2011 data are similar to
those observed in 2012, up to small differences expected from the
lower pile-up conditions and changes in the OFC optimisation procedure.

\begin{figure}%
  \centering
  \includegraphics[width=\columnwidth]{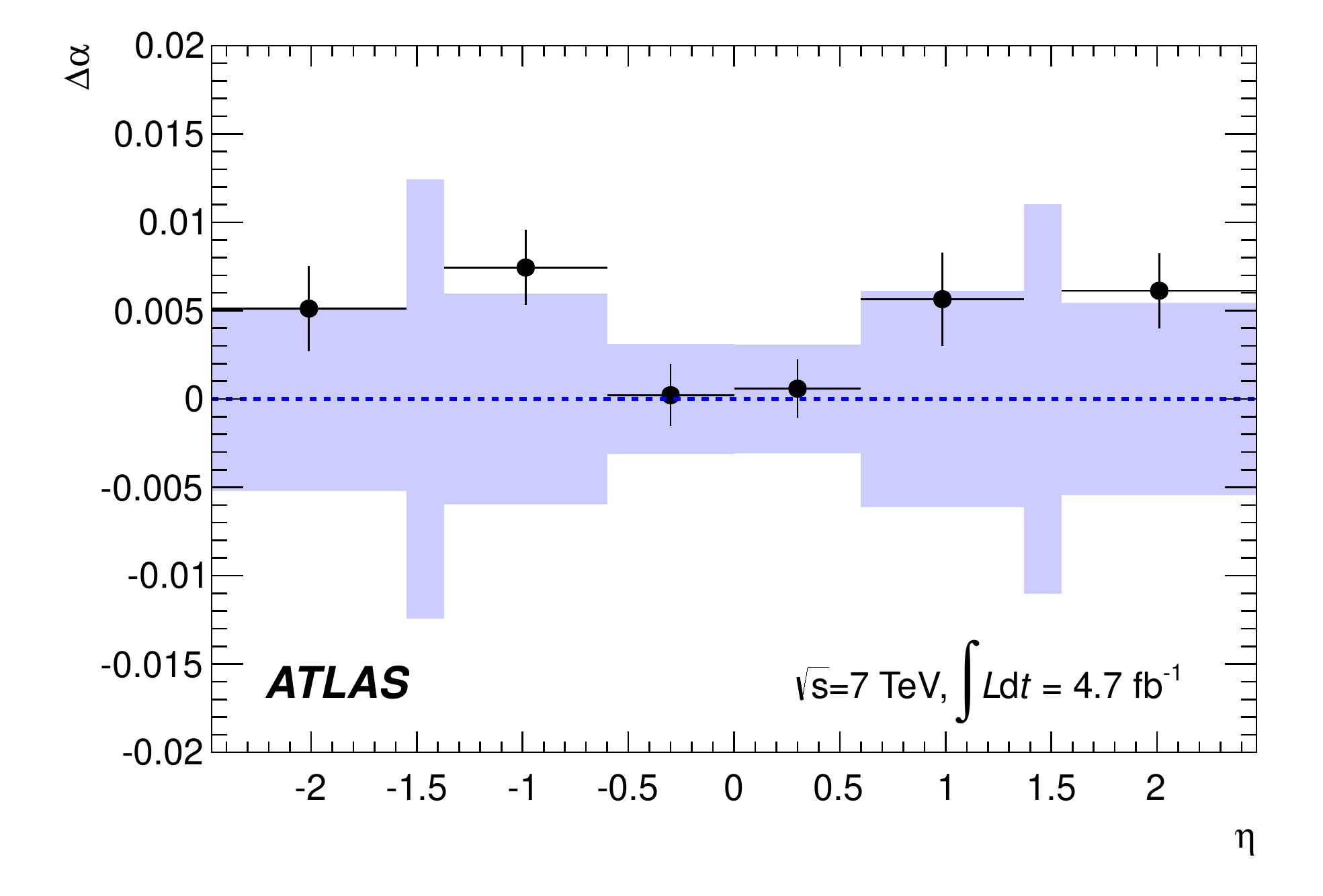}
  \caption{Energy scale factors $\Delta\alpha$ obtained after $Z$-
    based calibration from the \Jpsi~sample as a function of the electron pseudorapidity, using
    2011 data. The band represents the calibration systematic uncertainty. The error bars on the data points
    represent the total uncertainty specific to the
    $J/\psi\rightarrow ee$ analysis.
    \label{fig:jpsi_z_uniformity2011}}
\end{figure}

The energy scale uniformity obtained using $J/\psi\rightarrow ee$
events, after application of the full calibration chain, is shown in
Fig.~\ref{fig:jpsi_z_uniformity2011}. As in 2012, the results agree
within uncertainties, although a  bias of one standard
deviation is observed for $|\eta|>0.6$. 

\begin{figure*}
  \centering
  \includegraphics[width=0.49\textwidth]{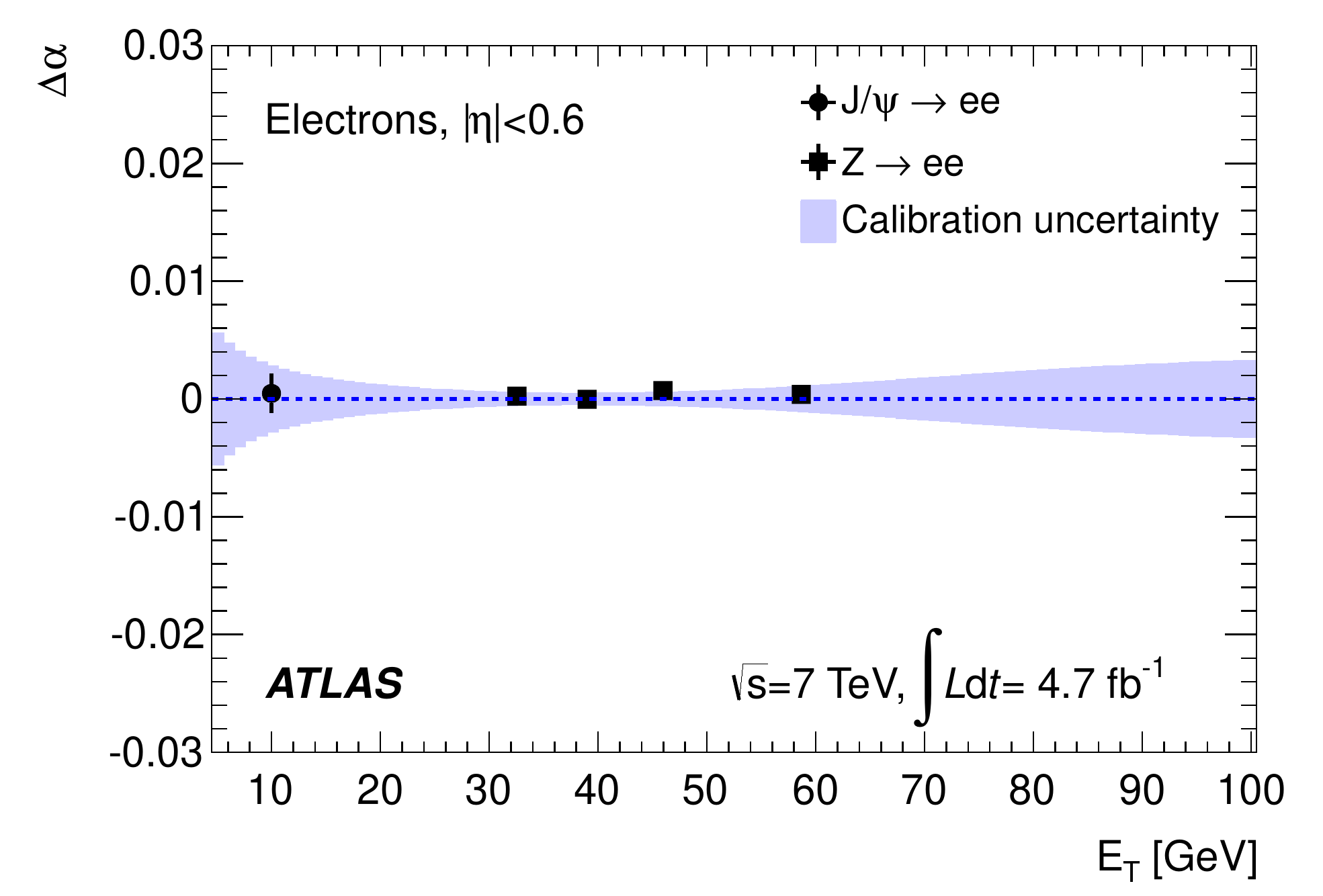}
  \includegraphics[width=0.49\textwidth]{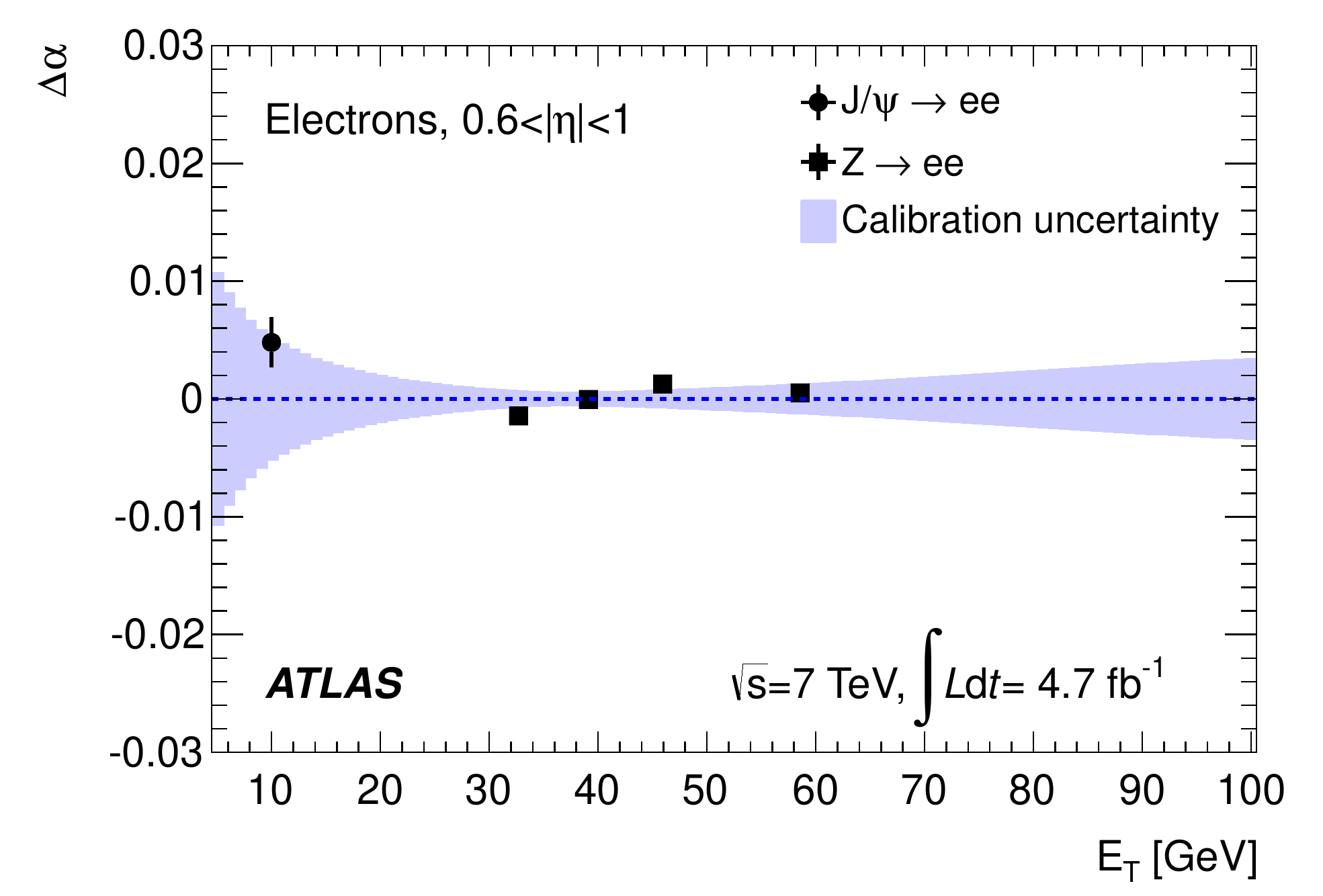} \\
  \includegraphics[width=0.49\textwidth]{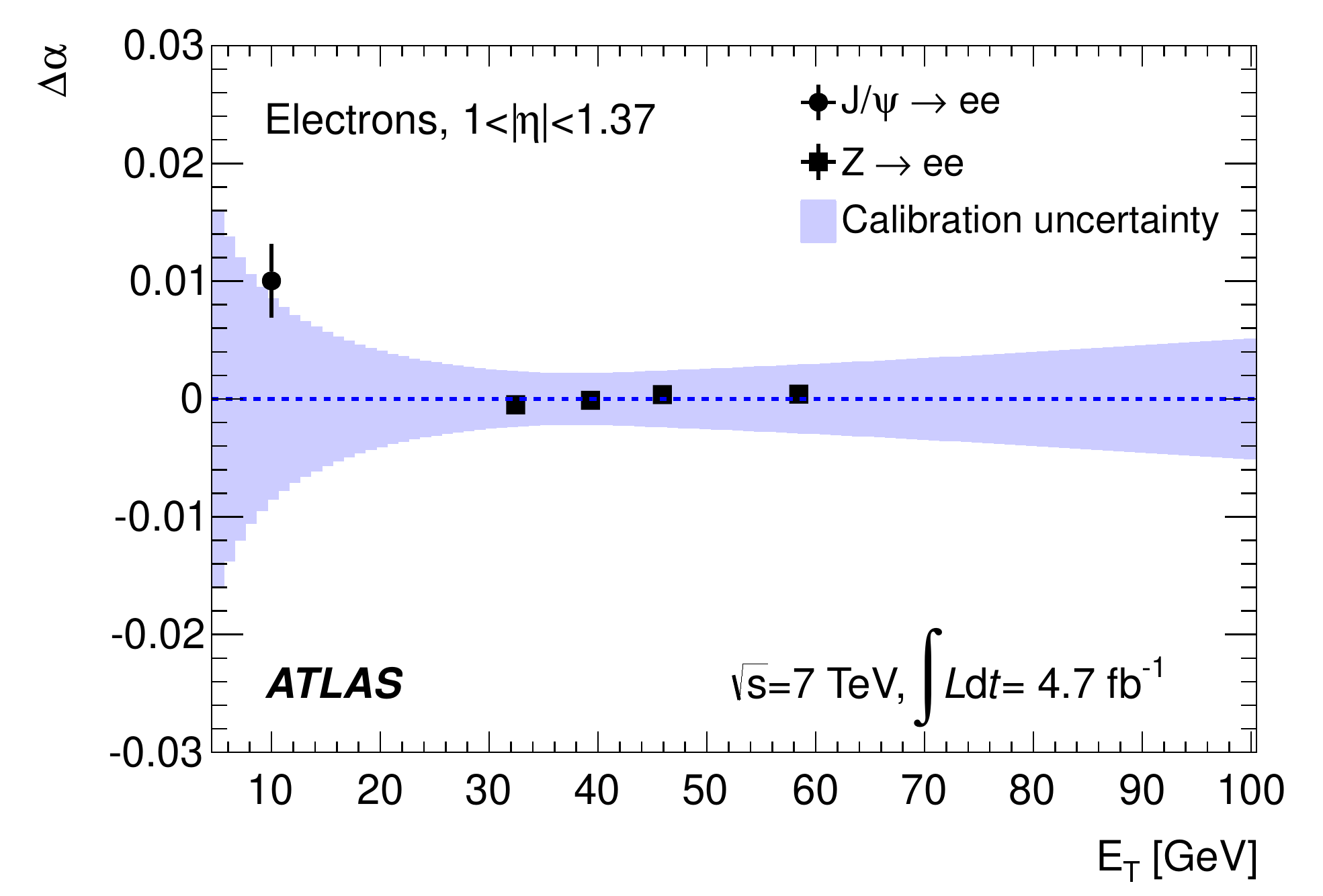}
  \includegraphics[width=0.49\textwidth]{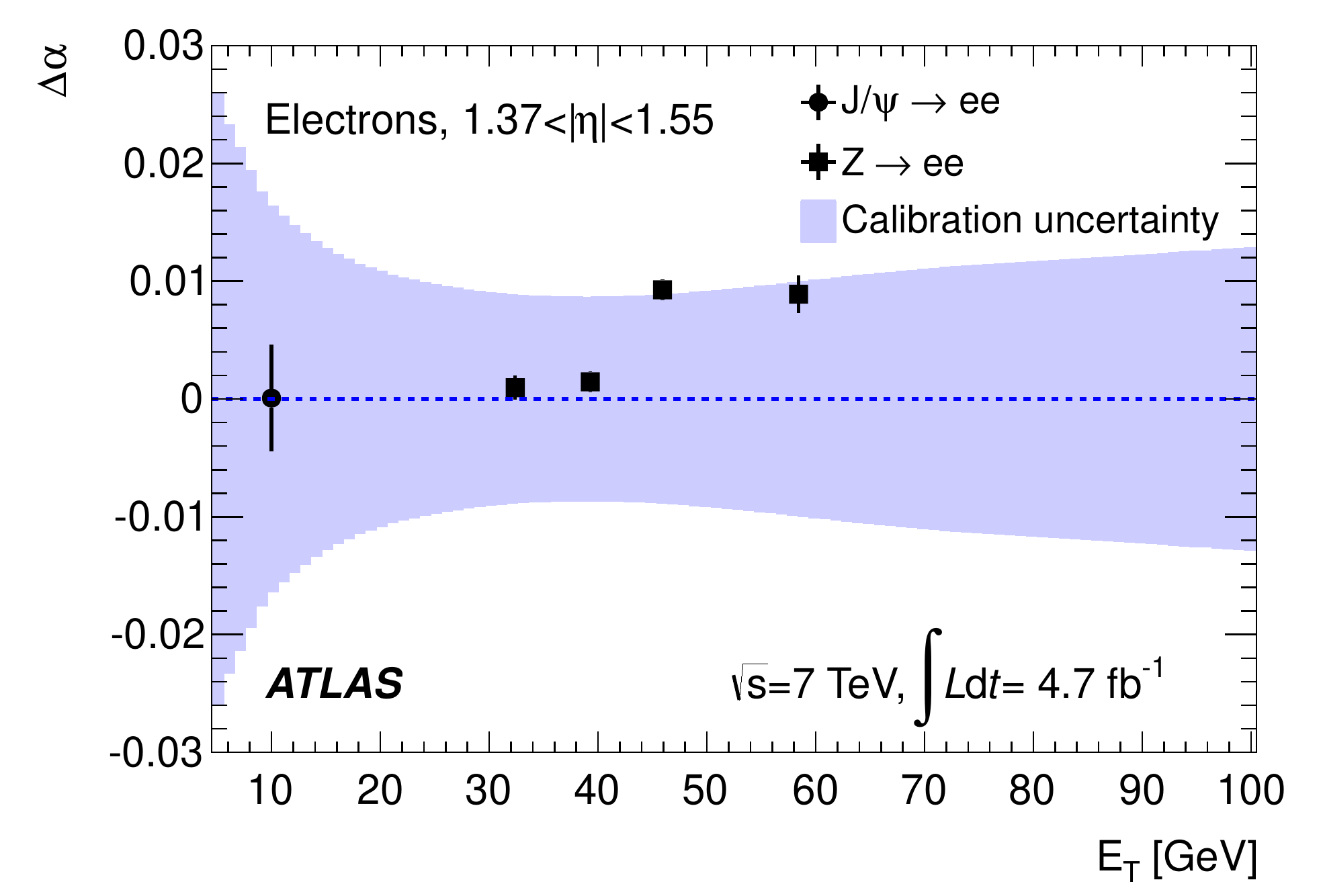} \\
  \includegraphics[width=0.49\textwidth]{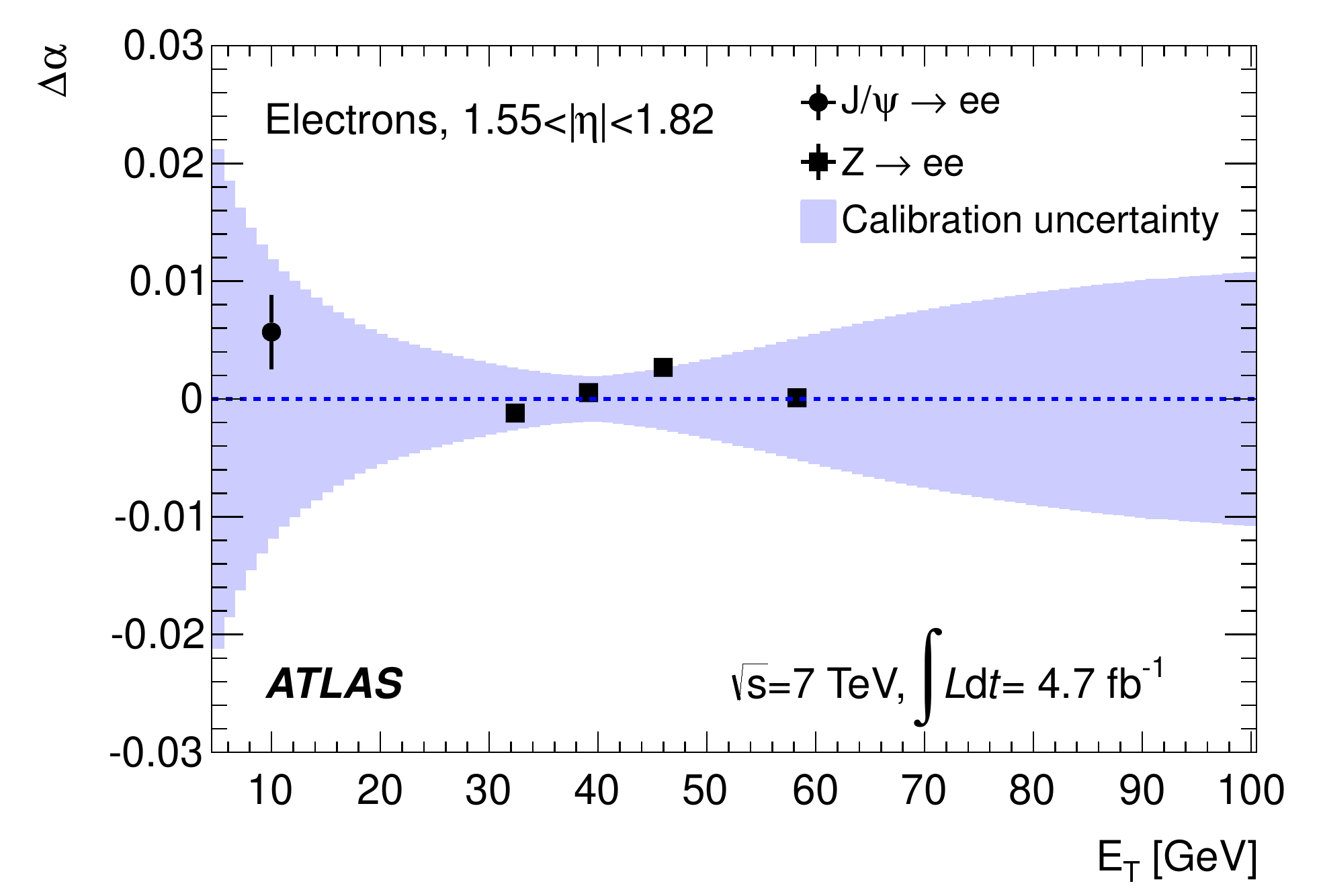}
  \includegraphics[width=0.49\textwidth]{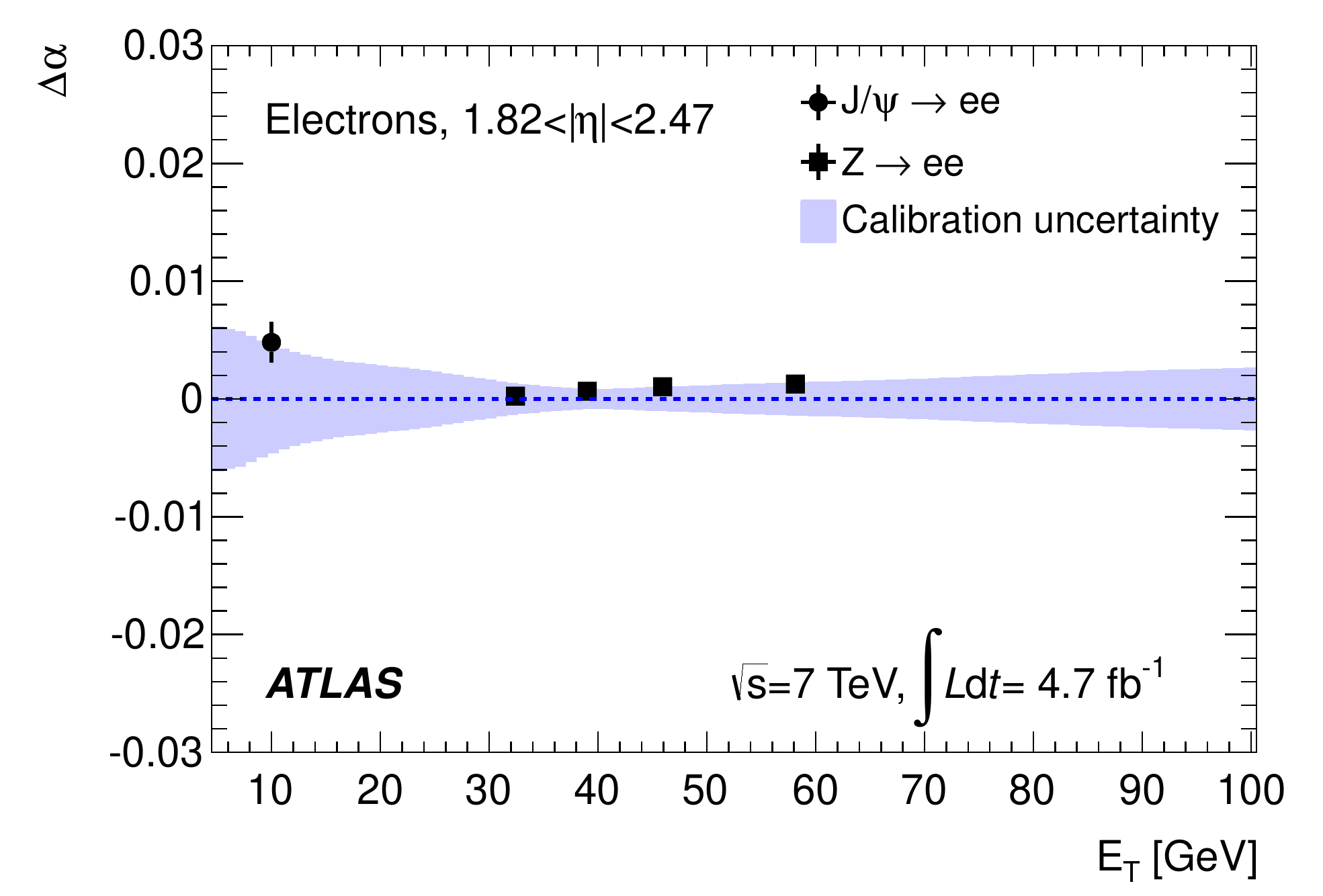}
  \caption{Energy scale factors $\Delta\alpha$ obtained after $Z$-based calibration from the \Jpsi\ and the
  \et-dependent $Z\rightarrow ee$ analyses, as function of \et,
  in different pseudorapidity bins and using 2011 data. The band represents the calibration systematic
  uncertainty. The error bars on the data points represent the total
  uncertainty specific to the cross-check analyses.\label{fig:jpsi_lin2011}}
\end{figure*}

The energy-dependent analysis is  
illustrated in Fig.~\ref{fig:jpsi_lin2011}. The $Z\rightarrow ee$
analysis is performed in four $E_{\mathrm{T}}$ bins, namely
27--35--42--50--100~GeV. For $1.37<|\eta|<1.52$, an observed energy dependence of the
energy scale of about 1\% leads to an increased systematic uncertainty
in this region; outside of the transition region, the results are
consistent with the $Z$-based calibration independently of $|\eta|$
and $\et$. 

\begin{figure*}
  \centering
  \includegraphics[width=0.49\textwidth]{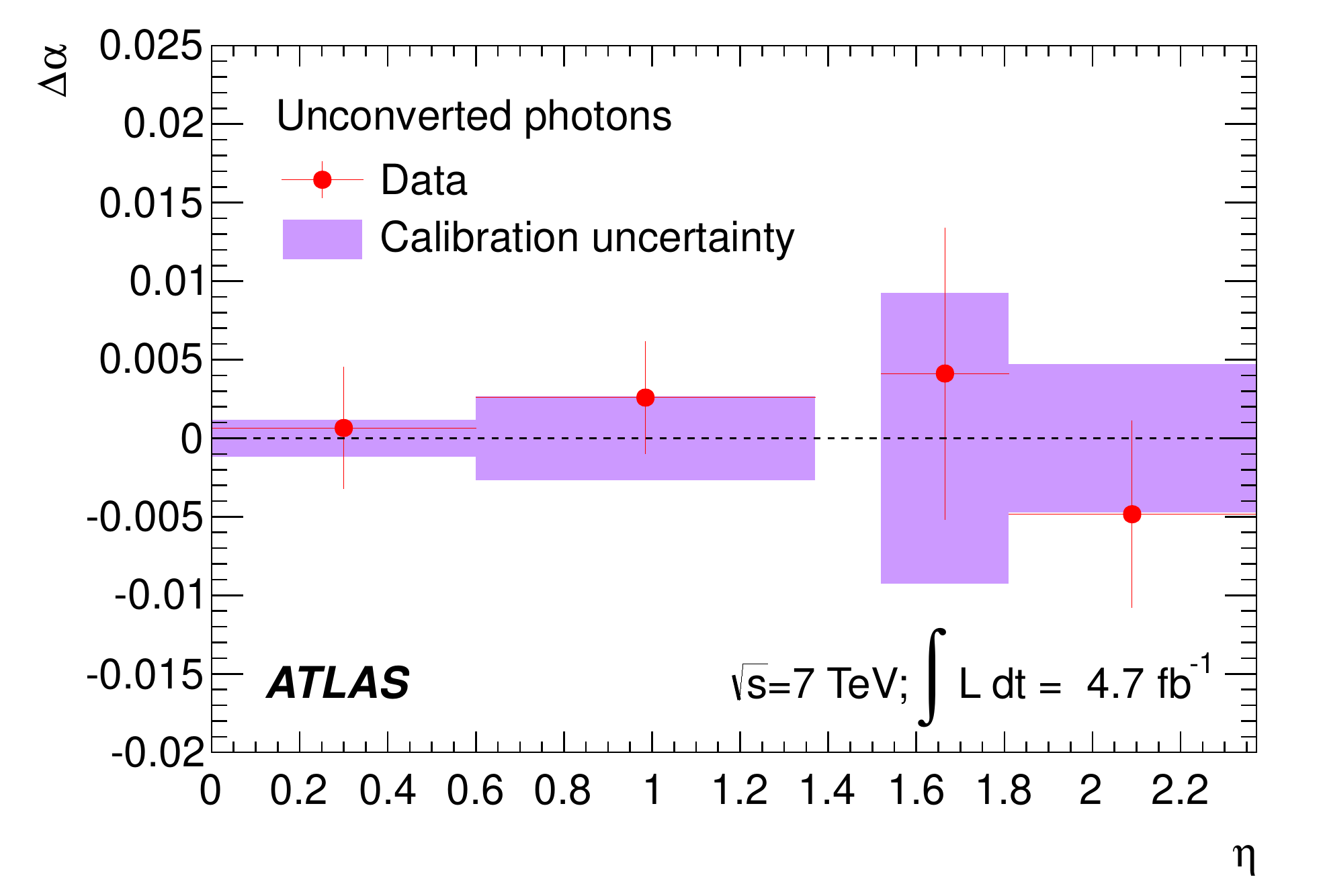}
  \includegraphics[width=0.49\textwidth]{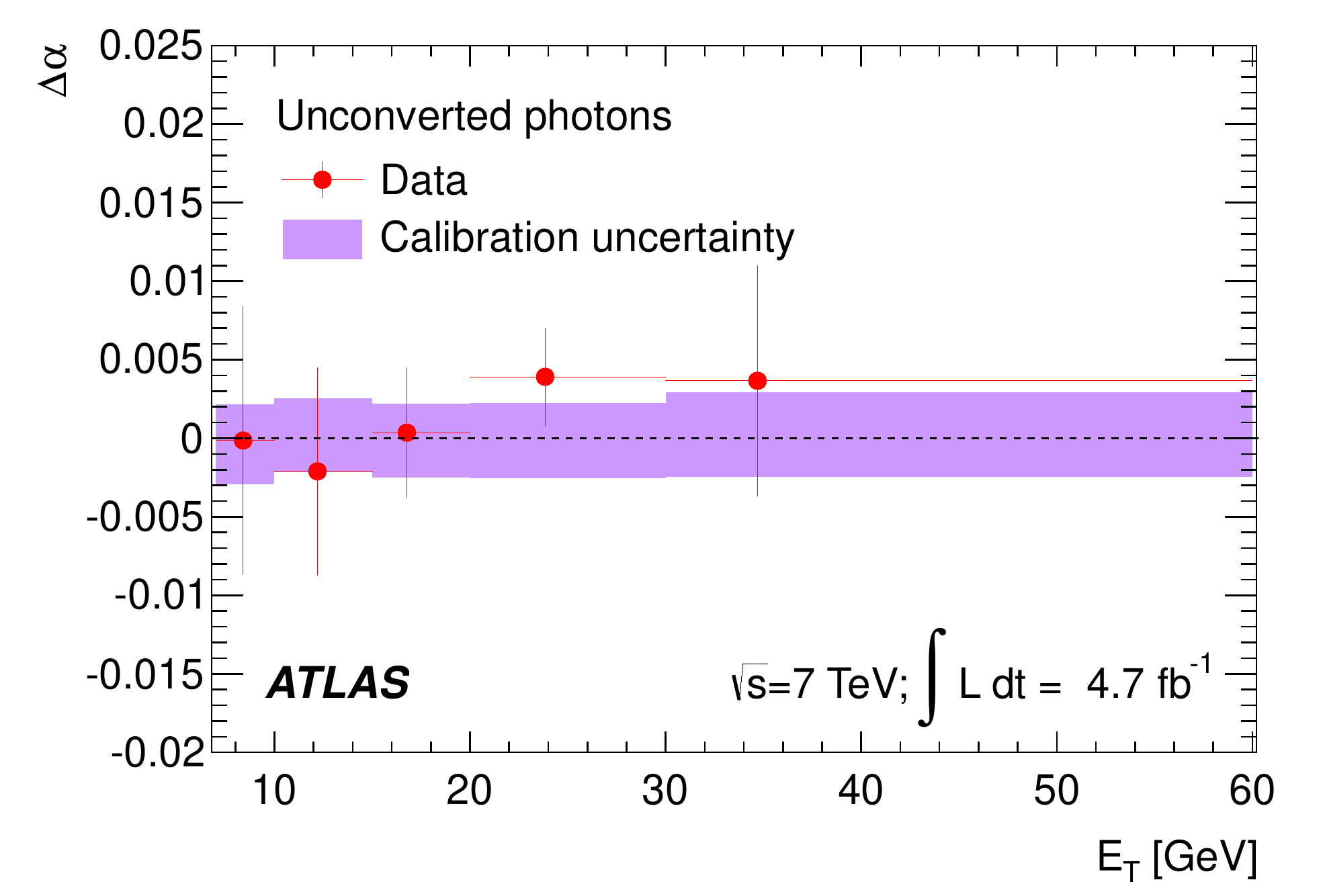} \\
  \includegraphics[width=0.49\textwidth]{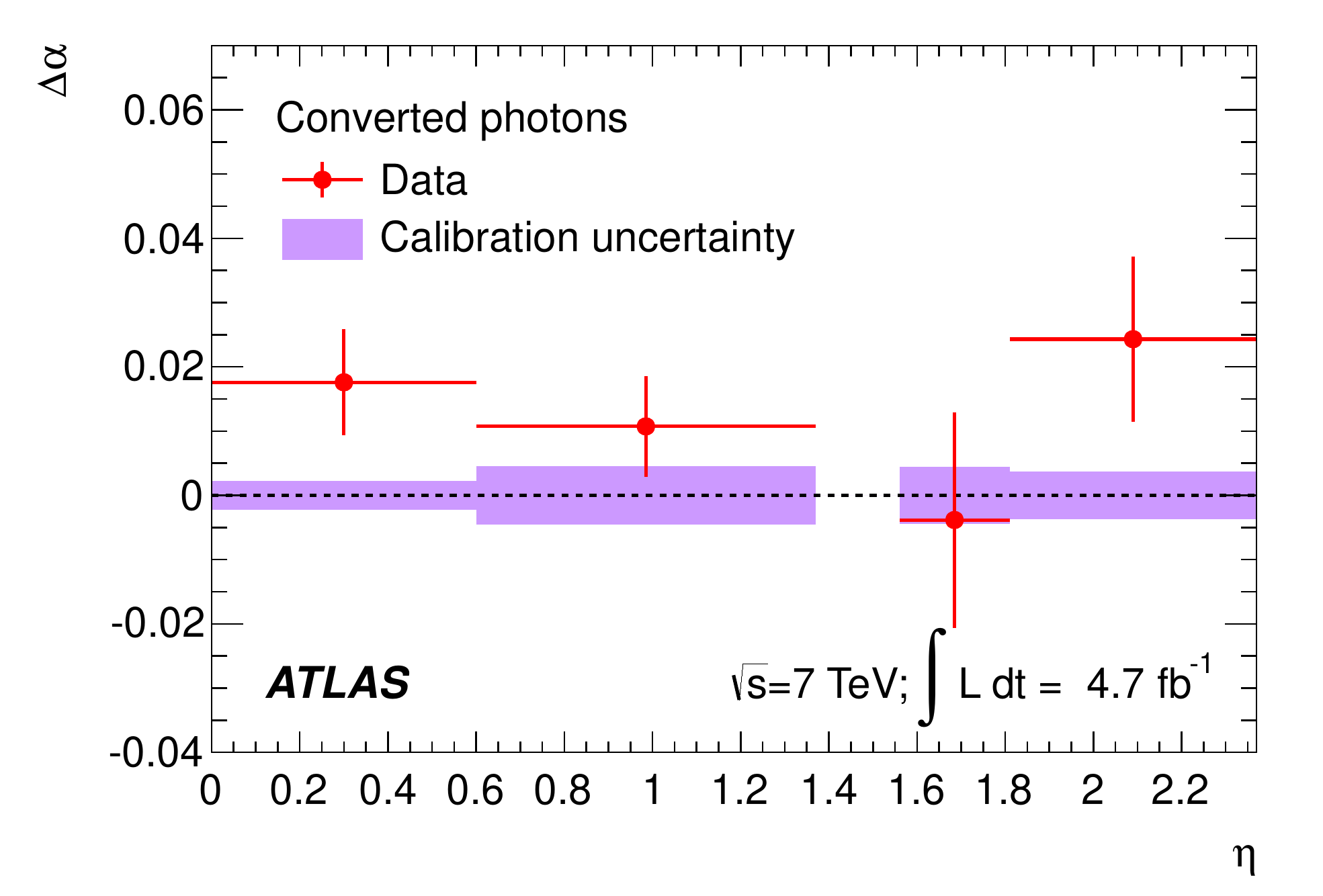}
  \includegraphics[width=0.49\textwidth]{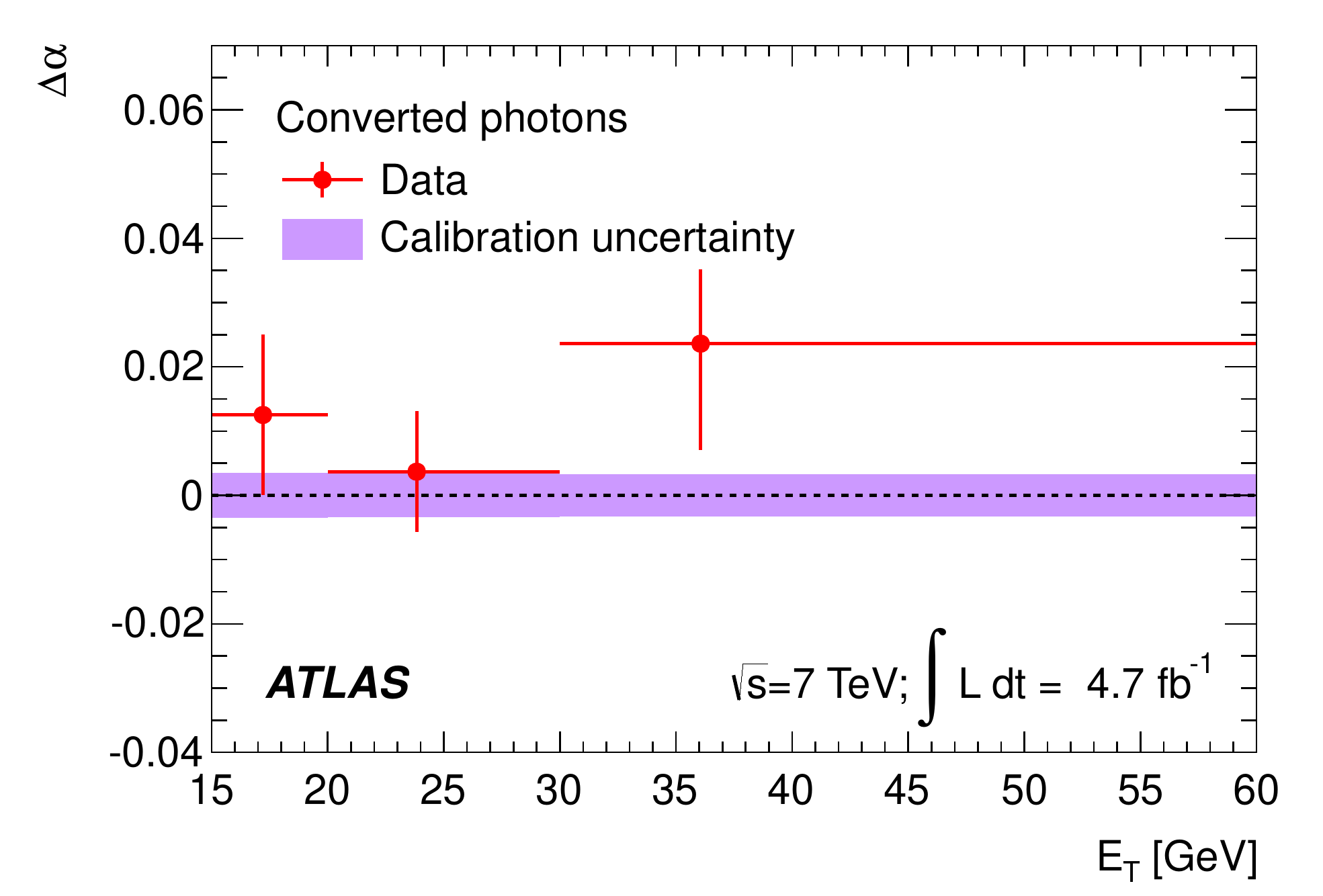}
  \caption{Combined photon scale factors $\Delta\alpha$ obtained
    after $Z$-based calibration for unconverted and converted
    photons, as a function of $|\eta|$ (left) and $E_{\mathrm{T}}$ (right) and
    using 2011 data. The band represents the calibration systematic 
    uncertainty. The error bars on the data points
    represent the total uncertainty specific to the
    $Z\rightarrow\ell\ell\gamma$ analyses.\label{fig:zllg2011}}
\end{figure*}

The radiative photon energy scale derived using $Z\rightarrow
\ell\ell\gamma$ is shown in Fig.~\ref{fig:zllg2011}. For unconverted
photons, the results agree with expectations. The discrepancy between
the energy scales obtained for converted photons and the $Z$-based
calibration is within two standard deviations.

\clearpage
\bibliographystyle{epjc}
\bibliography{calib_note}


\clearpage
\onecolumn
\begin{flushleft}
{\Large The ATLAS Collaboration}

\bigskip

G.~Aad$^{\rm 84}$,
B.~Abbott$^{\rm 112}$,
J.~Abdallah$^{\rm 152}$,
S.~Abdel~Khalek$^{\rm 116}$,
O.~Abdinov$^{\rm 11}$,
R.~Aben$^{\rm 106}$,
B.~Abi$^{\rm 113}$,
M.~Abolins$^{\rm 89}$,
O.S.~AbouZeid$^{\rm 159}$,
H.~Abramowicz$^{\rm 154}$
% ...
\bigskip
\\
% ...
$^{11}$ Institute of Physics, Azerbaijan Academy of Sciences, Baku, Azerbaijan\\
% ...
$^{84}$ CPPM, Aix-Marseille Universit{\'e} and CNRS/IN2P3, Marseille, France\\
% ...
$^{89}$ Department of Physics and Astronomy, Michigan State University, East Lansing MI, United States of America\\
% ...
$^{106}$ Nikhef National Institute for Subatomic Physics and University of Amsterdam, Amsterdam, Netherlands\\
% ...
$^{112}$ Homer L. Dodge Department of Physics and Astronomy, University of Oklahoma, Norman OK, United States of America\\
$^{113}$ Department of Physics, Oklahoma State University, Stillwater OK, United States of America\\
% ...
$^{116}$ LAL, Universit{\'e} Paris-Sud and CNRS/IN2P3, Orsay, France\\
% ...
$^{152}$ Institute of Physics, Academia Sinica, Taipei, Taiwan\\
% ...
$^{154}$ Raymond and Beverly Sackler School of Physics and Astronomy, Tel Aviv University, Tel Aviv, Israel\\
% ...
$^{159}$ Department of Physics, University of Toronto, Toronto ON, Canada\\
% ...
\end{flushleft}


\end{document}